\documentclass[12pt]{report}
\usepackage{graphicx}
\usepackage[dvips]{epsfig}

\newtheorem{theorem}{Theorem}
\newtheorem{lemma}{Lemma}
\newtheorem{proposition}{Proposition}
\newtheorem{conjecture}{Conjecture}

\newtheorem{definition}{Definition}
\newtheorem{corollary}{Corollary}

\begin{document}
\ifx\href\undefined\else\hypersetup{linktocpage=true}\fi 

\pagestyle{plain} 
\begin{titlepage}
\begin{center}
{\bf SEMICLASSICAL\thinspace \thinspace \thinspace \thinspace
STUDY\thinspace \thinspace \thinspace \thinspace OF\thinspace \thinspace
\thinspace \thinspace STATISTICAL\thinspace \thinspace \thinspace \thinspace
MEASURES\thinspace \thinspace \thinspace \thinspace OF}

{\bf SPECTRAL\thinspace \thinspace \thinspace \thinspace
FLUCTUATIONS\thinspace \thinspace \thinspace \thinspace IN\thinspace
\thinspace \thinspace \thinspace SOME}

{\bf PSEUDOINTEGRABLE\thinspace \thinspace \thinspace \thinspace BILLIARDS}
\end{center}

\vskip 1.0 in

\begin{center}
BY
\end{center}

\vskip .5 in

\begin{center}
{\bf HARISHKUMAR \thinspace DEVRAO \thinspace PARAB}
\end{center}

\vskip .5 in

\begin{center}
THEORETICAL PHYSICS DIVISION

BHABHA ATOMIC RESEARCH CENTRE

MUMBAI - 400 085
\end{center}

\vskip 1.0 in

\begin{center}
A THESIS SUBMITTED IN FULFILMENT OF

THE REQUIRMENTS OF THE DEGREE OF

DOCTOR OF PHILOSOPHY IN SCIENCE

IN THE SUBJECT OF

MATHEMATICS
\end{center}

\vskip .5 in

\begin{center}
UNIVERSITY OF MUMBAI

MUMBAI

July 1999
\end{center}
\end{titlepage}

\tableofcontents
\pagebreak
\pagestyle{plain}
\addcontentsline{toc}{chapter}{\numberline {}{Introduction}}

\textbf{{\Huge{\\ \\ \\ \\ \\ \\ Introduction}}}

\vskip 1in

Study of spectral fluctuations in a quantum dynamical system was initiated
by Wigner in 1956 \cite{0-wig}. He provided explanation of observed shortage
of close spacing in nuclear energy levels using statistical arguments. For
example, one can obtain precise information about levels ranging from number 
$N$ (of order $10^6$) to $N+n$ from neutron-capture experiments in heavy
nuclei. Theoretically, it is difficult to analyse levels of this order on
the basis of shell structure and collective or individual particle quantum
numbers. Wigner successfully shown that the highly exited states can be
understood by assuming a working hypothesis that all shell structure is
washed out and that no quantum numbers other than spin and parity remains
good via statistical theory of energy levels. The aim of statistical theory
is not to predict the detailed sequence of levels, but to describe the
general appearance and the degree of irregularity of the level structure in
the given quantum system.

This view led Wigner to surmise a possible spacing distribution based on the
assumption that matrix elements of the Hamiltonian matrix were unknown and
unknowable and are randomly distributed. Success of Wigner surmise in
explaining the spacing between energy eigenvalues resulted in further
development of a one of the tools to study statistical properties of levels,
known as Random Matrix Theory(see \cite{0-dy,0-mlm}).

The random matrix theory of Hamiltonian systems is based on the assumption
that we know very little about the dynamics of the given system except
certain symmetry properties. These symmetry properties impose restrictions
on the form of the Hamiltonian matrix. Based on the symmetry properties a
normalised probability distribution on the elements of the Hamiltonian
matrix can be obtained using a metric in the space of matrix elements which
is invariant under a similarity transformation. The joint probability
distribution is chosen to minimize information about the Hamiltonian matrix
and subjected to the condition that it is normalized to one and has elements
that remain finite.

For real symmetric Hamiltonians (e.g. systems having time reversal
symmetry,), the similarity transformation is orthogonal and the probability
distribution is said to be described by a Gaussian Orthogonal Ensemble (GOE)
of Hamiltonian matrices. For other interesting cases such as Hermitian and
real quaternion Hamiltonians, the similarity transformations are unitary and
symplectic respectively. The probability distribution of these is said to be
described by Gaussian unitary ensemble (GUE) and Gaussian Symplectic
Ensembles (GSE).

There are number of statistical properties of eigenvalues of random matrices
that are commonly used in the analysis of spectral properties. For, example
spacing distribution (or nearest neighbour spacing), the $\Delta _3-$%
statistic (or spectral rigidity) of Dyson and Mehta \cite{0-dy,0-mlm}.

Random matrix theory (RMT) have been successful in explaining universality
in spectral fluctuations (up to second order correlation) for different
classes of chaotic systems as above in an appropriate high energy limit. RMT
is expected to be applicable only on those time scales where the variables
associated with the classical dynamics are random enough, to fully randomize
the matrices, associated with the corresponding quantum operators. Moreover,
due to underlying assumptions on which framework of RMT is build, it is not
expected to shed any light on non-universal behaviour of spectral
fluctuations which is the characteristic of a given system.

A new dimension is added to the study of spectral properties of a quantum
system after the conjecture of Percival \cite{0-per} that there exist a {\it %
regular }and an {\it irregular }quantum spectra corresponding to the
integrable an non integrable classical dynamics. This indication that the
nature of classical dynamics can have its imprints in the behaviour of a
corresponding quantum system provided a new direction to study of
statistical properties of levels.

Classical Hamiltonian systems can display wide variety of dynamical features
from integrability to chaos. Before the works of Poincar\'e, classical
dynamics was viewed as a paradigm of regularity and predictability. The
works by Poincar\'e, and subsequently by Lyapunov, Birkhoff and others
opened the complicated world of classical dynamics. Within a few years after
these concept of determinism was abandoned from classical mechanics. The
concept about classical motion changed radically and led to birth of what is
nowadays called chaos. Broadly now, classical systems can be classified as
one showing complete integrability and others showing either complete chaos
or mixed behaviour.

For integrable (or regular) systems all constants of motion exist everywhere
in the phase space. A typical trajectory resides on a torus like constant
energy surface. On other hand one in case of completely chaotic systems no
constant of motion exist and a typical trajectory will occupy whole phase
space available. Most of the classical systems, however, belong to an
intermediate regime between these extremes, showing a mixed behaviour.
Typically, some of the trajectories will show regular behaviour winding
around tori. These trajectories are separated by the other orbits that
explore constant energy surface ergodically.

There are also many Hamiltonian systems that are just step away from the
complete integrability, known as pseudo-integrable systems. For these
systems constants of motion do not exist on a countable set of singular
points. These systems have $f$-dimensional ($f$ being dimension of the
system) sphere with $g$ handles as an invariant surface embedded in $2f$%
-dimensional phase space. Typical examples of pseudo-integrable systems are
rational of polygonal billiards, where a particle moves freely inside
polygonal enclosure whose each angle is of form $\pi /n$, reflecting
specularly from the walls. Results of some recent studies indicate that
generally one can expect coexistence of almost integrable and almost chaotic
regions in the mixed phase space. In this work we are concerned with
spectral properties of these systems.

The imprints of classical behaviour on a quantum system can best be
understood via semiclassical techniques which provides a information
regarding quantum dynamics in terms of properties of corresponding classical
system. Semiclassical techniques had been developed since the beginning of
the quantum mechanics. Due to easiness with classical mechanics the search
for connections between classical and quantum mechanics has been naturally
sought for. The WKB approximation was the first method built to extract
quantum-mechanical properties using information of classical orbits. The
recent interest in non-integrable systems further initiated development of
semiclassical techniques which are applicable for large class of classical
systems.

In this work we will consider dynamical systems as mentioned above known as
pseudointegrable billiards. Due to mathematical intractability of these
systems, few exact results are known. Numerical studies performed on the
energy spectra of these systems led to the belief that the spectral
statistic is intermediate between those for the Poisson ensemble and the GOE
of random matrices. However, the results on various measures of the spectral
statistics as analysed from the periodic orbit theory have been indicative
only and do not bring out a complete or explicit picture of underlying
correlations. Moreover, numerical studies have to deal with finite number of
levels only. There is thus possibility that numerical results may not
represent asymptotic trend.

Our aim here is to extend semiclassical formalism to obtain explicit
analytical expressions for various important measures of spectral
statistics. This will therefore can bring out dependence of spectral
properties on different system dependent parameters.
\vskip 1in
Plan of the thesis is as follows:

In chapter one we will discuss main issues involved in development of proper
semiclassical framework that can be applicable for wide class of systems.
The problem with WKB or similar methods developed earlier is that they were
applicable only for systems that are separable. Existence caustics prohibits
these methods to be useful beyond short time scale. Major improvement in
semiclassical methods has been taken place via Path Integrals which
ultimately leads to a semiclassical approximation to density of states in
terms of summation over periodic orbits of the classical system.

As the periodic orbits and their attributes such as stability, action etc.
becomes important in establishing a semiclassical approximation, study of
the nature of classical system in question becomes mandatory. In the case of
polygonal billiards there exist a large body of knowledge about classical
behaviour in the literature. Some of these are very helpful in the study of
periodic orbits and their various properties that are relevant to
semiclassical methods. These will be discussed in chapter two.

In chapter three we take up the first task \cite{0-hdp1}: detailed study of
periodic orbits in some pseudointegrable billiards. Exploiting geometrical
properties of the billiards considered we develop a modified version of
interval exchange maps, called as polar maps. From these maps we show how to
classify and enumerate all the periodic orbits of these billiards. We will
also quantify actions and other parameters required for the semicalssical
study.

In chapter four a very important attribute of periodic orbit i.e. growth
rate (with actions, periods or lengths) of periodic orbits is studied \cite
{0-hdp2}. We obtain exact (asymptotic) law for some pseudointegrable
billiards. We show that this law is quadratic in length and determine
constants associated with it. Some generalization of this law is also
suggested.

Using information gathered in chapter 3 and 4 we obtain explicit analytical
expression for two point correlation function and its fourier transform in
chapter five. Generally semicalssical study ends after establishing a sum
over periodic orbits. This does not bring out dependence of spectral
fluctuations on the various attributes of periodic orbits explicitly. We
here, go further to carry out this summation by converting it into a
integration via proper measure, which enables us to obtain analytical forms
for spectral fluctuation measures. This also brings out role played by
growth rate of periodic orbits explicitly.

In chapter six we use two point correlation function to obtain short and
intermediate range spectral measures such as spacing distribution, number
variance and spectral rigidity \cite{0-hdp3}. We will also compare our
results with some numerical experiments.

In the concluding chapter we discuss our results and outline some of the
future research activities in this field.

\chapter{Modern Approaches to Semi-Classical Quantization}

\section{Introduction:}

Semi-classical\thinspace \thinspace methods have been developed since the
beginning of quantum mechanics because of the importance of the models based
on classical particles and fields. Bohr and Sommerfeld initiated these
methods to quantize atomic systems. However their methods implicitly assumed
separability between the different internal degrees of freedom and hence
were of limited utility since most of the classical systems are not separable%
\cite{1-1}. For example even the three-body helium atom problem can not
conform to the Bohr-Sommerfeld scheme without further assumptions. There
were many attempts to resolve this problem e.g. see \cite{1-ern}. However,
major development in this direction had taken place with the advent of
Heisenberg's matrix mechanics and of Schr\"odinger's wave mechanics.

The Schr\"odinger wave equation of quantum mechanics turns out to be
extremely useful in the development of semi-classical theory since it allows
a discussion of the relation to classical mechanics along the lines similar
to the relation between wave optics and geometric optics, albeit in the
multidimensional phase space rather then in the three dimensional physical
space. This similarity was developed by Brillouin, Wentzel and Kramers for
one degree of freedom (well known as WKB approximation) based of well
developed mathematical framework of asymptotic methods of linear
mathematical physics\cite{1-1a,1-2}. For larger systems that are separable
and reducible to several one-degree of freedom systems variant of this
method (EWKB approximation) was developed.

Since semi-classical methods carry out the quantization on the basis of
knowledge of the classical motion, they are susceptible to the difficulties
caused by the non-linearities of Hamilton's equations of classical
mechanics. The inapplicability of standard WKB methods to non-linear systems
prompted further development of semi-classical methods\cite{1-3,1-4}. These
methods known as the periodic orbit quantization have been applied to a
large variety of systems. These methods basically use a semi-classical
approximations of the trace of the resolvent of the quantum Hamilton
operator. This trace formula, which establishes link between the quantum
operator and classical periodic orbits, can be used to obtain approximate
values of quantum eigenenergies for wide class of systems. The trace formula
which contains different terms depending on the nature and on the stability
of the periodic orbits. It not only improves the accuracy of semi-classical
methods but also provides a framework for the description of diffraction,
non-linear stability effects, bifurcation of periodic orbits etc.

In this thesis we will discuss the imprints of nature of classical
Hamiltonian system on the corresponding quantum mechanical system, in
particular on the collective properties of quantal eigenenergies. We will
use the framework of semi-classical methods for our study. Before coming to
the main theme, we will discuss in this chapter some important issues in
modern semi-classical methods.

\section{Short-wavelength Asymptotics of Schr\"odinger Equation:}

Consider the Cauchy problem with rapidly oscillating initial data for the
Schr\"odinger equation \cite{1-2,1-2a} 
\begin{equation}
\label{I-1}i\hbar \frac{\partial \psi ({\bf q},t)}{\partial t}=-\frac{\hbar
^2}{2m}\nabla ^2\psi ({\bf q},t)+V({\bf q})\psi ({\bf q},t) 
\end{equation}
\begin{equation}
\label{I-2}\psi \mid _{t=0}=A_0({\bf q})\exp (\frac i\hbar S_0({\bf q})) 
\end{equation}
where ${\bf q}\in R^n$, functions $V({\bf q}),S_0({\bf q})$ are real valued
and infinitely differentiable, $A_0({\bf q})$ is infinitely differentiable
with compact support. One can seek an asymptotic solution of the problem as $%
\hbar \rightarrow 0$and ${\bf q}\in R^n,\,\,0\leq t\leq T$. The equation (%
\ref{I-2}) can be regarded as the leading term of an asymptotic expansion in 
$\hbar $, in which $O(\hbar )$terms are neglected. It is also assumed that
an asymptotic solution of equation (\ref{I-1}) 
\begin{equation}
\label{I-3}\psi ({\bf q},t)=A({\bf q},t)\exp \left( \frac i\hbar S({\bf q}%
,t)\right) 
\end{equation}
which is valid only for the short elapsed time $t$, after which it should be
replaced by the sum of higher order terms in an asymptotic expansion. The
reason for this {\rm single term} break down is formation of caustic which
occur at time of order $(\hbar ^0)$. However for sufficiently longer time
above assumption breaks down altogether. We will elaborate these issues in
the following paragraphs.

Substituting equation (\ref{I-3}) in the Schr\"odinger equation and
neglecting terms of order $(\hbar )$ and higher, we get the time-dependent
Hamilton-Jacobi(H-J) equation for the action $S\left( {\bf q},t\right) $, 
\begin{equation}
\label{I-4}H\left( {\bf q},\frac{\partial S({\bf q},t)}{\partial {\bf q}}%
,t\right) +\frac{\partial S({\bf q},t)}{\partial t}=0 
\end{equation}
where $H$ is Hamiltonian and the canonical momentum ${\bf p}$ of $H({\bf q},%
{\bf p},t)$ is replaced by ${\bf p}=\partial S/\partial {\bf q}$. The order $%
(\hbar )$ term gives amplitude transport equation for $A$ or the continuity
equation for $\rho ({\bf q},t)=|A({\bf q},t)|^2$ as 
\begin{equation}
\label{I-5}\frac{\partial \rho \left( {\bf q},t\right) }{\partial t}+\frac
\partial {\partial {\bf q}}\left[ \rho ({\bf q},t)v({\bf q},t)\right] =0 
\end{equation}
where the velocity field $v=\partial H({\bf q},{\bf p},t)/\partial {\bf p}$
with ${\bf p}=\partial S/\partial {\bf q}$. One then solves equation (\ref
{I-4}) for $S({\bf q},t)$ and use this $S$ to solve equation (\ref{I-5}).

The H-J equation is typical of non-linear equations, hence has a bewildering
variety of solutions. Among them are the complete integrals which gives all
the trajectories of the system. As we are concerned with those solutions
that have a relationship to a quantum mechanical wave function, it is
desirable to find a way of defining $S$ such that it will be a single valued
function of its variables. The first step to do this is to use phase space
description of trajectories\cite{1-delos,1-perci}. The concept of {\sl %
Lagrangian Manifold} becomes important here.

\subsection{Lagrangian Manifold:}

The initial action $S_0({\bf q})$ and momentum field ${\bf p}_o=\partial
S_0/\partial {\bf q}$ can be viewed as a vector field on the $f${\it -}%
dimensional configuration space. This field represents the initial momenta
of the swarm of particles whose initial density is given by $\rho _0({\bf q}%
)=|A_0({\bf q})|^2$. This swarm of particles and vector field constitute the
classical and semi-classical interpretation of the initial action function
in configuration space. Since the momentum field imposes $f${\it \ }%
independent constraints on the $2f${\it \ }variables $({\bf q},{\bf p})$ of
the phase space, initial swarm of the particles lies on the $f${\it \ }%
-dimensional surface $\Lambda _0$. The surface $\Lambda _0$ is {\rm graph }%
of the function ${\bf p}={\bf p}_0({\bf q})$ i.e. the set of points in phase
space of the form $({\bf q},{\bf p}_0({\bf q}))$. This surface satisfy
properties of {\sl Lagrangian manifold,} $\Lambda ${\sl ,} the formal
definition of which is as follows:

\begin{definition}
{\sl Lagrangian manifold }is a {\it f-dimensional }surface $\Lambda $ in 
{\it 2f-dimensional }phase space such that $\forall ({\bf q},{\bf p})\in
\Lambda $ and for $\forall \,\,\delta \overline{z}\equiv (\delta {\bf q}%
,\delta {\bf p})$ tangent to $\Lambda $ at $({\bf q},{\bf p})$ representing
a small displacement , the action of the symplectic form (i.e. differential 
2-form) on all $\delta \overline{z}_1,\delta \overline{z}_2\in \{\delta 
\overline{z}\}$ defined as 
\begin{equation}
\label{I-6}w(\delta \overline{z}_1,\delta \overline{z}_2)\equiv \delta 
\overline{z}_1\cdot \underline{\underline{J}}^{-1}\cdot \delta \overline{z}%
_2\equiv \delta {\bf p}_1\cdot \delta {\bf q}_2-\delta {\bf p}_2,\delta {\bf %
q}_1=0
\end{equation}
where\thinspace \thinspace $\underline{\underline{J}}$\thinspace \thinspace
is unit symplectic matrix.
\end{definition}

All the curves in the 2-dim. phase space are {\sl Lagrangian manifolds}. The
concept is therefore really needed for multidimensional problems.\ It can be
immediately seen that surfaces ${\bf q}=const.$ or ${\bf p}=const.$ are $%
\Lambda $. Since the value of symplectic form of (\ref{I-6}) is invariant
under canonical transformations every $\Lambda $ is a constant ${\bf q}$ or
constant ${\bf p}$ surface in some set of canonical co-ordinates. The graph
of any curl-free momentum field is surface $\Lambda $. The converse is
however, not true, since $\Lambda $ may contain points at which derivatives
in the curl-free condition $(i.e.\partial {\bf p}_i/\partial {\bf q}_j)$ are
not defined. Such points are generally associated with caustics.

The {\it f-}dimensional vectors ${\bf q}$ and ${\bf p}$ can be regarded as
smooth functions of $u=(u_{1,}u_2,......,u_f)$, ''co-ordinates'' labelling
of $\Lambda $. The variables ${\bf q}(u)$ are locally invertible if Jacobian 
$|\partial {\bf q}/\partial {\bf p}|$ do not vanish. Then and only then one
can define ${\bf p}({\bf q})={\bf p}(u({\bf q}))$, a function of ${\bf q}$
on $\Lambda $. If above Jacobian vanishes then%
$$
\frac{\partial p_i}{\partial q_j}=\sum_k\frac{\partial p_i}{\partial u_k}%
\cdot \frac{\partial u_k}{\partial q_j} 
$$
will behave badly. In $1-$dimension since $\partial p_i/\partial u_k\neq 0$
it will diverge. In higher dimensions some of the eigenvalues of $[\partial 
{\bf p}/\partial u]$ may vanish at the same place where some of the
eigenvalues of $\left[ \partial {\bf q}/\partial u\right] $ also vanish,
resulting in the complicated behaviour. A phase space vector $\delta z$
tangent to $\Lambda $ can be written in the $u-$co-ordinates as $\partial
z=\left( \frac{\partial {\bf q}}{\partial u}\delta u,\frac{\partial {\bf p}}{%
\partial u}\delta u\right) $. If the matrix $[\partial {\bf q}/\partial u]$
is singular, then for all $\delta u\neq 0$ and $\delta q=0$ , $\delta z$ has
vanishing components in the $q$-components resulting in a caustic in the
configuration space. The order of caustic or the number of null eigenvectors 
$\delta u$, is given by co-rank of the matrix $[\partial {\bf q}/\partial u]$%
. The order of caustic as well as set of points for which $\partial
q/\partial u=0$ are independent of the choice of $u$ on $\Lambda $. The
important theorem about $\Lambda $ is

\begin{theorem}
A region of $\Lambda $ which is free of singular points if projected on
configuration space gives curl-free momentum field.\cite{1-2,1-2a}
\end{theorem}

Since flow preserves the symplectic structure (cf. Liouville Theorem) $%
\Lambda $ evolves into another $\Lambda ^{^{\prime }}$ under time evolution.
Any function $S({\bf q})$ which satisfy ${\bf p}({\bf q})=\partial
S/\partial {\bf q}$ on $\Lambda $ will now be called {\it generating
function }of $\Lambda $.

\begin{theorem}
On a region of $\Lambda $, free of singular points, one can define unique
generating function up to additive constant.\cite{1-2a}
\end{theorem}

This additive constant is usually associated with phase conventions in
semi-classical applications. For $\Lambda $, having region of singular
points, one can divide $\Lambda $ into sub-regions which extend up to and
separated by the caustics as shown in Fig.1.1. 
\begin{figure}[htbp]
\begin{center}
\framebox{\epsfig{file=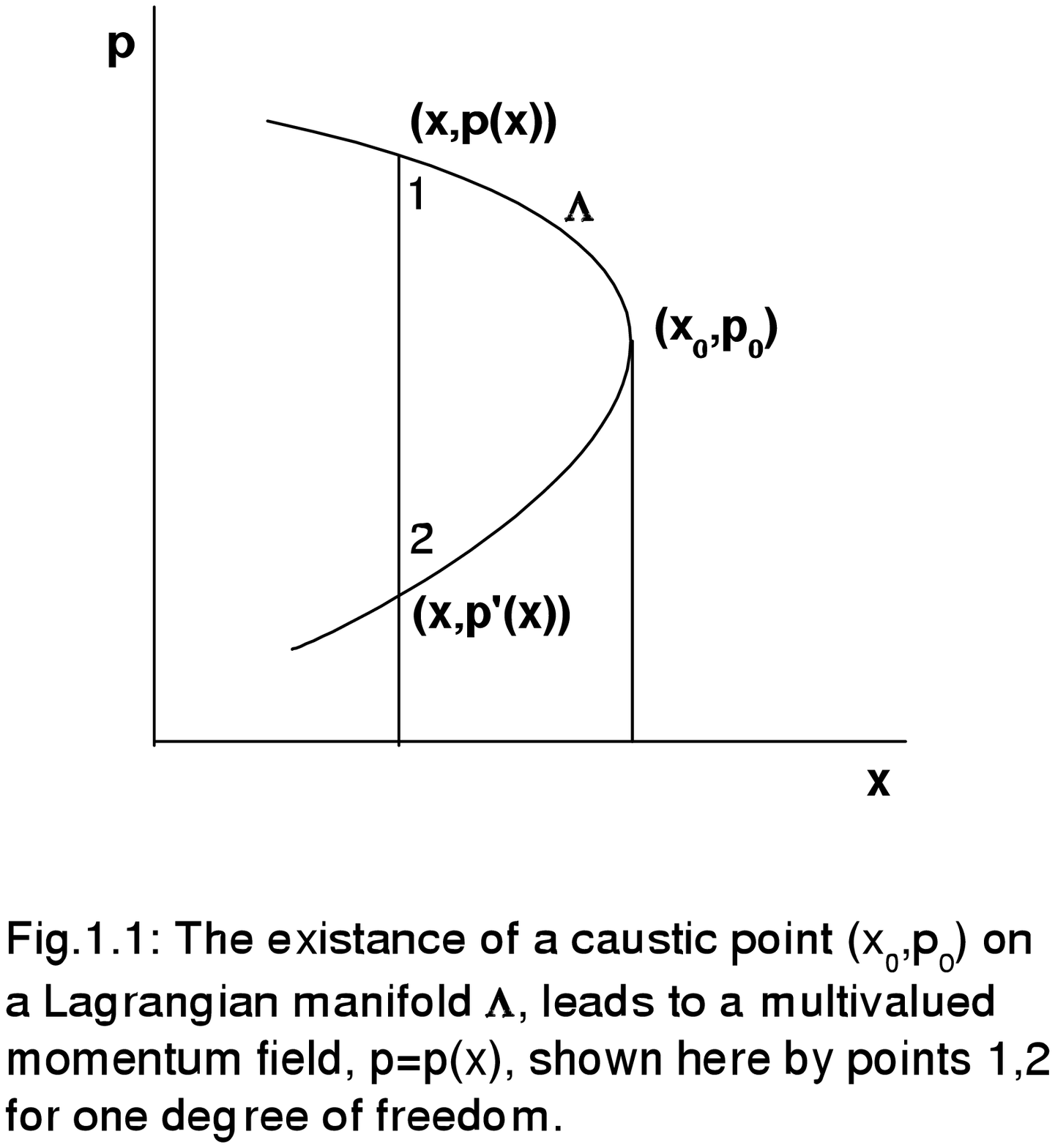,bbllx=75,bblly=198,bburx=540,bbury=700,height=6in}}
\end{center}
\end{figure}
Each such region corresponds
to a distinct branch of the momentum field ${\bf p}_b({\bf q})$ and
associated generating function $S_b({\bf q})$ up to its own additive
constant. One can define action function $S({\bf q},{\bf p})$ on $\Lambda $
itself, as the line integral $\int p\cdot dq$ along the contour belonging to
the $\Lambda $ and taken relative to an arbitrary initial point, see Fig.
1.2. This can be done by demanding that the different functions $S_b({\bf q}%
) $ approach one another at the caustic dividing branches. In this way one
can also link some or all of the additive constants together. The function $%
S({\bf q},{\bf p})$ will be multi-valued not only due to caustics but also
due to non-trivial (e.g. {\it multi-connectedness) }topology of $\Lambda $.
\begin{figure}[htbp]
\begin{center}
\framebox{\epsfig{file=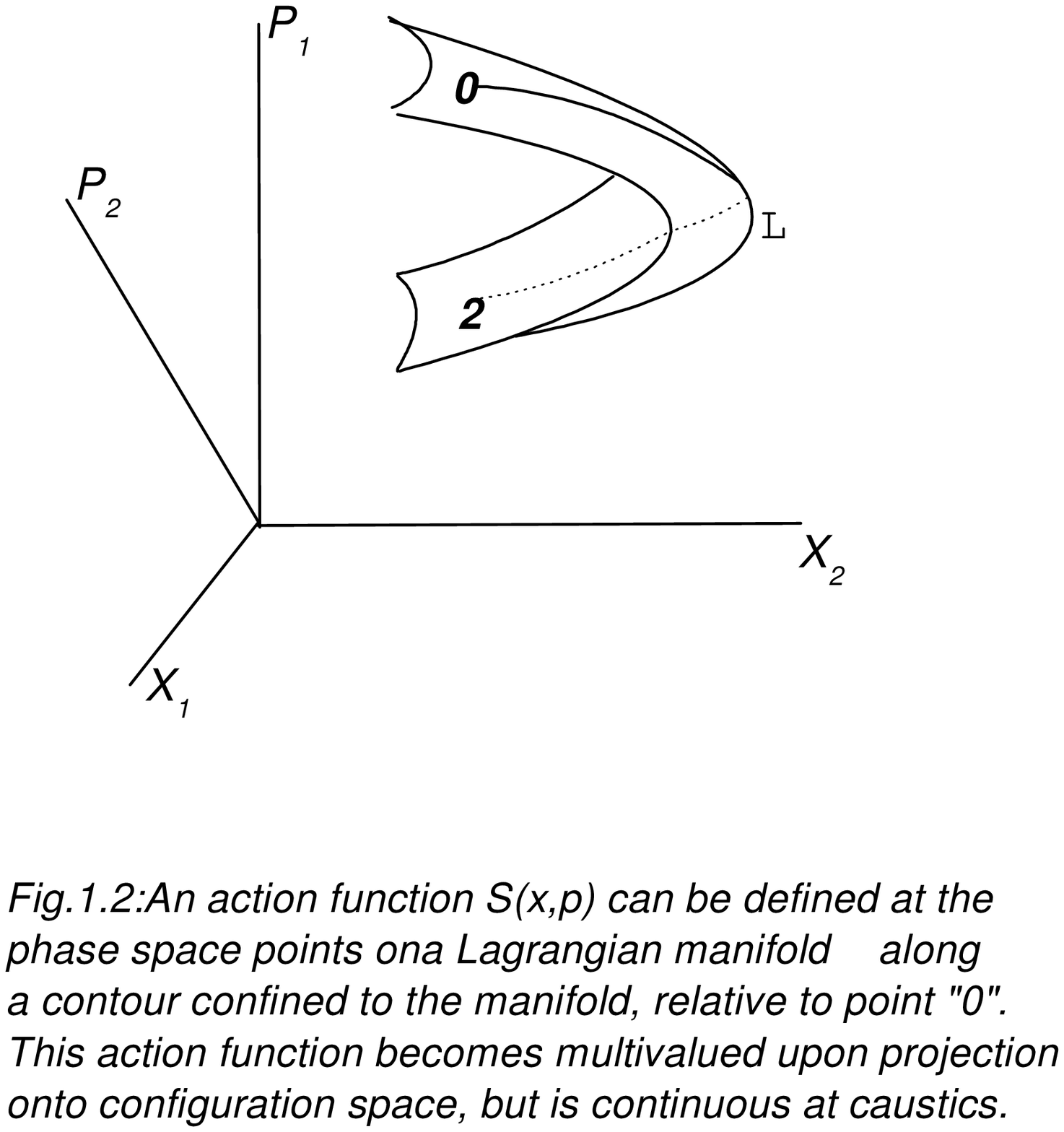,bbllx=80,bblly=170,bburx=560,bbury=700,height=6in}}
\end{center}
\end{figure}
The generating function in general is specific to the choice of co-ordinates
Some $\Lambda $ consist entirely of caustic points (e.g. $x=const.$ in
1-dimension). Such manifolds do not have generating function $S(x)$.
However, with respect to other co-ordinate systems one can define a
generating function for such surfaces. The locations of the caustic points
are relative to the representation being used. This fact plays important
role in the WKB\ theory. It is possible to cover every $\Lambda $
(Fig.1.3) with overlapping regions such that every region is caustic free in
some representation obtained from a commutating mixture of ${\bf q}$'s and $%
{\bf p}$'s.
\begin{figure}[htbp]
\begin{center}
{\epsfig{file=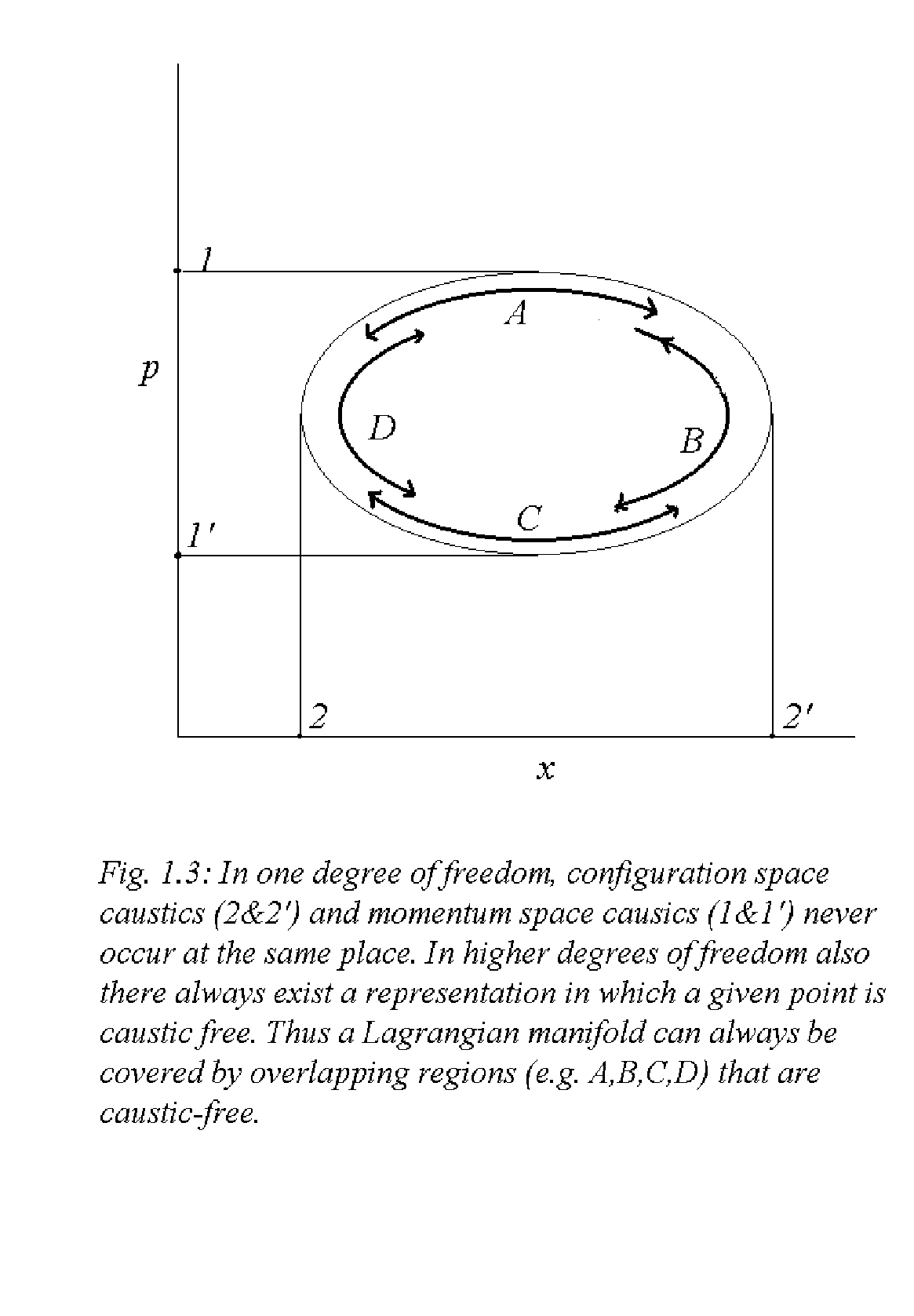,bbllx=80,bblly=170,bburx=560,bbury=700,height=6in}}
\end{center}
\end{figure}
A whole family of wave functions, such as a complete set of commutating
observables, will reproduce a whole family of $\Lambda $ manifolds
parameterized by some set of parameters $\lambda =(\lambda _{1,}\lambda
_{2,}\lambda _{3,........,}\lambda _f).$ Thus phase space is foliated up
into $f$- parameter family of $f$-dimensional $\Lambda $ manifolds. The
function $S({\bf q},\lambda )$ is one of the generating functions of
canonical transformations in classical mechanics.

\subsection{Solution of H-J and Amplitude Transport Equations:}

Now consider an initial Lagrangian manifold $\Lambda ^{^{\prime }}$, free of
singular points, obtained from initial action $S({\bf q}^{^{\prime
}},t^{^{\prime }})=S_0({\bf q}^{^{\prime }})$ and ${\bf p}({\bf q}^{^{\prime
}},t^{^{\prime }})=\partial S({\bf q}^{^{\prime }},t^{^{\prime }})/\partial 
{\bf q}^{^{\prime }}.$ Each point of $\Lambda ^{^{\prime }}$ evolves under
Hamilton's equations, mapping into another $\Lambda ^{^{\prime \prime }}$ at
time $t^{^{\prime \prime }}$. If $\Lambda ^{^{\prime \prime }}$ is also free
of singular points then one can define a generating function $S({\bf q}%
^{\prime \prime },t^{\prime \prime })$, which is also a solution of H-J
equation, as 
\begin{equation}
\label{I-7}S({\bf q}^{\prime \prime },t^{\prime \prime })=S({\bf q}^{\prime
},t^{\prime })+R({\bf q}^{\prime \prime },t^{\prime \prime }:{\bf q}^{\prime
},t^{\prime }) 
\end{equation}
where $R({\bf q}^{\prime \prime },t^{\prime \prime }:{\bf q}^{\prime
},t^{\prime })=\int\limits_{{\bf q}^{^{\prime }},t^{^{\prime }}}^{{\bf q}%
^{^{\prime \prime }},t^{^{\prime \prime }}}[{\bf p}\cdot d{\bf q}-Hdt]$ is
Hamilton's principal function. As $t^{^{\prime \prime }}\longrightarrow
t^{^{\prime }},$one gets $R\longrightarrow 0$ and $S({\bf q}^{\prime \prime
},t^{\prime \prime })\longrightarrow S({\bf q}^{\prime },t^{\prime })$ thus
satisfying required initial conditions.

Since the particle density is conserved one can write $\rho ({\bf q}^{\prime
\prime },t^{\prime \prime })d{\bf q}^{^{\prime \prime }}=\rho ({\bf q}%
^{\prime },t^{\prime })d{\bf q}^{^{\prime }}$, and since $\rho =|A|^2$, we
get 
\begin{equation}
\label{I-8}A({\bf q}^{\prime \prime },t^{\prime \prime })=A({\bf q}^{\prime
},t^{\prime })\left| \det \frac{\partial {\bf q}^{^{\prime }}}{\partial {\bf %
q}^{^{\prime \prime }}}\right| ^{1/2}. 
\end{equation}

Here the absolute sign in this equation required only when one considers
many branches.

Thus the WKB solution to the Cauchy problem can be written as 
\begin{equation}
\label{I-9}\Psi ({\bf q}^{\prime \prime },t^{\prime \prime })=A({\bf q}%
^{\prime },t^{\prime })\left[ \det \frac{\partial {\bf q}^{\prime }}{%
\partial {\bf q}^{\prime \prime }}\right] ^{1/2}\exp \left\{ \frac i\hbar
\left[ S({\bf q}^{\prime },t^{\prime })+R({\bf q}^{\prime \prime },t^{\prime
\prime }:{\bf q}^{\prime },t^{\prime })\right] \right\} 
\end{equation}
This expression diverges when ${\bf q}^{\prime \prime }$ is at caustic (i.e. 
$\det \partial {\bf q}^{\prime }/\partial {\bf q}^{\prime \prime }$
diverges) of $\Lambda ^{\prime \prime }$. This is illustrated in Fig. 1.4
for one degree of freedom. This divergence represents a non uniformity in
the variables $({\bf q}^{\prime \prime },t^{\prime \prime })$ of $\Psi $ in $%
\hbar $, since the error in $\Psi ,\longrightarrow 0$, like $\hbar $, as $%
\hbar \longrightarrow 0$ for fixed $({\bf q}^{\prime \prime },t^{\prime
\prime })$, however it goes to infinity for fixed $\hbar $ as $({\bf q}%
^{\prime \prime },t^{\prime \prime })$ approaches caustic.
\begin{figure}[htbp]
\begin{center}
{\epsfig{file=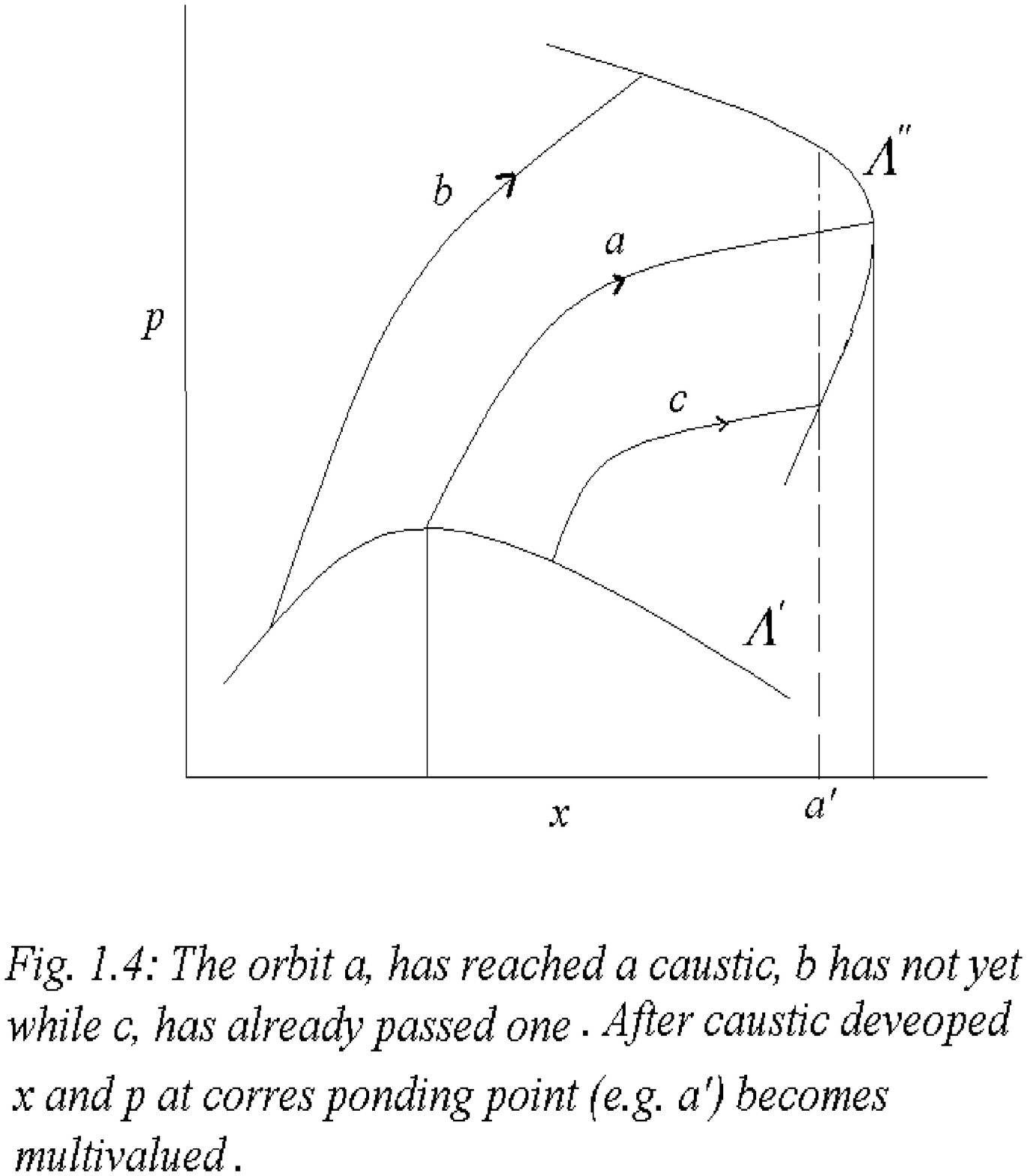,bbllx=80,bblly=170,bburx=560,bbury=700,height=6in}}
\end{center}
\end{figure}
Though the solution near caustic is not valid one can continue the solution
through the divergence. However, function ${\bf q}^{\prime }({\bf q}^{\prime
\prime },t^{\prime },t^{\prime \prime })$ is then multi-valued. By passing
through caustic the determinant changes sign via divergence. Thus $\left[
\det \frac{\partial {\bf q}^{\prime }}{\partial {\bf q}^{\prime \prime }}%
\right] ^{1/2}$ becomes imaginary after first caustic pass. One can absorb
this imaginary part into a phase factor as $\exp (-i\eta \pi /2)$, forcing $%
A $ to be positive. The $\eta $ here is well known as {\em Maslov index}.
Considering multiplicity of orbits arriving at ${\bf q}^{\prime \prime }$ at
time $t^{\prime \prime }$, one can write 
\begin{equation}
\label{I-10}\Psi ({\bf q}^{\prime \prime },t^{\prime \prime
})=\sum\limits_bA_b({\bf q}^{^{\prime \prime }},t^{^{\prime \prime }})\exp
\left\{ \frac i\hbar S_b({\bf q}^{^{\prime \prime }},t^{^{\prime \prime
}})-i\eta _b\frac \pi 2\right\} . 
\end{equation}

To compute indices $\eta ^{\prime }s$ one needs to consider momentum-space
wave function which is Fourier transform of a configuration-space function.
The Fourier transform integral is evaluated by{\em \ stationary phase
approximation}. This leads to the momentum-space function 
\begin{equation}
\label{I-11}\Phi ({\bf p})=\exp \left( i\alpha \frac \pi 4\right) |\det 
\underline{\underline{M}}({\bf q})|^{-\frac 12}\exp \left\{ \frac i\hbar
\left[ S({\bf q})-{\bf q}\cdot {\bf p}\right] \right\} 
\end{equation}
where the symmetric matrix $\underline{\underline{M}}({\bf q})=\partial {\bf %
p}/\partial {\bf q}=\partial ^2S({\bf q})/\partial {\bf q}\partial {\bf q}$
and $\alpha $ is an integer given by the {\em index of inertia of} $%
\underline{\underline{M}}$ (i.e. number of positive eigenvalues minus number
of negative eigenvalues). If $\Lambda $ do not have singular points in say, $%
X-representation$ but do have a caustic in $P-representation$, then at the
momentum-space caustic $\det \underline{\underline{M}}({\bf q})\rightarrow 0$%
, hence $\alpha $ changes discontinuously across the caustic by $\pm 2\cdot
n $, since number of eigenvalues of $\underline{\underline{M}}$ which vanish
at caustic is same as the order $"n"$ of the caustic. Therefore relative
phase shift between the two branches will have form $\exp (-i\eta \pi /2)$,
and 
\begin{equation}
\label{I-12}\Phi ({\bf p})=\exp \left( i\alpha _0\frac \pi 4\right)
\sum\limits_b\widetilde{A}_b\left( {\bf p}\right) \exp \left\{ \frac i\hbar
\left[ \widetilde{S}_b({\bf p})-i\eta _b\frac \pi 2\right] \right\} 
\end{equation}
where $\alpha _0$ is index of inertia of one of the branches for which $\eta
=0$, and $\widetilde{S}_b({\bf p})=S_b({\bf q)}-{\bf q}\cdot {\bf p}$, $%
\widetilde{A}_b=\left( \widetilde{\rho }_b({\bf p})\right) ^{1/2}$ with $%
\widetilde{\rho }_b({\bf p})=\rho _b({\bf q})|\det \partial {\bf q}/\partial 
{\bf p}|$.

Thus the relative phase shifts between branches can be determined by
switching to a representation which is caustic free. Considering any two
branches the phases of the two branches of $\Phi _b({\bf p}^{\prime \prime
},t^{\prime \prime })$ can be written as$\eta $ $\widetilde{S}_b({\bf p}%
)-i\eta _b\pi /2+i\alpha _b\pi /4$. Demanding continuity of phases across
caustic in caustic free representation, we get 
\begin{equation}
\label{I-13}\eta _2=\eta _1+\frac{\alpha _2-\alpha _1}2. 
\end{equation}
The bounds on $\triangle \eta $, change in $\eta $ across a caustic is then $%
-n\leq \triangle \eta \leq +n$, $n$ being order of caustic. By extending
process, it is possible to associate total $\triangle \eta $ measured
between the end points of a curve segment on $\Lambda ,$ which may cross a
number of caustic and straddle several branches. The quantity $\triangle
\eta $ associated with a directed curve segment of $\Lambda $ is the Maslov
index of that segment. The index depends only on the geometry of $\Lambda $
and its' projection on configuration space. It also depends on the
representation chosen.

The method discussed in this section is simple and allows us to derive
several important expressions. However, caustics are causing such
difficulties that formulation has restricted applicability. The determinant $%
\det (\partial {\bf p}/\partial {\bf q})$ plays an important role in the
formulation which becomes singular at caustic. At caustic, the method of
matching mentioned above uses different equations than preceding ones. This
method is also difficult to apply for higher order terms in $\hbar $
expansion. For this reason it is advantageous to reformulate the problem in
terms of Path Integrals.

\section{Path Integrals and Semi-classical Limit}

An alternate semi-classical quantization technique {\it via }Feynman Path
Integral gained momentum due to problems discussed in the previous section.
The first hint regarding this comes from the work of Selberg\cite{1-sel} who
obtained a path integral formulation for a particle moving on a Riemann
surface of negative curvature. On the short time scales path integrals had
been used to study semi-classical integrable systems\cite{1-mor,1-cho}.
After overcoming the problem with caustics the semi-classical path integral
was extended to long times\cite{1-sch}. These efforts culminated in {\it the}
{\it trace formulae }of Gutzwillar, Balian and Bloch, as well as Berry and
Tabor \cite{1-3,1-4,1-5}. The trace formulas are among the very few
theoretical results of any generality that connects quantum mechanics with
classical mechanics {\it via }semi-classical approximation for quantal
density of states of a bound quantum system, in terms of a sum over the
periodic orbits of corresponding classical system. In this section we
briefly outline these developments.

\subsection{The resolvent operator and its' singularities}

Consider \cite{1-5a} time evolution of equation (\ref{I-1}), which is linear
in nature. Therefore we can define a linear operator $\hat U(t,t^{\prime })$
such that 
\begin{equation}
\label{II-1}\Psi _t\left( {\bf q}\right) =\hat U(t,t^{\prime })\Psi
_{t^{\prime }}({\bf q}) 
\end{equation}
The operator $\hat U$ known as evolution operator, obeys the Sch\"odinger
equation (\ref{I-1}), with the initial condition $\hat U(t^{\prime
},t^{\prime })=\hat I$, an identity operator on the space of state vectors.
The Hamiltonian defined as 
\begin{equation}
\label{II-2}\hat{H}_t=-\sum_{j=1}^f\frac{\hbar ^2}{2m_j}\nabla _j^2+V({\bf q}%
) 
\end{equation}
is a Hermitian operator $(\hat{H}=\hat{H}^{\dagger }),$ which implies $%
\partial _t\left\langle \Psi _t\left( {\bf q}\right) |\Psi _t\left( {\bf q}%
\right) \right\rangle =0$. Hence the evolution operator $\hat U$ is unitary $%
\left( \hat U^{\dagger }\hat U=I\right) .$ The formal solution for $\hat
U(t,t^{\prime })$ can be written as 
\begin{equation}
\label{II-3}\hat U(t,t^{\prime })=\hat I-\frac i{^\hbar
}\int\limits_{t^{\prime }}^tds\hat{H}_s\hat U(s,t^{\prime }) 
\end{equation}
In case of time-independent $\hat{H}_t$ one can solve this equation by
iterations to obtain the Dyson series for the operator which finally reduces
to 
\begin{equation}
\label{II-4}\hat U(t,t^{\prime })=\exp \left\{ -\frac{i\hat{H}(t-t^{\prime })%
}\hbar \right\} . 
\end{equation}
The evolution operator also obeys group property $\hat U(t,t^{\prime \prime
})\hat U(t^{\prime \prime },t^{\prime })=\hat U(t,t^{\prime }).$ (Using this
we can set $t^{\prime }=0$). One can introduce evolution operators for {\tt %
``}forward{\tt ''} and {\tt ``}backward{\tt ''} propagation in time such as 
\begin{equation}
\label{II-5}\pm \Theta (\pm t)\hat U(t)=\hat U^{\pm }(t), 
\end{equation}
where $\Theta (t)$ is the Heaviside step function (i.e. $\Theta (t)=1\forall
t>0,\Theta (t)=0\forall t<0$). The form of (\ref{II-4}) is indication of
conjugacy of energy and time variables as involved in Fourier transform.
Thus one can describe a system either in the time-domain or in the energy
domain. Both of these description are related by a Fourier transform.

The Fourier transform of the $\hat U^{\pm }(t)$ is 
\begin{equation}
\label{II-6}
\begin{array}{ccc}
\hat G_{\pm }(E) & = & -\frac i\hbar \int\limits_{-\infty }^\infty dt\exp
\left\{ 
\frac{itE}\hbar \right\} \hat U^{\pm }(t) \\  & = & \mp \frac i\hbar
\int\limits_0^\infty dt\exp \left\{ \pm \frac{it(E-\hat H)}\hbar \right\} 
\end{array}
\end{equation}
where $E$ is the eigenvalue of the stationary Schr\"odinger equation $\hat
H|\Psi (E)\rangle =E|\Psi (E)\rangle $, obtained from taking Fourier
transform of (\ref{I-1}). The Integral (\ref{II-6}) does not exist, hence
Green operator does not have meaning as an operator-valued functions, but
can be meaningful as operator-valued distribution. We can redefine $\hat
G_{\pm }(E)$ by replacing $E\rightarrow E\pm i\eta ,\eta >0$ (i.e. by
analytical continuation in complex $E$ surface), then 
\begin{equation}
\label{II-7}\hat G_{\pm }(E)=\lim _{\eta \rightarrow 0_{\pm }}(E\pm i\eta
-\hat H)^{-1}
\end{equation}
The object $\hat G(z)=(z-\hat H)^{-1}$ is known as resolvent of $\hat H$.
The operators $\hat G_{\pm }(E)$ are now called as forward and backward
Green operators. The resolvent in bounded, except for the values of $z$
which corresponds to the spectrum of Hamiltonian. Since, eigenvalues of the
Hamiltonian, being a Hermitian operator , are real and $\Theta (t)+\Theta
(-t)=1$, one can write evolution operator {\it via} inverse Fourier
transform of $\hat G(z)$ as a contour integral 
\begin{equation}
\label{II-8}\hat U^{\pm }(t)=\frac 1{2\pi i}\int\nolimits_{C^{\pm }}dz\exp
\left\{ \frac{-i+z}\hbar \right\} \hat G(z).
\end{equation}
For $t>0\,\,(t<0)$, the contribution of the contour $C^{+}\left(
C^{-}\right) $ is zero. The resolvent has two types of singularities on the
real axes. The discrete spectrum of $\hat H$ results in poles on the real
axis below an energy threshold $E_0$. The Continuous spectrum on other hand
results in the branch points (in general of order two, since energy is
related to momentum by relation $E=E_{c\;}+p^2/2m$) at real energies\ $%
\left\{ E_c\right\} .$ Thus spectral decomposition of the evolution operator
can be carried out in terms of a sum over bound states $|\Psi _b\rangle $
and sum over continuous states $|\Psi _C(E)\rangle ,$ of different channels $%
C$, with eigenvalues $\exp \left\{ -\frac{iE_bt}\hbar \right\} ,\exp \left\{
-\frac{iEt}\hbar \right\} $ respectively, as\cite{1-6a} 
\begin{equation}
\label{II-9}\hat U(t)=\sum\limits_b|\Psi _b\rangle \exp \left\{ -\frac{iE_bt}%
\hbar \right\} \langle \Psi _b|+\sum\limits_C\int\limits_{E_c}^\infty
dE|\Psi _C(E)\rangle \exp \left\{ -\frac{iEt}\hbar \right\} \langle \Psi
_C(E)|
\end{equation}

The eigenfunctions of $\hat{U}$ are also the eigenfunctions of $\hat{H}$.
The energy eigenvalues are then poles of the Green operator, and hence also
of its trace \cite{1-6}. Considering bounded systems only we can write the
expression for the level density (or density of states) $d(E)$ as 
\begin{equation}
\label{II-10}
\begin{array}{ccc}
d(E) & \equiv \sum\limits_{n=0}^{\infty} \delta (E-E_n)\ \ & = \textsf{Tr} \delta (E-\hat{H}) \\ 
 & -\frac 1\pi \lim _{\eta \rightarrow
0_{+}}\textsf{Im}\quad \textsf{Tr} & \frac 1{E+i\eta -\hat{H}} 
\end{array}
. 
\end{equation}
The level density is derivative of the staircase function defined as the
number of eigenvalues below energy $E:N(E)=Number\left\{ E_n<E\right\}
=\sum_n^\infty \Theta (E-E_n).$ The singularities of the level density on
the real axis give eigenvalues of the discrete spectrum of $\hat{H}.$ It is
difficult to establish exact quantum mechanical expression for the evolution
operator. The path integral formulation is proved to be quite useful in
establishing an expression for the evolution operator and a propagator.

\subsection{A Propagator and Path Integral}

In position representation equation\ (\ref{II-1}), can be written as 
\begin{equation}
\label{II-11}
\begin{array}[t]{ccc}
\Psi _t({\bf q}) & = & \langle q|\hat U(t,t^{\prime })|\Psi _{t^{\prime
}}\rangle \\  
& = & \int dq^{\prime }\,\,\,\langle q|\hat U(t,t^{\prime })|q^{\prime
}\rangle \langle q^{\prime }|\Psi _{t^{\prime }}\rangle \\  
& = & \int dq^{\prime }K(qq^{\prime }:tt^{\prime })\Psi _{t^{\prime
}}(q^{\prime }) 
\end{array}
\end{equation}
where $K(qq^{\prime }:tt^{\prime })=\langle q|\hat U(t,t^{\prime
})|q^{\prime }\rangle $ is propagator, which gives wave function at time $t$
and position $q$ once the wave function at $t^{\prime },q^{\prime }$ is
known. It may be noted that equation (\ref{II-11}) is in fact the expression
of Huygens principle of the wave propagation. The propagator satisfies
Schr\"odinger equation (\ref{I-1}) along with initial condition 
\begin{equation}
\label{II-12}\lim _{t\rightarrow t^{\prime }}K(qq^{\prime }:tt^{\prime
})=\delta (q-q^{\prime }). 
\end{equation}
It also follows from the group composition properties of the evolution
operator and definition of the propagator that the propagator also satisfies
the semi-group property {\it e.g.} 
\begin{equation}
\label{II-13}K(qq^{\prime }:tt^{\prime })=\int dq^{\prime \prime
}K(qq^{\prime \prime }:tt^{\prime ^{\prime }})K(q^{\prime \prime }q^{\prime
}:t^{\prime \prime }t^{\prime }) 
\end{equation}
Thus transition from $(q^{\prime },t^{\prime })\longrightarrow (q,t)$
involves all the possible points, hence all possible paths. This notion of
propagation over all possible path is very important in the path integral
formulation of quantum mechanics\cite{1-7a}.

Consider a transition between two space-time points $(q_i,t_i)$ and $%
(q_f,t_f)$. Divide the time interval $\tau =t_f-t_i$ in $N$ equal pieces of
length $\Delta t=\tau /N$. The semi-group property implies (with $%
q_N,t_N=q_f,t_f$ and $q_0,t_0=q_i,t_i$) 
\begin{equation}
\label{II-14}K(q_fq_i:t_ft_i)=\int \cdot \int dq_1dq_2\cdot \cdot \cdot
dq_{N-1}\prod\limits_{n=0}^{N-1}K(q_{n+1}q_n:\Delta t) 
\end{equation}
The paths entering in the above equation are not necessarily classical
trajectories, and in general continuous but non-differentiable. Formally,
these paths are similar to those of Brownian motion, which are Markovian
Processes. Using equation(\ref{II-4}), the propagator for the Hamiltonian of
form (\ref{II-2}), over a small time interval $\Delta t$ can be written as 
\begin{equation}
\label{II-15}K(q_{n+1}q_n:\Delta t)=\langle q_{n+1}|\exp \{-\frac{i\hat{H}%
\Delta t}\hbar \}|q_n\rangle . 
\end{equation}
Using identity $\exp \{\hat{A}+\hat{B}\}=\exp \{\hat{A}\}\exp \{\hat{B}%
\}\exp \left\{ -\frac 12[\hat{A},\ \hat{B}]\right\} \cdot \cdot \cdot $,
where commutator $[\hat{A},\ \hat{B}]=\hat{A}\hat{B}-\hat{B}\hat{A}$ and
dots represents higher order commutators, one can write 
\begin{equation}
\label{II-16}\exp \{-\frac{i\hat{H}\Delta t}\hbar \}\simeq \exp \{-\frac{%
i\Delta t\hat{T}}\hbar \}\exp \{-\frac{i\Delta t\hat{V}}\hbar \}\left(
1+\frac 12\left( \frac{\Delta t}\hbar \right) ^2[\hat{T}\ \hat{V}]\right) 
\end{equation}
In the limit $N\longrightarrow \infty ,\quad \Delta t$ becomes small enough
so that $O(\Delta t^2)$ terms can be neglected. Therefore, equation (\ref
{II-15}) becomes%
$$
K(q_{n+1}q_n:\Delta t)=\langle q_{n+1}|\exp \{-\frac{i\hat{T}\Delta t}\hbar
\}|q_n\rangle \exp \{-\frac{i\Delta t\hat{V}}\hbar \}, 
$$
which can be rearranged in form%
$$
K(q_{n+1}q_n:\Delta t)=\left( \frac m{2\pi i\hbar \Delta t}\right) ^{-\frac
f2}\exp \left\{ \frac{i\Delta t}\hbar \left[ \frac m2\left( \frac{{\bf q}%
_{n+1}-{\bf q}_n}{\Delta t}\right) ^2-V\right] \right\} . 
$$
Thus in limit $N\rightarrow \infty ,$ equation (\ref{II-14}) becomes 
\begin{equation}
\label{II-17}K({\bf q}_N{\bf q}_0:\Delta t)=\lim _{N\rightarrow \infty
}\int \cdots \int \left( \frac m{2\pi i\hbar \Delta t}\right) ^{-\frac
12}\prod_{j=1}^{N-1}d{\bf q}_j\left( 2\pi \hbar \Delta t\right) ^{-\frac
f2}\exp \left\{ \frac i\hbar \int\limits_{t_i}^{t_f}dt\textsf{ L}({\bf q},%
{\bf \dot q})\right\} 
\end{equation}
where, {\sf L}$({\bf q},{\bf \dot q})=T-V,$ is the classical Lagrangian. In
the phase of above equation we have Hamilton's Principal Function 
\begin{equation}
\label{II-18}R[{\bf q}(t)]{\sf =}\int\limits_{t_1}^{t_f}\textsf{ L}({\bf q%
},{\bf \dot q})dt. 
\end{equation}
The physical path (or classical trajectory) will be realized only if one
extrmizes $R[{\bf q}].$

We are now interested in the semi-classical limit of equation(\ref{II-17}).
The semi-classical limit is obtained {\rm via }method of stationary phase
approximation, which amounts to an expansion about the classical path.

\subsection{Stationary Phase Approximation:}

The kernel of the propagator (\ref{II-17}) is complex exponential of the
type ($\exp (\frac{iR}\hbar )$). The stationary phase approximation \cite
{1-7} is an asymptotic approximation of the Feynman Path Integral where $%
\hbar $ is considered as a small parameter.

Consider a simple example, i.e. the evaluation of following integral 
\begin{equation}
\label{II-19}F(\frac 1\hbar )=\int\limits_{-\infty }^\infty dx\,\,\exp
\left( \frac{if(x)}\hbar \right) 
\end{equation}
When $\hbar \rightarrow 0,$ the exponential becomes highly oscillatory
function of $x$ and the integral becomes nearly zero. However there are
points where the oscillations stop. These points are located where the
variation of $f(x)$ slow down. Therefore phase become stationary at these
points, i.e. $f^{\prime }(x_c)=0.$ The set of points $\{x_c\}$ are called
set of stationary points. The idea is to find stationary points in the
domain of integration, including boundaries, and calculate their
contribution to the integral separately.

This is done by expanding $\exp \left( \frac{if(x)}\hbar \right) $ around
each of the stationary point $x_c$. Then equation (\ref{II-19}) can be
written as 
\begin{equation}
\label{II-20}F(\frac 1\hbar )=\int\limits_{-\infty }^\infty dx\,\,\exp
\left( \frac i\hbar \left[ f(x_c)+\frac{f^{\prime \prime }(x_c)}%
2(x-x_c)^2+\cdot \cdot \cdot \right] \right) . 
\end{equation}
In most of the examples $f^{\prime \prime }(x)\neq 0,$ and gives dominant
contribution hence, one can neglect cubic or higher order terms. If
integrand is analytically continued into the complex plane of $x$, one needs
to consider complex critical points. Integral can then be performed by a
steepest-descent method. Tunneling is one of the physical phenomenon in
which complex critical points are important.

By considering only quadratic terms equation (\ref{II-20}) can be
transformed into an imaginary Gaussian integral, which can be evaluated to
give 
\begin{equation}
\label{II-21}F(\frac 1\hbar )=\sqrt{\frac{2\pi \hbar i}{f^{\prime \prime
}(x_c)}}Exp\left( \frac{if(x_c)}\hbar \right) 
\end{equation}

\vskip .5in

Turning back to equations(\ref{II-17}) and (\ref{II-18}), the stationary
points (or paths) are solutions of equation%
$$
\delta R\left( {\bf q}(t)\right) =0, 
$$
which therefore are the classical trajectories satisfying boundary condition 
${\bf q}(0)={\bf q}_0,{\bf q}(t)={\bf q}$, as well as Newton's equations. We
can expand action $W$ around the critical path (i.e. classical orbit) ${\bf q%
}^c\left( t\right) $ 
$$
R[{\bf q}^c+\delta {\bf q}]=R[q^c]+\frac{\delta R}{\delta {\bf q}}[{\bf q}^c%
]\delta {\bf q+}\frac{\delta ^2R}{2!\delta {\bf q}^2}[{\bf q}^c]\delta {\bf q%
}\delta {\bf q}+{\bf .....} 
$$

It may be recalled that the type of extremum for the classical orbit and
hence linear stability of the orbit depends on the nature of second
variation. Thus the information about the stability of the classical orbit
enters into the semi-classical framework via second variation of the action
functional. The second variation of the action is given by 
\begin{equation}
\label{II-21a}\delta ^2R[f]=\int\limits_0^tdt\delta {\bf q}^{\top }\left(
t\right) \left( {\cal L}[f]\right) \delta {\bf q}(t);\,\,\,\,\,\,\,\,\,\,%
\delta {\bf q(}0{\bf )=}\delta {\bf q}(t)=0 
\end{equation}
where ${\cal L}=-\frac{d^2}{dt^2}\delta {\bf q-}\partial _{qq}^2V[f]\delta 
{\bf q.}$ A solution of $\delta ^2R[f]=0$ is a Jacobi field along the
classical path. The equation of motion for the variation under this
Lagrangian ${\cal L}$ is of the form of Jacobi-Hill equation. With
stipulated boundary conditions, we have a Sturm-Liouville problem for the
operator $\hat{{\sf D}}{\sf =}-\underline{1}\frac{d^2}{dt^2}{\bf -}\partial
_{qq}^2V$ , over the time interval $(0,t)$. The operator $\hat{{\sf D}}$ is
real symmetric so that it has real eigenvalues $\mu _n$ corresponding to
real eigenfunction ${\bf u}_n$, forming a complete basis on which it is
possible to expand any variation satisfying boundary conditions as 
\begin{equation}
\label{II-22}\delta {\bf q=}\sum\limits_{n=0}^wa_n{\bf u}_n 
\end{equation}
with $\hat{{\sf D}}{\sf \cdot }{\bf u}_n=\mu _n{\bf u}_n,\,\,\,\,{\bf u}%
_n(0)={\bf u(}t{\bf )=0\,\,,\,\,\,}$and $\int_0^tdt{\bf u}_m^{\top }(t)\cdot 
{\bf u}_n(t)=\delta _{nm}$. The second variation becomes diagonal in this
new basis. 
\begin{equation}
\label{II-23}\delta ^2R=\frac 12\sum\limits_{n=1}^\infty \mu _na_n^2 
\end{equation}
The nature of the classical path depends on the sign of the quadratic form
of the second variation, which in turn depends on the number of negative
eigenvalues of the operator $\hat{{\sf D}}$. In general variation of
eigenvalues is either negative or zero for solutions of equation (\ref{II-22}%
). Therefore with increasing time interval $\mu _n$ will decrease crossing
zero at time say $T_n.$ The equation $\hat{{\sf D}}{\sf \cdot \delta }{\bf q=%
}0$ admits a nontrivial solutions at this times satisfying {\it b.c. }$%
\delta {\bf q}(T_n)=0.$ At other times no such solution exist. The conjugate
points corresponds to the times $T=T_n.$

Both sides of equation (\ref{II-22}) are then differentiated with respect to 
$T$ and the integral (\ref{II-21a}) can be solved.

There exist several classical paths ${\bf q}_l\left( t\right) $ that goes
from ${\bf q}_0$ to ${\bf q}$ during time $t$, and each of these stationary
solution contribute to the propagator. From above comments we can write
equation (\ref{II-17}) as 
\begin{equation}
\label{II-24}K({\bf q\,\,\,q}_0\,:t)=\sum\limits_l\left( \frac m{2\pi i\hbar
\Delta t}\right) ^{\frac{Nf}2}\exp \left\{ \frac i\hbar R({\bf q}%
_l^c)\right\} \int d^{(N-1)f}[\delta {\bf q}]\exp \left\{ \frac i{2\hbar
}(\partial _{qq}^2R)\delta {\bf q}\delta {\bf q}\right\} . 
\end{equation}
The higher order terms are obtained in \cite{1-5a}. The matrix of the second
derivatives $\underline{D}=\partial _{qq}^2R[{\bf q]}$ is a $(N-1)f\times
(N-1)f$ matrix given by%
$$
\underline{D}=\frac 1{\Delta t}\left[ 
\begin{array}{cccccc}
A_1 & -1 & 0 & 0 & \cdots & 0 \\ 
-1 & A_2 & -1 & 0 & \cdots & 0 \\ 
0 & -1 & A_3 & -1 & \cdots & 0 \\ 
\vdots & \vdots & \vdots & \vdots & \ddots & \vdots \\ 
0 & 0 & 0 & \cdots & -1 & A_{N-1} 
\end{array}
\right] 
$$
where $A_n$ are $f\times f$ matrices given by $A_n=A_{\alpha \beta }(n\Delta
t)=2\delta _{\alpha \beta }-\Delta t^2\partial _{\alpha \beta }^2V$. In the
limit $\Delta t\rightarrow 0$ the matrix $\underline{D}$ is related to the
Jacobi-Hill operator $\hat{D}.$ The propagator can be obtained evaluating
all the Gaussian integrals and their moments. Finally one gets 
\begin{equation}
\label{II-25}K({\bf q\,\,\,q}_0\,:t)=\sum\limits_l\left( \frac m{2\pi i\hbar
}\right) ^{\frac f2}\left| \det _{ij}\frac{\partial ^2R_l}{\partial
q_0^i\partial q^j}\right| ^{1/2}\exp \left\{ \frac i\hbar R_l({\bf q}_0\,\,%
{\bf q};t)-\frac{i\pi \nu _l}2\right\} . 
\end{equation}
where $\nu _l$ is the Morse index.

As mentioned earlier a Fourier transform establishes the bridge between the
energy and time domain, the energy Green function can be obtained by taking
Fourier transform of equation (\ref{II-25}): 
\begin{equation}
\label{II-26}G({\bf q},{\bf q}_0)=\frac 1{i\hbar }\int\limits_0^\infty
dt\exp (\frac{iEt}\hbar )K({\bf q},{\bf q}_0;t). 
\end{equation}
This integral can again be evaluated using stationary phase approximation.
Here the stationary phase condition is $\partial _t[Et+R_l]=0$. If we define
the reduced action $S_l({\bf q}_0,{\bf q;}E)=Et+R_l({\bf q}_0,{\bf q};t({\bf %
q}_0,{\bf q};E))=\int_{q_0}^q{\bf p\cdot }d{\bf q}$ we obtain a quantity
independent of time, $\partial _tS_l=0$. This stationary phase condition
picks up all the classical trajectories $l$ from ${\bf q}_0$ to ${\bf q}$ at
given energy $E$. The following steps are same as above and as a result we
get 
\begin{equation}
\label{II-27}G({\bf q}_0,{\bf q;}E)=\frac 1{i\hbar (2\pi i\hbar
)^{(f-1)/2}}\sum_l\left| \det D\right| ^{1/2}\exp \left\{ \frac{iS_l({\bf q}%
_0,{\bf q;}E)}\hbar -\frac{i\nu _l\pi }2\right\} 
\end{equation}
where 
$$
D=\left[ 
\begin{array}{cc}
\frac{\partial ^2S}{\partial q_0\partial q} & \frac{\partial ^2S}{\partial
q_0\partial E} \\ \frac{\partial ^2S}{\partial E\partial q} & \frac{\partial
^2S}{\partial E^2} 
\end{array}
\right] . 
$$

Finally, to obtain semi-classical density of states as in equation(\ref
{II-10}), we take trace of above Green function. The contributions to the
trace mainly comes from two sources: (i)very short paths for which the
propagator is delta function \cite{1-bm}. In fact these are equilibrium
points of the system. (ii) periodic orbits of non-zero length. The
contribution of paths that are closed but not periodic, is negligible as a
result of destructive interference among themselves. The former leads to
Thomas-Fermi term for the average density of states 
\begin{equation}
\label{II-28}d_{av}(E)=\frac 1{\hbar ^f}\int d{\bf p}d{\bf q}\delta (E-H(%
{\bf p,q)).} 
\end{equation}
The second source leads to the oscillatory contribution for the density of
states, which in general form can be written as 
\begin{equation}
\label{II-29}d_{osc}(E)=\sum\limits_l\sum\limits_rA_{l,r}(E)\cos \left(
r\left( S_l(E)-\alpha _l\right) \right) 
\end{equation}
where $l,r$ denotes the primitive periodic orbits and their repetitions
respectively, and $S_l,\alpha _l$ denotes action and phase. The amplitudes $%
A_{l,r}$ depend on two aspects of the periodic orbit-whether they are stable
and whether they are isolated. The total density of states is sum of these
contributions, 
\begin{equation}
\label{II-30}d(E)=d_{av}(E)+d_{osc}(E). 
\end{equation}

\section{Density of States for Pseudo-integrable Billiards}

In the pseudo-integrable billiards almost all(in sense of Lebesgue measure)
periodic orbits are marginally stable and non-isolated (i.e. occur in
bands.). The density of states is given by \cite{1-rb} 
\begin{equation}
\label{III-1}d(E)=\frac m{2\pi \hbar ^2}Re\sum\limits_i\int \int d{\bf q}%
H_0^{(1)}\left( \frac{l_i}\hbar \sqrt{2mE}\right) \exp (i\alpha _i\pi ). 
\end{equation}
where $m$ is mass of the particle(billiard ball) and $l$ is length of the
periodic orbits. The index $\alpha _i$(half of Maslov indices $\nu $)
represents number of specular reflections at the boundary of the billiard.
And $H_0^{(1)}$ is a Bessel Function of third kind (Hankel function). Since
periodic orbits form continuous families on the invariant surface and each
orbit in the family has the same value of $l_i,$ independent of ${\bf q,}$
integration in (\ref{III-1}) is trivial. The oscillatory contribution to the
density of states in the asymptotic form ($\hbar \rightarrow 0$) is then
given by 
\begin{equation}
\label{3-2}d_{osc}(E)=\left( \frac m{2\pi ^2\hbar ^2}\right)
^{3/4}E^{-1/4}\sum\limits_i\frac{A_i}{l_i^{1/2}}\cos \left( \frac{l_i}\hbar 
\sqrt{2mE}+(\alpha _i-\frac 14)\right) , 
\end{equation}
where $A_i$ represents area of the bands of the periodic orbits. The
summation here is over all primitive periodic orbits and their repetitions.
The average contribution from zero length periodic orbit gives the Wyle area%
\cite{1-bf} contribution to the level density 
\begin{equation}
\label{3-3}
\begin{array}{c}
d_{av}(E)= 
\frac{mA_R}{2\pi \hbar ^2}Re(A_RH_0^{(1)}(0))+higher.order.terms. \\ =\frac{%
mA_R}{2\pi \hbar ^2}J_0(0)+h.o.t.=\frac{mA_R}{2\pi \hbar ^2}+h.o.t. 
\end{array}
\end{equation}

where, $A_R$ is configuration space area of the billiard and higher order
terms include corrections from boundary and corners of the configuration
space.

Integrated density of states or mode number $N(E)=Number\left\{
E_n<E\right\} =\sum_n^\infty \Theta (E-E_n)$ is then given by 
\begin{equation}
\label{III-4}N(E)=\int\limits_0^EdE^{\prime }\,\,\,d(E^{\prime }) 
\end{equation}

Once the complete information about classical variables such as $l,A$ is
obtained, equations (\ref{3-2}) and (\ref{3-3}) can be used to study various
spectral fluctuation properties in which we are interested.

\vskip .5in

In this chapter, we have attempted to give brief sketch of the developments
in semi-classical techniques taken place during last few decades. In
particular the problem begun by Selberg of finding semi-classical properties
of a quantum particle moving on a constant negative curvature Riemann
surface, has enjoyed a flurry of new activity in recent years. An extensive
development of concepts by Gutzwiller and other workers have resulted in
many applications in laser spectroscopy, Rydberg states of atoms or
molecules, electronic semiconductor devices etc. The success for the path
integral approach is in the fact that in all these systems periodic orbits
can be classified, their actions computed, and the path integrals can be
summed. Many techniques has been developed to carry out this task. The
Gutzwiller trace formula, however, gives only leading order terms in the $%
\hbar $ expansion. Higher order terms also has been computed recently by
Gaspard\cite{1-5a}.

\chapter{Billiards - Classical Dynamics}

\section{Introduction}

Billiards are dynamical systems corresponding to the motion of a point like
particle in a bounded domain $Q,$ which is a compact Riemannian manifold
with piecewise smooth boundary $\partial Q.$ We assume $Q\subset Q_0,$ $Q_0$
being a closed $C^\infty \,$Riemannian manifold. The boundary $\partial Q$
consist of a finite number of smooth compact $C^\infty $ sub-manifolds $%
\partial Q_1,\partial Q_2,\cdot \cdot \cdot \partial Q_r\,$ of co-dimension $%
1.$ The points of the boundary ${\bf q}{\cal \in }\partial Q$ are singular
of order $n$ if $\partial Q$ is only $C^n$ at ${\bf q}$. The points which
are not singular are called regular and represented by a set $\partial 
\widetilde{Q}=\partial Q\setminus \{{\bf q}\ {\rm of\ order}\ <\infty \}.$

The particle reflect from the boundary according to law of elastic
reflections. The motion between reflections corresponds to the geodesic flow 
$G_t,$ associated to a Hamiltonian $H_0$ with $Q_0$ as configuration space.
The reflection of the particle trajectories from a singular point of the
boundary is not well defined. Being of measure zero (in Lebesgue sense)
these trajectories are of little importance from the point of view of
ergodic theory. These trajectories may play important role in semi-classical
theories, rendering diffraction effects. Before turning to specific types of
billiards we are interested, few relevant terminologies, definitions are
given below.

\subsection{Phase space}

The phase space of the billiards is the set of all tangent vectors of fixed
length, $M=\{{\bf x}=({\bf q,v})|{\bf q}\in Q,{\bf v}\in S^{d-1}({\bf q)\}}$
where $d=\dim Q$ and $S^{d-1}({\bf q)}$ is the unit sphere of dimension $d-1$
over {\bf q}. The reflections of ${\bf x}\in \partial \widetilde{\widetilde{M%
}}=\{{\bf x}=({\bf q,v)}\in M|{\bf q}\in \partial \widetilde{\widetilde{Q}}%
\} $ where $\partial \widetilde{\widetilde{Q}}=\partial Q\backslash \{{\bf q}%
{\rm \ of\ order\,\,0\},}$ are defined by $R:({\bf q,v)\rightarrow }({\bf %
q,v-}2\left\langle {\bf n}_{{\bf q}}{\bf ,v}\right\rangle {\bf n}_{{\bf q}%
}). $ Here, ${\bf n}_{{\bf q}}$ is the unit inward normal to $\partial Q$ at 
${\bf q.}$ Let $\pi :M\rightarrow Q({\rm equivalently,\,\,}{\bf %
x=(q,v)\longmapsto q})$ be the natural projection on $Q.$ Since $\pi ^{-1}(%
{\bf q})={\bf q\otimes }S^{d-1}({\bf q})$, we have $\dim M=2d-1.$ For ${\bf %
x\in }M,$ the point ${\bf q}=\pi ({\bf x})$ is said to be carrier of ${\bf x}
$. $M$ possesses a natural involution sending each point ${\bf x=}({\bf q,v)}%
\in M$ into a point ${\bf x}^{\prime }=({\bf q,-v)}\in M.$

\begin{definition}
The measure $\mu $ on $M$ is defined as 
\begin{equation}
\label{2-1}d\mu =d\rho ({\bf q})d\omega ({\bf q})
\end{equation}
where $d\rho ({\bf q})$ is the element of volume generated by the Riemannian
metric into $Q$ and $d\omega ({\bf q})$ is the Lebesgue measure into $%
S^{d-1}({\bf q})\ni {\bf v.}$
\end{definition}

The flow $G_t$ of the billiard corresponds to a vector field $\{G_t({\bf x}),%
{\bf x}\in M_0\}$, where $G_t({\bf x})$ is a tangent vector to $M_0$ at $%
{\bf x}$. ($M_0$ being unit tangent bundle over $Q_0$. The flow $G_t$ thus
determines the motion of a particle with unit velocity along geodesic lines.
We define flow $\Phi _t$at the moment of a reflection from boundary $%
\partial Q$ as follows 
\begin{equation}
\label{2-2}\Phi _{\tau ({\bf x})}=\left\{ 
\begin{array}{cc}
G_{\tau ({\bf x})}, & {\rm if}\,\,\,\,\,{\bf x}\notin M_{-} \\ G_{\tau ({\bf %
x})}\circ R, & {\rm if\,\,\,\,\,}{\bf x}\in M_{-} 
\end{array}
\right. 
\end{equation}
where, $M_{-}=\{{\bf x}=({\bf q,v)}\in \partial \widetilde{\widetilde{M}}%
\mid \left\langle {\bf n}_{{\bf q}},{\bf v}\right\rangle \leq 0\}$ and $\tau
\left( {\bf x}\right) $ is the nearest strictly positive moment of a
boundary reflection of the trajectory issued from ${\bf x}$.

Let $N_{ij}$ be the set of all interior points ${\bf x\in }M$ such that the
segment of the geodesic line in the direction of ${\bf x}$ intersect $%
\partial Q$ on $\partial Q_i\cap \partial Q_j$. Denote by $N^{\left(
1\right) }$, the set of all points ${\bf x\in }M$ which will be contained in 
$\cup _{i\neq j}N_{ij}$ at some step of construction of the geodesic flow.
Denote by $N^{\left( 2\right) }$, a set all ${\bf x}$ for which process of
construction of a geodesic leads to an infinite number of reflection in
finite time. Then

\begin{definition}
If for almost every (in the sense of measure $\mu $) ${\bf x\in }M^{\prime
}=M\setminus (N^{\left( 1\right) }\cup N^{\left( 2\right) })$, we have a
geodesic segment of a finite length with end point located at a regular
point, the billiards are said to be proper.
\end{definition}

We will consider here proper billiards only. Dynamics of billiards is of
Hamiltonian nature. However, due to the reflections on the boundary its
dynamics cannot be completely described by a Hamiltonian, unlike a particle
moving under the influence of a conservative force. There are many examples
of dynamical system that can equivalently be described in a billiard. For
example system of two particles of masses $m_1,m_2$ moving in a unit
interval $[0,1],$ bouncing off from boundary as well as from each other
elastically is equivalent to a triangular billiard. Another example is of
three hard rods sliding along a frictionless ring and making elastic
collisions which is also equivalent to a billiard on triangular table\cite
{2.1}. One can also deform a Hamiltonian system mathematically into a
billiard.\cite{2.1a,2.1b,2.1c}

The billiards can exhibit all features of dynamical systems, from
integrability to chaotic behaviour. The behaviour of course depends on the
geometry of the billiard table. Billiards can be classified according to
their behavioural pattern as: 1) Hyperbolic billiards or dispersing
billiards, e.g., polygons with smooth obstacles\cite{2.2}, some billiards
with convex boundary\cite{2.3}, 2)Elliptical billiards with strictly
positive curvature convex tables\cite{2.4}. 3) Parabolic billiards, e.g.,
polygonal billiards. Here we shall concentrate on polygonal billiards in the
Euclidean plane.

\section{Polygonal billiards}

Let $P$ be closed, connected, non-self intersecting polygon in the Euclidean
plane, $P\subset {{\sf R\hspace*{-0.9ex}\rule{0.15ex}{1.5ex}\hspace*{0.9ex}}}%
^2$, whose boundary $\partial P$ consist of a finite number of line
segments(edges), we denote them in arranged order $\partial P_1,\partial
P_2\cdot \cdot \cdot \partial P_r$ (some times we will use notations $%
a_1,a_2..$etc. for convenience) such that $\partial P_{i\pm 1}$ are
neighbour of $\partial P_i$. The points $\partial P_i\cap \partial P_{i\pm
1} $ are vertices of $P$. We thus have a set of regular points $\partial 
\widetilde{P}=\partial \widetilde{\widetilde{P}}=\{{\bf q\in }\partial P|%
{\bf q}\notin \partial P_i\cap \partial P_{i\pm 1}\}$. It is obvious that $%
{\bf q\in }\partial P_i\cap \partial P_{i\pm 1}$ are zero order singular
points. We denote its phase space by $M\left( P\right) $. At each regular
point ${\bf q}\in \partial P_i$, the unit normal vector is same and shall be
denoted by ${\bf n}_i$.

Consider the one-parameter group $\{T^t\}$ of transformations on $M$ (for 
{\bf \linebreak}${\bf x=(q,v)\in M}^{\prime },-\infty <t<\infty $) defined
as follows: $T^t{\bf x}_o$ is obtained by starting at ${\bf q}_o$ and
drawing a continuous path inside $P$ consisting of straight line segments
and of total length $|t|$ and ending at ${\bf x}_t=({\bf q}_t,{\bf v}_t)$.
The straight line segments (except first and last) begin and end on $%
\partial P$ and is called as the link of trajectory a given trajectory. The
direction change at the boundary in passing from one link to the next is
made in accordance with laws of elastic reflections. This is done via a map $%
\sigma _i:S^1\longrightarrow S^1$ at each point ${\bf x=}({\bf q,v}),{\bf q}%
\in \partial P_i$ which acts according to the formula $\sigma _i{\bf v}={\bf %
v-}2\left\langle {\bf n}_i,{\bf v}\right\rangle {\bf n}_i$ If path hits
vertex we stop trajectory there. The set of the form $\pi (\{T^t{\bf x:}%
-\infty <t<\infty ,{\bf x}\in M^{\prime }\})$ represents configuration
trajectory of $P$. If trajectories hit vertices of $P\,$ both in the future
and in the past then trajectory is finite, we call it a {\it generalized
diagonal} of $P.$

The procedure of unfolding polygons can be applied now (see Fig. 2.1). The
configurational trajectory $\pi (\{T^t{\bf x}\})$ has vertices on $\partial
P_i$. Starting from any reference point $({\bf q}_0,{\bf v}_0)$ on the
trajectory, let trajectory has successive vertices on the boundary segments
of $a_1,a_2,a_3...$of $P$. Then reflections of $P$ with respect to these
faces transform broken line to the trajectory into a straight line
intersecting the polygons $P,P_{a_1},P_{a_1,a_2}.....$. Here, $%
P_{a_1,a_2...a_i}$ is the polygon obtained by reflecting $P$ with respect to
the sides $a_1,a_2,...a_i$. The velocity of the motion on the part of
trajectory after $k^{th}$ reflection is ${\bf v}_k=(\sigma _{a_k}\sigma
_{a_{k-1}}...\sigma _{a_1}){\bf v}_0$. Let $G_P$ be the sub-group of the
isometry group of the unit circle $S^1$ generated by the isometries $\sigma
_1\sigma _2...\sigma _r$. In the analysis of billiard dynamics, the
singularities produced by the vertices play a major role. The vertex angles
and the relative lengths of the edges are important characteristics of $P$
as far as dynamics is concerned. If all the angles of $P$ are commensurable
with $\pi $ then we call $P$ a {\it rational billiard}. The group $G_P$ is
finite for rational billiards.
\begin{figure}[htbp]
\begin{center}
{\epsfig{file=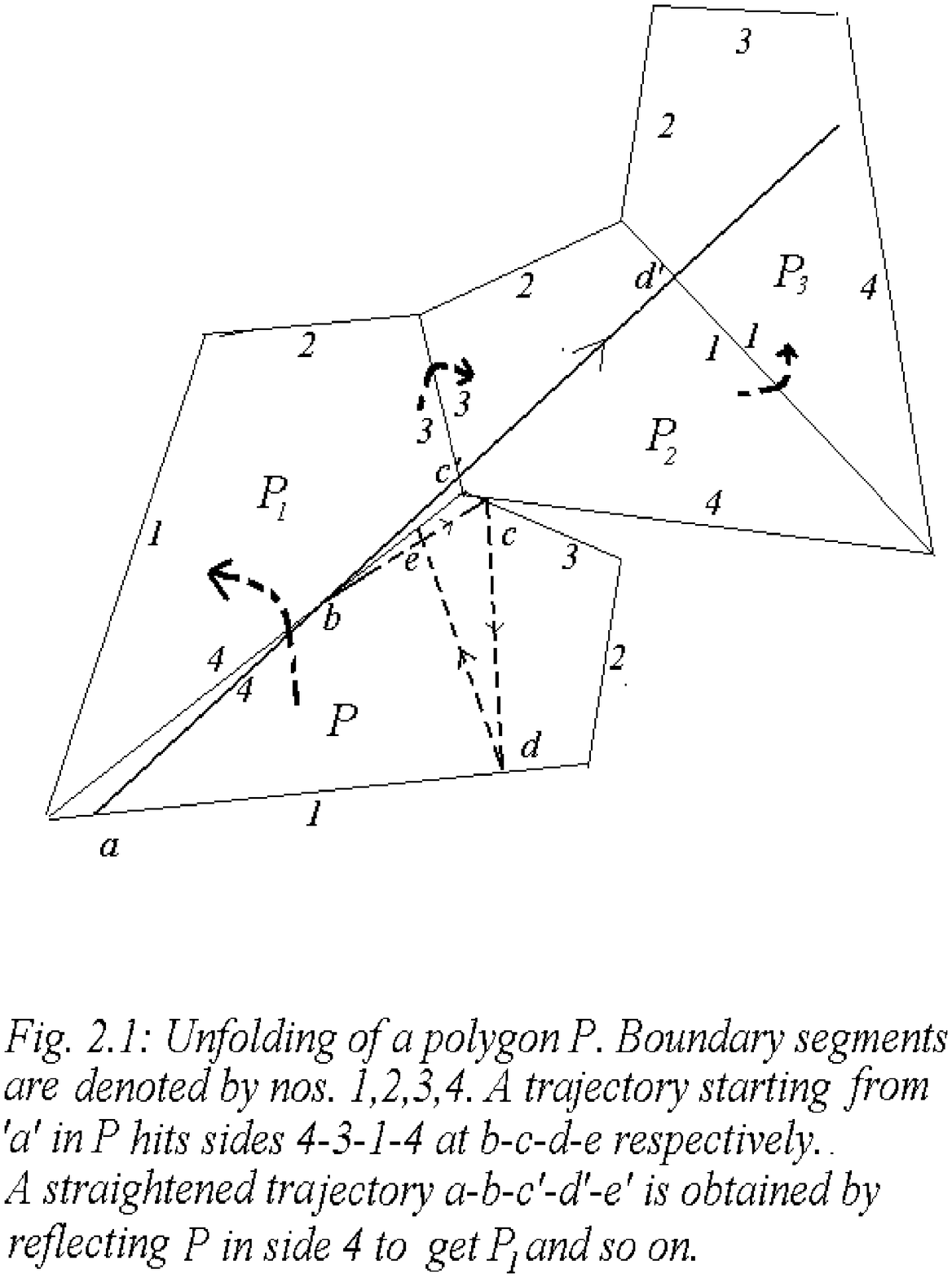,bbllx=80,bblly=170,bburx=560,bbury=700,height=6in}}
\end{center}
\end{figure}
The problem of billiard dynamics deals with behaviour of the billiard
trajectory. The question mainly falls into two categories, one concerning
statistics which belong to ergodic theory and other concerning the topology
of trajectories. Here we are more interested in the topological properties
then the ergodic properties of the polygonal billiards. Most of the
information we need for semi-classical study can be extracted from
techniques developed in these areas. As shown in the previous chapter
semi-classical properties of a system depends mainly on the periodic orbits.
We shall therefore simply state some well known facts about ergodic
properties of polygonal billiards. And discuss topological properties that
mainly concerns about periodic orbits of polygonal billiards.

The billiard in a typical polygon is ergodic \cite{2.5}. A prevailing
opinion in the mathematical community is that polygonal billiards are never
mixing, but this has not been established yet. However there is a conjecture
about existence of {\it weakly mixing }polygonal billiards\cite{2.5a} which
is stronger property than ergodicity but weaker than mixing. The rational
polygonal billiards are however, proved to be not mixing . Some of the
important facts relevant to our work are presented below.

\begin{proposition}
The set of generalized diagonals of $P$ is countable.\cite{2.6}
\end{proposition}

{\it Proof: }Let $X$ be the set of polygons obtained by unfolding along all
billiard trajectories in $P$. Then $X$ belongs to the set of polygons $%
\{gP:g\in G_P\}$ which is countable because the group $G_P$ has finite
number of generators and hence countable. This implies that set of pairs of
vertices of any two polygons in $X$ is also countable.$\bullet $

\begin{theorem}
For any ${\bf q}\in P\subset {{\sf R\hspace*{-0.9ex}\rule{0.15ex}{1.5ex}%
\hspace*{0.9ex}}}^2$ and almost all (w.r.t. Lebesgue measure) ${\bf v}\in S^1
$ the closer of the configurational trajectory of the point ${\bf x}=({\bf %
q,v})$ with respect to the billiards flow $\{T^t\}$ contains at least one
vertex of the polygon.
\end{theorem}

Let $\gamma $ be periodic trajectory (of length ${\it L}$)in $P$ with $m$
links $\gamma _0,\gamma _1,\cdot \cdot \cdot \gamma _{m-1}$ such that $%
\gamma _0$ is the reflection of $\gamma _{m-1}$. Unfolding of $\gamma $ then
results in set of polygons $P_0=P,P_1,\cdot \cdot \cdot P_{m-1},P_m$. Choose
a link $\gamma _0$ and let $l$ be the line through $\gamma _0$. The element $%
g\in G_P$ that moves $P$ into $P_m$, also moves $\gamma _0$ into $\widetilde{%
\gamma }_m$ ({\it tilde }here represents link in the unfolded polygon that
corresponds to the link in original polygon) which belongs to the same line $%
l$ as $\gamma _0$. For example in Fig 2.1, $\gamma $ is $abcd..$, links $%
\gamma _0,\gamma _1...$ are simply segments $ab,bc...$ so on and links in
the unfolded polygons are $bc^{\prime },c^{\prime }d^{\prime }$ etc.

Thus $g$ preserves the line $l$, hence $g$ is either a parallel translation
along $l$ preserving orientation, or a sliding reflection with axis $l$
reversing the orientation. Since $gP=P_m$, $g$ is the product of $m$
reflections, $m$ is even if $g\,$ preserves orientation and odd if reverses
orientation.

\begin{corollary}
If $m$ is even, $\gamma $ extends to a band of periodic trajectories of
length ${\it L}$ parallel to $\gamma $. Both boundaries are unions of the
generalized diagonals. If $m$ is odd, every trajectory $\gamma ^{\prime }$
starting close to $\gamma $ and parallel to it comes back after $m$
reflections to the same edge, at the same distance from $\gamma $ and in the
same direction but on the opposite side of $\gamma $.
\end{corollary}

{\it Proof: }It is convenient to denote $\gamma $ by $\gamma _0$. unfolding $%
P$ along $\gamma _0$ we obtain: 1)the sequence $P=P_0,P_1,\cdot \cdot \cdot
,P_m,$ 2)the line $l_0$ and 3) the motion $g$ such that $P_m=gP$. Let the
point $x_0$ on the edge $a$ of $P$ be the starting point of $\gamma _0$ and
let $y_0=gx_0$ be the corresponding point on the edge $b=ga$ of $P_m$. The
periodicity of $\gamma _0$ implies that $l_0$ goes from $x_0$ to $y_0$.

Let $\gamma $ be the trajectory starting at $x\in a$ close to $x_0$ and
parallel to $\gamma _0$. If $\gamma $ is closes enough to $\gamma _0$,
unfolding along $\gamma $ we obtain the line $l$ through $x$ parallel to $%
l_0 $ and passing by the same sequence $P=P_0,P_1,\cdot \cdot \cdot ,P_{m-1}$
of polygons. Let $l$ intersect $b$ at $y$.

If $\gamma _0$ is even, by previous discussion, the quadrangle formed by $%
a,b,l_0,$ and $l$ is a parallelogram, thus $y=gx$, i.e. $\gamma $ comes back
to x in the same direction. Hence,$\gamma $ is periodic. Now start moving $%
\gamma _0$ to the right parallel to itself. We then get $\gamma
_t,\,\,\,t\geq 0$. For small $t$, $\gamma _t$ is periodic of the same length
as $\gamma _0$. Increasing $t$ we come to the moment $t=t_1$ such that $%
\gamma _1=\gamma _{t_1}$, hits the vertex of $P$. Since $\gamma _{t_1}$ is
the limit of the periodic trajectory, it is also a generalized diagonal or
union of such. The same argument works for $t\leq 0$.

When $\gamma _0$ is odd, the quadrangle above is a trapezoid and the point $%
x_1=g^{-1}y$ on the side a where $\gamma $ returns after $m$ reflections is
symmetric to $x$ with respect to $x_0$.$\bullet $

\begin{theorem}
In any rational polygonal billiard there is at most a finite number of prime
periodic orbits of odd periods.
\end{theorem}

Consider convex rational polygon $P$ with $r$ vertices. Its interior angles $%
\alpha _1,\alpha _2,\cdot \cdot \cdot ,\alpha _r$ being commensurable with $%
\pi $, can be written in form $\alpha _i=(k_i/n)\pi $, where $n\geq 1$ and
the greatest common divisor of the numbers $n,k_{1,}k_2,\cdot \cdot \cdot
k_r $ is $1$. Then we have following lemma for group $G_P$.

\begin{lemma}
The group $G_P$ is isomorphic to the symmetry group of a regular polygon
with $n$ vertices.
\end{lemma}

{\it Proof:} Note that in the velocity space $S^1$ the composition of the
symmetries corresponding to the sides bounding the angle $\alpha _i$ is a
rotation $R_{2\alpha _i}$ of the circle $S^1$ by an angle $2\alpha _i=\left(
2k_i/n\right) \pi $. Hence, for any family of integers $s_0,s_1,\cdot \cdot
\cdot ,s_r,$ the group $G_P$ contains the rotation by an angle $\alpha
=\left( s_0n+s_1k_1\cdot \cdot \cdot +s_rk_r\right) \left( 2\pi /n\right) $.
But it follows from our assumptions that there exist a family $s_0,s_1,\cdot
\cdot \cdot ,s_r,$ for which $s_0n+s_1k_1\cdot \cdot \cdot +s_rk_r=1$. Hence
the group $G_P$ contains the rotation $R_{2\pi /n}$ and all rotations by
angles which are multiples of $2\pi /n$. Moreover, the group $G_P$ contains $%
n$ symmetries with respect to the axes of the form $R_{2\pi /n}l,k=0,1,\cdot
\cdot \cdot ,n-1$, where $l$ is any of the axes of symmetry of $\sigma
_1,\sigma _2,\cdot \cdot \cdot ,\sigma _r$. The transformations indicated
above generate the symmetry group of a regular polygon of $n$ sides.$\bullet $

The group $G_P$ is thus isomorphic to the {\it dihedral} group $D_n$. The
group $G_P$ therefore has $2n$ elements and the circle $0\leq \theta \leq
2\pi $ is divided by the action of $G_P$ into $2n$ intervals $\pi
(i-1)/n\leq \theta \leq \pi i/n,\,\,\,i=1,\cdot \cdot \cdot ,2n$. Every $%
\theta $, $0\leq \theta \leq 2\pi $ is equivalent by action of $G_P$ to a
unique $\theta _1,\,\,\,\,\,0\leq \theta \leq \pi /n,\,$ so that set of
orbits of $G_P$ is parametrized by $[0,\pi /n]$. The orbit of every $\theta $%
, $0\leq \theta \leq \pi /n$ has $2n$ elements, the orbits of $\theta =0$
and $\theta =\pi /n$ have $n$ elements each.

\section{Invariant Surface of Polygonal Billiards}

\begin{theorem}
If $G_P$ is a finite group, then the billiards in the polygon $P$ are not
ergodic. Moreover, to each orbit of the natural action of the group $G_P$ on 
$S^1$ (i.e., to the set $\Omega =\Omega ({\bf v}_0)=\{g{\bf v}_0\in S^1:g\in
G_P\}$), corresponds to the set $A_\Omega $, invariant with respect to $%
\{T^t\}$, consisting of all ${\bf x=(q,v})\in M$ such that ${\bf v}\in
\Omega $. \cite{2-sin}
\end{theorem}

{\it Proof: }Suppose ${\bf x}=({\bf q,v})\in A_\Omega ,{\rm i.e.},{\bf v}=g_0%
{\bf v}_0,\,\,\,g_0\in G_P$. Then any $t\in {{\sf R\hspace*{-0.9ex}%
\rule{0.15ex}{1.5ex}\hspace*{0.9ex}}},$ we have $({\bf q}_t{\bf ,v}_t)=T^t%
{\bf x}\in A_\Omega .$ Since the group $G_P$ is finite we can find a set of
orbits of the group $G_P:C\subset S^1/G_P$ whose measure differs from zero
or one. The set \thinspace $A=\cup _{\Omega \in C}A_\Omega $ is invariant
w.r.t. $\{T^t\}$ and $\mu (A)$ differs from zero or one.$\bullet $

This theorem states that for a finite group $G_P,$ only a finite number of
directions may be obtained when we move along billiards trajectories from
the given initial direction.

It may be noted that any vector ${\bf v}\in S^1$ can equivalently expressed
in terms of angle $\theta $. From the discussion of previous section for any 
$\theta ,\,\,\,\,0\leq \theta \leq \pi /n,$ the set $R_\theta \equiv
A_\Omega $ is invariant under the billiard flow. The invariant surface $%
R_\theta $ are {\it level surfaces }of the function $\psi =M(P)\rightarrow
[0,\pi /n]$, defined as $\psi (x,\eta )=\theta $, where $0\leq \theta \leq
\pi /n$ and $\eta =g\theta $\cite{2.6,2-fox,2-keane,2-rb}. The function $%
\psi $ is independent of the Hamiltonian $H$ that gives rise to the flow of
the billiard and is a constant of motion.

\subsection{Construction of the invariant surface}

The invariant surfaces $R_\theta $ for $0\leq \theta \leq \pi /n$ are
isomorphic to a surface $R$ which can be geometrically constructed the
polygon $P$ (see Fig. 2.2 for an example). To construct $R$ which is made up
of $2n$ copies of $P,$ say $P_1,P_2\cdot \cdot \cdot ,P_{2n}$, choose $%
\theta _1,\,\,0\leq \theta _1\leq \pi /n$ and denote $\theta _2,\theta
_3,\cdot \cdot \cdot \theta _{2n}$ the elements of the $G_P$-orbit of $%
\theta _1$ in the natural order. Now take polygon $P_1$ remove it from list
of $2n$ copies of $P_1$ and consider any trajectory in direction $\theta _1$%
, it hits a side say $a_{ij}$ of $P_1$. Reflect $P_1$ on the plane about $%
a_{ij}$, join the reflected $P_1$(say $P_{1^{^{\prime }}}$) to the figure
under construction (i.e. $R$) resulting in a combined figure $%
P_1+P_{1^{^{\prime }}}=P_1^{^{\prime }}$ and remove a copy of $%
P_{1^{^{\prime }}}$ from set of $2n$ copies of $P$. Repeat the whole process
with $P_1^{^{\prime }}$, for trajectory in the same direction $\theta _1$
but not necessarily along the same line. Each time trajectory meets a side $%
a_{ij}$, carry out reflection. If $P_{i^{\prime }}$ is still in the list of $%
2n$ polygons join it to the figure under construction and remove it from the
list. If $P_{i^{\prime }}$ is not in the list then identify side $a_{ij}$
and corresponding side of $P_{i^{\prime }}$ in the $P_i^{^{\prime }}$ which
are already drawn and transfer a trajectory to $P_{i^{\prime }}$. Repeat the
procedure until all the $2n$ polygons are exhausted from the list, and
resulting in a polygonal surface $R$. In the case of overlapping between $%
P_i^{^{\prime }}$ we consider them belonging to the different copies of the
plane. There are many possible surfaces $R$ for a given $P$.
\begin{figure}[htbp]
\begin{center}
{\epsfig{file=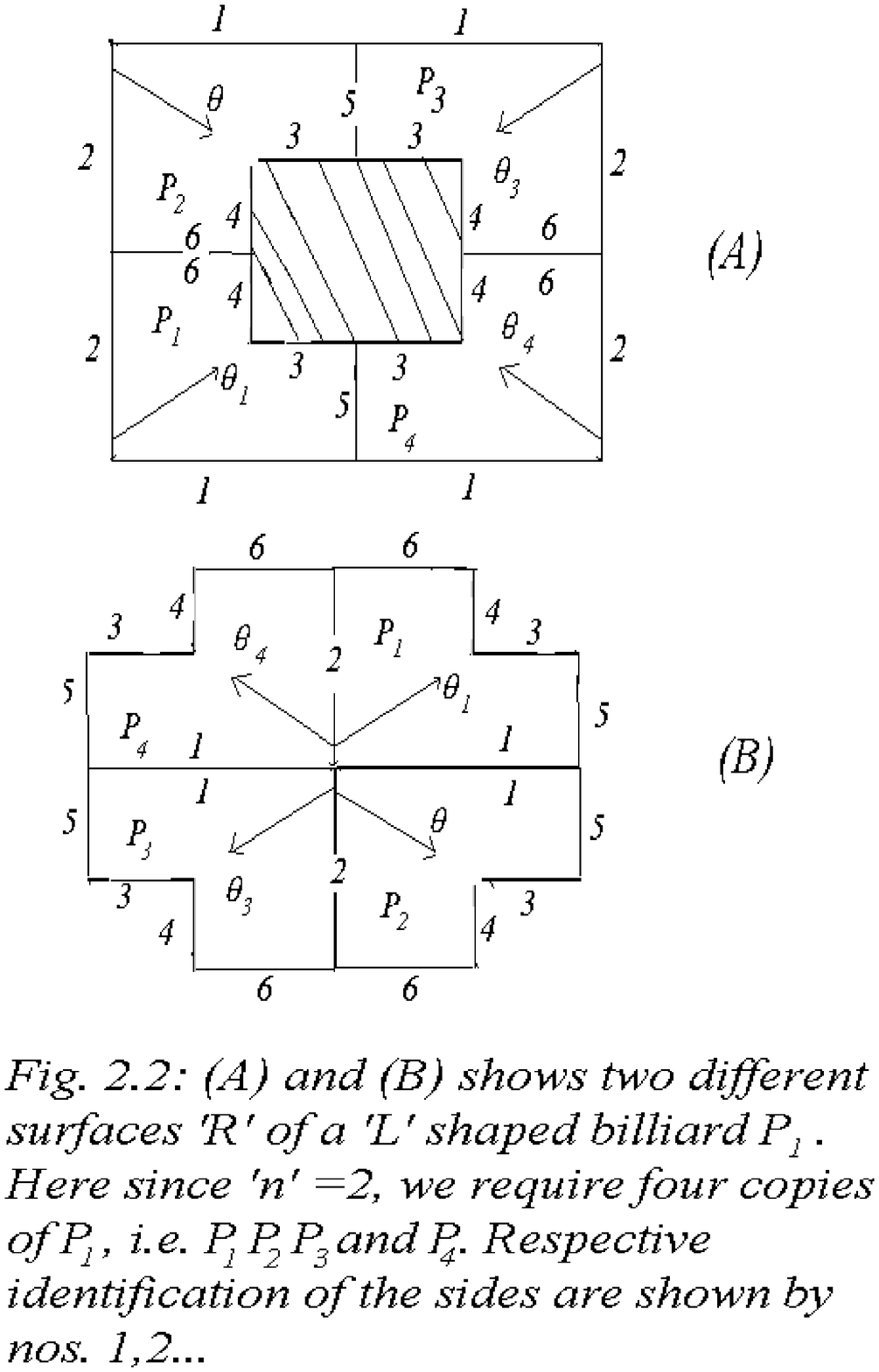,bbllx=80,bblly=170,bburx=560,bbury=700,height=6in}}
\end{center}
\end{figure}
The polygon $R$ has an even number of sides and they are divided in pairs,
each side of any pair differs from the other one by a parallel translation.
The billiard flow have simple realization on $R$. Consider a trajectory in
any direction $\theta $, suppose it meets boundary segment $a$ of $R$. Since
sides of $R$ comes in a pair, let $b$ be pair of $a$. The trajectory meeting 
$a$ gets instantly transferred to $b$ by the parallel translation that
identifies $a$ and $b$. The trajectory then starts anew in the same
direction $\theta $ from the side $b$.

If each identified pair glued together one gets the invariant surface $R$(
we keep here same notation $R$ for simplicity reason). It is therefore clear
that $R$ is a closed surface without any boundary. It can also be shown that 
$R$ is orientable. The gluing or identifying sides preserve the orientation
of the side.

\subsection{Topology of the invariant Surface}

The topological type of a closed orientable surface $R$ is determined by its
genus $g(R)$. The surface of genus $g$ looks like a pretzel with $g$ holes
or equivalently a sphere with $g$ handles.

Again, let $\pi m_i/n_i,\,\,\,i=1,...,r,$ be the vertex angles between the
sides of $P$. From discussion above one can see that each vertex $a_i$ of
\thinspace $P$ with angle $\pi m_i/n_i$ gives rise to $n/n_i$ singular
points $a_{ij}$ of the flow on $R$. Each $a_{ij}$ has $2m_i$ equally spaced
throngs and the flow lines in a small neighbourhood of such singular point
are shown in Fig.(2.3). Singular points of this type are called
multisaddles, the flow lines coming into and going from the singular point
are called the incoming and the outgoing separatrices. Thus a vertex $a_i$
gives rise to $n/n_i$ multisaddles on $R$ with $m_i$ number of incoming and
outgoing separatrices each. Varying $\theta $ does not change the position
of multisaddles $a_{ij}$ but uniformly rotates the separatrices around $%
a_{ij}$.
\begin{figure}[htbp]
\begin{center}
{\epsfig{file=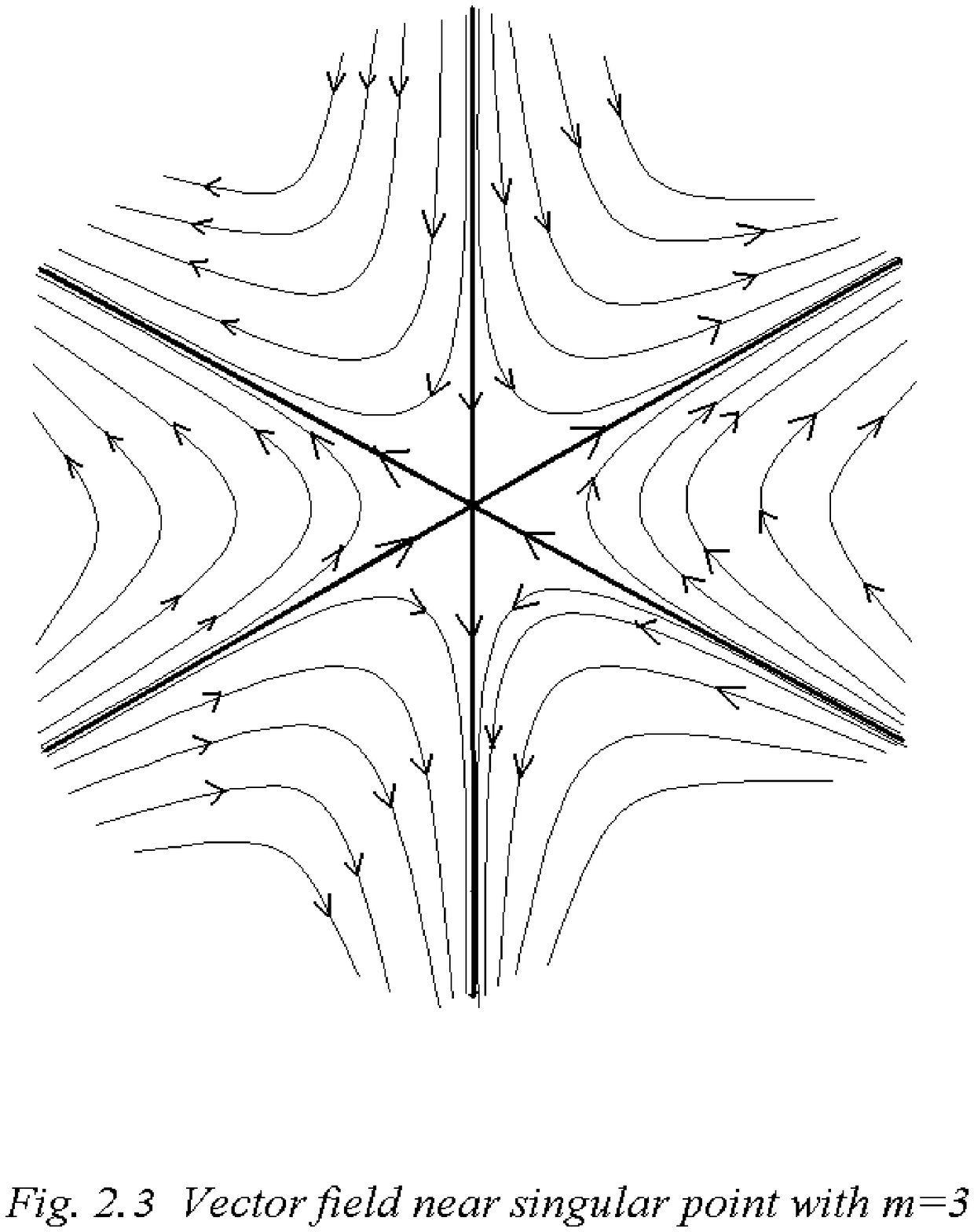,bbllx=80,bblly=170,bburx=560,bbury=700,height=6in}}
\end{center}
\end{figure}
The index of a multisaddle $a$ with $2m$ separatrices is equal to $m-1$. The
index formula for the Euler number gives 
\begin{equation}
\label{2-3}\chi (R)=\sum_{i=1}^r\frac n{n_i}(1-m_i)=2-2g(r) 
\end{equation}
.and hence the genus of the surface $R$ is 
\begin{equation}
\label{2-4}g(R)=1+\frac n2\sum_{i=1}^r\frac{m_i-1}{n_i}. 
\end{equation}
Thus the topology of $R$ is thus determined by the angles of $P$. The
surfaces $R_0,R_{\pi /n}$ are called exceptional invariant surfaces. Their
topology, is not determined by the angles of $P$\cite{2-kat}.

The topology of the invariant surface determines whether the system is
completely integrable or not. We will now discuss this in a dynamical sense.

\section{Integrability and beyond:}

The dynamics of classical Hamiltonian system is completely described by the
Hamilton's equations of motion%
$$
\dot q_i=\frac{\partial H}{\partial p_i};\,\,\,\,\,\,\,\,\,\,\,{\rm and}%
\,\,\,\,\,\,\,\,\,\,\,\dot p_i=-\frac{\partial H}{\partial q_i} 
$$
where symbols have their usual meaning i.e. $H$ is the Hamiltonian and $%
(q_i,p_i)$ are the canonical coordinate-momentum pair.

Consider a system of $f$ degrees of freedom with the Hamiltonian $H.$ The
Hamiltonian induces the flow $\Phi _t$ on the phase space $M$. The dynamical
system is integrable (in the sense of Liouville)\cite{2-arn,2-arn1} if there
exists a canonical transformation such that the new momenta are the
constants of motion. A necessary and sufficient condition for this
transformation to exist is that it should be possible to find $f$ analytic
functions $\left\{ F_i\right\} _{i=1...f}$ on $M$ satisfying following
conditions :

\begin{itemize}
\item  The Poisson bracket, $\left\{ F_i,H\right\} =0$ for all i, which
implies $F_i(\Phi _tx)=F_i(x)$ for all $x\in M,$

\item  all $F$ are in involution with each other i.e. $\left\{
F_i,F_j\right\} =0\,$ for all $i,j=1...f$,

\item  all $grad(F_i)$ are linearly independent.
\end{itemize}

Here the Poisson bracket $\left\{ A,B\right\} =\sum\limits_{i=1}^f(\frac{%
\partial A}{\partial q_i}\frac{\partial B}{\partial p_i}-\frac{\partial A}{%
\partial p_i}\frac{\partial B}{\partial q_i})$, and $grad=(\nabla _q,\nabla
_p)$. One can then define a level set of the functions $\left\{ F_i\right\} $
as $M_k=\left\{ x\in M|F_i(x)=k_{i,}\,\,\,\,i=1...f\right\} $. This level
set is invariant under flow $\Phi _t$. furthermore if $M_k$ is compact and
connected ,then it is diffiomorphic to the $f-$dimensional torus\cite{2-arn1}%
. The vector field $\left\{ {\bf V}_i\right\} ;\,\,\,{\bf V}_i=\left( \nabla
_pF_i,-\nabla _qF_i\right) $ is linearly independent, tangent to $M_k$, and
commutating. According to the Noether's theorem \cite{2-noe,2-reich} these
constants of motion (or isolating integrals) result from the symmetries of
the dynamical system.

Polygonal billiards are examples of dynamical systems where there exist one
or two constants of motion depending upon the angles and ratio of the sides
of the polygon. For some rational polygons with a rational ratio of their
sides, there exist two constants of motion satisfying all integrability
conditions except one i.e. $\left\{ F_i,F_j\right\} \neq 0$ at the vertices
of the polygon. Thus the vector field ${\bf V}$ becomes singular at
countable number of points in the phase space. Hence it is not possible to
obtain global action-angle variables or constants of motions. Following
Richens and Berry\cite{2-rb}, we call these systems a pseudo-integrable.

\section{Birkhof-Poincar\'e Maps and Interval Exchange Maps}

The Birkhof-Poincar\'e map is first return map for billiards. For simplicity
let $P$ be simply connected billiard. We enumerate the vertices of $P$
counterclockwise $A_o,...A_{n-1},A_n=A_0$. Denote by $%
a_i=[A_{i-1},A_i],i=1,....,n$ the edges of $P$. The boundary $\partial P$
with length coordinate $x$ is isomorphic to the circle of perimeter equal to
the length($L$) of $\partial P$. The set $\Omega \subset $ $M(P)$ of vectors
with footpoints in $\partial P$ can be parameterized by coordinates $0\leq
x\leq L$ and $-\pi /2\leq \theta \leq \pi /2$. The vector $(x,\theta )$ has
foot points in $x\in a_i\subset \partial P$ and angle $\theta $ between it
and the inner normal to $a.$ Set $L_k=\left| a_1\right| +\cdot \cdot \cdot
+\left| a_k\right| ,\,\,\,k=0,...,n$. The coordinates $\theta $ in $%
(x,\theta )$ is not well defined for $x=L_k$ because these $x\,$ are
vertices of $P$ and the angle $\theta $ can be measured with respect to any
of the two normals. The set $\Omega $ with points excluded is isomorphic to
the cylinder $[0,L)\times [-\pi /2,\pi /2]$ with deleted intervals $%
L_0\times [-\pi /2,\pi /2],.....,L_{n-1}\times [-\pi /2,\pi /2]$ (see Fig.
2.4).
\begin{figure}[htbp]
\begin{center}
{\epsfig{file=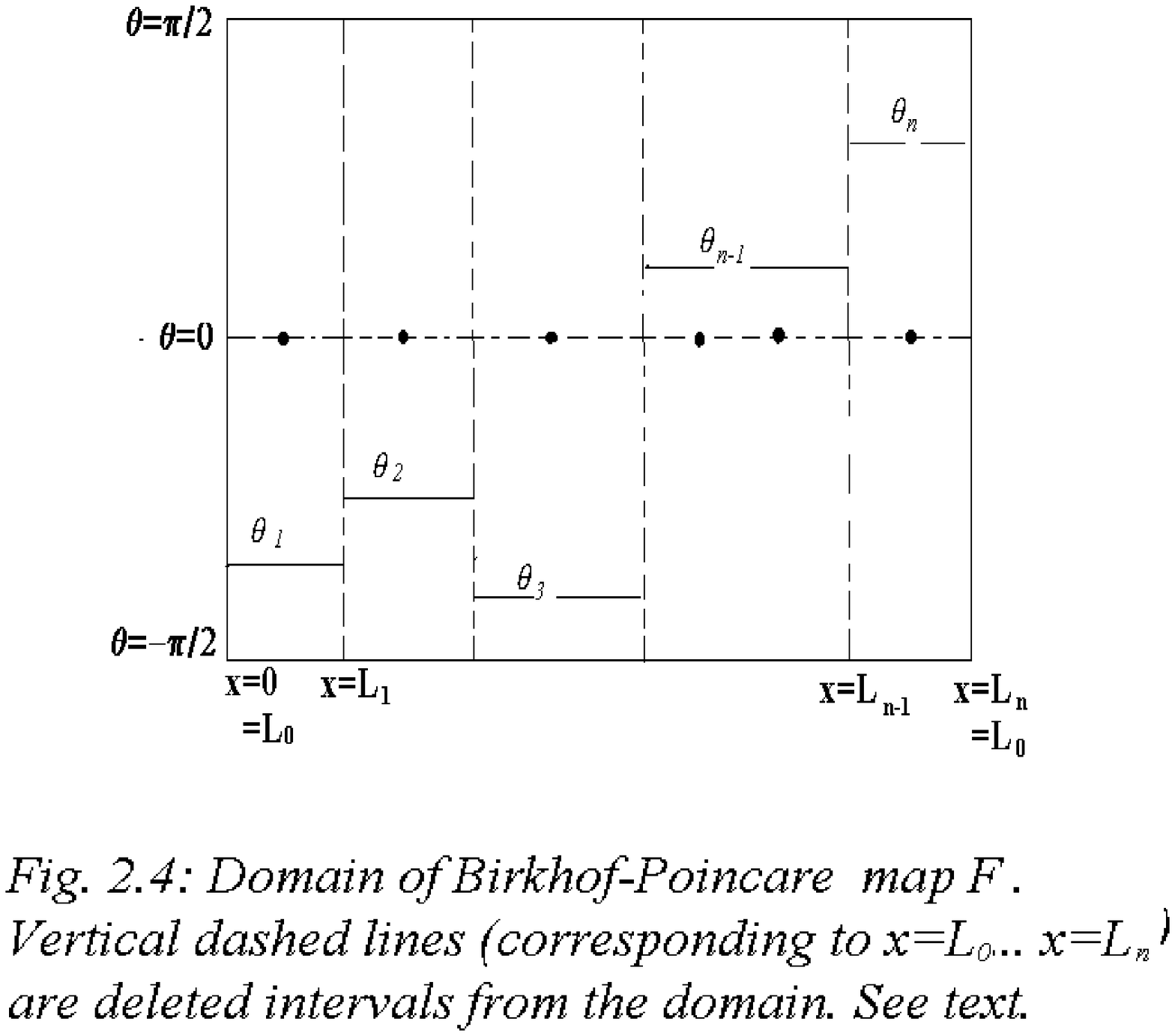,bbllx=80,bblly=170,bburx=560,bbury=700,height=6in}}
\end{center}
\end{figure}
The Birkhof-Poincar\'e map $F:\Omega \rightarrow \Omega $ is defined as
follows: consider a trajectory from $x$ in direction $\theta ,$ $(x,\theta
)\in \Omega $. When trajectory hits $\partial P$ the first time and bounces
off, it determines another point $(y,\eta )=F(x,\theta )$ to $\Omega .$ The
mapping $F$ is not well defined on the deleted intervals $L_0\times [-\pi
/2,\pi /2],.....,L_{n-1}\times [-\pi /2,\pi /2]$ as well as on the
boundaries $\theta =\pm \pi /2$. Fix an edge $a_i=[A_{i-1},A_i]$ and a
vertex $A_{j,}\,\,j\neq i-1,i$. Points $(x,\theta ),\,\,\,\,x\in a_i,\,$
such that the ball goes to the corner A$_j$ form a curve in $\Omega $ on
which map $F$ is not well defined. Each rectangle of $\Omega $ is divided by
these curves into the domains of continuity of $F$. Thus, the set of
discontinuities of $F$ is the union of a finite number of curves in $\Omega $%
, $F$ is obviously invertible and $F^{-1}$ is the Birkhof-Poincar\'e map for
the billiards with time reversed. The $F$ -invariant Lebesgue measure is $%
\cos \theta d\theta dx$. The properties of the billiard flow are easily
translated into the properties of the mapping $F$.

\subsection{Interval exchange Map For Rational Billiards}

Since the billiard flow $G^t$ decomposes into the family $G_\theta ^t$ of
flows, $0\leq \theta \leq \pi /N,$ the Birkhof-Poincar\'e map $F$ also
decompose into the one parameter family $F_\theta :\Omega _\theta
\rightarrow \Omega _\theta $ mappings, where $\Omega _\theta =\Omega \cap
R_\theta $ is set of vectors with foot points on $\partial P$ with
directions equivalent to $\theta $ and $F_\theta =F\mid \Omega _\theta $.
The family $\{F_\theta \}$ is the family of interval exchanges \cite
{2-bold,2-sin,2-kh}.

Suppose the space $I$ is semi-interval $[0,1)$ and $\xi =(I_1....I_r)$ is a
partition of $I$ into $r\geq 2$ disjoint semi-intervals $%
I_1=[0,d_1),....,I_r=[d_{r-1},1),\,\,\,0<d_1<...<d_{r-1}<1$. Let $\omega
=(\omega _1,...\omega _r)$ be permutation of the number $\{1,...r\}$.

\begin{definition}
Suppose the transformation $T:I\rightarrow I$ is a translation $T_{\alpha
_i}x=x+\alpha _i({\rm mod}\,\,\,1)$on each of the semi-intervals $I_i$(the
number $\alpha _i$ depends on $i$) and exchanges the semi-intervals
according to permutations $\omega $ i.e. The semi-intervals $%
T\,\,\,I_i=T_{\alpha _i}I_i=I_i^{^{\prime }}$ adhere to each other in the
order $I_{\omega _1}^{^{\prime }}...I_{\omega _r}^{^{\prime }}:$ then $T$ is
said to be interval exchange transformation corresponding to the partition $%
\xi \,$ and the permutation $\omega $.
\end{definition}

Thus if $I$ is piece of wire then the transformation $T$ essentially is the
cutting of $I$ into pieces $I_1....I_r$, rearranging them according to
permutation $\omega $ and welding them together again. Identifying $I$ with
the circle we can think of $T$ as an interval exchange on the circle. An
exchange of two intervals is given be one number $0<\alpha <1\,$ and it is
simply rotation by angle $\alpha $. It is clear that interval exchanges are
invertible transformations of $I$ preserving the Lebesgue measure $\rho $
and the numbers $\alpha _1....\alpha _r$ are well defined (mod $1\,$ ) by
the pair $(\xi ,\omega ).$

Now to see that the family $\{F_\theta \}$ is the family of interval
exchanges recall that in coordinates $(x,\theta )$ the set $\Omega $ is $%
[0,L)\times (-\pi /2,\pi /2)$ and $F(x,\theta )=(y,\eta )$ where $\eta $
locally depends only on $\theta $. The map $F$ perseveres $dm=\sin \theta dx$
which is interpreted as the mass element carried by the flow in direction $%
\theta $. Fix a direction $\theta $ and for each side $a_i$ of $P$ let $-\pi
/2<\theta _{i_1}<...<\theta _{i_N}<\pi /2$ be the directions $D_N$%
-equivalent to $\theta $ (they depend on $i\,$ because $a_i$ determines
angle of reference). The set $\Omega _\theta $ is the union $a_1\times
\left\{ \theta _{1_1}...\theta _{1_N}\right\} \cup ...\cup a_p\times \left\{
\theta _{p_1}...\theta _{p_N}\right\} $ of horizontal intervals. The mapping 
$F$ preserves the set $\Omega _\theta $ which is union of $pN$ horizontal
intervals (they can be glued into one interval) and the length element $dm$
on $\Omega _\theta .$ The positive orientation of $P$ induces orientation of 
$\Omega _\theta $ (each interval oriented from left to right), the map $F$
reverses the orientation. Thus, the restriction $F_\theta $ of $F$ to $%
\Omega _\theta $ is an interval exchange with the flipping of intervals.
Multiplying $F_\theta $ by the trivial orientation reversing map $%
J:(x,\theta )\rightarrow (L-x,\theta )$ we obtain honest interval exchange.
Dividing by total length $m_\theta $ of $\Omega _\theta $ which is 
$$
m_\theta =\left| a_1\right| (\sin \theta _{1_1}+...+\sin \theta
_{1_N})+...+\left| a_p\right| (\sin \theta _{p_1}+...+\sin \theta _{p_N}) 
$$
we normalize $F_\theta $ to an interval exchange on $[0,1)$ (with flipping).
For a fixed polygon $P$ the parameter of $F_\theta $ i.e. the number of
exchanged intervals, the permutation $\omega $ and the length of interval
depends only on $\theta .$ The obvious upperbound on the number of exchanged
intervals is $p^2N.$

The interval exchange maps has been used to prove many statistical and
topological properties of polygonal billiards\cite{2-sin,2-kh}. Our main
interest is in enumeration, classification of the periodic orbits. In the
next chapter we will use modified form of interval exchange to carry out
this task in case of some specific examples.

\chapter{Periodic Orbits in Some Pseudo-integrable Billiards}

\section{Introduction}

In this chapter we take up first task to study spectral fluctuations of
pseudo-integrable billiards within semi-classical framework, i.e. to
enumerate and classify all the periodic orbits of some typical
pseudo-integrable billiards and also define their respective actions. We
shall consider $\pi /3$-rhombus billiard for this purpose. This specific
example serves as a paradigm model as there are hardly any general results
known in the literature. Furthermore the methodology we develop here, being
based on interval exchange transformation, may be used to enumerate and
classify periodic orbits in many rational polygonal billiards.

\section{Periodic orbits of the $\pi /3$-rhombus billiard}

The $\pi /3$-rhombus billiard is an example of an almost integrable system.
As discussed before, billiard flow for a particle inside an almost
integrable polygon is called an almost integrable billiard. The most
fascinating mathematical questions are related to the periodic orbits of
these billiards. For complete analytical semi-classical study of such
systems one needs to enumerate and classify all the periodic orbits and also
able to compute actions of periodic orbits of the given system. The example
where the enumeration and classification of periodic orbits is analytically
carried out are very rare. Apart from the trivial enumeration of the orbits
for a separable barrier billiard there is no instance where a complete study
exists.

\subsection{Enumeration}

To begin with, let us briefly recapture how the motion of a particle inside $%
\pi /3$-rhombus shaped encloser can be visualized as motion on an equivalent
barrier billiard \cite{3.1,3-par}. It is simple to see (ref. Fig. 3.1) that
after three successive reflections ($A\rightarrow B\rightarrow C$) of the
rhombus $A$ around a vertex of angle $2\pi /3$, the rhombus returns upon
itself but with reversed orientation with vertices $2$ and $4$ interchanged.
In other words, we obtain the final configuration of vertices as if we have
reflected the rhombus about the shorter diagonal of the rhombus joining the
vertices $1$ and $3$. If we continue the reflections, it will take exactly
three more, or equivalently, another reflection about the shorter diagonal,
for the rhombus to identify itself with original orientation. In this
picture, due to double-valuedness of configuration of vertices per direction
( by direction we mean one of the three directions the rhombus is facing in
Fig. 3.1), one can visualize three rhombus-orientations in Fig. 3.1 on one
sheet(or,plane) and the subsequent three orientations (required to obtain
the original configuration of vertices in rhombus-A) on another sheet(or,
plane). One can visualize a trajectory of particle reflecting from a wall of
the rhombus by letting the particle move straight and appropriately
reflecting the rhombus about the wall. It is this way of analysing that
turns out to be more fruitful and hence the discussion on the tessellation
of plane by rhombi. Due to its equivalence to the Riemann surface of $%
z^{1/2},z$ being a complex variable, we notice that the two sheets discussed
above are joined along straight lines (The complex counterparts are branch
cuts) that cannot be crossed; we call these barriers. Furthermore, as we
have seen above, going to the next plane is to compensate for a phase $\pi $%
, the trajectory must reflect from the barrier.
\begin{figure}[htbp]
\begin{center}
{\epsfig{file=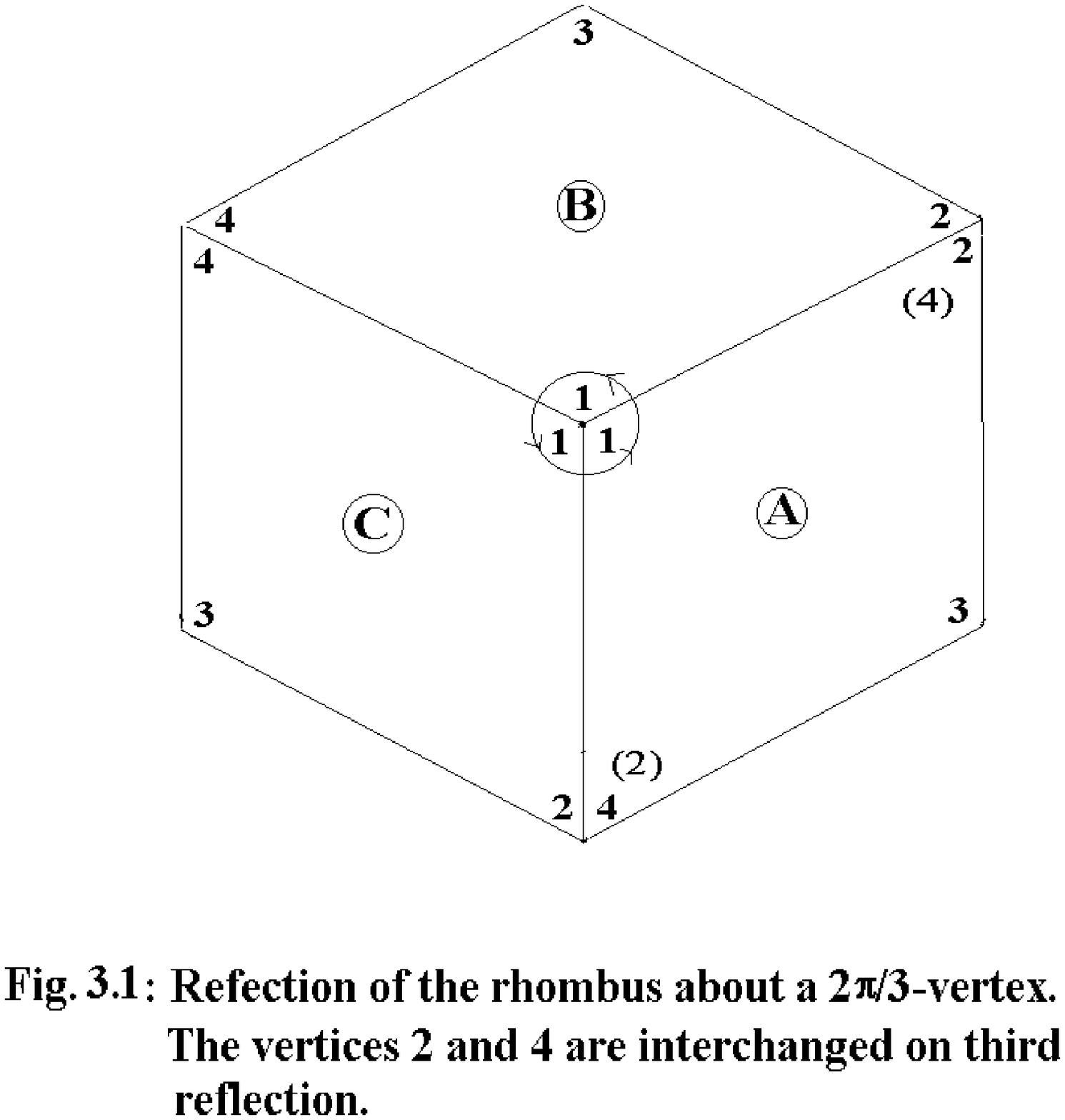,bbllx=80,bblly=170,bburx=560,bbury=700,height=6in}}
\end{center}
\end{figure}
Alternatively, after three reflections, we can reflect third rhombus back
onto itself (hence compensating for the phase in the third step), i.e. the
fourth rhombus comes to lie under the third rhombus. If we continue
reflections now, the sixth rhombus will come to lie under the first rhombus.
In this picture the point in Fig. 3.2 will become a (monkey) saddle point.
\begin{figure}[htbp]
\begin{center}
{\epsfig{file=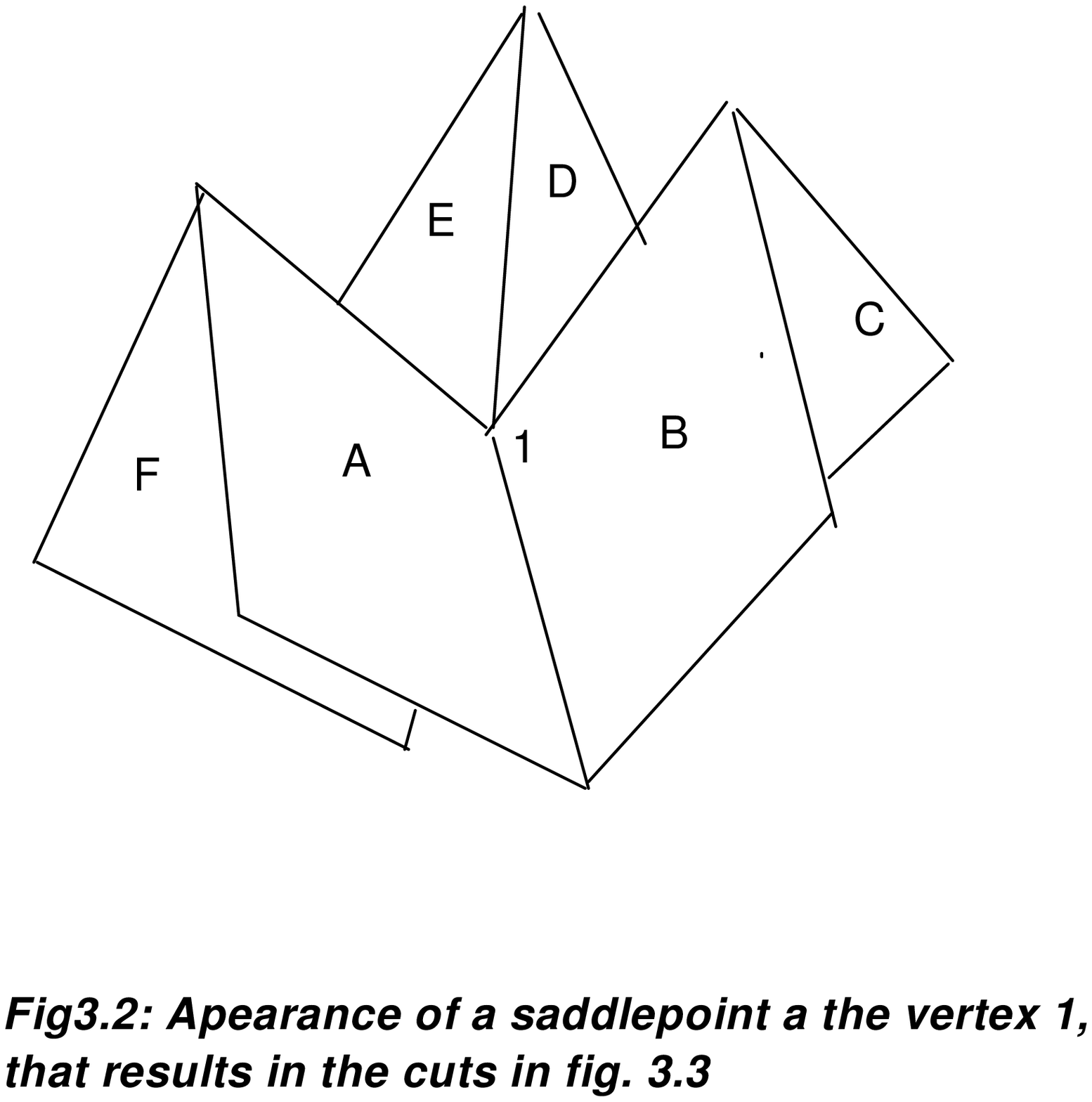,bbllx=80,bblly=170,bburx=560,bbury=700,height=6in}}
\end{center}
\end{figure}
Continuing the process of reflections, we can get planes connected by cuts
between the saddle points. This construction, projected onto two dimensions,
entails an orbit looking like a zig-zag line. We can now construct the
fundamental region using six replicas of rhombus, and subsequently
tessellate the two-dimensional plane by staking the fundamental regions side
by side, exploiting the translational symmetry. On doing so, we will
generate a barrier billiard shown in Fig 3.3 with two sets of planes (call
them top and bottom) interspersed with each other.

Now concentrate our attention on the barrier billiard where the barrier to
gap ratio is two. the barrier constituted of contributions from two rhombi
and hence, there are two distinguishable sub-barriers giving rise to a
single barrier of length twice that of gap. Classification of the periodic
orbits had been carried out for a barrier billiard with barrier to gap ratio
equal to unity\cite{3.2}. It must be noted that the barrier billiard
corresponding to the rhombus problem is more general than the barrier
billiard studied in \cite{3.2}. since the barrier to gap ratio is two and
moreover the barriers appear in an oblique manner at an angle of $\pi /3$.

First, we observe that bifurcations of the orbits take place at the two ends
of the barriers and at the center of the barrier (e.g. one example of such
bifurcation is shown in Fig. 3.3, where two neighbouring trajectories on the
opposite side of a trajectory $OABC..$bifurcate at end of a barrier near
vertex $C$). This is due to the fact that each half of the barrier is
contributed from two different rhombi in the fundamental region, and the
point of bifurcation actually corresponds to a vertex.

The single connected surface is made up of two planes - a top and bottom.
Under the covering of the surface by fundamental regions (double hexagons),
the surface divides into alternate arrays of both planes containing
barriers. Obviously, it does not matter which plane is called top(or
bottom). This argument allows us to choose an origin which, for obvious
reasons, dictated by symmetry of the barriers, is chosen to be the center of
the barrier, denoted by $O$ in Fig.3.3. Calling the length of a side of the
rhombus by $L$, the barrier length is $2L$ and gap length is $L$. On the
vertical axis, the perpendicular distance between adjacent arrays of a
(top/bottom) plane is $\sqrt{3}L$. Since the factor of $\sqrt{3}$ is common
in the vertical axis, we choose to measure the length in this direction in
terms of $\sqrt{3}$, thereby making the ordered pairs labeling the points
purely consisting of integers, $(q,p)$. For instance, a point $O^{\prime }$
in the Fig. 3.3 will be labeled by $\left( 3,2\right) $.
\begin{figure}[htbp]
\begin{center}
{\epsfig{file=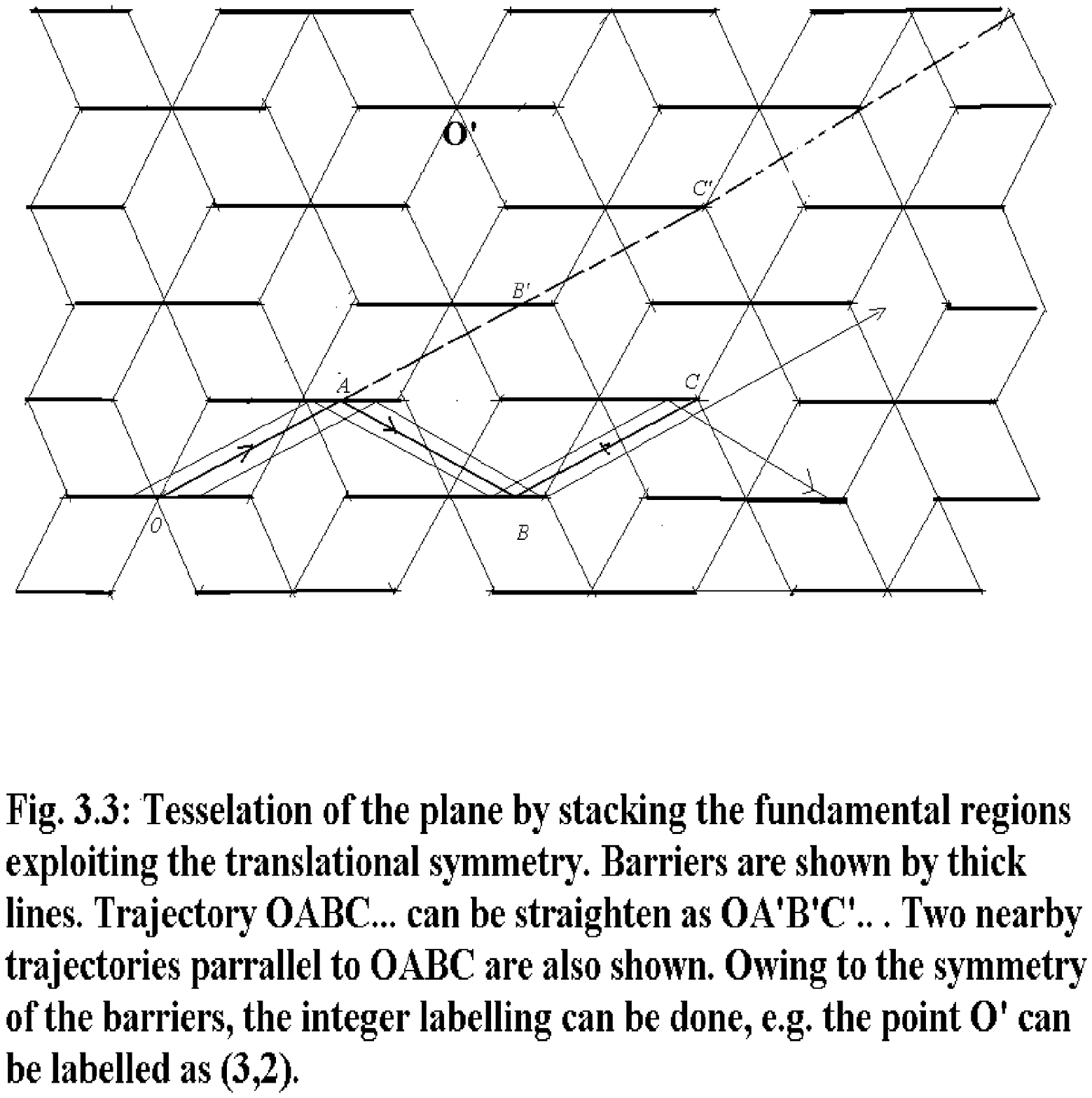,bbllx=80,bblly=170,bburx=560,bbury=700,height=6in}}
\end{center}
\end{figure}
A typical trajectory on this surface will be made up of alternate motions in
the top and the bottom planes via barriers (i.e. when a trajectory hits a
barrier on a upper plane it simply appears on the lower plane and vice
versa). Since every plane consist of an identical array of barriers, the
trajectory starting at an angle with the plane from an initial point and
ending on an equivalent point on the same plane constitutes a periodic orbit
(see e.g. Fig. 3.4). Instead of following the zig-zag path, we can unfold
the trajectory into an exactly equivalent straightened version as shown in
Fig. 3.3. As stated above this straightened trajectory will lie on both
planes, crossing planes at the barriers. Subsequently, we must decide which
directions lead to periodic orbits.
\begin{figure}[htbp]
\begin{center}
{\epsfig{file=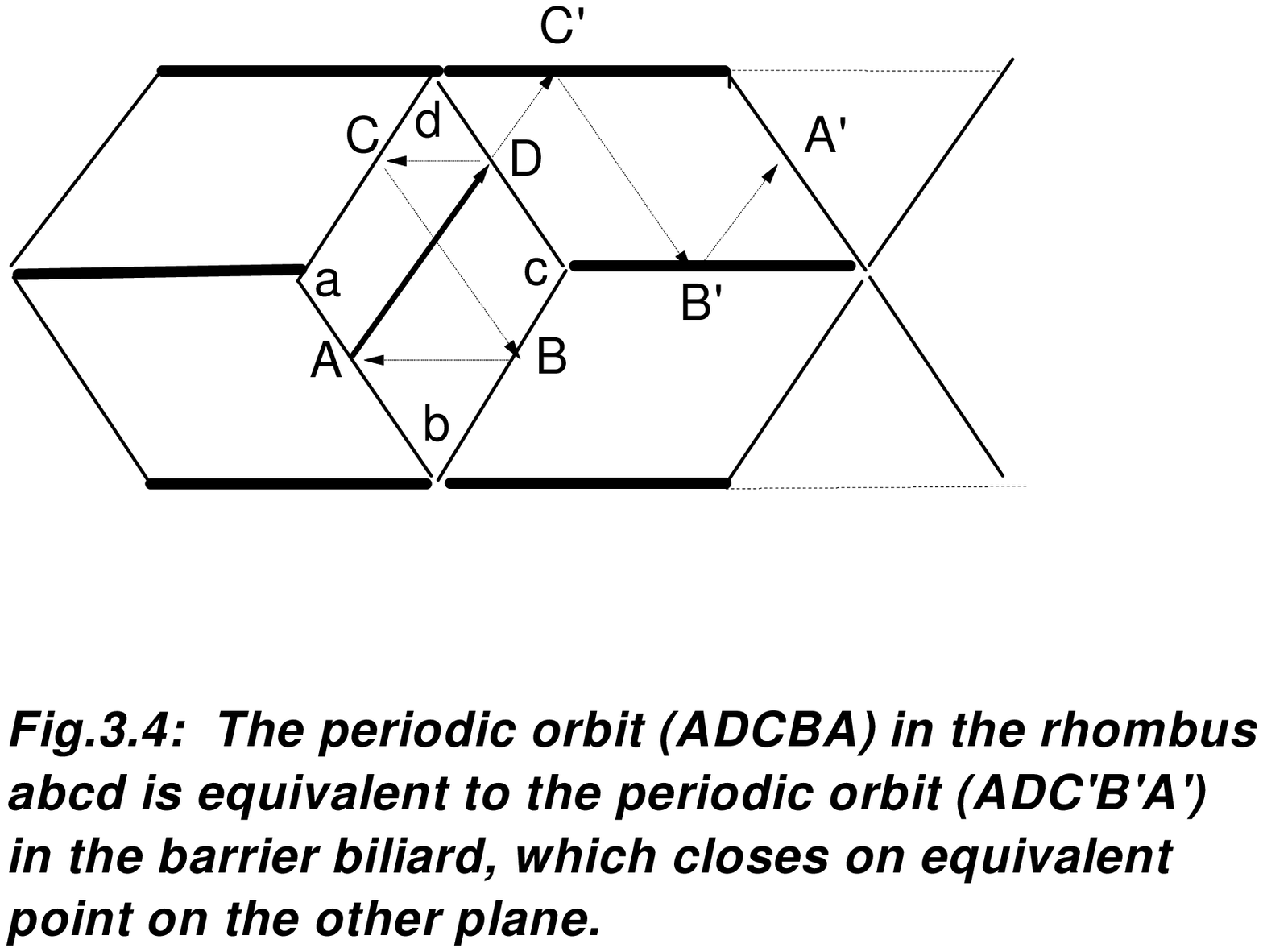,bbllx=80,bblly=170,bburx=560,bbury=700,height=6in}}
\end{center}
\end{figure}
By virtue of the integer labeling, it is clear that all those directions
that end on (integer, integer) ordered pair correspond to periodic orbits.
Leaving apart a factor of $\sqrt{3}$ in the vertical direction, these
directions correspond to rational gradients on each (top, bottom) plane.
Also, starting from origin, we must consider only those end points such that 
$q$ and $p$ are co-prime, since only such pairs results in primary periodic
orbits. If $q$ and $p$ are not co-prime but have g.c.d. $k$, then orbit
ending at $(q,p)$ represents $k^{th}$ repetition of a primary periodic orbit
ending at $(q/k,p/k)$. By the symmetry of the barriers on the plane, we need
to restrict ourselves to an upper-half region only. Further taking care of
geometry of our system, we restrict further to either $p\leq q$ or $q\leq
3p, $ obviously then, three classes emerge, {\it viz. }$(q,p)$ can be ({\it i%
})(odd,odd), or, ({\it ii})(odd,even), or, ({\it iii})(even,odd). Next, we
have to classify the number of bands or families of periodic orbits that
correspond to each direction.

For equal barrier to gap ratio, it was shown that any trajectory with
irrational gradient can be approximated arbitrarily well by trajectories
with rational gradients, utilizing Klein's string construction or the
continued fraction expansion. However, those trajectories never close in
position and momentum both, rather they form a curious zig-zag path, for
quadratic irrational gradients, has a fractal dimension. For the barrier in
our case, same holds. Hence, we conclude that the trajectories with any
irrational gradient do not close. It, therefore follows that if we take into
account all the rational gradients avoid over-counting and classify
different families/bands, we would have enumerated all the periodic orbits.

We shall now use modified form of interval exchange transformations and name
it polar construction for the reasons of clarity and easy generalization.

\subsection{Polar Construction}

Exploiting the periodicity of barrier-gap-barrier-gap... string, we wrap the
basic string of skewed sets of barrier-gap-barrier on two circles
representing two different planes. Each circle has three basic divisions,
coming out of two barriers(from two rhombi, joined together) and a gap. Each
division now corresponds to an angle $2\pi /3.$ The fact that barriers on
two planes are stacked in a skewed manner is accounted for giving
appropriate phase difference between equivalent points on two circles (see.
Fig. 3.5). A $(q,p)$ direction can be represented these circles by the
following procedure.

Divide each segment of the inner (outer) circle into $p$ parts, after fixing
the origin at the point joining the two sub-barrier segments. The origin of
the outer (inner) will be at an angle of $\pi q/3p$ from the origin of the
inner (outer) circle moving along the circular arc in a definite sense (we
use anticlockwise movement). After fixing the origin, the outer circle has
to be divided into $3p$ parts of equal length with one division at the
origin. Since there are two gradients, positive and negative, we adopt the
convention of marking outside (inside) of the circle as a representative of
positive (negative) gradient.

To follow a trajectory, we start with an arbitrary point on one of the $p$
sub-segments of the barrier segment on the right of the origin, on the inner
circle. Next point will be on the outer circle just at the same distance
(number of sub-segments) from the origin (of the outer circle) as the
previous point was from its respective origin. As is clear from earlier
discussion, these points must alter between the outer and the inner circles.
The following point comes on the inner circle $q$ segments away from the
original one, and the next on the outer circle $q$ segments away from the
earlier one, and so on. Going on in this way, after a finite number of
points, we will reach the starting point on the inner circle and that would
make one periodic point. Although this is generally valid, in some cases it
may not lead to the minimal length of the orbit. Such would be the case
when, exactly after half the number of reflections, the trajectory will
close, i.e., reach a point corresponding to the same respective sub-segment
as it started with, on the other circle. At first sight, it might seem
erroneous to consider this as a periodic orbit. However, it may be recalled
that though the procedure of erecting barriers and subsequent polar
construction is a way leading to an easy classification of orbits; however,
more fundamental idea is the straightening of a trajectory-reflecting the
domain about the edge on which the particle is incident. For the cases where
the length of an orbit turns out to be double by polar construction, one can
easily see that using the domain-reflection method, one gets the correct
length. Thus, without ambiguity, the trajectory in the polar construction
must be considered periodic even if it seems to be closing on the other
plane (see, for example illustration in Fig. 3.4). With this clarification,
we need only think in terms of the polar construction. Different sequences
of barriers and gap correspond to the different orbits.

\subsection{Classification}

As we have seen above, the most elementary classification is in $(q,p)$
being $\left( odd,odd\right) ,$ $\left( odd,even\right) \,\,\,$and $\left(
even,odd\right) $. Having set the origin at the center of the barrier, the
trajectory sets off in some rational direction and reaches either a center
of some barrier or end (left or right) of some barrier. It is rather obvious
to see that for each of the three cases written above there are subclasses
which we shall call: the center-to-center(CC) case and the center-to
edge(CE) case. Our Procedure of classification is in following steps:

(i) using the polar construction, we depict the trajectory on the circles,
with an opening and an ending point on one of the sub-segments of the
segment. We go on to the other to other sub-segments of the same segment,
exploring the positive and negative gradients till both sides of the circles
are filled. In all, we must fill $12p$ points.

(ii)check if the orbit has already closed a half-way on the outer circle or,
equivalently, on the other plane.

With these steps in mind, we now take up each class separately and classify
the bands of periodic orbits in full.

{\bf Case 1: Odd-Odd CE}

We first describe through a simple example as to how we would arrive at
general conclusions. Our approach would be to make a conclusion based on
empirical data obtained by ''brute force''. At the end, we will provide with
a rationale supporting and explaining the conclusion obtained.

Let us consider the case of $(q,p)=(1,3).$ The corresponding polar
construction is shown in Fig.3.5. As can be seen, point $1$ and point $10$
identify with each other, forming a periodic orbit after six bounces. Also,
there is an orbit with negative gradient. It should be noted that all the
sub-segments are visited by just these two orbits. The orbit close a
half-way of $(3q,3p),$on the other plane and, there are two bands of orbits.
\begin{figure}[htbp]
\begin{center}
{\epsfig{file=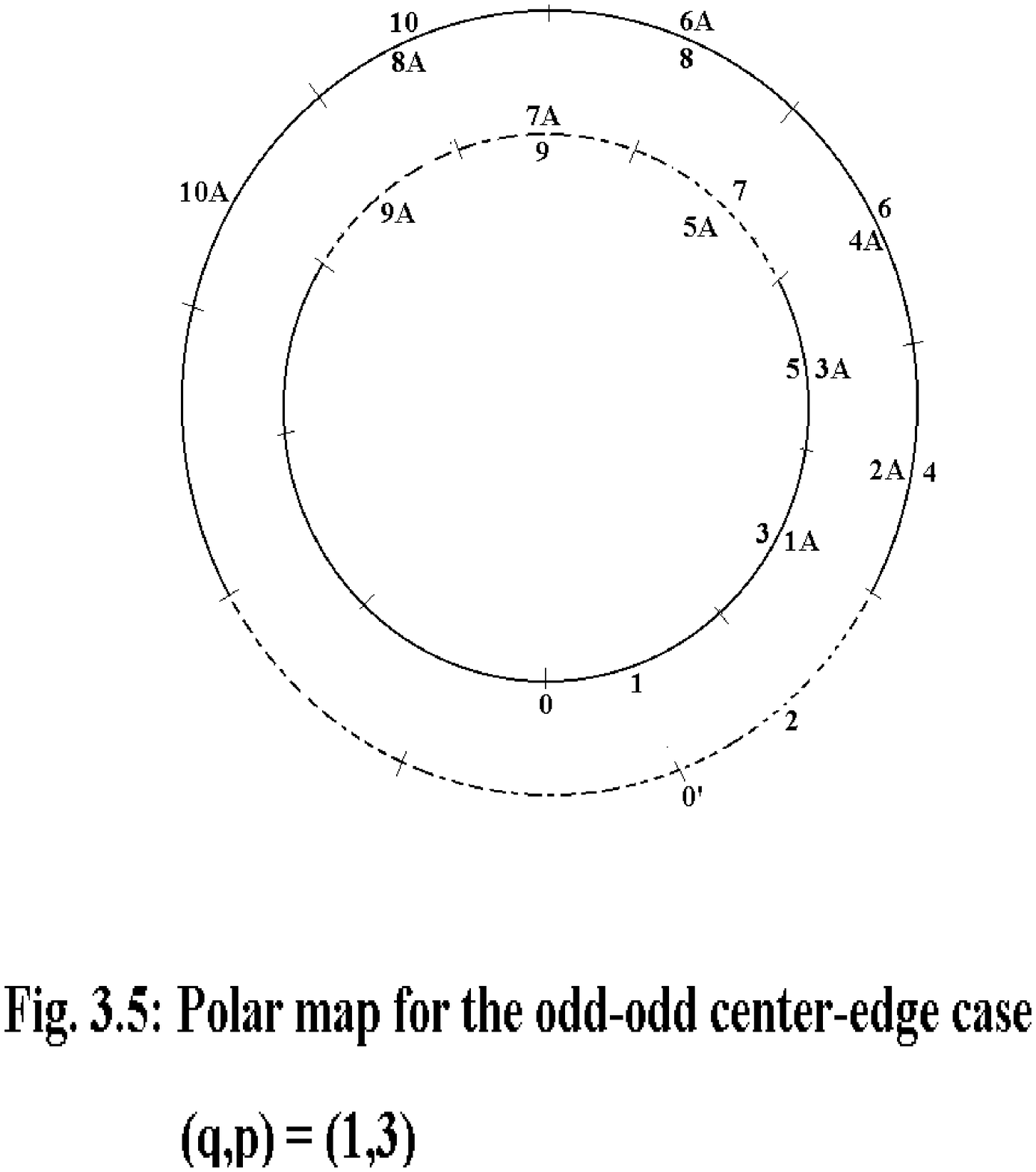,bbllx=80,bblly=170,bburx=560,bbury=700,height=6in}}
\end{center}
\end{figure}
Drawing the polar construction for other odd-odd CE cases, it can be seen
that there are only two bands of orbits as seen above.

Now we come to discuss the rationale behind this classification of orbits.
Due to the polar construction, each sub-segment is an arc of angle $2\pi /3p$%
. Translating the formation of periodic orbit by the polar construction into
an equation, we trivially get 
\begin{equation}
\label{3.1}N(2\pi /3p)q=2\pi M 
\end{equation}
where $N$ denotes the number of sub-segments and $M$ denotes the number of
rotations by $2\pi $. Henceforth, we call $N$ by the ''crossing index'' and $%
M$ by the ''rotation number'' eq.(\ref{3.1}) is simply

\begin{equation}
\label{3.2}Nq=3Mp 
\end{equation}

where $M,q,p$ are positive integers. Since this is a CE case, $q$ is not a
multiple of three, i.e. $q\neq 3l,l\varepsilon Z.$ Thus, the only way in
which eq.(\ref{3.2}) can be satisfied is if $q=M$ and $3p=N.$ Note that $M$
and $N$ will be odd as both $q$ and $p$ are odd. The crossing-index on one
circle is $3p$ implying that the total crossing- index is $6p$ after which
the orbit closed. In all, there are $12p$ sub-segments and hence there are
exactly two bands of periodic orbits.

In general, the crossing-index is given by

\begin{equation}
\label{3.3}N=[q,3p]/q 
\end{equation}

and the rotation number is given by

\begin{equation}
\label{3.4}M=[q,3p]/3p 
\end{equation}

where $[a,b]$ denotes the lowest common multiple of $a$ and $b$. Trivially,
for the CE case, $N=3p$ and $M=q,$for the CC case, $N=p$ and $M=q/3$.

{\bf Case 2: Odd-Odd CC}

Firstly, all the rational directions pointing toward an infinite number of
avenues correspond to the bands of periodic orbits. Points $(q,p)$
corresponding to avenues are of the form $(3(2k-1),1),k\epsilon Z_{+}.$ For
each direction there will be two bands corresponding to positive and
negative gradients.

We go over to a representative of a general case, viz., $(q,p)=(3,5)$. The
polar construction is depicted in Fig. 3.6. There are three strings of
points corresponding to three bands of periodic orbits. Including the
opposite gradients, there are four distinct bands in all. The string
starting with point $1$ and ending with point $11$ closes after six
reflections. Same is the case for $1B-11B$ orbit. The opposite gradient
counterpart of $1-11$ $(1B-11B)$ is equivalent to itself, starting from the
outer circle (at $6(6B)$). So these strings give us two bands of periodic
orbits. Care must be taken for the periodic orbit starting with the point $%
1A $. The orbit closes at $6A$ as this is the subsequent corresponding to $%
1A $ on the inner circle and the gradient matches. The orbit$1A-6$A closes
after four reflections and occurs in a band. Taking the opposite gradient,
we get two bands of periodic orbits here.
\begin{figure}[htbp]
\begin{center}
{\epsfig{file=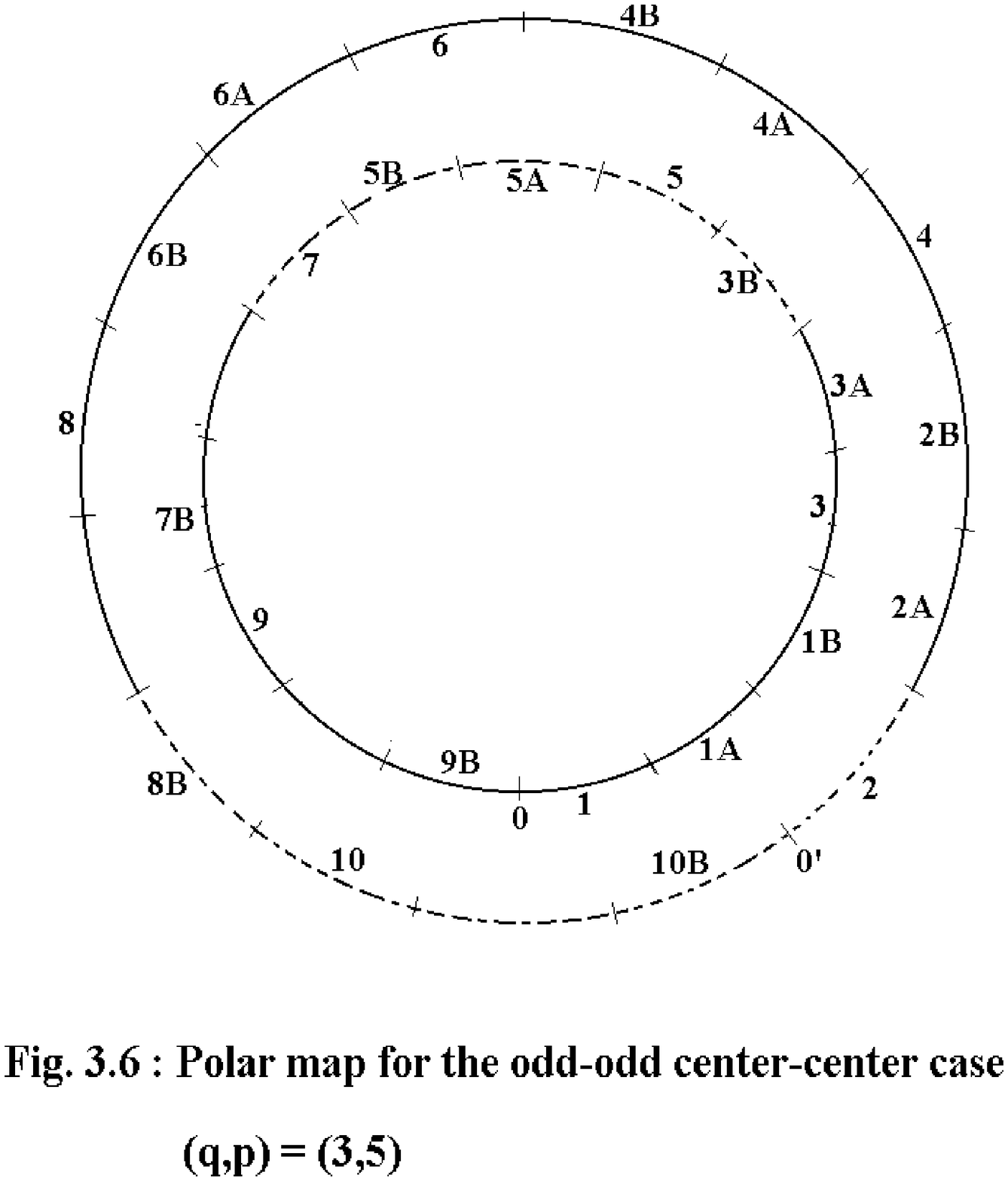,bbllx=80,bblly=170,bburx=560,bbury=700,height=6in}}
\end{center}
\end{figure}
For other odd-odd CC cases, we obtain the same results as in the above
mentioned case. Let us now see the reason for the occurrence of four bands
of periodic orbits in this case.

From eqs (\ref{3.3}) and (\ref{3.4}), $M=q/3$ and $N=p$. For each orbit, the
crossing-index will be $2p$ accounting for the other circle also. In total,
there are $12p$ segments and it clearly follows that there must be six bands
of periodic orbits. Subtracting the two equivalent bands, we are left with
four bands.

{\bf Case 3: Odd-Even CE}

Consider $(q,p)=(1,2)$. The polar construction is drawn in Fig. 3.7. There
is a bifurcation in the band of trajectories starting with point $1$, and
further continue the two bands, primed and unprimed, to eventually close at $%
13^{\prime }$ and $13$ respectively. This feature, which can be succinctly
put as bifurcation of the vector fields at vertices and continuation of
trajectories in the form of bands (a signature of zero Liapunov exponent),
is typical of pseudo-integrable system. The points $1$ and $6^{\prime }$
are, indeed, identical. Orbits emanating from $1$ and $6\prime $ ($1A$ and $%
6A^{\prime }$) will be the same . Consequently, on allowing the opposite
gradients, we get just two bands of periodic orbits.
\begin{figure}[htbp]
\begin{center}
{\epsfig{file=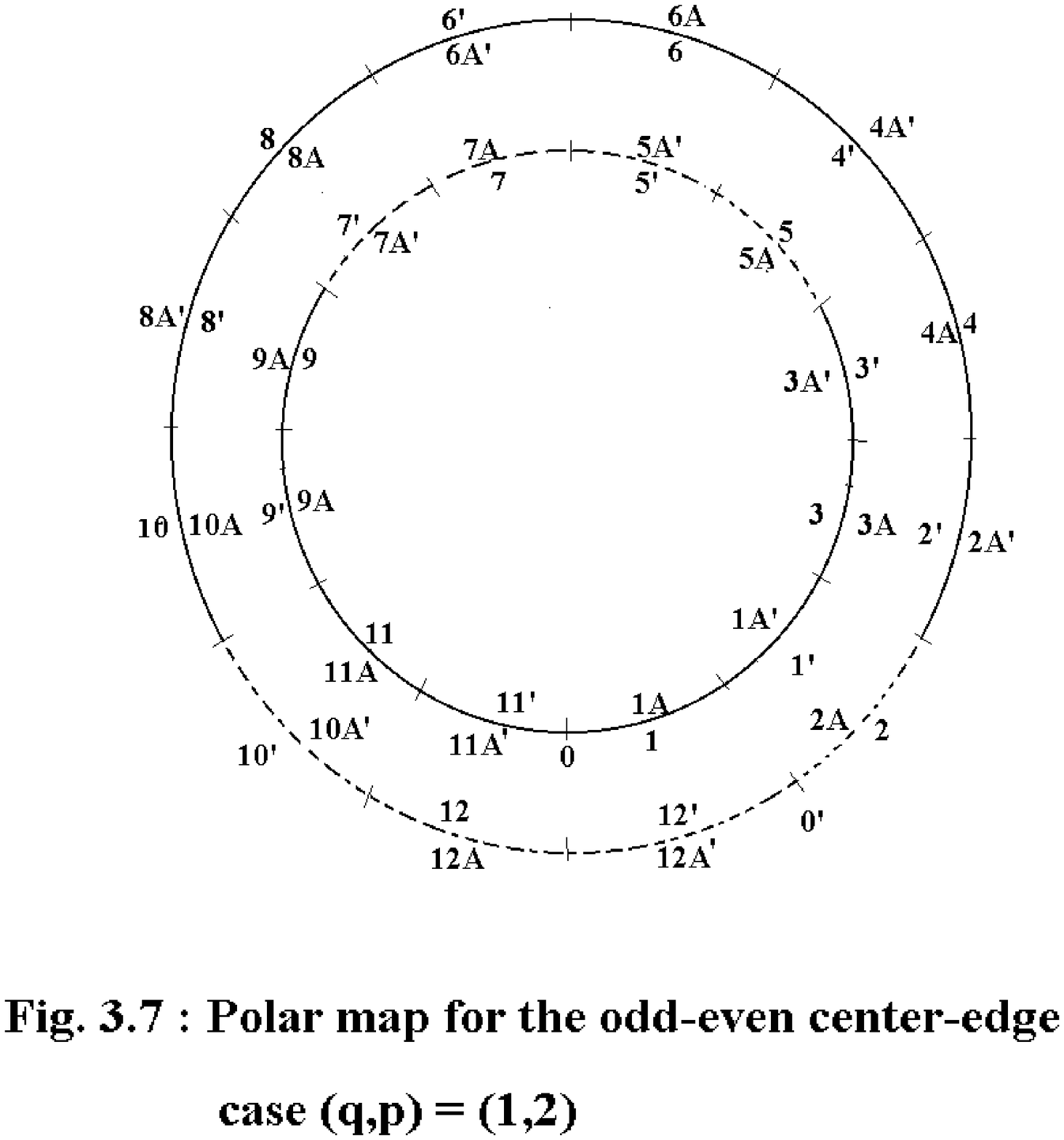,bbllx=80,bblly=170,bburx=560,bbury=700,height=6in}}
\end{center}
\end{figure}
All other examples of this class give rise to same number of bands and the
orbit-types are also similar. Of course, the lengths and other details will
be different.

As the arguments for justification follow on the same lines, we do not
repeat them for this and further cases.

{\bf Case 4: Odd-Even CC}

For this case, consider $(q,p)=(3,2)$, polar construction for the same is
drawn in Fig. 3.8. Due to bifurcations, we have drawn double the marks,
thus, making explicit that we have to fill $24p$ points in all. As can be
seen from the diagram, there are four bands of periodic orbits: an orbit
corresponding to the string $12...9$; another orbit corresponding to the
string $1A2A...9A$; and a positive-negative gradient pair, $1^{\prime
}2^{\prime }...5^{\prime }$ (plus the opposite gradient). It is interesting
to see that bifurcations of these vector fields take place at all possible
places, viz. $Z_1,Z_2$and $Z_3$.
\begin{figure}[htbp]
\begin{center}
{\epsfig{file=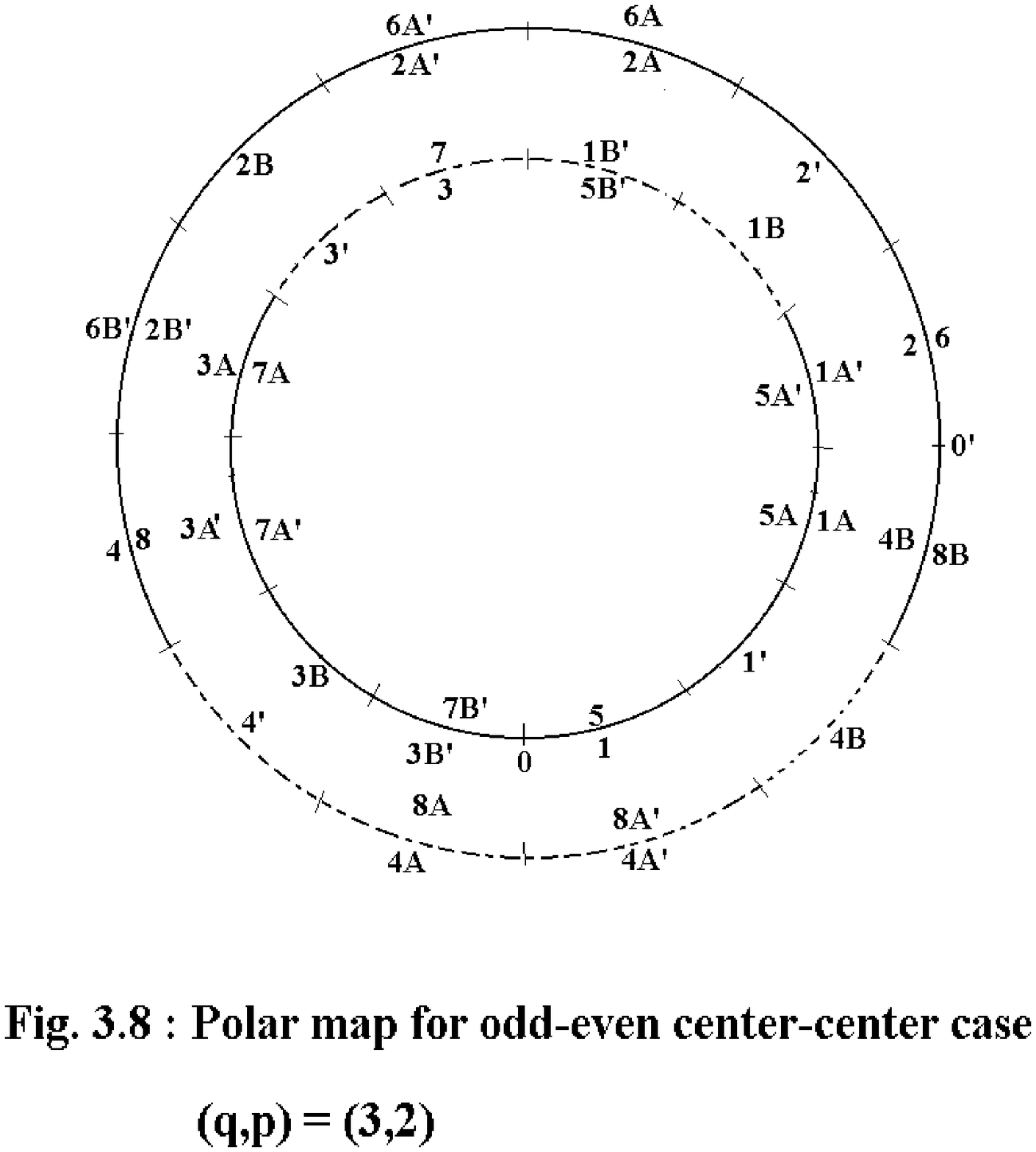,bbllx=80,bblly=170,bburx=560,bbury=700,height=6in}}
\end{center}
\end{figure}
Considering other odd-even cases, one can easily see that the same
conclusions about the number of bands etc. will hold, leaving apart the
finer details.

{\bf Case 5: Even-Odd CE}

Consider $(q,p)=(2,1)$, the polar construction is shown in Fig. 3.9. There
are two bands of periodic orbits. In fact, the periodic orbit formed with
opening point as the negative-gradient-equivalent of the point $1$ is
identical to the orbit $1A2A...7A$, starting from $2A.$ Thence, strings $%
1\,\,2...7$ and its negative gradient counterpart are the two bands in this
case.
\begin{figure}[htbp]
\begin{center}
{\epsfig{file=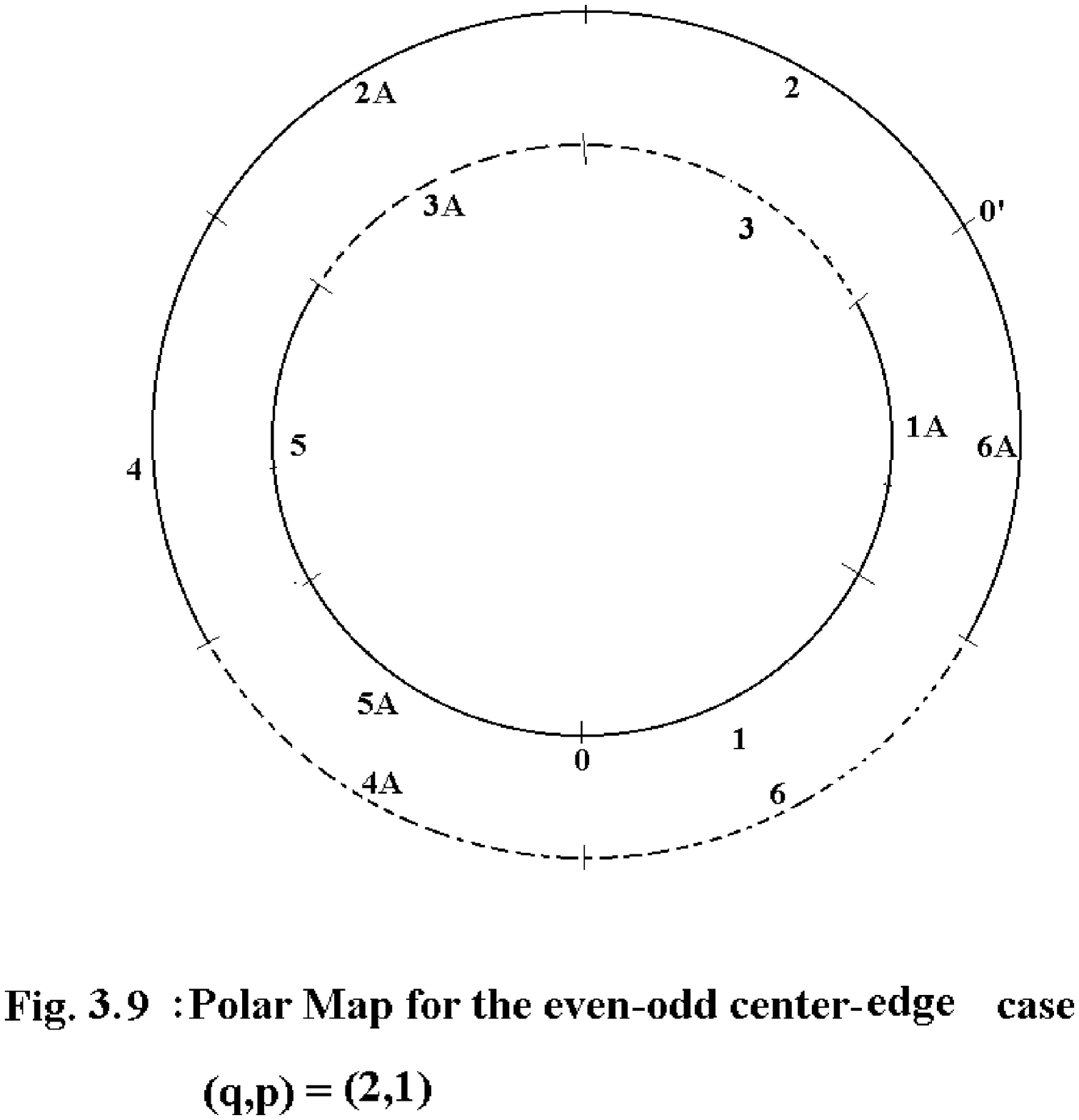,bbllx=80,bblly=170,bburx=560,bbury=700,height=6in}}
\end{center}
\end{figure}
{\bf Case 6: Even-Odd CC}

A general case can be studied through the example of $(q,p)=(6,5)$, polar
construction being depicted in Fig. 3.10. Although the diagram is getting
rather complicated, in the same manner as discussed earlier, it can be
concluded that there are four bands of periodic orbits. General validity of
this conclusion can be verified without undue hardship.
\begin{figure}[htbp]
\begin{center}
{\epsfig{file=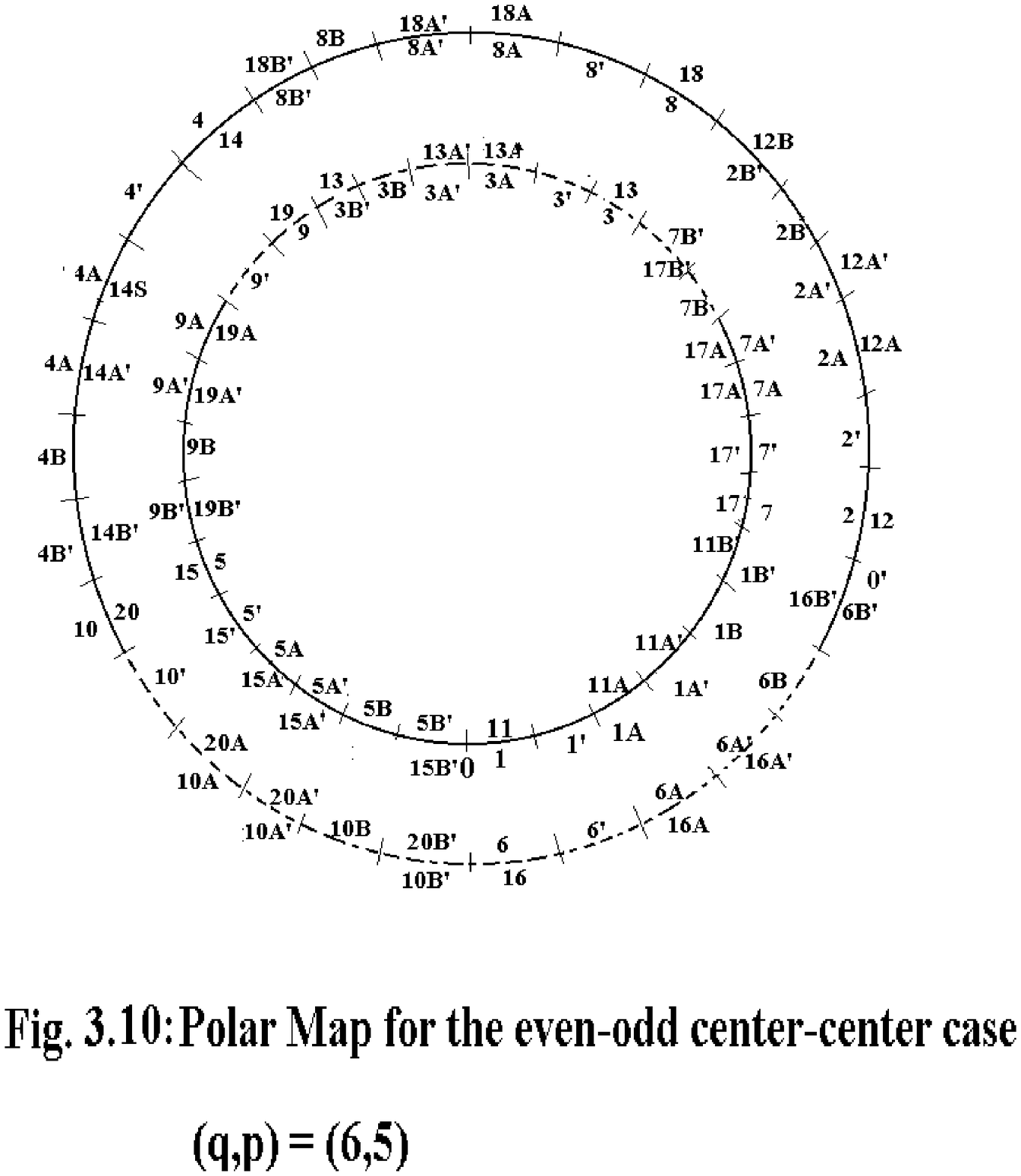,bbllx=80,bblly=170,bburx=560,bbury=700,height=6in}}
\end{center}
\end{figure}
The results of this section, along with the lengths of the periodic orbits
and the phase space areas of the bands in which they occur are summarized in
Table I. This concludes our discussion on the enumeration and classification
of the periodic orbits in $\pi /3-$rhombus billiard. The periodic orbits of
other billiards such as Hannay-McCraw billiard \cite{3.2} or rectangle
billiard with slit at the center can also be enumerated and classified in
the same way.

\section{The H-M Billiard}

The particle moves here in a configuration space where a hard line-segment
barrier is placed at the center parallel to one of the sides of the periodic
cell of side length $L$, and the length of the barrier is half of the side
length(i.e. $L/2$). The invariant integral surface of this system has genus
equal to two $(g=2)$. The fundamental region consist of two cells, due to
presence of a barrier (see, Fig. 3.11). The opposite boundaries of the
fundamental region are identified topologically resulting in one handle of a
sphere. The barriers in each cell of the fundamental region has to be
identified topologically which results in another handle of the sphere. By
stacking the domains of the billiard side by side in both the orthogonal
directions, one obtains an infinite lattice of barriers and gaps, with
barrier to gap to ratio unity. One can label the end points of barriers by
integer-pairs which form lattice points. It can be easily seen that the
straightened version of a rational gradient $(=|p/q|)$ trajectory will
initially meet lattice point $(q,p)$ and then repeat itself by meeting
lattice points $(mq,mp)$ where $m\in Z.$ On the other hand, the irrational
gradient trajectory will never visit any lattice point though it will come
arbitrarily close to many lattice points, hence will never be periodic. Thus
the periodic orbits in the system are the ones which hit any lattice point $%
(q,p)$ in this array of barriers, the gradient of such trajectories will be
given by $|p/q|$. By above arguments, we need to consider only the pairs$%
(q,p)$ such that $q$ and $p$ are co-prime since they only give a primitive
periodic orbits, and points $(mq,mp),$ where $m\in Z$, gives $m$ repetitions
of a primitive periodic orbit corresponding to $(q,p)$. Each such $(q,p)$
gives different number of bands or families of periodic orbits, depending on
whether the pair is odd-odd $(o,o)$ or even-odd $(e,o)$ or odd-even $(o,e)$.
The length of periodic orbit in a given family corresponding to a lattice
point $(q,p)$ is given by

$$
l=cL\sqrt{\left( q^2+p^2\right) } 
$$
where $c$ depends on the number of families of periodic orbits. The periodic
orbits can now be enumerated and classified exactly same way as done in the
case of the $\pi /3$-rhombus billiard\cite{3.2}. The result is summarised in
the table 3.2.
\begin{figure}[htbp]
\begin{center}
{\epsfig{file=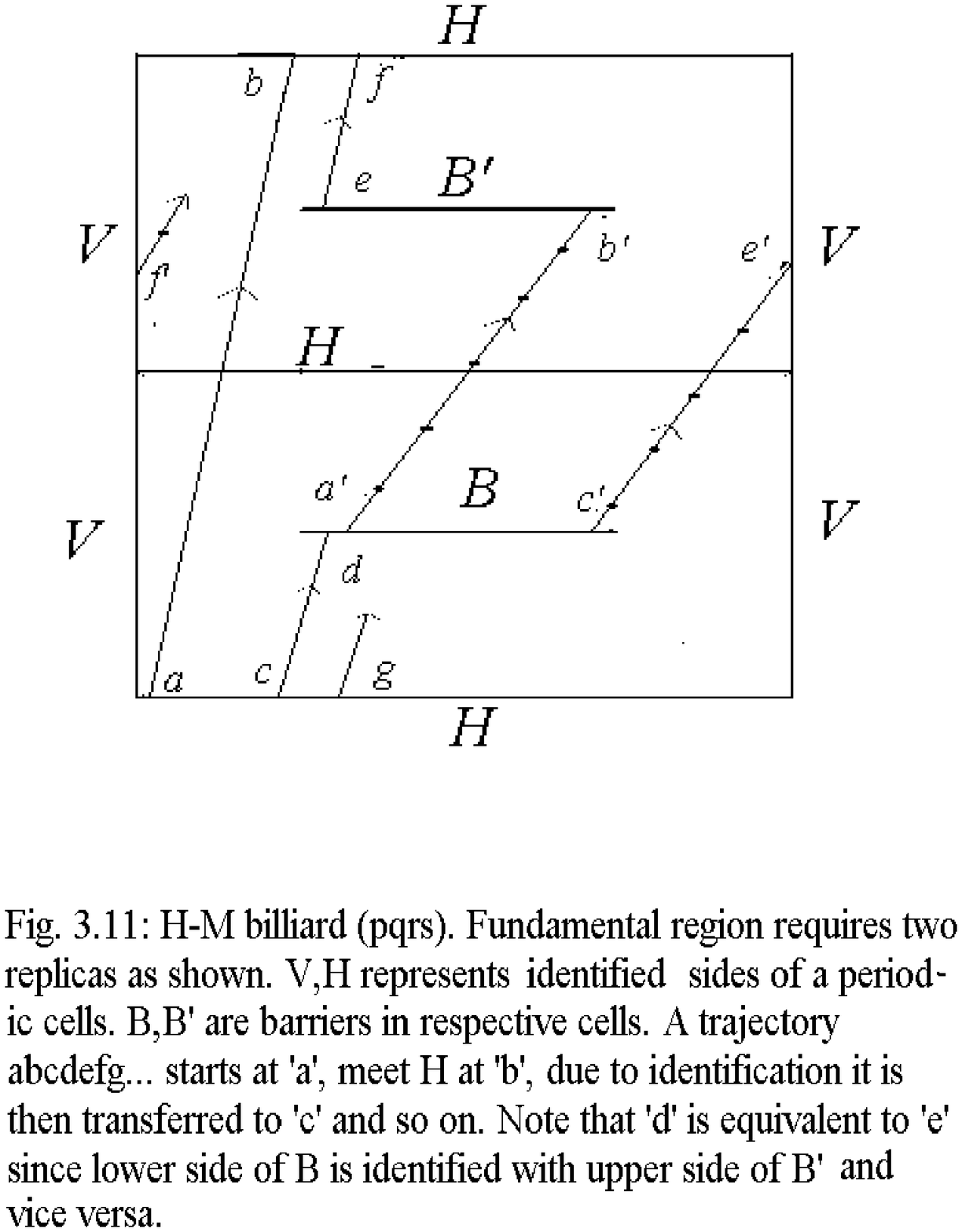,bbllx=80,bblly=170,bburx=560,bbury=700,height=6in}}
\end{center}
\end{figure}
\section{The Single Slit Rectangle billiard}

Another simple example of pseudo-integrable billiard is a simple
modification of H-M billiard, where instead of periodic cells one can
consider a linear barrier of length $L$ placed at the center of the
rectangle $(L,2L),$ parallel to longer side. This simple modification
results in the pseudo-integrable billiard whose invariant surface is
topologically equivalent to a sphere with three handles(genus, $g=3$). The
fundamental region now consist of four replicas of the configuration space
instead of two as in the H-M\ billiard. Opposite sides of the fundamental
region are identified as well as each pair of barrier lying one above each
other has to be identified separately which results in genus $3$ surface
(see Fig. 3.12). By stacking the domains of the billiard side by side in
both the orthogonal directions, one obtains an infinite lattice of barriers
and gaps, with barrier to gap to ratio unity, which is exactly same as the
lattice of barriers and gaps one obtains in the case of H-M\ billiard. One
can proceed in the same way as before to label lattice point. The periodic
orbits can then be enumerated and classified using polar maps. The result is
summarised in table 3.3.
\begin{figure}[htbp]
\begin{center}
{\epsfig{file=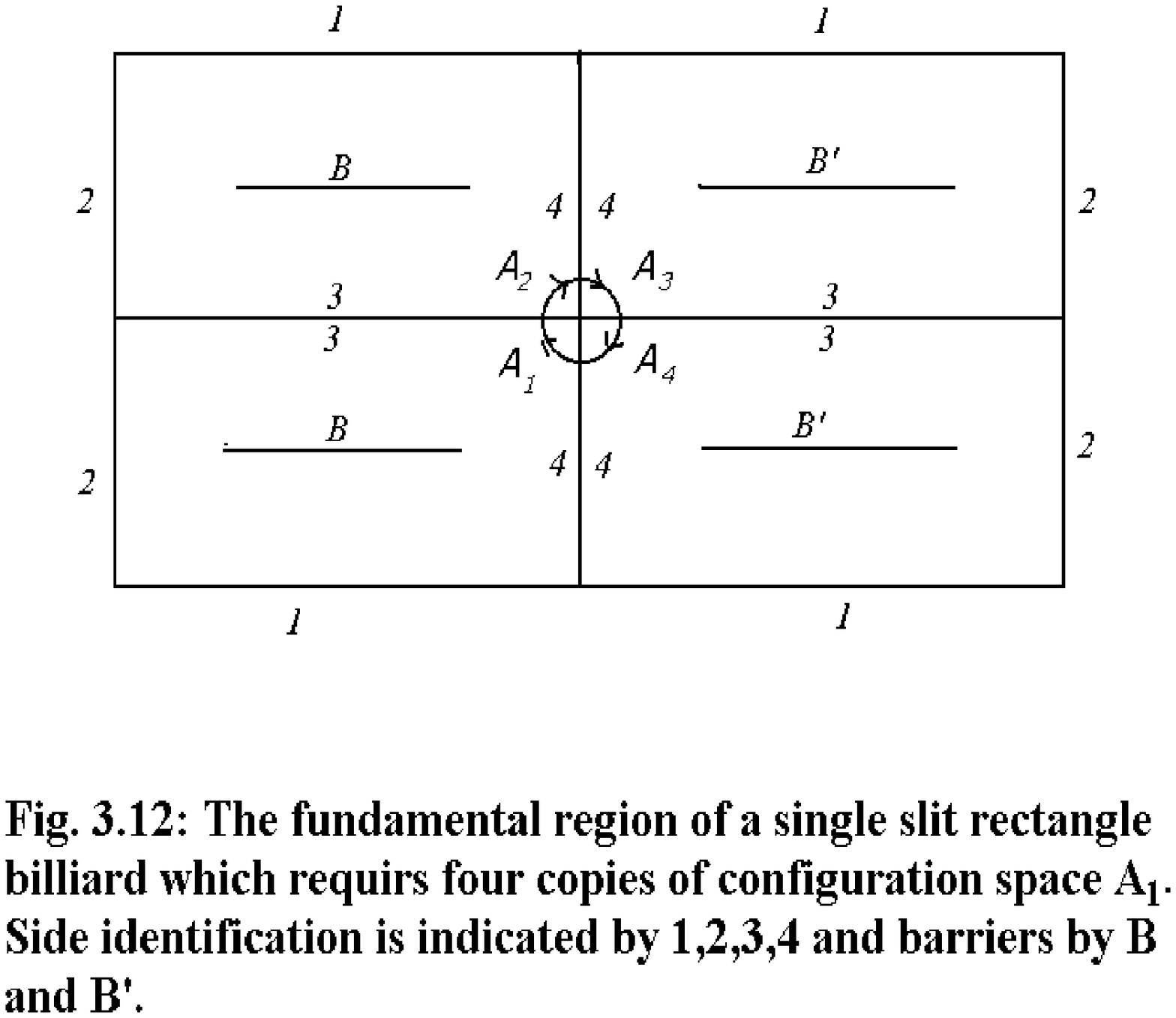,bbllx=80,bblly=170,bburx=560,bbury=700,height=6in}}
\end{center}
\end{figure}
In this chapter we have developed a methodology to enumerate and classify
periodic orbits of some pseudo-integrable billiards. The same methodology
can be used to study large number of rational polygonal billiards in
particular one can apply technique to rectangular billiard with many
slits(or barriers) or L-shaped billiards.

\newpage

\begin{table} 
  \begin{tabular}{|l|l|l|l|l|l|} \hline
     {\bf Type}  &  {\bf Classes}  &  {\bf No. of } & {\bf Closing } & {\bf Length} & {\bf Band}  \\ 
                         &                          & {\bf Families}&{\bf Point}& & {\bf Areas} \\  \hline
    Center-center& & & & & \\   
& & & & & \\    
 &  odd-odd       &  1-1 & $(q/2,p/2)$ &$Lf_{pq}/2$&$A_R$ \\
& & & & & \\ 
          &     &  2  &$(q,p)$ & $Lf_{pq}$ & $2A_R$\\ \hline
& & & & & \\  
   &  odd-even   &  1-1 & $(q,p)$ &$Lf_{pq}$& $A_R$ \\
& & & & & \\  
         &     &  2  & $(2q,2p)$ & $2Lf_{pq}$ & $2A_R$\\    \hline
& & & & & \\ 
    &  even-odd   &  1-1 & $(q,p)$ &$Lf_{pq}$& $A_R$ \\
& & & & & \\  
         &     &  2  & $(2q,2p)$ & $2Lf_{pq}$ & $2A_R$\\  \hline
& & & & & \\   
Center-edge& & & & & \\
& & & & & \\  
 &  odd-odd   &  1-1 & $(3q/2,3p/2)$ &$3Lf_{pq}/2$&$3 A_R$ \\  \hline
& & & & & \\   
  &  odd-even   &  1-1 & $(3q,3p)$ &$3Lf_{pq}$&$3 A_R$ \\ \hline
& & & & & \\  
  &  even-odd   &  1-1 & $(3q,3p)$ &$3Lf_{pq}$&$3 A_R$ \\   \hline
 \end{tabular}  
\caption{Summary of the results obtained in this chapter about 
periodic orbits of $\pi/3$-rhombus billiard. The symbol $f_{pq}=(q^2+3p^2)^{1/2}$, 
$L$ and $A_R=3^{1/2}L^{2}/2$ are length and area of the billiard. Note that total band 
area occupied by periodic orbits assigned to a single lattice point is always $6A_R$ (i.e. the
phase space area of the billiard). }
\newpage
\end{table}

\begin{table} 
  \begin{tabular}{|l|l|l|l|l|} \hline
    {\bf Classes}  &  {\bf No. of } & {\bf Closing } & {\bf Length} & {\bf Band}  \\ 
                             & {\bf Families}&{\bf Point}& &                        {\bf Areas} \\  \hline   
 & & & & \\    
   odd-odd       &  1 & $(4q,4p)$ &$4Lf_{pq}/2$&$2A_R$ \\ \hline
 & & & & \\ 
    even-odd$^{1}$    &  1-1 &$(q,p)$ & $Lf_{pq}$ & $A_R/2$\\ \hline
 & & & & \\  
    even-odd$^{2}$   &  1 & $(2q,2p)$ &$2Lf_{pq}$& $A_R$ \\ \hline
 & & & & \\  
     odd-even     &  2 & $(2q,2p)$ & $2Lf_{pq}$ & $A_R$\\    \hline
  \end{tabular}
  \caption{Summary of results for single H-M billiard, The symbol $f_{pq}=(q^2+p^2)^{1/2}$, 
$L$ and $A_R=L^{2}$ are length and area of the billiard.  Again here  total band 
area occupied by periodic orbits assigned to a single lattice point is always $2A_R$ (i.e. the
phase space area of the billiard). }
\end{table}
\begin{table} 
  \begin{tabular}{|l|l|l|l|l|} \hline
    {\bf Classes}  &  {\bf No. of } & {\bf Closing } & {\bf Length} & {\bf Band}  \\ 
                             & {\bf Families}&{\bf Point}& &                        {\bf Areas} \\  \hline   
 & & & & \\    
   odd-odd       &  2 & $(4q,4p)$ &$4Lf_{pq}/2$&$2A_R$ \\ \hline
 & & & & \\ 
    even-odd$^{1}$    &  2 &$(2q,2p)$ & $2Lf_{pq}$ & $A_R$\\ \hline
 & & & & \\  
    even-odd$^{2}$   &  2 & $(2q,2p)$ &$2Lf_{pq}$& $A_R$ \\ \hline
 & & & & \\  
     odd-even     &  2  & $(4q,4p)$ & $4Lf_{pq}$ & $2A_R$\\    \hline
  \end{tabular}
  \caption{Summary of results for single slit rectangle billiard, The symbol $f_{pq}=(q^2+p^2)^{1/2}$, 
$L$ and $A_R=2L^{2}$ are length and area of the billiard.  Here also  total band 
area occupied by periodic orbits assigned to a single lattice point is always $6A_R$ (i.e. the
phase space area of the billiard). }
\end{table}

\chapter{Growth Rate of Periodic Orbits}

\section{Introduction}

An important characteristic of a dynamical system is the asymptotics of the
number of periodic orbits(or growth rate of periodic orbits) having lengths
less than or equal to $l$. We denote this function by $F(l)$. From the point
of view of application, semi-classical theory of spectral statistics can be
carried out only after knowing the exact form of this law. The problem
addressing the distribution of the periodic orbits with the period of the
orbits is discussed in this chapter. The answer to the problem is known only
for the chaotic dynamical systems which possess only unstable, isolated
periodic orbits. In such cases growth rate of periodic orbits is known to be
exponential\cite{4-pary1,4-pary2}. e.g. For so called axiom A systems $F(l)$
is given by 
$$
F(l)=\exp \left\{ \kappa l\right\} /\kappa l\,\,\,\,\,\,\,\,{\rm as}%
\,\,\,\,\,\,l\rightarrow \infty 
$$
where $\kappa >0$ is the topological entropy of the system. For the
pseudo-integrable and their special cases, the almost integrable billiards(A
polygon $P$ is called almost integrable if the group $G_P$ generated by the
reflections in the side of $P$ is a discrete subgroup of the group of
motions of the Euclidean plane $R^2$), there are some conjectures and
incomplete results.

For example a theorem due to \cite{4-kat,4-gut} states that in case of any
arbitrary polygon $P,$ the function $F(l)$ grows slower than any
exponential, as $l\rightarrow \infty $. By Corollary (2.1), $F(l)$ is
bounded above by the number of generalized diagonals of length less than $l$%
. The sub-exponential growth rate of the number of generalised diagonals is
estimated from above by the entropy of the billiard, which is zero \cite
{4-sin,4-gal}. The above theorem is the only known upper bound on the
periodic orbits in general polygons. As for a lower bound, no result has
been established yet. On the other hand, periodic orbits in rational
polygons have efficient bounds from below and from above. The proofs of
these estimates use the theory of holomorphic quadratic differentials on
Riemann surface. We state here a theorem \cite{4-mas1,4-mas2} regarding this,

\begin{theorem}
Let $P$ be a rational polygon. Then there are positive constants $0<c_1<c_2$
such that for all sufficiently large values of $l$, we have 
\begin{equation}
\label{4-1}c_1l^2\leq F(l)\leq c_2l^2.
\end{equation}
\end{theorem}

These quadratic bounds are not likely to hold for general (irrational)
polygons. The expectation is that, for general polygons there are polynomial
bounds on $F(l)$.

\begin{conjecture}
Let $P$ be an arbitrary polygon. then there exist positive constants $c_1,c_2
$ and integers $1\leq n_1\leq n_2$ such that for sufficiently large $l$ we
have 
\begin{equation}
\label{4-2}c_1l^{n_1}<F(l)<c_2l^{n_2}.
\end{equation}
\end{conjecture}

For arbitrary rational polygons condition (\ref{4-1}) may well give best
possible estimates on the $F(l).$ However, there are non-trivial examples
when the asymptotics of $F(l)$ can be computed exactly.

Our investigations on some models are carried out analytically with
comparisons shown with the numerical results. We sharpen the existing
theorem on proliferation law for almost-integrable systems in a significant
manner. Our analysis enables us to give a general law of proliferation of
the periodic orbits in the pseudo-integrable billiards.

If we consider an integrable system, $\Delta $ corresponding to an almost
integrable system, $P$ and let $g$ be the genus of the surface $R$
corresponding to $P$, denote by $\mid \Delta \mid $ and $\mid P\mid $ the
respective areas of $\Delta $ and $P$, Gutkin \cite{4-gut2} proved that
there exists a constant $c_1$ such that

\begin{equation}
\label{4-3}F(l)=c_1\frac{\pi g}{|p|}l^2+0(l). 
\end{equation}

The constant $c_1\varepsilon [1,\frac{\mid p\mid }{|\Delta |}].$

\section{ Law of Proliferation of Periodic Orbits in Pseudo-integrable
Billiards}

Our results in the previous chapter clearly showed that there are countable
number of families of periodic orbits. By a family of periodic orbit, we
mean an isolated trajectory closing after an odd number of reflections, or a
band of trajectories closing after an even number of reflections. We have
also seen that the periodic trajectories only occur in bands for the $\pi
/3- $rhombus billiard due to its equivalence with the barrier billiard. The
number of families of periodic orbits of length less than $x$ is finite for
any $x;$we shall call this number by Counting Function, $F(x)$ now on.

In the subsections below, the law for an almost integrable and a
pseudo-integrable billiard \cite{4-harish1,4-harish2} is derived.

\subsection{The $\pi $/$3$-Rhombus Billiard}

We now present number-theoretic arguments to obtain $F(x)$ for the $\pi $/$3$%
-rhombus billiard. Subsequently, we shall discuss and compare our results
with the above-mentioned results by Katok \cite{4-kat} and Gutkin \cite
{4-gut2}. The length of periodic orbits in the given family corresponding to
the lattice point $(q,p)$ where $q,p$ are co-prime is given by 
\begin{equation}
\label{4-4}l=c_2L\sqrt{(q^2+3p^2)} 
\end{equation}
where $c_2$ depends on family of periodic orbits as seen in the Table 3-1 of
chapter (3). If $l<x$ for a given family of points $(q,p)$ should be counted
in $F(x)$. We can draw a circle of radius $l_1(=q_0L)$ then, all points $%
(q,p)$ having family of periodic orbit with length $c_2l_1\leq x$ (or $%
l_1\leq x/c_2$) should be considered for the calculation of $F(x)$.
Referring to Fig. 4.1, area of a quarter circle is ($\pi l_1^2/4$) and the
area of the square OABC is ($l_1^2$). We shall denote the integer
(fractional) part of a number by [...](\{...\}). The number of lattice
points in OABC is 
\begin{equation}
\label{4-5}
\begin{array}{ccc}
N & = & (q_0+1)[ 
\frac{(q_0+1)}{\sqrt{3}}] \\  & = & (q_0+1)^2/\sqrt{3}-(q_0+1)[\frac{(q_0+1)%
}{\sqrt{3}}] 
\end{array}
\end{equation}
\begin{figure}[htbp]
\begin{center}
{\epsfig{file=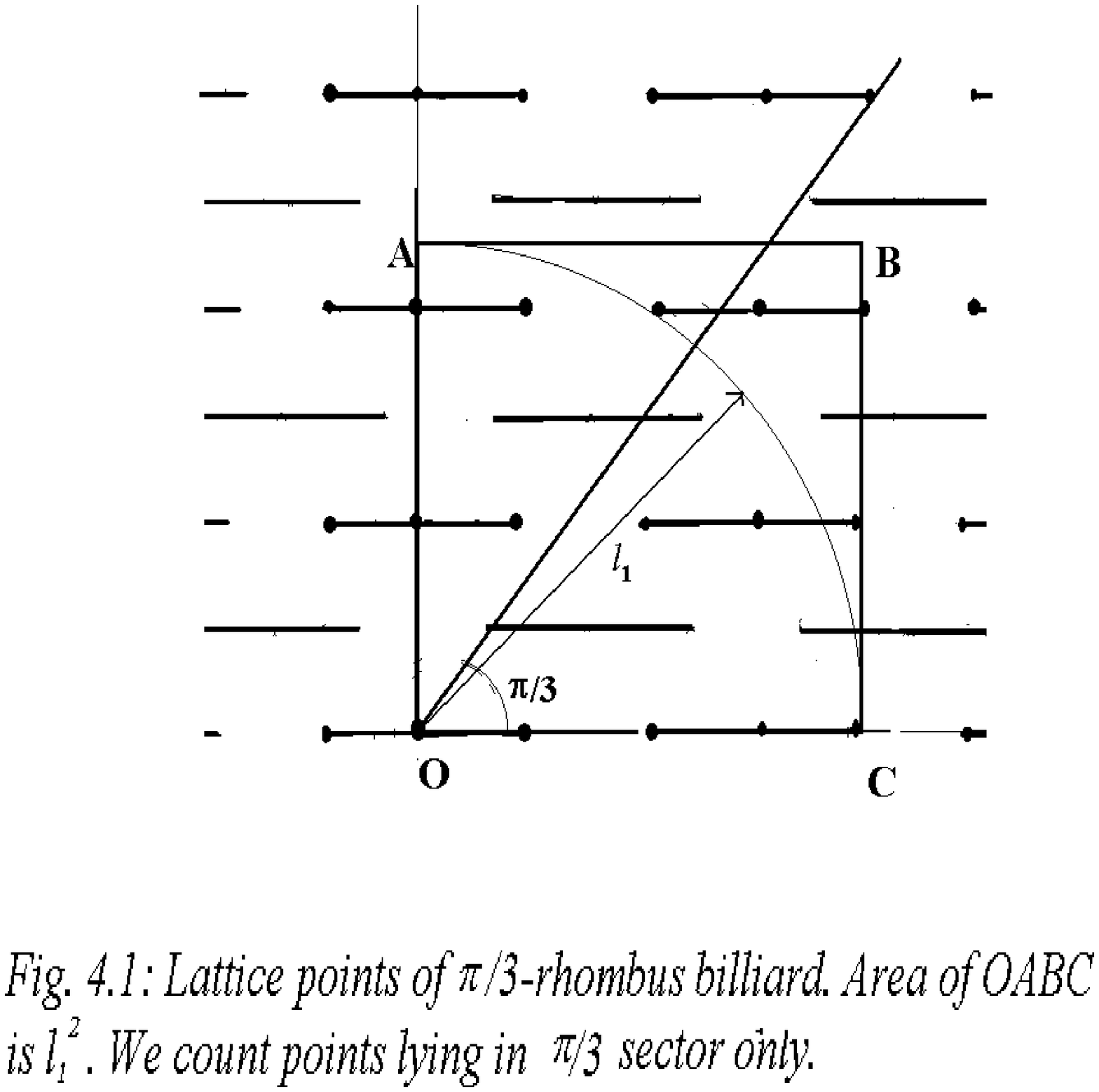,bbllx=80,bblly=170,bburx=560,bbury=700,height=6in}}
\end{center}
\end{figure}
On an average, $[(q_0+1)/\sqrt{3}]$ is 1/2 (obviously); hence

\begin{equation}
\label{4-6}N\sim \frac{q_0^2}{\sqrt{3}}+\frac{(2q_0+1)}{\sqrt{3}}-\frac{%
(q_0+1)}2 
\end{equation}
Therefore,number of lattice points in a quarter circle is

\begin{equation}
\label{4-7}N_q=N(\pi l_1^2/4)/l_1^2=\frac{\pi N}4 
\end{equation}
Since the probability that two randomly chosen numbers are co-prime is ($%
6/\pi ^2$) \cite{4-sch}, the number of co-prime lattice points is

\begin{equation}
\label{4-8}N_c=(\frac 6{\pi ^2})(\frac{\pi N}4)\sim \frac{\sqrt{3}l_1^2}{%
2\pi L}+\frac{A\sqrt{3}}{4\pi } 
\end{equation}

where 
\begin{equation}
\label{4-9}A=(4-\sqrt{3})l_1/L+2-\sqrt{3}. 
\end{equation}
For reasons discussed in the previous section, we are concerned in finding
the number of points in a ($\pi /3$) sector, the area of which is two-thirds
that of a quarter circle. Hence, for ($\pi /3$)sector, 
\begin{equation}
\label{4-10}N^{\prime }(l_1)\sim \frac{l_1^2}{\sqrt{3}\pi L^2}+\frac A{2%
\sqrt{3}\pi } 
\end{equation}

Taking only the dominant contribution ($O(l_1^2)$), with the help of Table
(3.1) (cf. chapter 3), we can write for the number of periodic orbits whose
length is $\leq x$ as 
\begin{equation}
\label{4-11}
\begin{array}{ccc}
F(x) & \sim & 2(P_{oocc}N_{oocc}(2x)+P_{oocc}N_{oocc}(x)+P_{oecc}N_{oecc}(x)
\\  
&  & +P_{oecc}N_{oecc}(x/2)+P_{eocc}N_{eocc}(x)+P_{eocc}N_{eocc}(x/2) \\  
&  & +P_{ooce}N_{ooce}(2x/3)+P_{ooce}N_{ooce}(x/3)+P_{ooce}N_{ooce}(x/3)) 
\end{array}
\end{equation}

where ($P_{oocc}$) is the probability that given co-prime lattice point is
of odd-odd, centre-centre type and so on, $N_{oocc}(l)$ is total number of
odd-odd, centre-centre type co-prime lattice points contained in the sector.
Out of the four points, only one is odd-odd (or even-odd or odd-even), also
two out of three points are of centre-edge type and one is of centre-centre
type. Therefore, we have

$$
P_{oocc}=P_{oecc}=P_{eocc}=\frac 19\,\,\,\,\,, 
$$

$$
P_{ooce}=P_{oece}=P_{eoce}=\frac 29. 
$$

Thus we can write

$$
F(x)=2((4+1+1+\frac 14+1+\frac 14)(\frac{x^2}{9\sqrt{3}\pi L^2})+(\frac
49+\frac 19+\frac 19)(\frac{2x^2}{9\sqrt{3}\pi L^2})) 
$$

or

$$
=\left( \frac{53}{27\sqrt{3}\pi L^2}\right) \,\,x^2. 
$$
In terms of $|P|$ this can be written as 
\begin{equation}
\label{4-12}F(x)=0.049\,\,733\left( \frac{2\pi }{|P|}\right) x^2. 
\end{equation}

Of course, apart from the dominant term that is quadratic in $x$, there will
be terms of $O(x)$ and $O(1)$. It is, however, important to note here that
term of $O(x)$ is not related to the orbits periodic after an odd number of
reflections ( hence, isolated). These terms only present more exact
expression for $F(x)$ arising from the above arguments. Their origin is in
the points contributing to $F(x)$ lying on the boundary of the sector. On
the same lines as above, terms of $O(x)$ and $O(1)$ are found to be $%
(26/81\pi )(4\sqrt{3}-3)(x/L)$ and $(12/27\pi )(2\sqrt{3}-3)$ respectively.
Thus the counting function is of form

\begin{equation}
\label{4-12a}F(x)=\left( \frac{53\sqrt{3}}{81\pi L^2}\right) x^2+\left( 
\frac{26}{81\pi L}\right) (4\sqrt{3}-3)x+\left( \frac{12}{27\pi }\right) (2%
\sqrt{3}-3). 
\end{equation}

\subsection{The H-M Barrier Billiard:}

We now present our calculation for the Hannay-McCraw billiard \cite{4-hm}.
Recall the discussion about this billiard in the previous chapter where we
have argued that the periodic orbits in this system are the ones which hit
any lattice point $(q,p)$ in this array of barriers, the gradient of such
trajectories will be given by $|p/q|$. Therefore we consider only the pairs$%
(q,p)$ such that $q$ and $p$ are co-prime since they only give a primitive
periodic orbits, and points $(mq,mp),$ where $m\in Z$, gives $m$ repetitions
of a primitive periodic orbit corresponding to $(q,p)$. Each such $(q,p)$
gives different number of bands or families of periodic orbits, depending on
whether the pair is odd-odd $(o,o)$ or even-odd $(e,o)$ or odd-even $(o,e)$.
The length of periodic orbit in a given family corresponding to a lattice
point $(q,p)$ is given by 
$$
l=cL\sqrt{\left( q^2+p^2\right) } 
$$
where $c$ depends on the number of families of periodic orbits. It can be
seen from table (3.2) of chapter 3, that, for $(q,p)$ as \cite{4-hm}

\begin{equation}
\label{4-13}c=,\left\{ 
\begin{array}{ccc}
4 & \textsf{one band, closing at }(4q,4p) & \textsf{for }(o,o) \\ 2,2 & \textsf{%
two bands, each closing at }(2q,2p) & \textsf{for }(o,e) \\ 1,1,2 & \textsf{two
bands closing at }(q,p)\textsf{ and one at }(2q,2p) & \textsf{for }(e,o) 
\end{array}
\right. 
\end{equation}
If $l\leq x$ for a given family of points $(q,p),$the contribution from this
family of $(q,p)$ should be counted in $F(x).$ Drawing a quarter circle of
radius $l_1$ in the quadrant under consideration, all points $(q,p)$ having
family of periodic orbits with length $cl_1\leq x$ (or $l_1\leq x/c$) must
be considered for the calculation of $F(x)$. The quarter circle is inscribed
in a square OABC with side length $l_1$ ( Fig.4.2). thus the area of this
square is $l_1^2$ and the area of the quarter circle OAC is $\pi l_1^2/4$.
We shall denote the integer (fractional) part of a number by [ ] (\{\}). The
number of lattice points in the square OABC is given by

$$
N=\left( \left[ \frac{l_1}L\right] +1\right) ^2\left( \left( \frac{l_1}%
L-\left\{ \frac{l_1}L\right\} +1\right) ^2\right) 
$$
Taking fractional part of $1_1/L$, on an average as $1/2$, we can write

$$
N=\left( \frac{l_1}L\right) ^2+\left( \frac{l_1}L\right) -\frac 14 
$$
Then the number of lattice points in a quarter circle is just

\begin{equation}
\label{4-14}N_q=\left( \frac N{l_1^2}\right) \left( \frac{\pi l_1^2}4\right)
=\frac{\pi N}4 
\end{equation}
\begin{figure}[htbp]
\begin{center}
{\epsfig{file=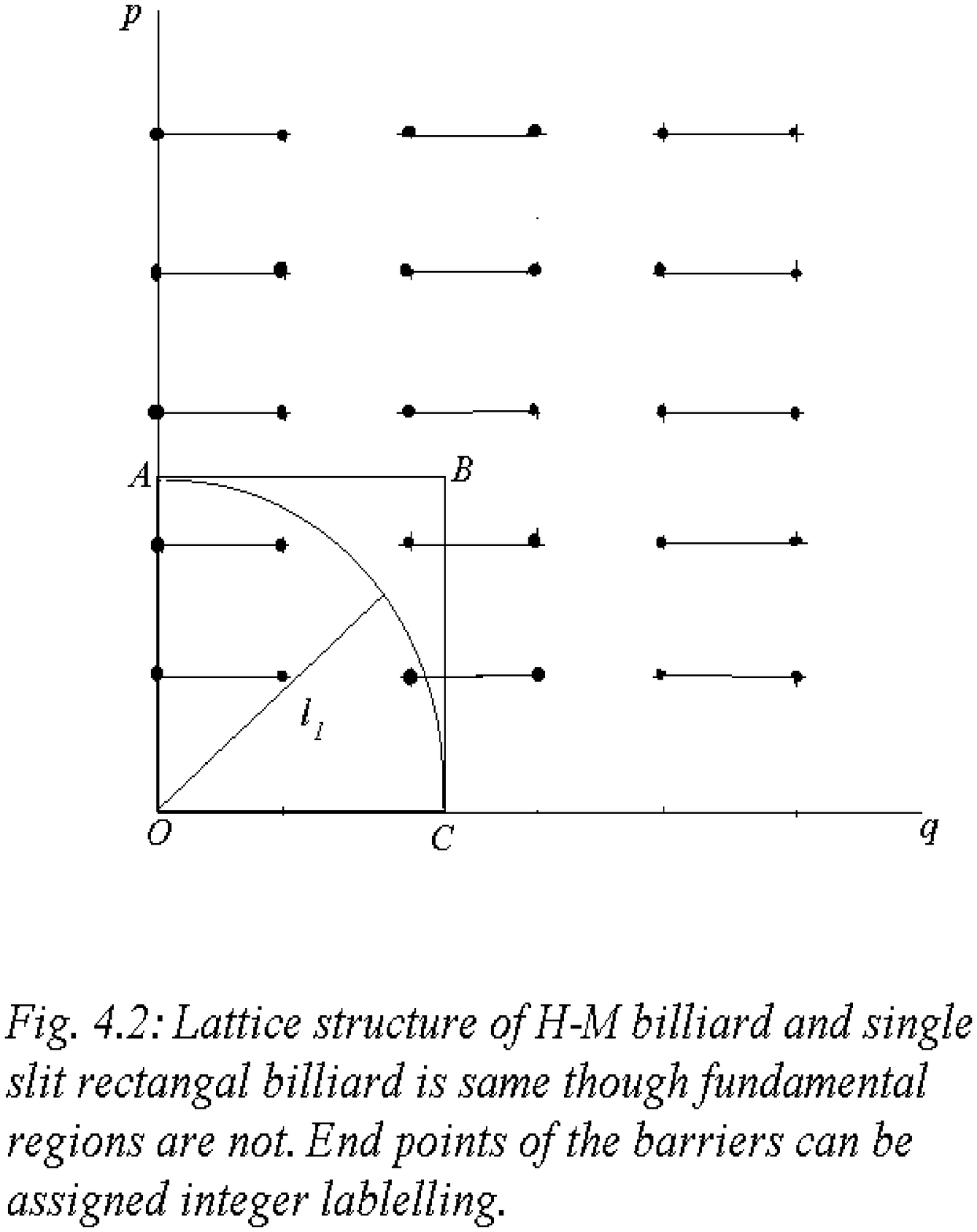,bbllx=80,bblly=170,bburx=560,bbury=700,height=6in}}
\end{center}
\end{figure}
Since the probability that two randomly chosen numbers are co-prime is $%
\left( 6/\pi ^2\right) $ the number of co-prime lattice points in quarter
circle is

\begin{equation}
\label{4-15}
\begin{array}{ccc}
N_c & = & \left( \frac 6{\pi ^2}\right) \left( 
\frac{\pi N}4\right) \\  & = & \left( \frac 3{2\pi L^2}\right) l_1^2+\left(
\frac 3{2\pi L}\right) l_1+\frac 3{8\pi } 
\end{array}
\end{equation}

The counting function can now be written explicitly as (using equation(\ref
{4-13}))

\begin{equation}
\label{4-16}F(x)=P_{oo}N_{oo}\left( \frac x4\right) +2P_{oe}N_{oe}\left(
\frac x2\right) +2P_{oe}N_{oe}\left( x\right) +P_{eo}N_{eo}\left( \frac
x2\right) 
\end{equation}

where again e.g.,$P_{oo}$ is the probability that a given co-prime lattice
point is of $(odd,odd)$ type and $N_{oo}(x)$ is the total number of odd-odd
co-prime lattice points contained in the quarter circle of radius $x(=N_c).$
Trivially,

\begin{equation}
\label{4-16a}P_{oo}=P_{oe}=P_{eo}=\frac 13 
\end{equation}
thus 
\begin{equation}
\label{4-17}
\begin{array}{ccc}
F(x) & = & \frac 13\left[ N_c\left( \frac x4\right) +3N_c\left( \frac
x2\right) +2N_c\left( x\right) \right] \\  
&  & \left( \frac{45}{32\pi L^2}\right) x^2+\left( \frac{15}{8\pi L}\right)
x+\frac 3{4\pi } 
\end{array}
\end{equation}
This is the asymptotic law of proliferation of periodic orbits for system
under consideration. How fast actual $F(x)$ converges to equation(\ref{4-17}%
) depends on the rate of convergence of $P_{oo},P_{oe},P_{eo}$ and $N_c$ in
accordance with equations(\ref{4-16a}) and (\ref{4-15}) respectively. It can
be easily seen that $P$ converges rapidly to $1/3$. In Table (4.1), we
compare the actual number of co-prime pairs with the results obtained by
equation(\ref{4-15}). It can be seen that (even) at $x=50$, the $\%$
difference between the numerical and analytical results is only $2\%$. For
similar reasons, we get an equally remarkable agreement in the case of the $%
\pi /3$-rhombus billiard as seen in Fig.(4.3).
\begin{figure}[htbp]
\begin{center}
{\epsfig{file=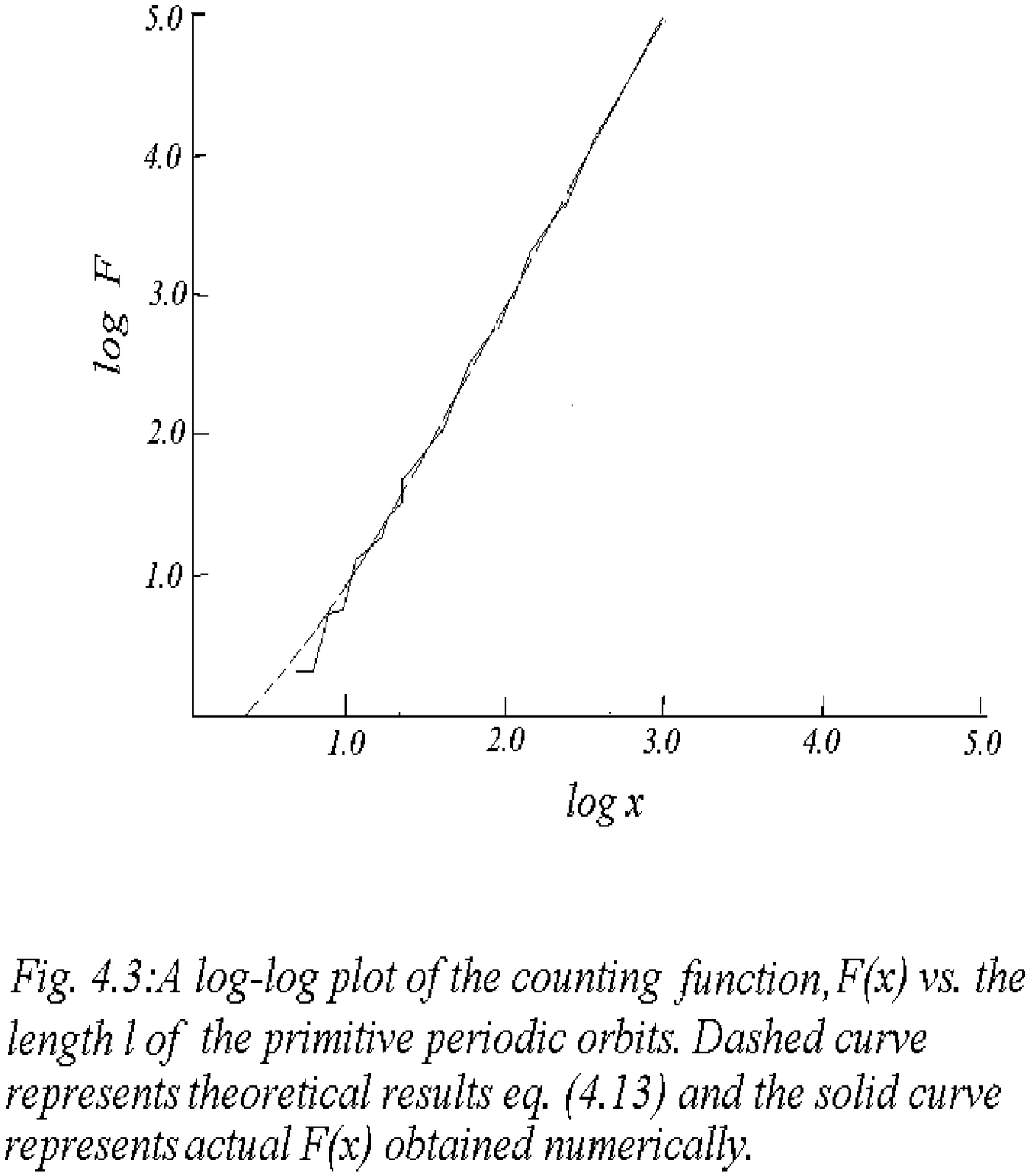,bbllx=80,bblly=170,bburx=560,bbury=700,height=6in}}
\end{center}
\end{figure}
\subsection{The Single Slit Rectangle Billiard}

This billiard has been already described in the last chapter. Again we
consider only the pairs$(q,p)$ such that $q$ and $p$ are co-prime since they
only give a primitive periodic orbits, and points $(mq,mp),$ where $m\in Z$,
gives $m$ repetitions of a primitive periodic orbit corresponding to $(q,p)$%
. Each such $(q,p)$ gives different number of bands or families of periodic
orbits, depending on whether the pair is odd-odd $(o,o)$ or even-odd $(e,o)$
or odd-even $(o,e)$. The length of periodic orbit in a given family
corresponding to a lattice point $(q,p)$ is given by

$$
l=cL\sqrt{\left( q^2+p^2\right) } 
$$
where $c$ depends on the number of families of periodic orbits, which is now
changed due to a simple modification of H-M billiard to single slit
rectangle, and is now given by (ref. Table 3.3 of chapter 3) 
\begin{equation}
\label{4-18}c=,\left\{ 
\begin{array}{ccc}
4 & \textsf{two bands, each closing at }(4q,4p) & \textsf{for }(o,o) \\ 4 & 
\textsf{two bands, each closing at }(4q,4p) & \textsf{for }(o,e) \\ 2,2 & \textsf{%
four bands each closing at }(2q,2p) & \textsf{for }(e,o) 
\end{array}
\right. . 
\end{equation}
Remaining analysis is same and will not be repeated here. Proliferation law
of periodic orbits can be deduced similarly and will be of same form $%
ax^2+bx+c$.

\vskip .5in

The similarity of equations (\ref{4-12a}) and (\ref{4-17}) indeed suggest an
immediate generalization. However, first of all, let us discuss the reason
underlying the difference between the quadratic coefficient obtained by us
and the one by Gutkin in the theorem paraphrased earlier in this section.
The calculations show that this difference is of an order or so. The reason
is as follows. In considering the number of lattice points formed by
stacking fundamental region of the corresponding integrable system, the
condition of co-primality was not taken into account by Gutkin. As explained
earlier an orbit labeled by a pair $(mq,mp)$, where $m\in Z$, gives $m$
repetitions of a primitive periodic orbit corresponding to point $(q,p)$
where $q$ and $p$ are co-prime. Hence, ignoring the co-primality condition
leads to an over counting of the periodic orbits. Further, due to symmetry
in the tessellated two dimensional plane, calculation need to be performed
for ($\pi /4$)sector in the single slit or H-M billiard and for ($\pi /6$)
sector in the $\pi /3$-rhombus billiard. In general, of course, for a domain
with a discrete symmetry of order $N$, only a ($\pi /N$) sector needs
consideration. Finally, one must note that a basic difference between the
lattice generated by fundamental polygonal billiard and corresponding
integrable system, lies in the incomplete tessellation of the plane by
non-integrable billiards. For instance, the barriers are of zero width and
finite length in the two examples considered in this section. It is this
structure that enables us to completely classify the orbits $via$ integer
labeling. The relative weight$(P_{oo},P_{oe},P_{eo})$ in H-M billiard and $%
(P_{oocc},P_{ooce})$ etc. in the ($\pi /3$)-rhombus billiard for different
types of co-prime lattice points differ in different system and lead to a
different quadratic coefficient. Hence, to give a general formula for the
law of proliferation of periodic orbits exactly demands a complete
enumeration and classification of periodic orbits. Although this important
question cannot be answered today, a general recipe in the following is
presented which comes very close to an exact formula for quadratic
coefficient.

\section{Proliferation of Periodic Orbits Considering Repetitions}

We now discuss the nature of this law if repetitions of the orbits are also
counted. this finds application in the semi-classical theory of spectral
statistics. If we follow on the similar lines as above, we obtain the
asymptotic behaviour detailed below.

If we are counting repetitions of the primitive orbits of length, $l<x$,
then the primitive orbits of length between $x$ and $x/2$ will not be
repeated; the primitive orbits with length $l;$ $x/2<l<x/3$ will be repeated
once; the primitive orbits with length $x/3<l<x/4$ will be repeated twice
and so on. Thus taking into account these repetitions, one can write an
expression for number of {\tt `}effective{\tt '} co-prime lattice points
within a quarter circle of radius $x,N(x)$ as,

\begin{equation}
\label{4-19}N_r(x)=N_c(x)-N_c\left( \frac x2\right) +2\left( N_c\left( \frac
x2\right) -N_c\left( \frac x3\right) \right) +...+n\left( N_c\left( \frac
xn\right) -N_c(1)\right) 
\end{equation}

Here $n$ is the largest integer less than $x$, we have neglected $N(l)(l<1)$
since there are no periodic orbits of length less than $1$ in the system we
have considered. Equation (\ref{4-19}) can be rewritten as

\begin{equation}
\label{4-20}N_r(x)=N_c(x)+N_c\left( \frac x2\right) +N_c\left( \frac
x3\right) +...+N_c\left( \frac xn\right) -nN_c(1) 
\end{equation}
Then, the modified counting function, $F_r(x),$ for say H-M billiard becomes 
\begin{equation}
\label{4-21}F_r(x)=\frac 13\left[ N_r\left( \frac x4\right) +3N_r\left(
\frac x2\right) +2N_r\left( x\right) \right] 
\end{equation}
substituting equation(\ref{4-20}) in equation(\ref{4-21}), we get 
\begin{equation}
\label{4-22}
\begin{array}{lll}
F_r(x) & = & \left( \frac 13\right) \left( \left( 
\frac{A(n_4)}{16}+\frac{3A(n_2)}2+2A(n)\right) \left( \frac 3{2\pi }\right)
x^2\right. \\  &  &  \\  
&  & \left. +\left( \frac{B(n_4)}4+\frac{3B(n_2)}2+2B(n)\right) \left( \frac
3{2\pi }\right) x-\left( n_4+n_2+n\right) \left( \frac 3\pi \right) \right) 
\end{array}
\end{equation}
where $n_4$ and $n_2$ are the largest integers less than $x/4$ and $x/2$
respectively. $A$ and $B$ are given by 
$$
A(n)=\sum\limits_i\frac
1{i^2};\,\,\,\,\,\,\,\,\,\,\,\,\,\,\,\,\,\,\,\,\,\,\,\,\,B(n)=\sum\limits_i%
\frac 1i 
$$
Asymptotically ($n\rightarrow \infty $), $A(n)=\pi ^2/6$ and $B(n)=\log
(n)+\gamma $ where ($\gamma $) is the Euler-Mascheroni constant, equal to $%
0.5772157....$. The proliferation law with repetitions for other billiards
considered above can similarly be deduced.

The rate of convergence of actual $F_r(x)$ to equation(\ref{4-22}) depends
upon the rate of convergence of actual $N_r(x)$ to equation(\ref{4-20}). In
Table(4.1), we compare actual number of 'effective' co-prime pairs $N_r(x)$
with that given by equation (\ref{4-20}). Note that the $\%$ difference
between actual 'effective' co-prime numbers and those obtained from equation(%
\ref{4-20}) at $x=50$, is $5\%$ ( it may be recalled that this figure is
almost 2.5 times the one observed for $N_c$). It is for this reason that the
convergence to the quadratic law (equation(\ref{4-22})) is much slower if
one considers repetitions.

To conclude, we have shown analytically that the rate of proliferation of
the periodic orbits is exactly quadratic (in length) asymptotically. We have
seen that this is in complete agreement with extensive numerical
calculations on two model pseudo-integrable systems. The reason underlying
the asymptote, $ax^2+bx+c,$ to the counting function is clearly related to
the tessellation of the two dimensional plane by the fundamental region of
the billiards.

\section{Generalization}

It is well known that a rational polygon can periodically tile a surface
everywhere flat, in the sense of null Gaussian curvature, except at isolated
vertex points of singular negative curvature. A periodic structure that
tiles the almost everywhere flat surface may consist of several polygons and
hence space can be assigned distinct labels (albeit complicated) taking
account of different periodicities, in a spirit similar to the one presented
above. To enumerate distinct primitive periodic orbits, one needs a
condition analogous to the co-primality condition required by the two
systems discussed above, since out of all lattice points lying on the same
line of a given slope only one will give a primitive periodic orbit. Let us
denote the probability of the ''co-primality condition'' to be satisfied by
distinct labels $P_c$. Furthermore, the classification entailing each
distinct label will give rise to relative weights in which the orbits will
be distributed, let us denote it by $P_j$($j$ denote classes). For a polygon
with symmetry group of order $N$, the points to be considered will be
restricted to a $\pi /N$ sector. This number can be written as 
$$
N_L=\sum\limits_i\left( \alpha _ix^2+\beta _ix+\gamma _i\right) /N 
$$
where the summation is over all the periodicities. And number of
''co-prime'' points are $N_{LC}(x)=P_CN_L(x)$. For each class of periodic
orbit for which the weight is $P_j,$ there may be $k$ types of periodic
orbits closing at length $\xi _{kj}x$. We assume that $k\,$ remains constant
for given class. With this the counting function becomes%
$$
F(x)=\sum\limits_jP_j\sum\limits_kN_{LC}\left( \frac x{\xi _{kj}}\right) 
$$
and hence coefficient of $F(x)=ax^2+bx+c$ are%
$$
\begin{array}{ccc}
a & = & \frac{P_C}N\sum\limits_i\alpha _i\sum\limits_jP_j\sum\limits_k\xi
_{jk}^{-2} \\ b & = & \frac{P_C}N\sum\limits_i\beta
_i\sum\limits_jP_j\sum\limits_k\xi _{jk}^{-1} \\ c & = & \frac{P_C}%
N\sum\limits_i\gamma _i\sum\limits_jP_j 
\end{array}
$$

Hence, counting function will quadratic and convergence will depend on both $%
P_C$ and $P_L$. However, If $k\,$ is not constant but depends on $x$ then
above analysis will have to be modified and $F(x)\,$ may not be quadratic.

\newpage

\begin{table}\centering
\begin{tabular}{|l|l|l|l|l|l|l|}\hline
\(x\) & \(N_{c}\) &  \(N_{c}\)&    \(/diff\)    &    \(N_{r}\)   &   \(N_{r}\)   &   \(/diff\)\\
         &                   & \(eq.(\ref{4-15})\)& \(Col. 2,3\)&     &\(eq.(\ref{4-20})\)&\(Col.5,6\)\\\hline
3.0     &  2  &     5      &      150.0        &    2        &    4      &       100.0\\
4.0     &  6  &     9      &      50.0          &     6       &     9      &50.0\\
5.0&10&14&40.0&12&17&41.6\\
6.0&16&19&18.7&18&23&27.7\\
7.0&20&26&30.0&26&34&30.7\\
8.0&28&34&21.4&36&46&27.7\\
9.0&36&42&16.6&48&59&22.9\\
10.0&46&52&13.0&62&75&20.9\\
20.0&190&200&5.2&280&313&11.8\\
30.0&424&443&4.4&654&707&8.1\\
40.0&764&782&2.3&1188&1264&6.3\\
50.0&1192&1217&2.0&1876&1977&5.3\\
60.0&1718&1747&1.6&2720&2848&4.7\\
70.0&2330&2372&1.8&3724&3876&4.1\\
80.0&3048&3093&1.4&4884&5063&3.7\\
90.0&3876&3910&.8&6208&6406&3.2\\
100.0&4770&4822&1.1&7684&7912&2.9\\
200.0&19088&19193&.55&31066&31593&1.7\\
300.0&42972&43114&.33&70162&71009&1.2\\
400.0&76402&76585&.24&124974&126145&.94\\
500.0&119372&119604&.19&195484&197005&.78\\\hline
\end{tabular}
\vskip .5in
\caption{ Comparision between actual and theoretical values for $N_{c}$ and $N_{r}$}
\end{table}

\chapter{Two-Point Cluster Function and Form Factor: A Diagonal Approximation
}

\section{Introduction}

We Consider the discrete spectrum $E_1,E_2,E_3,\cdot \cdot \cdot $ of a $f$%
-dimensional bound quantum system. As mentioned in the chapter (1) one can
define a ''global'' function \thinspace $N(E)\,$ which gives number of
levels less than or equal to $E$ on the set of this discrete levels. The
function $N(E)\,$ is obviously a staircase like function, since it jumps by
unity at the occurrence of an eigenvalue $E_i$ and remains constant
otherwise. Of course if spectrum consist of degeneracies then jump in $N(E)$
at the particular degeneracy will not be unity but equal to number of
degenerate levels at particular point on energy axis. Typical functional
nature of $N(E)\,$ is therefore non-analytical. When we approximate $N(E)$
by a semi-classical expression we use approximate analytical form for the
exact staircase. We use same symbol $N(E)$ for both actual staircase and its
semiclassical approximation, mostly we will be talking about semiclassical
approximation below.

As mentioned earlier it is possible to separate semi-classical approximation
to $N(E)\,$ in a smooth part $N_{av}(E)$ and the fluctuating part $%
N_{osc}(E) $ (i.e. $N(E)=N_{av}(E)+N_{osc}(E)$). In the particular case of a
particle (of mass $m=1/2)$ in polygonal box (or polygonal billiard) of area $%
A_R,$ $N_{av}(E)$ is given by 
\begin{equation}
\label{5-1}N_{av}(E)=\frac{A_RE}{4\pi \hbar ^2}-\frac{L_D\sqrt{E}}{4\pi
\hbar }+C 
\end{equation}
where $L_D$ is length of the parameter of the boundary, and $C$ is a
constant containing complex information on the geometrical and topological
properties of the billiard.

This average behaviour may be eliminated in order to characterize and
compare the fluctuation patterns of different systems whose corresponding $%
N_{av}(E)$ is not same. For this purpose it is convenient to {\tt ``}unfold%
{\tt ''} the original spectrum $\{E_i\}$ through the mapping $E\longmapsto r$
\begin{equation}
\label{5-2}r_i=N_{av}(E_i). 
\end{equation}
The effect of (\ref{5-2}) is that the sequence $\{r_i\}$ has on the average
a constant mean spacing equal to unity, irrespective of the particular form
of the function $N_{av}(E)$. By construction $r_i\simeq i-1/2(i=1,2,...)$
and the departures 
\begin{equation}
\label{5-3}\delta _i=r_i-(i-1/2) 
\end{equation}
of $r_i$ from its average value $i-1/2$ are the level fluctuations. As a
result of unfolding the spectrum in a way mentioned above it is obvious that
the average part of level density of unfolded spectrum $d_{av}(y)$ is unity.

\section{Spectral Fluctuation Measures}

The level fluctuations can be characterized in a systematic way using the $k$%
-point correlation functions and measures derived from them. The $k$-point
correlation function is defined as \cite{5-dys,5-mlm} 
\begin{equation}
\label{5-4}R_k(x_1,...,x_k)=\frac{n!}{(n-k)!}\int\limits_{-\infty }^\infty
...\int\limits_{-\infty }^\infty P_n(x_1,...x_n)dx_{k+1}...dx_n. 
\end{equation}
where $P_n$ is a joint probability density function of the levels $%
\{x_i\}_{i=1...n}$ and $1\leq k\leq n$. The $R_k$ is thus a probability
density of finding a level(regardless of labeling) around each of the
points $x_1...x_k,$ the positions of the remaining levels being unobserved.
Each function $R_k;\,\,\,\,k>1$ contains terms of various kinds describing
the grouping of $k$ levels into various subgroups or clusters.

In practice it is convenient to introduce $k$-level cluster functions
obtained from $R_k$ by subtracting out the lower-order correlation terms, as 
\begin{equation}
\label{5-5}T_k(x_1...x_k)=\sum\limits_G(-1)^{k-m}(m-1)!\prod%
\limits_{j=1}^mR_{G_j}(x_t,\textsf{with }k\textsf{ in }G_j). 
\end{equation}
Here $G\,$ stands for any division of the indices $[1,...,k]$ into subgroups 
$[G_1,...,G_m].$ For example, 
\begin{equation}
\label{5-6}
\begin{array}[t]{lll}
T_1(x) & = & R_1(x) \\ 
{\rm and} &  &  \\ 
T_2(x_1,x_2) & = & -R_2(x_1,x_2)+R_1(x_1)R_1(x_2) 
\end{array}
\end{equation}
Measuring the energies in the units of the mean level spacing $\lambda =1$
and introducing the variables $y_i=x_i/\lambda $, (this is nothing but the
unfolding mentioned earlier) the cluster functions can be written as 
\begin{equation}
\label{5-7}Y_k(y_1,...y_k)=\lim _{n\rightarrow \infty }\lambda
^nT_n(x_1,...x_k). 
\end{equation}
The $Y_k$'s are then well defined and finite everywhere. The cluster
functions being isolated from the effects of the lower correlations,
vanishes as the separation \linebreak $r=\mid r_1-r_2\mid $ becomes larger
and larger. Among these two-level cluster function $Y_2$ is of prime
importance. When the collection of levels are treated as a classical Coulomb
gas, $Y_2$ defines the shape of the neutralizing charge cloud induced by
each particle around itself \cite{5-mlm}. As mentioned earlier many
fluctuation measures can now be expressed in terms of the $Y_{k\textsf{ }}$'s.
We shall discuss this at appropriate place.

Consider now one point measure, the energy level density function 
\linebreak
$d(y)=\sum\limits_i\delta (y-y_i)$, where $\delta $ is usual Kronecker's
delta function. The average with respect to the probability distribution
function $P(y_1...y_n)$ (so called ensemble average) of the level density
function is given by 
\begin{equation}
\label{5-8}
\begin{array}{lll}
\langle \langle d(y)\rangle \rangle & = & \int\limits_{-\infty }^\infty
dy_1\cdot \cdot \cdot \int\limits_{-\infty }^\infty dy_nd(y)P(y_1...y_n) \\  
&  &  \\  
& = & n\int\limits_{-\infty }^\infty dy_1\cdot \cdot \cdot
\int\limits_{-\infty }^\infty dy_n\delta (y-y_1)P(y_1...y_n) \\  
&  &  \\  
& = & R_1(y)\,\,\,\,\,\,\,\,=\,\,\,\,\,\,\,\,\,\,T_1(y) 
\end{array}
. 
\end{equation}
Also the density-density correlation function (for energy) is defined as 
$$
\begin{array}[t]{lll}
\langle \langle d(y)d(y^{\prime })\rangle \rangle & = & \sum\limits_{i=1}^n%
\sum\limits_{j=1}^n\int\limits_{-\infty }^\infty dy_1\cdot \cdot \cdot
\int\limits_{-\infty }^\infty dy_n\delta (y-y_i)\delta (y^{\prime
}-y_j)P(y_1...y_n) \\  
&  &  \\  
& = & n\sum\limits_{i=1}^n\int\limits_{-\infty }^\infty dy_1\cdot \cdot
\cdot \int\limits_{-\infty }^\infty dy_n\delta (y-y_1)\delta (y^{\prime
}-y_1)P(y_1...y_n) \\  
&  &  \\  
&  & \,\,\,+n(n-1)\sum\limits_{i\neq j=1}^n\int\limits_{-\infty }^\infty
dy_1\cdot \cdot \cdot \int\limits_{-\infty }^\infty dy_n\delta (y-y_1)\delta
(y^{\prime }-y_2)P(y_1...y_n) \\  
&  &  \\  
& = & \delta (y-y^{\prime })R_1(y)+R_2(y,y^{\prime }) 
\end{array}
$$
Hence, from equation (\ref{5-6}), we get 
\begin{equation}
\label{5-9}\langle \langle d(y)d(y^{\prime })\rangle \rangle =\delta
(y-y^{\prime })R_1(y)+R_1(y)R_1(y^{\prime })-T_2(y,y^{\prime }). 
\end{equation}
Since the ensemble average is same as the spectral average(see e.g.\cite
{5-hak}), which is defined as (for level density) 
\begin{equation}
\label{5-10}\langle d(y)\rangle =\frac 1\eta \int\limits_{y-\eta /2}^{y+\eta
/2}dx\,\,\,\,d(x), 
\end{equation}
we can write in an asymptotic limit 
\begin{equation}
\label{5-11}\langle d(y)\rangle =R_1(y)=Y_1(y)=d_{av}(y)=1 
\end{equation}
and 
\begin{equation}
\label{5-12}\langle d(y)d(y^{\prime })\rangle =\delta (y-y^{\prime
})+1-Y_2(y,y^{\prime }). 
\end{equation}
Thus two-point correlation function $R_2(y_1,y_2)$ or cluster function $%
Y_2(y_1,y_2)$ are related to the spectral average of density-density
correlation function as 
\begin{equation}
\label{5-13}R_2(y,y^{\prime })=\langle d(y)d(y^{\prime })\rangle -\delta
(y-y^{\prime }) 
\end{equation}
and 
\begin{equation}
\label{5-14}Y_2(y,y^{\prime })=1+\delta (y-y^{\prime })-\langle
d(y)d(y^{\prime })\rangle . 
\end{equation}
Equations (\ref{5-13}) and (\ref{5-14}) are the important relation to
develop semi-classical understanding of spectral fluctuation properties in
which we are interested. Since we can write density of states as sum of
average and fluctuating part ($d(y)=d_{av}(y)+d_{osc}(y)$) the
density-density correlation function becomes%
$$
\begin{array}{lll}
\langle d(y)d(y^{\prime })\rangle & = & \langle
(d_{av}(y)+d_{osc}(y))(d_{av}(y^{\prime })+d_{osc}(y^{\prime }))\rangle \\  
&  &  \\  
& = & d_{av}(y)d_{av}(y^{\prime })+d_{av}(y)\langle d_{osc}(y^{\prime
})\rangle \\  
&  &  \\  
&  & +d_{osc}(y)\langle d_{av}(y^{\prime })\rangle +\langle
d_{osc}(y)d_{osc}(y^{\prime })\rangle , 
\end{array}
$$
since $\langle d_{osc}(y)\rangle =\langle d_{osc}(y^{\prime })\rangle \simeq
0,$ and also $d_{av}(y)=1,$ we finally get 
\begin{equation}
\label{5-15}\langle d(y)d(y^{\prime })\rangle =1+\langle
d_{osc}(y)d_{osc}(y^{\prime })\rangle . 
\end{equation}
Using this, equations (\ref{5-13}) and (\ref{5-14}) can be rewritten as 
\begin{equation}
\label{5-16}R_2(y,y^{\prime })=\langle d_{osc}(y)d_{osc}(y^{\prime })\rangle
+1-\delta (y-y^{\prime }) 
\end{equation}
and 
\begin{equation}
\label{5-17}Y_2(y,y^{\prime })=\delta (y-y^{\prime })-\langle
d_{osc}(y)d_{osc}(y^{\prime })\rangle . 
\end{equation}
These are the equations that we will use below to study spectral fluctuation
properties as we can substitute semi-classical expression obtained in
chapter\ (1) for $d_{osc}(y)$. Note, existence of $\delta (y-y^{\prime })$
in eqs. (\ref{5-16}) and (\ref{5-17}). Since $R_2$ or $Y_2\,$ must be free
of $\delta $-function, density-density correlation function also must
contain similar $\delta $-function. Indeed this is due to the
self-correlation of a level.

All the relevant information to evaluate semi-classical periodic orbit sum
is obtained in the previous two chapters. For an easy reference we will
reproduce briefly those results in the following section. Further the
example, we consider here ($\pi /3$-rhombus billiard) also possesses a
symmetry with respect to both the diagonals. This results in modification of
the semi-classical expression for the density of states, since one can now
work with decomposed state space due to symmetry. These issues are also
discussed in the following section before we take up study of statistical
measures.

\section{Semi-classical Density of States}

As stated in chapter (1) the oscillatory part of density of states in case
of pseudo-integrable is given by\cite{5-pjr-mvb} 
\begin{equation}
\label{5-18}d(E)=\frac m{2\pi \hbar ^2}Re\sum\limits_i\int \int d{\bf q}%
H_0^{(1)}\left( \frac{l_i}\hbar \sqrt{2mE}\right) \exp (i\alpha _i\pi ). 
\end{equation}
where all symbols are as explained in chapter (1). Here we take $m=1/2$,
then by taking real part and in the asymptotic limit ($\hbar \rightarrow 0$)
we get (note that $\alpha $'s are integers) 
\begin{equation}
\label{5-19}d_{osc}(E)=\left( \frac 1{8\pi ^3\hbar ^3}\right)
^{1/2}E^{-1/4}\sum\limits_i\frac{A_i}{l_i^{1/2}}\cos \left( \frac{l_i}\hbar 
\sqrt{E}+\pi (\alpha _i-\frac 14)\right) . 
\end{equation}

We now {\tt ``}unfold{\tt ''} the energy using only first term of $N_{av}$
for the reasons of simplicity (as a trade off between simplicity and
accuracy), through mapping $E\rightarrow r_0:$%
\begin{equation}
\label{5-20}r_0=\frac{A_RE}{4\pi \hbar ^2}. 
\end{equation}
This means unfolded levels will not have mean spacing equal to unity exactly
but only approximately. In fact from equation (\ref{5-1}) it is easy to see
that mean level spacing after unfolding is less then unity by a factor ($%
1-L_D/(\sqrt{4\pi A_Rr_0})+C/r_0)$. Since for semi-classical analysis is
valid for large value of $r_0$, it is obvious that unfolding levels using
just a leading term of $N_{av}(E)$ $\,$ will not introduce serious error in
the analysis for large value of $r_0$.

From equation (\ref{5-20}) we can see that%
$$
\frac{dr_0}{dE}=\frac{A_R}{4\pi \hbar ^2}\simeq d_{av}(E), 
$$
and hence, 
\begin{equation}
\label{5-21}d(r_0)=d(E)\left( \frac{dr_0}{dE}\right) ^{-1}\simeq
1+d_{osc}(E)\left( \frac{dr_0}{dE}\right) ^{-1}=1+d_{osc}(r_0). 
\end{equation}
The oscillatory part of the density of states (eq.(\ref{5-19})) can then be
written in terms of rescaled energies as 
\begin{equation}
\label{5-22}d_{osc}(r_0)=\left( \frac{r_0^{-1/4}}{\pi ^{3/4}A_R^{3/4}}%
\right) \sum\limits_i\frac{A_i}{l_i^{1/2}}\cos \left( l_i\sqrt{\frac{4\pi r_0%
}{A_R}}+(\alpha _i-\frac 14)\pi \right) . 
\end{equation}

As mentioned above when a quantum dynamical system possesses a symmetry, its
state space may be decomposed into subspaces of definite symmetry type. The
Schr\"odinger equation can then be restricted to these subspaces resulting
in a symmetry- projected spectrum. It is interesting to know how these
symmetries influence the fluctuation properties of the whole spectrum.

\subsection{Symmetry-projected Green's function}

Let $\hat{H}\,$ be a quantum Hamiltonian for a system of $f$ degrees of
freedom, invariant under a discrete group $G.$ We will assume that $G$ acts
classically as a group of Euclidean point transformations ( i.e. combination
of translations rotations, and reflections). The quantum mechanical action
of $G$ in the ${\bf x}$ representation is given by 
\begin{equation}
\label{5-22a}\hat U(g)|{\bf x}\rangle =|g\cdot {\bf x\rangle } 
\end{equation}
where $g$ is an element of $G$ and $g\cdot {\bf x}$ denotes its action on $%
{\bf x}$. By virtue of this symmetry, $\hat{H}$ may be restricted to a
subspace which is invariant under the symmetry and whose states transform
according to an irreducible representation of $G$. The projection onto an
invariant subspace is given by 
\begin{equation}
\label{5-22b}P_m=\frac{d_m}{|G|}\sum\limits_{g\in G}\chi _m(g)\hat U^{\top
}(g), 
\end{equation}
where, $\chi _m(g)\,$ is the character of the $m^{th}$ irreducible
representation of dimension $d_m$, and $|G|\,$ is the order of group $G$\cite
{5-tin}.

A semi-classical approximation for $G_m(E)$, the trace of the
symmetry-projected Green's function is defined as 
\begin{equation}
\label{5-22c}G_m(E)={\rm Tr}[P_m(E-\hat{H})^{-1}]. 
\end{equation}
The poles of $g_m(E)$ are the energy eigenvalues of symmetry $m$ and the
pole strength give the eigenvalue degeneracies. One can now follow steps
given in chapter (1) to obtain the periodic orbit sum formula. The
symmetry-projected Green's function, $G_m({\bf x,x}^{\prime };E)=\langle 
{\bf x}|P_m(E-\hat{H})^{-1}|{\bf x}^{\prime }\rangle $ can be obtained by
restricting to an invariant subspace only which is given by\cite{5-rob} 
\begin{equation}
\label{5-22d}G_m({\bf x,x}^{\prime };E)=\frac{d_m}{|G|}\sum\limits_{g\in
G}\chi _m(g)G(g\cdot {\bf x,x}^{\prime };E). 
\end{equation}
It may be noted that the full semi-classical Green's function involves sum
over orbits between ${\bf x}^{\prime }$ and ${\bf x}$, its symmetry
projection involves a larger family of orbits. The sum over classical orbits
is taken for orbits that begin at ${\bf x}^{\prime }$ and end at $g\cdot 
{\bf x}$, a point related to ${\bf x}$ by symmetry. These orbits can be
associated in a one-to-one fashion with orbits on a symmetry reduced phase
space.

A general form of symmetry (discrete) projected oscillatory part of density
of states can be obtained from the corresponding trace formula\cite{5-rob}.
In the case of pseudo-integrable polygonal billiards it reduces to (in an
unfolded spectrum) 
\begin{equation}
\label{5-22e}d_{osc}^m(r)=d_m\left( \frac{r^{-1/4}}{\pi ^{3/4}A_R^{3/4}}%
\right) \sum_j\frac{A_j}{l_j^{1/2}}\cos \left( l_j\sqrt{\frac{4\pi r}{A_R}}%
-\frac \pi 4+\mu _j\pi \right) \chi _m(g_p^m). 
\end{equation}
Here, the terms $l_p,A_p$ are length and band area of the periodic orbits in
the primitive polygon that tessellate full phase space. The sum over all $m$%
, of equation (\ref{5-22e}) reproduces equation (\ref{5-22}).

The spectrum of the $\pi /3$-rhombus billiard is composed of superposition
of four different modes due to system symmetries \cite{5-db-srj}.
Semi-classically, it is not possible to distinguish all four modes \cite
{5-pjr-mvb} separately. However, it is possible to separate equilateral
triangle modes(odd parity classes) where wave function is zero on the
shorter diagonal and pure rhombus modes(even parity classes) where gradient
of wave function is zero on the shorter diagonal. This can be done by
incorporating and exploiting the symmetry of the system in a semi-classical
treatment, to obtain symmetry projected spectrum.

We will follow treatment of \cite{5-rob} to study effect of symmetries on
the fluctuation properties of the spectrum. We take longer diagonal along $y$%
-axis and shorter diagonal along $x$-axis. $\pi /3$-rhombus billiard has
symmetry group $C_2\bigotimes C_2$, consisting of reflections about $x$ and $%
y$ axes. The elements of the group can be denoted by $x^ay^b$, where $x$ and 
$y$ represent commuting reflections about a respective axis. The primitive
cell $V$ \cite{5-rob} is then half equilateral triangle having sides half of
longer (${\cal L}_y$,) and shorter (${\cal L}_x$) diagonals and one side of
the rhombus ${\cal L}$. The group has four one dimensional ($d_m$ in (\ref
{5-22e}) is therefore 1) irreducible representations, labeled by numbers $%
(p,q)$. These numbers describe parities under reflections about $x$ and $y$
axes respectively. For states, even under $x$ reflection, $p$ = $0$, and odd
under $x$ reflection $p$ = $1$. Similarly, $q$ takes values $0$ or $1$
depending on parity under $y$ reflection. The characters of the group can be
easily deter$\min $ed in this case and are given as 
\begin{equation}
\label{5-22f}\chi _{p,q}(g_l^n)=\exp \{-\pi n(pa_l+qb_l)\} 
\end{equation}
where, $a_l,b_l$ are number of times $l^{{\rm th}}$ primitive periodic orbit
bounces from ${\cal L}_x$ and ${\cal L}_y$ respectively, $n$ counts
repetitions of periodic orbits and $g_l^n$ in group element $x^{a_l}y^{b_l}$%
. Pure rhombus modes are therefore represented by $p=0;q=0$ or $1$ and the
equilateral triangle modes by $p=1;\ q=0$ or $1$.

Recalling the comments about the Maslov indices in the chapter (1), one can
see that the character of the group actually play a role of the Maslov
indices $\mu $ in this case. Thus, for the pure rhombus modes Neumann
boundary condition is applied to the shorter diagonal ${\cal L}_x$ and then
bounces on the same are not taken into the consideration. For equilateral
triangle modes however, since one have Dirichlet boundary condition on $%
{\cal L}_x$, bounces on it should be counted.

From the polar constructions we have used \cite{5-s-h} in chapter (3) to
deter$\min $e and classify the periodic orbits $a_l$ and $b_l$ can be deter$%
\min $ed easily. It can be seen that $b_l$ is always even, hence $qb_l$ does
not play any role here. This is the basic reason that one can not separate
these two parity classes (belonging to integrable and pseudo-integrable
parts) further into the four different modes. Also it turns out that for all
the centre-edge orbits, \ $a_l$ \ is even. However, for all the
centre-centre orbits there are three bands of periodic orbits each occupying
area of $4A_R$ ($A_R$ is area of $V$). Two of these bands bounces odd number
of times on ${\cal L}_x$ and one band bounces even number of times (i.e. $%
a_l $ is odd for two bands and even for one band). The same is true for side 
${\cal L}$, since total number of bounces on three sides of $V$ are always
even. Using all these facts discussed above, equation (\ref{5-22e}) becomes 
\begin{equation}
\label{5-22g}d_{osc}^m(r)=\left( \frac{r^{-1/4}}{\pi ^{3/4}A_R^{3/4}}\right)
\sum_j\frac{A_j}{l_j^{1/2}}\cos \left( l_j\sqrt{\frac{4\pi r}{A_R}}-\frac
\pi 4+\mu _j\pi -pa_j\pi \right) 
\end{equation}
Note, repetition index $n$ is absorbed in $a_j$ here. $A_j,l_j$ and $A_R$
represent respectively the band area of periodic orbits of primitive cell $V$%
, length of the periodic orbits and area of half equilateral triangle. The
Maslov indices, $\mu _j$ here counts bounces only on the side ${\cal L}$ of
the primitive cell $V$. Thus for the pure rhombus modes since $p=0$, $\mu _j$
will play important role, while for equilateral triangle modes ($p=1$), $%
(\mu _j-a_j)$ is always even for all the periodic orbits. If therefore, $\mu 
$'s are neglected one will end up in erroneous results.

Using above discussion and following the steps of \cite{5-rob} we write
density of states for the pure rhombus modes as 
\begin{equation}
\label{5-22h}d_{osc}^m(r)=\left( \frac{2r^{-1/4}}{\pi ^{3/4}A_R^{3/4}}%
\right) \sum_j\frac{A_j}{l_j^{1/2}}\cos \left( l_j\sqrt{\frac{4\pi r}{A_R}}%
-\frac \pi 4\right) 
\end{equation}
for centre-edge orbits. For all centre-edge points, $A_j$ will be $12A_R$
i.e. same as phase space area. And for centre-centre orbits we write 
\begin{equation}
\label{5-22i}
\begin{array}{lll}
d_{osc}^m(r) & = & \left( 
\frac{2r^{-1/4}}{\pi ^{3/4}A_R^{3/4}}\right) \sum_j^a\frac{A_j}{l_j^{1/2}}%
\cos \left( l_j\sqrt{\frac{4\pi r}{A_R}}-\frac \pi 4\right) \\  &  &  \\  
&  & +\sum_j^e\frac{A_j}{l_j^{1/2}}\cos \left( l_j\sqrt{\frac{4\pi r}{A_R}}%
-\frac \pi 4\right) -\sum_j^o\frac{A_j}{l_j^{1/2}}\cos \left( l_j\sqrt{\frac{%
4\pi r}{A_R}}-\frac \pi 4\right) 
\end{array}
\end{equation}

Here, superscripts $a,e,o$ over summation sign represents all repetitions,
even repetitions and odd repetitions of the primitive periodic orbits. This
is simple to understand, since centre-edge points do not result in periodic
orbits of different lengths in case of the rhombus there contribution will
be same as the rhombus or in the half equilateral triangle. However, note
that each centre-centre point results in two kinds of bands of the periodic
orbits one having the same length as that of the periodic orbits in the half
equilateral triangle and other having length twice of the same. These orbits
do not belong to the half equilateral triangle. The density of states for
equilateral triangle mode has same form as (\ref{5-22h}). It is simple to
check that adding these contributions we do get density of states for the
complete rhombus. We shall now turn our attention to study of two-level
cluster function.

\section{Two-Level Cluster Function}

In this section we shall obtain a closed analytical expression for two level
cluster function in pseudo-integrable billiards using semi-classical
framework. To do this it is clear from equation (\ref{5-17}) that we need to
evaluate semi-classical product $\langle d_{osc}(y)d_{osc}(y^{\prime
})\rangle ,$ (i.e. spectral average of density-density correlation function.

\subsection{Density-density correlation function}

As defined above density-density correlation function (oscillatory part) is
given by $d_{osc}(y)d_{osc}(y^{\prime })$. We consider correlation of levels
around $r_0$ i.e. we will put $y=r_0+r_1$ and $y^{\prime
}=r_0+r_2:r_1,r_2<<r_0$ then using equation (\ref{5-22}) correlation
function becomes%
$$
\begin{array}{lll}
d_{osc}(r_0+r_1)d_{osc}(r_0+r_2) & = & \frac{r_0^{-1/2}}{\pi ^{3/2}A_R^{3/2}}%
\sum_{j,k}\frac{A_jA_k}{l_j^{1/2}l_k^{1/2}}\left\{ \cos \left( \sqrt{\frac{%
4\pi }{A_R}}\left( l_j\sqrt{r_0+r_1}+\pi \left( \alpha _j-\frac 14\right)
\right) \right) \right. \\  &  &  \\  
&  & \times \left. \cos \left( \sqrt{\frac{4\pi }{A_R}}\left( l_k\sqrt{%
r_0+r_2}\right) +\pi \left( \alpha _k-\frac 14\right) \right) \right\} 
\end{array}
$$
or by using simple trigonometrical identities 
\begin{equation}
\label{5-25a}
\begin{array}{clll}
& d_{osc}(r_0+r_1)d_{osc}(r_0+r_2) & = & \frac{r_0^{-1/2}}{2\pi
^{3/2}A_R^{3/2}}\sum_{j,k}\frac{A_jA_k}{l_j^{1/2}l_k^{1/2}} \\  &  &  & 
\left\{ \sin \left( 
\sqrt{\frac{4\pi }{A_R}}\left( l_j\sqrt{r_0+r_1}+l_k\sqrt{r_0+r_2}\right)
+\kappa _{+}\right) \right. \\  &  &  & \left. +\cos \left( \sqrt{\frac{4\pi 
}{A_R}}\left( l_j\sqrt{r_0+r_1}-l_k\sqrt{r_0+r_2}\right) +\kappa _{-}\right)
\right\} 
\end{array}
\end{equation}
where $\kappa _{\pm }=\pi (\alpha _j\pm \alpha _k),$ terms due to Maslov
indices(i.e. number of bounces on the billiard wall). In the case of
polygonal billiards it can be seen that the number of bounces of a particle
on wall is always even if boundary conditions on the all segments of the
boundary are same either Dirichlet or Neumann\cite{5-gut}. Hence, we will
drop them from our analysis until we consider an example where they becomes
important (these are the cases where boundary conditions are mixed i.e. on
some segment it is Dirichlet, on the others it is Neumann).

It is easy to see that due to local averaging the first term in the above
equation will be negligibly small even when statistically significant number
of levels are considered. The second term also vanishes, unless%
$$
\left| \sqrt{\frac{4\pi }{A_R}}\left( l_j\sqrt{r_0+r_1}-l_k\sqrt{r_0+r_2}%
\right) \right| \,<1 
$$
The reason for this is that in the semi-classical limit this term will
oscillates rapidly\cite{5-mvb,5-ha-ozo} (it is well known that if argument
of $\cos $ or $\sin $ function is large, any small change in it will result
in rapid oscillations). Due to these rapid oscillations, spectral averaging
will result in negligibly small contributions unless above condition is
satisfied.

Since $r<<r_0$ we can approximate $\sqrt{r_0\pm r}\sim \sqrt{r_0}\left( 1\pm
r/2r_0\right) $ to write above condition as 
\begin{equation}
\label{5-26}\left| \sqrt{\frac{4\pi r_0}{A_R}}\left( \delta +\frac
r{2r_0}l\right) \right| \,\,<1 
\end{equation}
where for a simplicity we assume with out loss of any generality $%
l_j=l+\delta /2$ $l_k=l-\delta /2$, and $r_{1\left( 2\right) }=+(-)r/2$.
Note the presence of correlation range {\tt `}$r=|r_1-r_2|${\tt '} in the
deno$\min $ator of the second term, which appears because we are dealing
with density-density correlation function directly and not with it's Fourier
transform as generally done. The equation (\ref{5-26}) indicates that one
can separate the correlation function (\ref{5-25a}) in two parts as 
$$
d_{osc}(r_0+r_1)d_{osc}(r_0+r_2)=\sum\limits_{l_j=l_k}+\sum\limits_{l_j\neq
l_k}. 
$$
The first part is known as diagonal part ($\delta =0$) and other part is
called as off-diagonal part ($\delta \neq 0$).

Now let us take close look at condition (\ref{5-26}) which can be written as%
$$
\left| \delta +\frac r{2r_0}l\right| \,\,<\sqrt{\frac{A_R}{4\pi r_0}} 
$$
This is equivalent to following conditions 
\begin{equation}
\label{5-26a}
\begin{array}[t]{ll}
{\rm for\,\,\,\,}0<\delta : & 0<\delta < 
\sqrt{\frac{A_R}{4\pi r_0}}-\frac r{2r_0}l \\  &  \\ 
{\rm for\,\,\,\,}\delta <0{\rm \,\,\,and\,\,\,\,}|\delta |<\frac{rl}{2r_0}:
& \frac r{2r_0}l- 
\sqrt{\frac{A_R}{4\pi r_0}}<|\delta |<\frac r{2r_0}l \\  &  \\ 
{\rm for\,\,\,\,}\delta <0{\rm \,\,\,and\,\,\,\,}|\delta |>\frac{rl}{2r_0}:
& \frac r{2r_0}l<|\delta |<\frac r{2r_0}l+\sqrt{\frac{A_R}{4\pi r_0}} 
\end{array}
\end{equation}
One can see that in $\lim \,\,r_0\rightarrow \infty $, (or equivalently $%
\hbar \rightarrow 0$),: $\delta \rightarrow 0,$ hence, off-diagonal
contributions may not be important in the asymptotic limit and one can
safely rely on the diagonal approximation only. Generally, one is interested
in the asymptotic properties of spectral fluctuations only.

In practice, however, one always works with finite $r_0$, in fact most of
the numerical studies deals with at most few thousands of levels. It should
be emphasized that in such cases off-diagonal contributions may not be
negligible since finite number of levels are considered.

When $\delta =0$ (i.e. $l_j=l_k$) this condition becomes 
\begin{equation}
\label{5-27}l<\frac 1r\sqrt{\frac{A_Rr_0}\pi }=\frac{l_H}{2\pi r} 
\end{equation}
where $l_H=\sqrt{4\pi A_Rr_0}\,$ is {\tt ``}Heisenberg length{\tt ''} (which
corresponds to well known Heisenberg time $t_H=2\pi \hbar d_{av}(E),$via
relation $l_H=2t_H\sqrt{E}\,$).

The equation (\ref{5-27}) stipulates a condition for the validity of the
diagonal approximation. For any given $l_H$, it is sufficient to work with
diagonal terms for the length of periodic orbits satisfying above condition,
since then one can see that in general $\delta \sim 0$ for sufficiently
large number of levels. This condition is similar to the one obtained for
diagonal approximation for Fourier transform of $Y_2$ (known as the form
factor), except the presence of correlation range $r\,$ in the deno$\min $%
ator. It is obvious from this condition that for any fixed $l_H$ diagonal
approximation is strictly valid only at $r\sim 0$. For $0<<r\,$ and $%
l_H/2r<l $ the contribution of off-diagonal terms may not be insignificant.

Here, in this and next chapter we consider diagonal terms only. Our results
will therefore valid for $r<<1$.

\subsection{Diagonal approximation}

We can neglect first term of eq. (\ref{5-25a}). Since we are interested in
correlations over the number of few mean level spacings, much smaller than
total number of levels considered ($r<<r_0$) we can approximate $\sqrt{%
r_0\pm r}$ by $\sqrt{r_0}(1+r/2r_0)\,,$ then density-density correlation
function becomes 
\begin{equation}
\label{5-28}d_{osc}(r_0+r_1)d_{osc}(r_0+r_2)=\frac{r_0^{-1/2}}{2\pi
^{3/2}A_R^{3/2}}\sum_{j,k}\frac{A_j^2}{l_j}\cos \left( \frac{2\pi l_jr}{l_H}%
\right) , 
\end{equation}
where $r=|r_1-r_2|$ and $l_H=\sqrt{4\pi A_Rr_0}$, the Heisenberg length.

We have seen in chapter (3) that the periodic orbits occurring in the
pseudo-integrable polygonal billiards can be classified into different
families according to the nature of lattice points corresponding to the
periodic orbit as well as to the projected phase space area occupied by the
bands of periodic orbits. We have seen this via few examples of such
billiards. For these billiards density-density correlation function can be
written as 
\begin{equation}
\label{5-29}
\begin{array}{lll}
d_{osc}(r_0+r_1)d_{osc}(r_0+r_2) & = & \frac{r_0^{1/2}}{2\pi ^{3/2}A_R^{3/2}}%
\left\{ \sum_\alpha g_\alpha ^2A_\alpha ^2\sum_j\frac{\cos \left( \frac{2\pi
l_{\alpha ,j}r}{l_H}\right) }{l_{\alpha ,j}}+\right. \\  &  &  \\  
&  & \sum\limits_{\alpha ,\beta }\left( 1-\delta _{\alpha ,\beta }\right)
g_\alpha g_\beta A_\alpha A_\beta \sum_{j,k}\delta _{l_{\alpha ,j},l_{\beta
,k}}\frac{\cos \left( \frac{2\pi l_{\alpha ,j}r}{l_H}\right) }{l_{\alpha
,j}^{1/2}l_{\beta ,k}^{1/2}} 
\end{array}
\end{equation}
Here, subscripts $\alpha ,\beta $ denotes classes of bands of the periodic
orbits, $A_\alpha \ A_\beta $ are projective phase space area occupied by
the periodic orbits in the given class, $g_\alpha \ g_\beta $ denotes
degeneracy in the lengths of the periodic orbits. Other symbols have their
usual meaning. The second term on RHS takes care of systematic length
degeneracy occurring in different classes. However, we do not take into
account accidental degeneracy here. The summation over $j$ in the above
equation can be replaced by integration in the continuum limit, choosing a
proper measure. Heuristic arguments to obtain such measure are given below

\subsection{Sum$\min $g over all the periodic orbits}

Consider a length interval $(l,l+dl)$ such that $dl<<l$ but sufficiently
large enough to contain many events (by events here, we mean occurrence of a
periodic orbit of a length within the above interval). The proliferation law
of the periodic orbits, which gives the average number of periodic orbits of
length $\leq l$ for a class $\alpha $, has a form 
$$
F_\alpha (l)=a_\alpha l^2+O(l) 
$$
where $a_\alpha $ is system dependent constant. In our analysis we will
neglect $O(l)$ term. Number of periodic orbits having length between
(dropping subscript $\alpha $ temporarily), $l$ and $l+dl$, is given as $n$=$%
2aldl$. Hence the mean spacing between periodic orbits is $1/2al$. One can
divide interval $dl$ in $n$ cells each of the length equal to the mean
spacing. The length of the $i^{th}$ periodic orbit, $l_i\varepsilon \left(
l,l+dl\right) $ can be written as $l_i$=$l+i/2al+\eta _i$. Here, $\eta _i$'s
represent local fluctuations in the actual periodic orbit lengths from the
average lengths $l+i/2al$. It is therefore, reasonable to assume that $\eta
_i$'s are distributed symmetrically around zero. Most of the $\eta _i$' will
be closely distributed around zero with small fractions extending up to few
mean level spacings. The summation of form $\sum \cos ({\cal A}l)/l$ over j
for $l_j\varepsilon \left( l,l+dl\right) $ can be written as%
$$
\sum_{j=0}^{n-1}\frac{\cos ({\cal A}l_j)}{l{_j}}=\sum_{j=0}^{n-1}\frac{\cos
\left( {\cal A}(l+\frac j{2al}+\eta _j)\right) }l\left( 1-\frac j{2al^2}-%
\frac{\eta _j}l\right) 
$$
In our case ${\cal A}$=$\sqrt{\frac \pi {A_Rr_0}}r=2\pi r/l_H<<1$, hence $%
{\cal A}\eta _j\sim 0$. One can also notice that 2nd and 3rd terms in the
RHS bracket of above equation are $<<1$. Using these facts one can write
leading order terms of the above summation as%
$$
\sum\limits_{j=0}^{n-1}\frac{\cos \left( {\cal A}\left( l+\frac
j{2al}\right) \right) }{l_j}\simeq \frac{\cos ({\cal A}l)}l\sum_{j=0}^{n-\
1}\cos \left( \frac j{2al}\right) -\frac{\sin ({\cal A}l)}%
l\sum_{j=0}^{n-1}\sin \left( \frac j{2al}\right) . 
$$
Now summations on RHS can be easily performed by substituting, $%
n=2\,a\,\,l\,\,dl$ and using trigonometric relations (see e.g.\cite{5-grad}
page 30)%
$$
\begin{array}{lll}
\sum_{j=0}^{n-1}\frac{\cos \left( {\cal A}(l+\frac j{2al})\right) }{l_j} & 
\simeq & \frac{\cos ({\cal A}l)}l\cot (\frac{{\cal A}}{4al})\frac{Adl}2+%
\frac{\sin ({\cal A}l)}l\frac{Adl}2 \\  &  &  \\  
& \simeq & \frac{\cos ({\cal A}l)}l(2aldl)\sum_{k=0}^\infty (-1)^k\frac{%
2^{2k}B_{2k}}{(2k)!}\frac{{\cal A}}{4al}+\frac{\sin (Al)}l\frac{Adl}2 \\  &  
&  \\  
& \simeq & \frac{\cos ({\cal A}l)}l(2aldl)\left[ 1-\frac{{\cal A}^2}{64a^2l^2%
}+O(\frac{{\cal A}^3}{l^3})\right] +\frac{\sin ({\cal A}l)}l\frac{{\cal A}dl}%
2 
\end{array}
$$
Here, $B_{2k}$'s are Bernoulli's numbers. Neglecting terms of order $%
O(l^{-1})$ and higher as their contribution to summation is negligibly
small, we get 
\begin{equation}
\label{5-30}\sum_{j=0}^{n-1}\frac{\cos \left( A(l+\frac j{2al})\right) }{l_j}%
\simeq \frac{\cos (Al)}l2aldl 
\end{equation}
Same measure can be deduced from following simple relation%
$$
\begin{array}{ccc}
\sum_{j=0}^{n-1}\frac{\cos \left( Al_j\right) }{l_j} & = & \int 
\frac{\cos \left( Al\right) }l\sum\limits_{l_j}\delta (l-l_j)\,\,dl \\  & =
& \int \frac{\cos \left( Al\right) }l2aldl+\int \frac{\cos \left( Al\right) }%
ld_{osc}(l)dl 
\end{array}
$$
However it should be noted that 2nd term will have non-trivial contribution
unless ${\cal A}<<1$ as indicated above. This is the reason for which one
can not apply similar treatment to periodic orbit sum in density of
states(e.g. eq. (\ref{5-22})) or for off-diagonal terms as well.

\vskip .5in

Thus, returning to discussion on correlation function, using above results
the leading term in density- density correlation(\ref{5-29}) becomes 
\begin{equation}
\label{5-31}
\begin{array}{lll}
d_{osc}(r_0+r_1)d_{osc}(r_0+r_2) & = & \frac{r_0^{1/2}}{\pi ^{3/2}A_R^{3/2}}%
\left\{ \sum_\alpha g_\alpha ^2A_\alpha ^2a_\alpha I_{\alpha \beta }\right.
\\  &  & \left. +\sum\limits_{\alpha ,\beta }\left( 1-\delta _{\alpha ,\beta
}\right) g_\alpha g_\beta A_\alpha A_\beta \sum_{\mu \nu }\delta _{l_\mu
,l_\nu }I_{\mu \nu }\right\} 
\end{array}
\end{equation}

where 
\begin{equation}
\label{5-32}I_\alpha =\int_{l_{\min ,\alpha }}^{l_{\max ,\alpha }}dl\cos
\left( \frac{2\pi \,\,l\,\,r}{l_H}\right) 
\end{equation}
$I_{\mu \nu }$ can be similarly defined and $a_{\mu \nu }$ is appropriately
chosen coefficient. In the subsequent analysis this second term will not be
shown explicitly, though it will be included in the calculation of a
specific example. The choice of a measure enables us to write two-point
density correlation as 
\begin{equation}
\label{5-33}d_{osc}(r_0+r_1)d_{osc}(r_0+r_2)=\sum\limits_\alpha \kappa
_\alpha \left[ \frac{\sin \left( \frac{2\pi l_{\max ,\alpha }\,\,r}{l_H}%
\right) }{\pi r}-\frac{\sin \left( \frac{2\pi l_{\min ,\alpha }\,\,r}{l_H}%
\right) }{\pi r}\right] 
\end{equation}
where 
\begin{equation}
\label{5-34}\kappa _\alpha =\frac{a_\alpha g_\alpha ^2A_\alpha ^2}{\pi A_R} 
\end{equation}
and $l_H$ has usual meaning as above. In limit $l_{\max }\rightarrow \infty $
first term in the bracket represents contribution to the self-correlation
term (i.e., $\delta (r_1-r_2)$) of the levels. In writing the two-level
cluster function $Y_2$ we remove this self-correlation singularity as in eq.(%
\ref{5-19}). It may be noted that eq.(\ref{5-33}) indicates that $%
Y_2(r_1,r_2)$ is just a function of $r=|r_1-r_2|$. Hence, we write the
cluster function as 
\begin{equation}
\label{5-35}Y_2(r)=\sum\limits_\alpha \kappa _\alpha \frac{\left\langle \sin
\left( \frac{2\pi l_{\min ,\alpha }\,\,r}{l_H}\right) \right\rangle }{\pi r}%
. 
\end{equation}
Here, $\langle ...\rangle $ denotes spectral averaging as stated earlier.

\subsection{Taking spectral average}

The spectral average refers to all levels in an energy range that is
classically small, i.e. small in comparison with $r_0$, but large in
comparison with the mean level spacing, which is unity in an unfolded
spectrum or to be more precise large in comparison with the energy range
defined by outer energy scale defined by Berry \cite{5-mvb}. The outer
energy scale is given by $\hbar /T_{\min }$, where $T_{\min }$ is the period
of the shortest classical closed orbit. It may be noted that each classical
orbit with period $T$ causes deviation($\pm )$ of a level from the mean
position by order $\hbar /T$. Hence, outer energy scale corresponds to the
largest deviation of a level from the mean. In terms of number of mean level
spacing it is given by $r_{out}=\hbar d_{av}/T_{\min }\sim l_H/2\pi l_{\min
} $. In our case this range (say, $\sigma $) then roughly becomes $\sqrt{r_0}%
<<\sigma <<r_0$, since $l_H\sim O\left( \sqrt{r_0}\right) .$

We now consider spectral average of a function $f\left( {\cal A}r{\cal /}%
\sqrt{r_0}\right) $%
$$
\left\langle f\left( \frac{{\cal A}r}{\sqrt{r_0}}\right) \right\rangle
=\frac 1\sigma \int\limits_{-\sigma /2}^{\sigma /2}d\sigma ^{\prime
}\,\,f\left( \frac{{\cal A}r}{\sqrt{r_0+\sigma ^{\prime }}}\right) . 
$$
Since $\sigma ^{\prime }<<r_0$ we can approximate this as%
$$
\left\langle f\left( \frac{{\cal A}r}{\sqrt{r_0}}\right) \right\rangle
=\frac 1\sigma \int\limits_{-\sigma /2}^{\sigma /2}d\sigma ^{\prime
}\,\,f\left( \frac{{\cal A}r}{\sqrt{r_0}}\left( 1-\frac{\sigma ^{\prime }}{%
2r_0}\right) \right) . 
$$
Expanding $f$ in the Taylor series as%
$$
\begin{array}{c}
f\,\left( 
\frac{{\cal A}r}{\sqrt{r_0}}\left( 1-\frac{\sigma ^{\prime }}{2r_0}\right)
\right) =f\left( \frac{{\cal A}r}{\sqrt{r_0}}-\frac{{\cal A}^2r^2}{2r_0^2}%
\right) -\frac{{\cal A}r}{2r_0^{3/2}}f\,^{\prime }\left( \frac{{\cal A}r}{%
\sqrt{r_0}}-\frac{{\cal A}^2r^2}{2r_0^2}\right) \\ \left( \sigma ^{\prime }-%
\frac{{\cal A}r}{\sqrt{r_0}}\right) +... 
\end{array}
$$
If ${\cal A}r<<\sqrt{r_0}$ we can approximate spectral average as%
$$
\left\langle f\left( \frac{{\cal A}r}{\sqrt{r_0}}\right) \right\rangle
=\frac 1\sigma \int\limits_{-\sigma /2}^{\sigma /2}d\sigma ^{\prime }\left\{
\,\,f\left( \frac{{\cal A}r}{\sqrt{r_0}}\right) -f\,^{\prime }\left( \frac{%
{\cal A}r}{\sqrt{r_0}}\right) \left( \frac{{\cal A}r\sigma ^{\prime }}{%
2r_0^{3/2}}-\frac{{\cal A}^2r^2}{2r_0^2}\right) +..\right\} . 
$$
Which can be easily evaluated and we get%
$$
\left\langle f\left( \frac{{\cal A}r}{\sqrt{r_0}}\right) \right\rangle \sim
f\left( \frac{{\cal A}r}{\sqrt{r_0}}\right) +f\,^{\prime }\left( \frac{{\cal %
A}r}{\sqrt{r_0}}\right) \frac{{\cal A}^2r^2}{2r_0^2}+f\,\,^{\prime \prime
}\left( \frac{{\cal A}r}{\sqrt{r_0}}\right) \frac{{\cal A}^2r^2\sigma
^{\prime 2}}{3!\,2^3r_0^3}+... 
$$
It is thus obvious that second and higher order terms can be neglected
(since, $\sigma <<r_0$) if $f^{\prime }$ and higher derivatives are also
small. We will encounter here trigonometric functions (e.g. $\sin ,\cos $
and sin integrals) only.

\vskip .5in

Thus after spectral averaging the two-level cluster function becomes 
\begin{equation}
\label{5-36}Y_2(r)\simeq \sum\limits_\alpha \kappa _\alpha \frac{\sin \left( 
\frac{2\pi l_{\min ,\alpha }\,\,r}{l_H}\right) }{\pi r} 
\end{equation}

As per convention negative sign of $Y_2(r)$ represents positive correlation
and vice versa. For the finite spectrum occurring between $r_{\max }$ and $%
r_{\min }$ two-level cluster function is given by\cite{5-del,5-del1} 
\begin{equation}
\label{5-37}Y_2(r)=\delta (r)-\frac 1{\Delta r}\int\limits_{r_{\min
}}^{r_{\max }}dr_0\ \left\langle d_{osc}\left( r_0+r_1\right) d_{osc}\left(
r_0+r_2\right) \right\rangle 
\end{equation}
where $\Delta r$ = $r_{\max }-r_{\min }$. Note, that we have used here a
simple box function as a window function defined in \cite{5-del1} which
takes care of the finite number of levels and should not be misunderstood as
double averaging. Using eq.(\ref{5-36}), $Y_2(r)$ becomes 
\begin{equation}
\label{5-38}Y_2(r)\simeq \sum\limits_\alpha \kappa _\alpha \frac 1{\pi
r\Delta r}\int\limits_{r_{\min }}^{r_{\max }}dr_0\sin \left( \frac{2l_{\min
,\alpha }\,\,r}{l_H}\right) 
\end{equation}

Integration in (\ref{5-38}) is thus easy to carry out and $Y_2(r)$ can then
be written in simple analytical form as 
\begin{equation}
\label{5-39}
\begin{array}{ccc}
Y_2(r) & = & \sum\limits_\alpha 
\frac{\kappa _\alpha }{\pi \Delta r}\left\{ r_{\max }\left[ \frac{\sin
\left( \frac{2\pi l_{\min ,\alpha }\,\,r}{l_H}\right) }r+\frac{2\pi l_{\min
,\alpha }}{l_{H,\max }}\cos \left( \frac{2\pi l_{\min ,\alpha }\,\,r}{%
l_{H,\max }}\right) +\frac{4\pi ^2l_{\min ,\alpha }^2\,\,r}{l_{H,\max }^2}%
{\rm si}\left( \frac{2\pi l_{\min ,\alpha }\,\,r}{l_{H,\max }}\right)
\right] \right. \\  &  &  \\  
&  & -\left. r_{\min }\left[ \frac{\sin \left( \frac{2\pi l_{\min ,\alpha
}\,\,r}{l_H}\right) }r+\frac{2\pi l_{\min ,\alpha }}{l_{H,\min }}\cos \left( 
\frac{2\pi l_{\min ,\alpha }\,\,r}{l_{H,\min }}\right) +\frac{4\pi ^2l_{\min
,\alpha }^2\,\,r}{l_{H,\min }^2}{\rm si}\left( \frac{2\pi l_{\min ,\alpha
}\,\,r}{l_{H,\min }}\right) \right] \right\} 
\end{array}
\end{equation}
where, recall that $\kappa _\alpha $ is $A_\alpha ^2g_\alpha ^2a_\alpha /\pi
A_R$, $l_{_H,\max (\min )}$ is $\sqrt{4\pi A_R\,\,\,r_{\max (\min )}}$ and $%
{\rm si}(x)$ is sine integral $-\int_x^\infty dt\ \sin (t)/t$.

Apart from dependence on the energy window $(r_{\min },r_{\max })$ selected,
one can see that $Y_2(r)$ depends on two parameters (1) $l_{\min ,\alpha }$,
shortest periodic orbits in each class $\alpha $ (we will call it length
parameters) and (2)$\kappa _\alpha $, which is a measure of phase space area
occupied by bands of periodic orbits in class $\alpha $. Without referring
to any system, it is interesting to study how these parameters affect the
behaviour of $Y_2(r)$. Fig. 5.1 shows effect of variation in length
parameters where we have considered three classes of periodic orbits with
same value of parameter $\kappa $ and energy window. The length parameters
are, for curve a) 10,20,30, b)10,20,40, c)10,20,50, d)10,20,70. The $\kappa $%
's are same for all classes and curves: .3333 .3333 .3333. Number of levels
considered are 100-1000. Curve (g) is for G.O.E. As one increases highest
length parameter one gets more and more level repulsion. This indicates that
the presence of relatively longer periodic orbits will result in level
repulsion in the quantum spectrum. In Fig. 5.2 we show effect of variation
of parameters $\kappa $ for same values of length parameters.The $\kappa $
parameters are, for curve a) .6,.2,.2, b).2,.6,.2, c).2,.2,.6, d).1,.1,.8.
The ''lengths'' are same for all classes and curves:10,20,30. Number of
levels considered are 100-1000. One can see here that as weightage $\kappa $
increases for the longer length parameters, $Y_2(r)$ deviates from
Poissonian (i.e. $Y_2(r)$ =$0$) showing more and more negative correlations
or level repulsion. As far as the effect of energy window is concerned we
can see from above equation that as $r_{\max }\rightarrow \infty $ and $%
r_{\min }\ll r_{\max }$ so that $\Delta r\sim r_{\max }$, $Y_2(r)\sim \max
_\alpha (4\pi l_{\min ,\alpha }/\sqrt{r_{\max }})$, hence approaches a
Poissonian as $r_{\max }\rightarrow \infty $ in the diagonal approximation.
\begin{figure}[htbp]
\begin{center}
{\epsfig{file=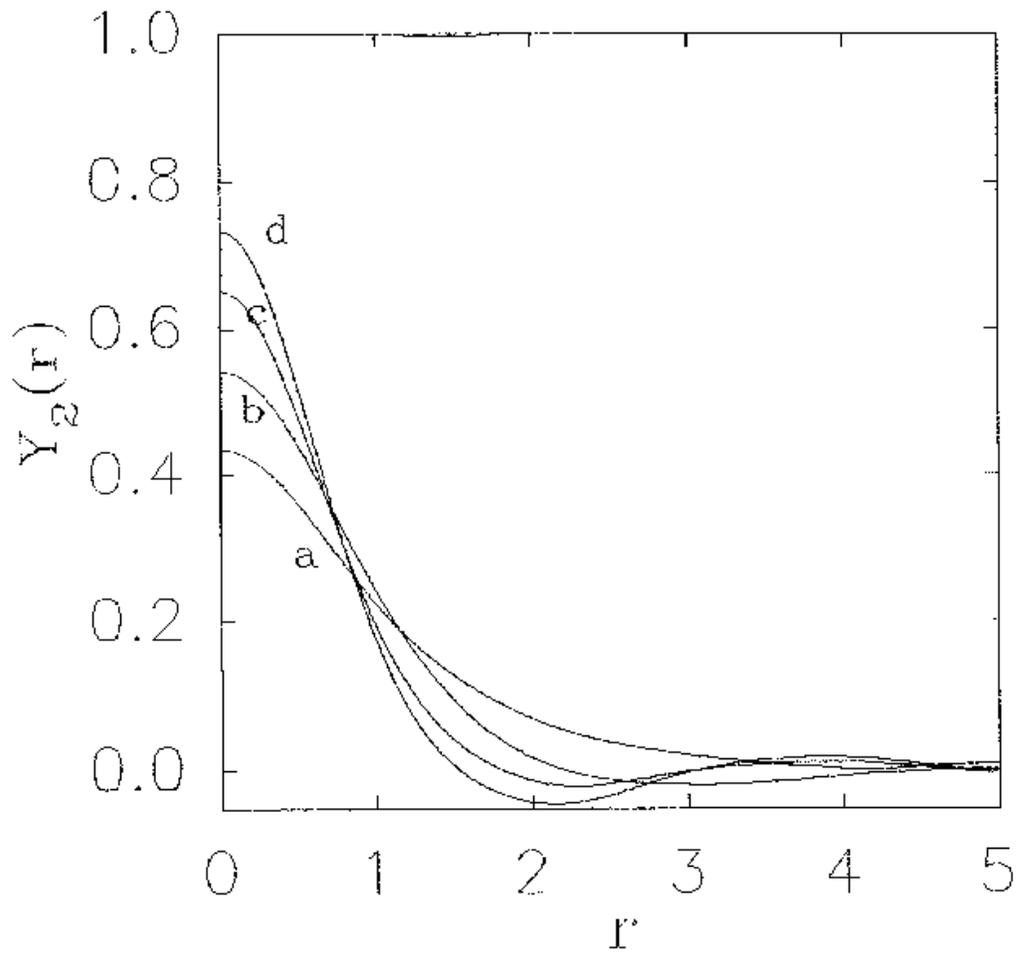,bbllx=80,bblly=170,bburx=560,bbury=700,height=6in}}
\caption{Two level cluster function: effect of variation in length parameters (see text)}
\end{center}
\end{figure}
\begin{figure}[htbp]
\begin{center}
{\epsfig{file=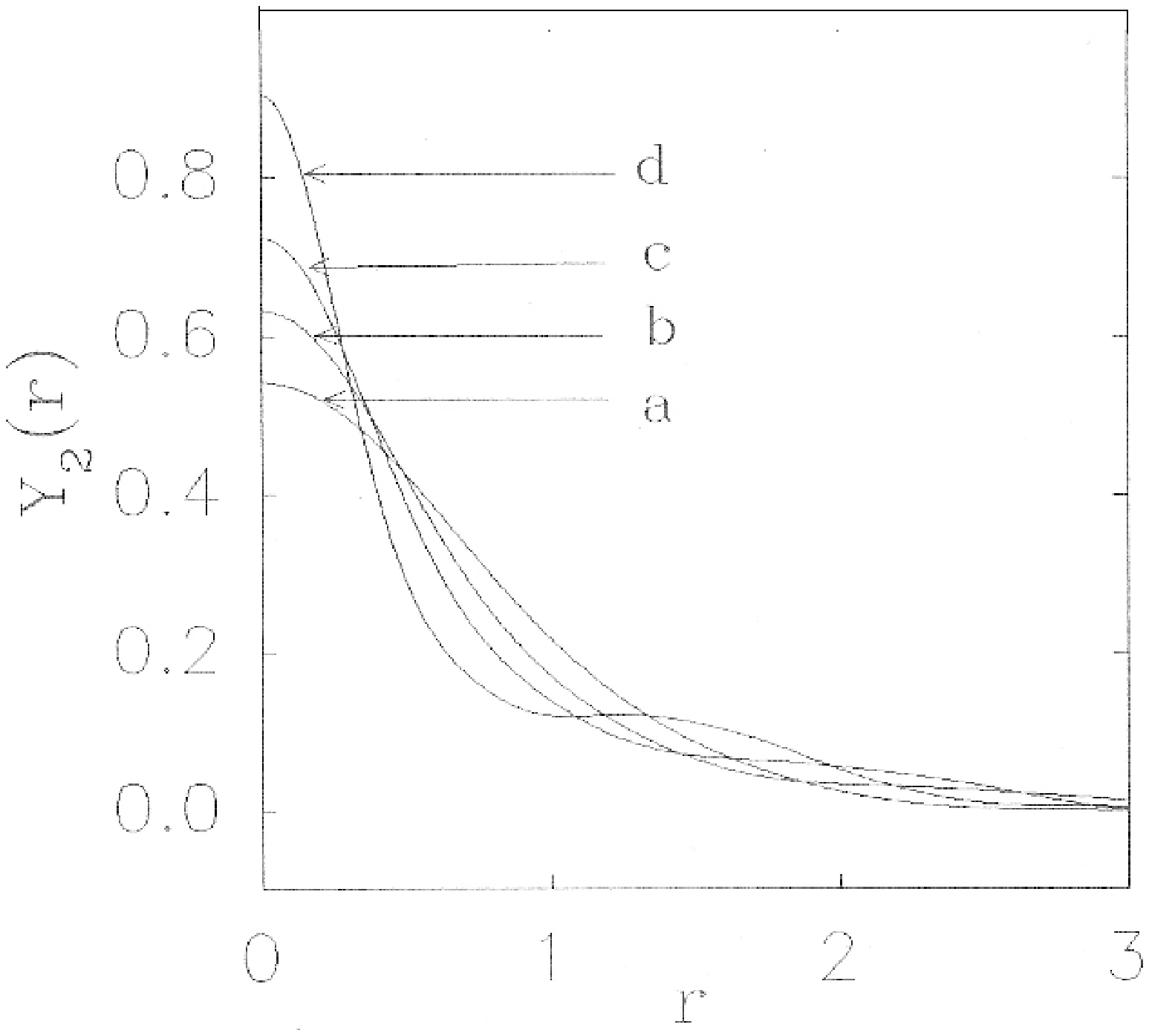,bbllx=80,bblly=170,bburx=560,bbury=700,height=6in}}
\caption{Two level cluster function: effect of variation
of parameters $\kappa $ (see text)}
\end{center}
\end{figure}
Rate of approach is to Poissonian is basically governed by $1/\sqrt{r_{\max }%
}$. It may be noted that $\max _\alpha (l_{\min ,\alpha })$ is also
important parameter since it will decide when one would expect Poisson like
results. For example in the case of $\pi /3$-rhombus billiard (pure rhombus
modes ) $\max _\alpha (l_{\min ,\alpha })=2\sqrt{93}{\cal L,}$ (${\cal L}$
being side length of the billiard), hence to see real asymptotic behaviour
one should consider number of levels must be at least a order of magnitude
larger than $1600$ levels. Unfortunately there exist few numerical results
in this range. Most of the numerical experiments can therefore be termed as
sub-asymptotic. In such cases as mentioned above off-diagonal contributions
may be significant. Any comparison of our results with numerical experiments
should be made by keeping this discussion in $\min $d.

We now consider two-point cluster function for $\pi /3$-rhombus billiard. We
reproduce relevant information in table 5.1 and 5.2. In Fig. 5.3 we show $%
Y_2 $ for complete $\pi /3$-rhombus billiard, including both even and odd
states that corresponds to two pure rhombus modes and two equilateral
triangle modes. In Fig 5.4 we consider two pure rhombus modes combinedly
(i.e. even states). Effect of number of levels considered is shown via
curves a)represents energy window $(47,370)$ b)$(420,740)$ c)$(47,743)$ and
d)$(1000,2000)$. It is evident from these figures that even states are more 
{\tt ``}away{\tt ''} from the Poissonian than the complete rhombus case.
This is obvious due to fact that complete rhombus billiard spectrum is
superposition of spectrums of four modes described above. One can also see
that as we increase number of levels $Y_2$ approach Poissonian.
Since spectral measures we consider here directly depend on $Y_2$ we expect
similar trend in case of these measures.
\begin{figure}[htbp]
\begin{center}
{\epsfig{file=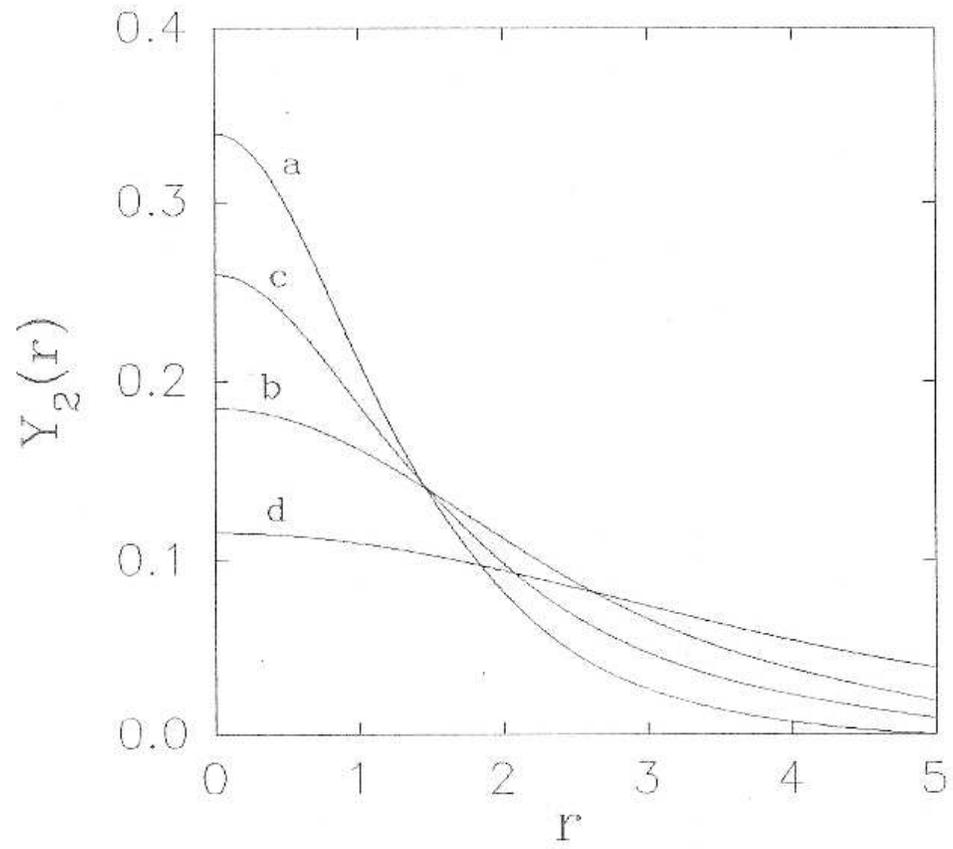,bbllx=80,bblly=170,bburx=560,bbury=700,height=6in}}
\caption{Two level cluster function: for complete $\pi /3$-rhombus billiard (see text)}
\end{center}
\end{figure}
\begin{figure}[htbp]
\begin{center}
{\epsfig{file=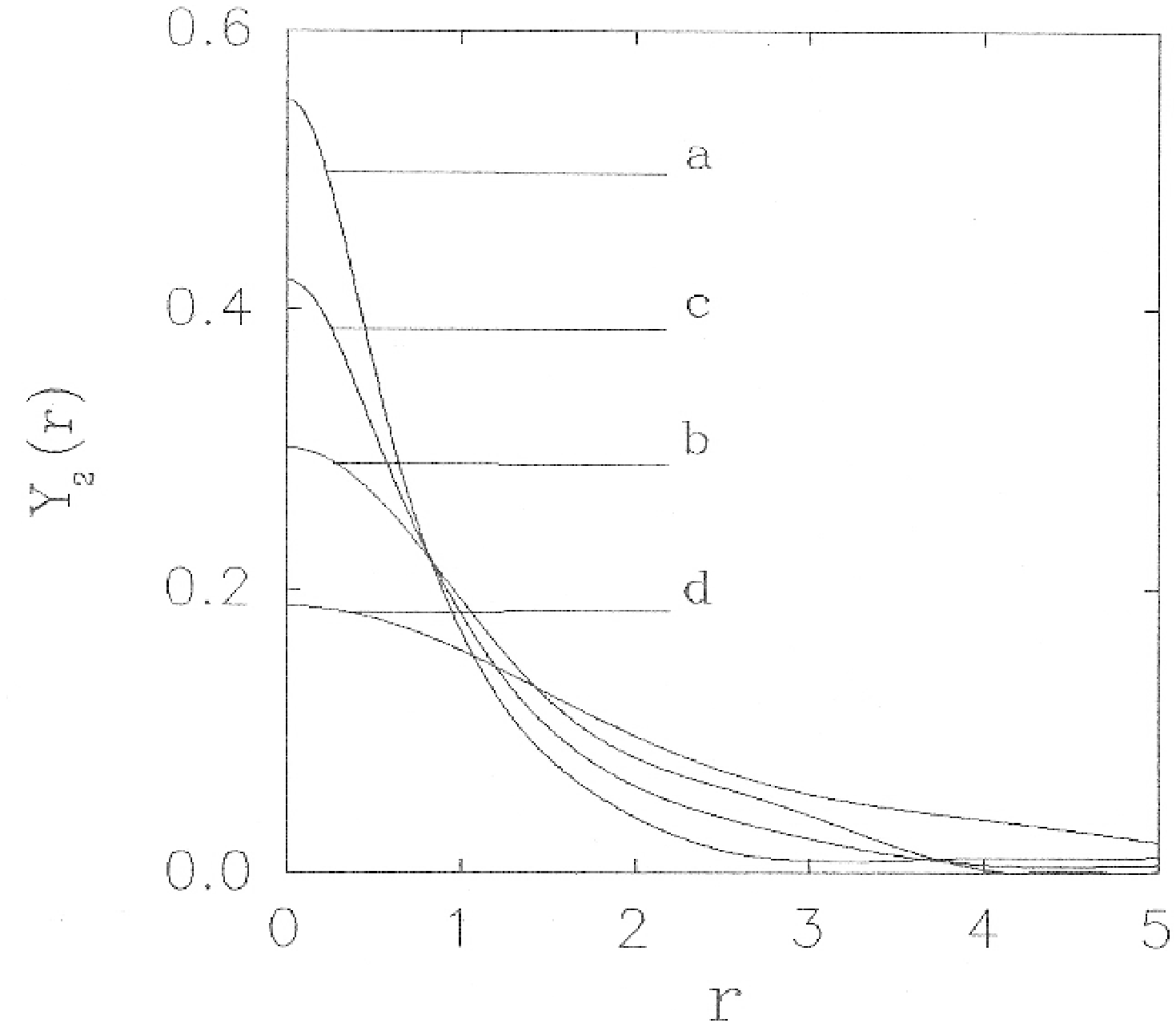,bbllx=80,bblly=170,bburx=560,bbury=700,height=6in}}
\caption{Two level cluster function: for $\pi /3$-rhombus billiard pure rhombus modes(see text)}
\end{center}
\end{figure}

\section{The Form Factor}

In this section we will consider one of the interesting quantity, the
Fourier transform of two-level cluster function, known as the Form Factor 
\cite{5-mlm}. It is well known that the Fourier transform of the density of
states gives important information about classical behaviour of the system,
i.e. it generates periodic orbit length spectrum of the classical system.
Similarly the form factor yields important information about periodic orbit
structure in the classical system. Before we elaborate this point, let us
derive a semi-classical expression for the form factor. We will use
definition of \cite{5-del} to derive expression for the form factor 
\begin{equation}
\label{5-40}b_2(\tau )=-\ 2\ \int\limits_0^\infty drY_2(r)\cos {(2\pi r\tau )%
} 
\end{equation}
where $\tau $ is a dimensionless time, related to $t$ by 
$$
\tau =\frac t{t_H}\ \ \,\,\,\,\,\,\,\,\,\,\,\,\,\,\,\,\,\,\,\,t_H=2\pi {%
\hbar }d_{av}. 
$$
Using (\ref{5-39}) the semi-classical expression for the form factor is
given by (since only finite numbers of levels are considered, we will
integrate above equation from $r=0$ to $b_\infty =\Delta {r}/2$, this will
not change our results since $Y_2(r)\sim 0$ for large $r$) 
\begin{equation}
\label{5-42}
\begin{array}{lll}
b_2{(\tau )} & = & -\sum\limits_\alpha \kappa _\alpha \left\{ 
\frac{r_{\max }}{\pi {\Delta }r}\left[ \left( 1-\frac{\gamma _{\alpha ,\max
}^2}{4\pi ^2\tau ^2}\right) \left[ {\rm Si}\left( \left( \gamma _{\alpha
,\max }-2\pi {\tau }\right) b_\infty \right) +{\rm Si}\left( \left( \gamma
_{\alpha ,\max }+2\pi {\tau }\right) b_\infty \right) \right] \right.
\right. \\  &  &  \\  
&  & + 
\frac{\gamma _{\alpha ,\max }}{\pi {\tau }}\sin {(2\pi {\tau }b_\infty )}%
\cos {\left( \gamma _{\alpha ,\max }b_\infty \right) } \\  &  &  \\  
&  & \left. + 
\frac{\gamma _{\alpha ,\max }^2}{\pi {\tau }}{\rm si}{\left( \gamma _{\alpha
,\max }b_\infty \right) }\left[ b_\infty \sin {\left( 2\pi {\tau }b_\infty
\right) }+\frac{\cos {\left( 2\pi {\tau }b_\infty \right) }}{2\pi {\tau }}%
\right] +\frac{\gamma _{\alpha ,\max }^2}{4\pi {\tau ^2}}\right] \\  &  &  
\\  
&  & - 
\frac{r_{\min }}{\pi {\Delta }r}\left[ \left( 1-\frac{\gamma _{\alpha ,\min
}^2}{4\pi ^2\tau ^2}\right) \left[ {\rm Si}\left( \left( \gamma _{\alpha
,\min }-2\pi {\tau }\right) b_\infty \right) +{\rm Si}\left( \left( \gamma
_{\alpha ,\min }+2\pi {\tau }\right) b_\infty \right) \right] \right. \\  &  
&  \\  
&  & + 
\frac{\gamma _{\alpha ,\min }}{\pi {\tau }}\sin {\left( 2\pi {\tau }b_\infty
\right) }\cos {\left( \gamma _{\alpha ,\min }b\right) } \\  &  &  \\  
&  & \left. \left. +\frac{\gamma _{\alpha ,\min }^2}{\pi {\tau }}{\rm si}{%
\left( \gamma _{\alpha ,\min }b_\infty \right) }\left[ b_\infty \sin {\left(
2\pi {\tau }b_\infty \right) }+\frac{\cos {\left( 2\pi {\tau }b_\infty
\right) }}{2\pi {\tau }}\right] +\frac{\gamma _{\alpha ,\min }^2}{4\pi {\tau
^2}}\right] \right\} 
\end{array}
\end{equation}
where $\gamma _{\alpha ,\max (\min )}=2\pi l_{\alpha ,\min }/l_{H,\max (\min
)}$. As $b_\infty $ becomes large enough, then we can approximate sine
integrals as follows, 
\begin{equation}
\label{5-42a}{\rm Si}(\pm x)\simeq {\pm }\frac \pi 2{\mp }\frac{\cos (x)}x,\;%
{\rm si}(x)\simeq -\frac{\cos (x)}x. 
\end{equation}
Substituting above relations in (\ref{5-42}) we get, for $\tau <\min _\alpha
(\gamma _{\alpha ,\max }/2\pi )$ 
\begin{equation}
\label{5-43}b_2(\tau )=-\sum_\alpha \kappa _\alpha -\sum_\alpha \kappa
_\alpha \{\frac{r_{\max }\sin (\gamma _{\alpha ,\max }b_\infty )-r_{\min
}\sin (\gamma _{\alpha ,\min }b_\infty )}{\pi {\Delta }r}\}\frac{\sin (2\pi {%
\tau }b_\infty )}{\pi {\tau }b_\infty }. 
\end{equation}
And for $\tau \gg \max _\alpha (\gamma _{\alpha ,\max }/2\pi )$ 
\begin{equation}
\label{5-44}b_2(\tau )=-\sum_\alpha \kappa _\alpha \left\{ \frac{r_{\max
}\cos (\gamma _{\alpha ,\max }b_\infty )-r_{\min }\cos (\gamma _{\alpha
,\min }b_\infty )}{\pi {\Delta }r}\right\} \frac{\cos (2\pi {\tau }b_\infty )%
}{\pi {\tau }b_\infty }+O(\frac 1{\tau ^2}). 
\end{equation}
Various important properties of the form factor has been discussed in \cite
{5-del,5-del1,5-cohen}. We bring out some new (not reported so far) features
of the form factor. In general, form factor $b_2(\tau )$ saturates to $0$
for $\tau \gg 1$ due to discrete nature of the spectrum. For $\tau \ll
1,b_2(\tau )\rightarrow -1$. In case of chaotic systems diagonal
approximation leads to monotonically increasing ($\sim \tau $) form factor.
The contribution of the off-diagonal terms however cancels this rise to
restore correct asymptotic behaviour of the form factor \cite{5-cohen}.

From (\ref{5-42}) one can see that there will be a spike at $\tau $ = $0$,
due to $\sin (2\pi {\tau }b)/\pi {\tau }b$ (which is $\sim \delta {(\tau )}$%
) factor on RHS. For $0<\tau <{\rm \min }_\alpha (\gamma _{\alpha ,\min
}/2\pi )$, $b_2(\tau )$ will be -1 since contribution of the same factor
becomes negligible and $\sum_\alpha \kappa _\alpha $ = 1. It should be noted
that above expressions for $b_2(\tau )$ have independent contributions from
the classes of periodic orbits in the diagonal approximation. The classes of
the periodic orbits are characterized by length of shortest periodic orbits (%
$l_{\min ,\alpha })$ that can be arranged in increasing order. For $\tau
\sim \min _\alpha (\gamma _{\alpha ,\min }/2\pi )$, corresponding
contribution from that family will decrease(in absolute sense) and
saturating ultimately to $0$ as $\tau $ becomes greater than $\gamma
_{\alpha ,\min }/2\pi $, resulting in net increase in the form factor. If
lengths of shortest periodic orbits in different classes are well separated
then $\tau $ may still be less than next $(\gamma _{\alpha ,\min }/2\pi )$,
the form factor will saturate at $-\sum_{\alpha ^{\prime }}{\kappa }_\alpha
>-1$, where summation is now taken over all classes except one(i.e.$\alpha $%
). This thus leads to a ''step'' like structure in the plots of the form
factor. If however, lengths of the shortest periodic orbits are not well
separated steps may merge into each other. The step structure is thus a
hallmark of existence of multiple, distinct '' characteristic time scales''
(hence also length scales) in these systems.

In fig. 5.5 and 5.6 we study effect of variations in the length parameters
and $\kappa $'s on the form factor. Again here choice of various parameters
is same as that in case of the two-point cluster function. Though general
conclusions that can be drawn are similar to that discussed in the previous
section, i.e., the existence of larger length parameters or more weightage
(via. $\kappa $ parameters) to higher length parameters shows deviation away
from Poissonian behaviour. Most important feature is imprints of classes of
periodic orbits. We have considered three classes for all the cases and the
form factor rises to zero in the same number of ''steps''. Existence of
steps in the form factor are thus signatures of the classes of bands of
marginally stable periodic orbits in the classical system. This feature is
apparent in the plots of the form factor in some of the works \cite
{5-del,5-del1} where systems, not far away from the integrability, like
kicked rotor or do$\min $o billiards are considered and in the examples that
we will consider in this paper. We wish to draw attention of readers to the
fact that the form factor saturates at $0$ in long $\tau $ limit and does
not rise monotonically as in case of chaotic system in the diagonal
approximation used above. This in turns implies that off-diagonal
corrections may not be considerable as far as the form factor is concerned.
As our study indicates the off-diagonal contributions may only affect the
form factor in the transition region where it rises from $-1$ to $0$. Effect
of the off-diagonal terms on step like behaviour is difficult to guess.
Recent approach(\cite{5-bogo}) to treat off-diagonal terms in terms of
diagonal contribution however, indicates that step like behaviour may be
preserved even after consideration of off-diagonal contributions. Secondly,
since diagonal form factor saturates at $0$ as expected, off-diagonal terms
may only change the transition region most likely flattening it.
\begin{figure}[htbp]
\begin{center}
{\epsfig{file=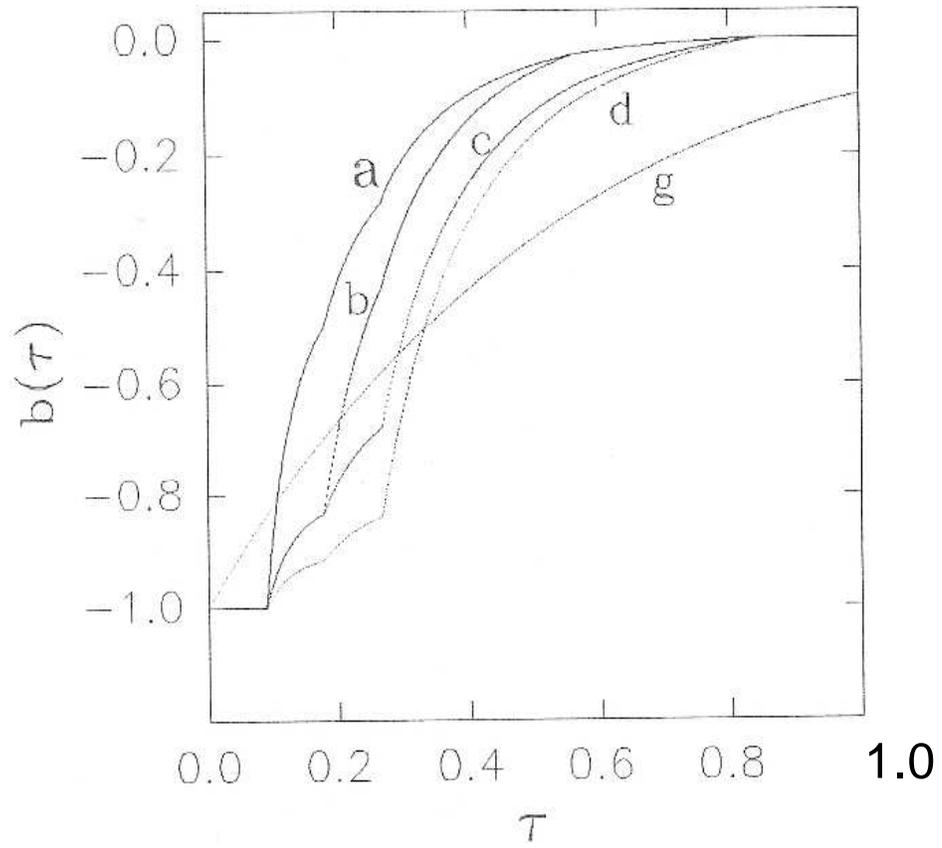,bbllx=80,bblly=170,bburx=560,bbury=700,height=6in}}
\caption{The form factor: Effect of variation of the length parameters (see text)} 
\end{center}
\end{figure}
\begin{figure}[htbp]
\begin{center}
{\epsfig{file=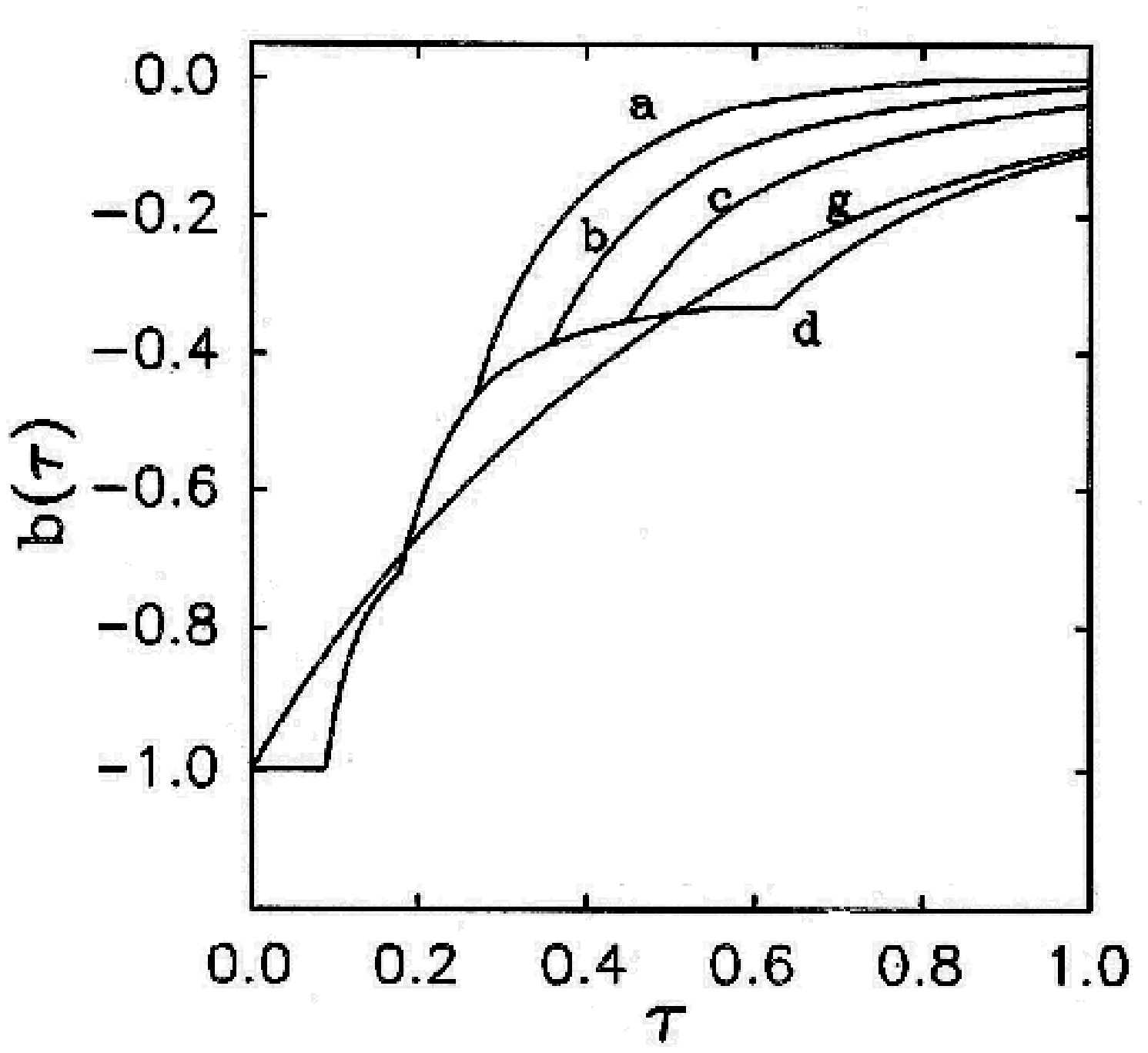,bbllx=80,bblly=170,bburx=560,bbury=700,height=6in}}
\caption{The form factor: Effect of variation of the $\kappa $ parameters (see text)} 
\end{center}
\end{figure}

Our analysis thus brings out a important feature of the form factor that has
not been paid attention so far. With increasing energy, rise in these steps
become sharper and sharper taking the form factor closer the Possonian one.
Also if differences in the length parameters are small, this step structure
may not be easily visible. This is evident from Fig. 5.7 and Fig. 5.8 where
we plot form factor for complete rhombus and for even states respectively.
Difference between two cases is self evident, rise is sharper in the former
case than the later case.

In this chapter we have obtained closed analytical expressions for two
interesting quantities that are used to study spectral fluctuation
properties using diagonal approximations. This is an important step which
should ultimately lead to complete understanding of spectral fluctuations by
including other contributions such as off-diagonal terms, diffraction
effects etc.
\begin{figure}[htbp]
\begin{center}
{\epsfig{file=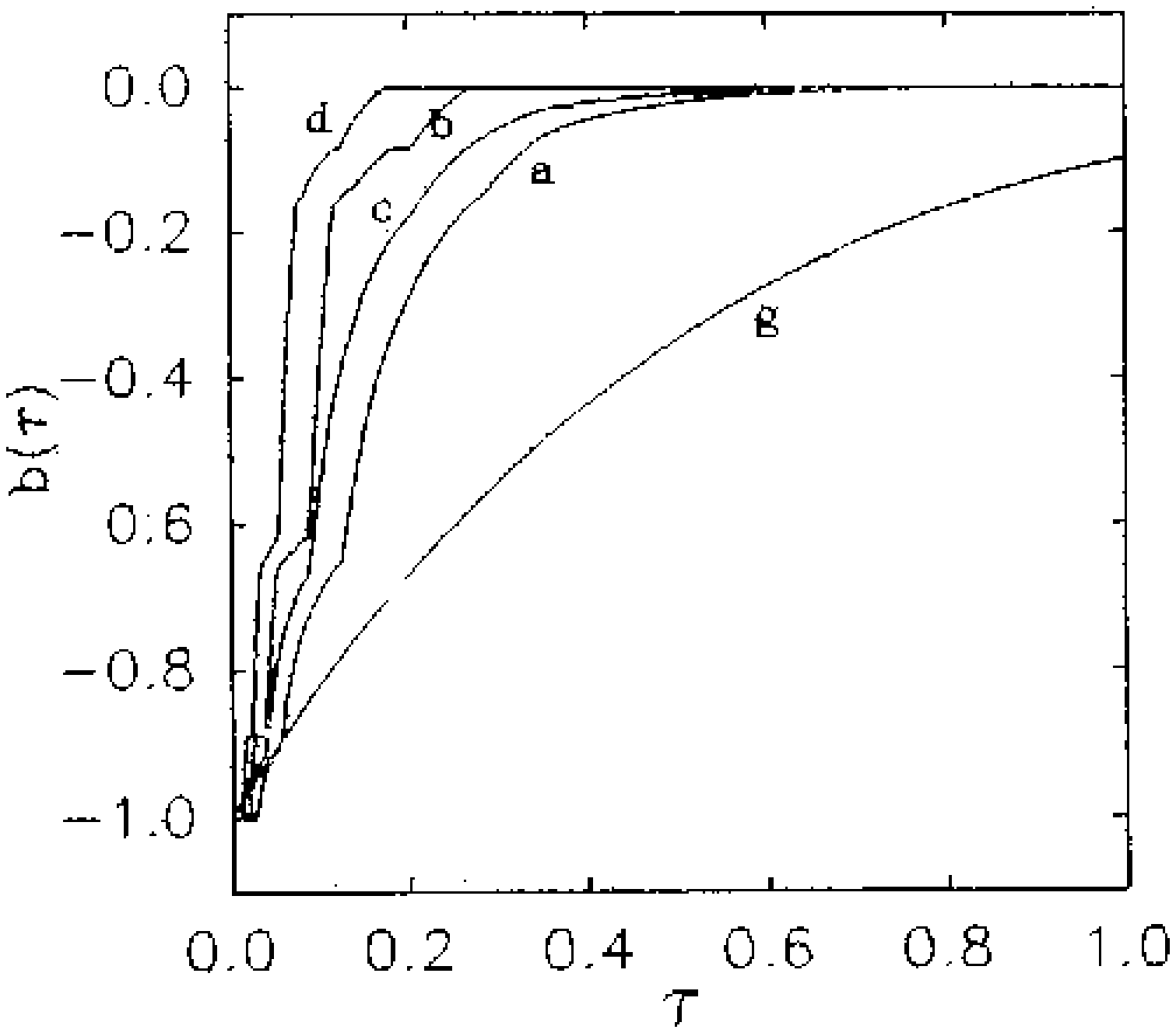,bbllx=80,bblly=170,bburx=560,bbury=700,height=6in}}
\caption{The form factor: For complete $\pi/3$- rhombus billiard (see text)} 
\end{center}
\end{figure}
\begin{figure}[htbp]
\begin{center}
{\epsfig{file=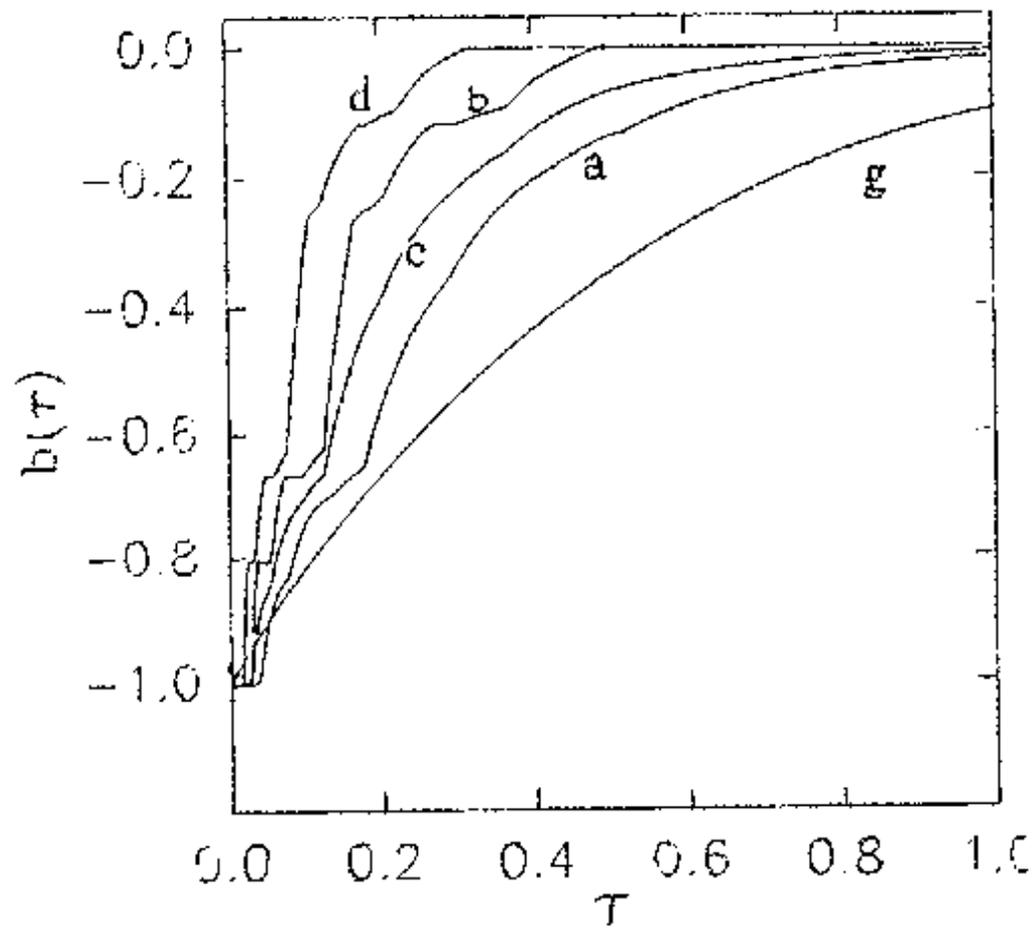,bbllx=80,bblly=170,bburx=560,bbury=700,height=6in}}
\caption{The form factor: For pure modes of $\pi/3$- rhombus billiard (see text)} 
\end{center}
\end{figure}

\begin{table}
\begin{tabular}{|l|l|l|l|l|l|}
\hline
Type&Class& Closing& Band&$\kappa_{\alpha}$&$l_{min}$\\
&& Point & Area&&  \\ \hline
&&&&&\\ 
{\bf centre-centre}&odd-odd&$(q/2,p/2)$&$A_R$&$\frac{1}{9}$&${\surd
3}{\cal L}$\\
&&&&&\\ 
&&$(q,p)$&$2A_R$&$\frac{2}{9}$&$2{\surd 3}{\cal L}$\\
\hline
&&&&&\\ 
&odd-even&$(q,p)$&$A_R$&$\frac{1}{36}$&${\surd 21}{\cal L}$\\
&&&&&\\
&&$(2q,2p)$&$2A_R$&$\frac{1}{18}$&$2{\surd 21}{\cal L}$\\
\hline
&&&&&\\ 
&even-odd&$(q,p)$&$A_R$&$\frac{1}{36}$&${\surd 39}{\cal L}$\\
&&&&&\\
&&$(2q,2p)$&$2A_R$&$\frac{1}{18}$&$2{\surd 39}{\cal L}$\\
\hline
&&&&&\\ 
{\bf centre-edge}&odd-odd&$(3q/2,3p/2)$&$3A_R$&$\frac{1}{3}$&$3{\surd
37}{\cal L}$\\
&&&&&\\
\hline
&&&&&\\ 
&odd-even&$(3q,3p)$&$3A_R$&$\frac{1}{12}$&$3{\surd 7}{\cal L}$\\
&&&&&\\
\hline
&&&&&\\ 
&even-odd&$(3q,3p)$&$3A_R$&$\frac{1}{12}$&$3{\surd 7}{\cal L}$\\
&&&&&\\
\hline
\end{tabular}
\caption{Summary of results for complete $\pi/3$-rhombus billiard.}
\end{table}
\begin{table}
\begin{tabular}{|l|l|l|l|l|l|l|}
\hline
Type&Class& Closing&Band&Repetitions&$\kappa_{\alpha}$&$l_{min}$\\
&& Point&Area&&&   \\
\hline
&&&&&&\\
{\bf centre-centre}&odd-odd&$(q/2,p/2)$&$4A_R$&all&$\frac{8}{18}$&${\surd
3}{\cal L}$\\
&&&&&&\\
&&&$8A_R$&even&$\frac{5}{36}$&$2{\surd 3}{\cal L}$\\
&&&&&&\\
&&&$8A_R$&odd&$-\frac{1}{4}$&${\surd 3}{\cal L}$\\
\hline
&&&&&&\\
{\bf centre-centre}&odd-even&$(q,p)$&$4A_R$&all&$\frac{2}{18}$&${\surd
93}{\cal L}$\\
&&&&&&\\
&&&$8A_R$&even&$\frac{5}{144}$&$2{\surd 93}{\cal L}$\\
&&&&&&\\
&&&$8A_R$&odd&$-\frac{1}{16}$&${\surd 93}{\cal L}$\\
&&&&&&\\
\hline
&&&&&&\\
{\bf centre-centre}&odd-even&$(q,p)$&$4A_R$&all&$\frac{2}{18}$&${\surd
39}{\cal L}$\\
&&&&&&\\
&&&$8A_R$&even&$\frac{5}{144}$&$2{\surd 39}{\cal L}$\\
&&&&&&\\
&&&$8A_R$&odd&$-\frac{1}{16}$&${\surd 39}{\cal L}$\\
\hline
&&&&&&\\
{\bf centre-edge}&odd-odd&$(3q/2,3p/2)$&$12A_R$&all&$\frac{1}{3}$&$3{\surd
37}{\cal L}$\\
&&&&&&\\
\hline
&&&&&&\\
&odd-even&$(3q,3p)$&$12A_R$&all&$\frac{1}{12}$&$3{\surd 61}{\cal L}$\\
&&&&&&\\
\hline
&&&&&&\\
&even-odd&$(3q,3p)$&$12A_R$&all&$\frac{1}{12}$&$3{\surd 19}{\cal L}$\\
&&&&&&\\
\hline
\end{tabular}
\caption{Summary  of  results  for $\pi/3$-rhombus billiard: Only
Pure Rhombus Modes. Here $A_R$ is area of $V$,  half  equilateral
triangle.}
\end{table}

\chapter{Short and Intermediate Range Spectral Fluctuation Measures}

\section{Introduction}

We shall obtain expressions for three important and popular spectral
fluctuation measures namely spacing distribution, number variance and
spectral rigidity using diagonal approximation for pseudo-integrable
billiards. Spacing distribution provides information about spectral
fluctuations on short energy scale i.e. on the scale of mean level spacing.
Other two measures are called as intermediate range spectral measures since
they provide information about fluctuations on few mean level spacing. In
this chapter

\section{Nearest Neighbour Spacing Distribution}

Nearest neighbour spacing (NNS) distribution is{\it \ the probability
density }$P(S)${\it \ of finding the nearest neighbour of a given level at }$%
r${\it , in the range }$r+S${\it \ to }$r+S+dS${\it .} As stated above $P(S)$
distribution of any given system enables us to study fine scale texture
(i.e., on the scale of mean level spacing) of the spectrum.

For Integrable systems spacing distribution is very well modeled by Poisson
distribution ($P(S)=\exp (-S)$) since the energy levels are random due to
number theoretic properties. On the other hand in case of chaotic systems
with time reversal symmetry (modeled by GOE of random matrix theory(RMT))
spacing distribution is given by the Wigner distribution function. 
\begin{equation}
\label{6-1}P(S)=\frac \pi 2S\exp \left( -\frac{\pi S^2}4\right) , 
\end{equation}
provided the level sequence has been normalized to unit mean level spacing.
RMT is expected to be applicable only on those time scales where the
variables associated with the classical dynamics are random enough, to fully
randomize the matrices, associated with the corresponding quantum operators.
Moreover, due to underlying assumptions on which framework of RMT is build,
it is not expected to shed any light on non-universal behaviour of spectral
fluctuations which is the characteristic of a given system.

Various formulas have been proposed to analyse $P(S)$ distribution for the
systems in the intermediate regime \cite{brody,mvb-rob,iz}. However, these
models are not applicable for the pseudointegrable systems because of basic
differences with other mixed systems.

The aim of this section is to develop a model for $P(S)$ distribution in
pseudointegrable systems. Once such model is developed it can be extended to
more general class of mixed system where almost regular and almost chaotic
states coexist. When the spectrum is unfolded to unit mean spacing of
neighbouring levels everywhere, one can use probabilistic arguments e.g.\cite
{hak,gutz} to develop expression for $P(S)$ distribution.

The conditional probability $g(S)dS$ of finding a level in the interval $%
(r+S,r+S+dS)$, given one at $r$ is related to $P(S)$ in the following way.
Choose a segment of length $\lambda $(integer), and divide it into small
intervals, all of the same length $\varepsilon $. We place $\lambda $
markers at random, independently of one another, with the probability $%
g(S)dS $, into the small intervals. The first marker above $0$ will hit any
particular small interval with probability $\varepsilon g(\xi )/\lambda $,
and miss any other small interval with probability $1-\varepsilon g(\xi
)/\lambda $, where $\xi $ is some coordinate inside the small interval in
question. Now we choose a contiguous interval of length $S$, and require the
probability that none of the markers fall within $S$, while there is a
marker in interval $(r+S,r+S+dS)$. We have to form the product of all the $%
1-\varepsilon g(\xi )/\lambda \,$ for $0<\xi <S$, and multiply with $%
\varepsilon g(S)/\lambda $. In the limit of small $\varepsilon $ or large $%
\lambda $ we find that 
\begin{equation}
\label{6-2}P(S)={\cal C}g(S)\exp \left( -\int_0^SdS^{\prime }g(S^{\prime
})\right) , 
\end{equation}
${\cal C}$ on RHS is used so as to satisfy condition $\int_0^\infty P(S)=1.$
Underlying assumptions in deriving this relation are: 1) Given a level at $r$%
, the probability that another level will be around $r+S$ is proportional to 
$S$ and does not depend on $r$. This assumption though applied for all $S$,
is valid only for small $S$. In other words, this assumption means (1)the
two point correlation function $R_2(r_1,r_2)$ is linear in $\mid r_1-r_2\mid 
$ and (2) the probabilities in various intervals of length $S/m$ obtained by
dividing $S$ into $m$ equal parts are mutually independent. In other words
three-point and higher correlations are negligibly small. Though both of
these the assumptions are inaccurate, above relation gives accurate result
for GOE of random matrix in terms of Wigner surmise, which indicates that
errors arising due to these assumption almost cancel each other \cite{mlm}.
It is, therefore reasonable to expect that above relation would yield good
approximation to $P(S)$ distribution in pseudointegrable billiards as for
such systems levels are expected to be less correlated than that in the
chaotic systems. Strictly speaking, $P(S)$ depends not only on two-point
correlations but on higher correlations too \cite{mlm,wig1}.

In case of homogeneous spectrum (i.e. density of states almost independent
of energy ), unfolded spectrum with unit mean spacing everywhere conditional
probability density $g(S)$ is nothing but the two- point correlation
function, or $g(S)$ =$1-Y_2(S)$ \cite{hak,bho-gi}. In terms of this the $%
P(S) $ distribution can be written as 
\begin{equation}
\label{6-3}P(S)={\cal C}\ e^{-S}\ (1-Y_2(S))\ e^{\int_0^SY_2(S^{\prime
})dS^{\prime }} 
\end{equation}
In our case ${\cal C}$ $=1$. Using expression for $Y_2(S)$ derived in the
last chapter and carrying out necessary algebra one can write NNS
distribution for pseudointegrable billiards as 
\begin{equation}
\label{6-4}P(S)={\cal A}(S)\ \exp ({\cal B}(S)-S) 
\end{equation}
where ${\cal A}(S)$ \ = \ $1-Y_2(S)$. And 
\begin{equation}
\label{6-5}
\begin{array}{ccc}
{\cal B}(S) & = & \sum \kappa _\alpha \left\{ 
\frac{r_{\max }}{\pi \Delta r}\left[ {\rm Si}(\gamma _{\alpha ,\max }S)+%
\frac{\sin (\gamma _{\alpha ,\max }S)}2+\frac{\gamma _{\alpha ,\max }S}2\cos
(\gamma _{\alpha ,\max }S)\right. \right. \\  &  &  \\  
&  & \left. 
\frac{\gamma _{\alpha ,\max }^2S^2}2{\rm si}(\gamma _{\alpha ,\max
}S)\right] \\  &  &  \\  
&  & - 
\frac{r_{\min }}{\pi \Delta r}\left[ {\rm Si}(\gamma _{\alpha ,\min }S)+%
\frac{\sin (\gamma _{\alpha ,\min }S)}2+\frac{\gamma _{\alpha ,\min }S}2\cos
(\gamma _{\alpha ,\min }S)\right. \\  &  &  \\  
&  & \left. +\left. \frac{\gamma _{\alpha ,\min }^2S^2}2{\rm si}(\gamma
_{\alpha ,\min }S)\right] \right\} 
\end{array}
\end{equation}
where {\rm Si}$(x)$ = $\pi /2+{\rm si}(x)$ and $\gamma _{\alpha ,\min (\max
)}=2\pi l_{\alpha ,\min }/l_{H,\min (\max )}$. The expression for $P(S)$ is
clearly normalized. For small $S$, neglecting terms of order $O(S^2)$ and
higher, one obtains 
\begin{equation}
\label{6-6}
\begin{array}{lll}
{\cal A}(S)\simeq {\cal A}s & = & -\sum_\alpha \kappa _\alpha 
\frac{2l_{\min ,\alpha }}{\sqrt{\pi A_R}\Delta r}\left[ \sqrt{r_{\max }}-%
\sqrt{r_{\min }}\right] \\  &  &  \\ 
{\cal B}(S)={\cal B} & = & \sum_\alpha \kappa _\alpha 
\frac{l_{\min ,\alpha }}{\sqrt{\pi A_R}\Delta r}\left[ \sqrt{r_{\max }}-%
\sqrt{r_{\min }}\right] S \\  & = & S- 
{\cal A}S \\  & {\rm thus} &  \\  
&  &  \\ 
P(S) & \simeq & {\cal A}\exp \left[ -{\cal A}S\right] 
\end{array}
\end{equation}
Therefore $dP(S)/dS\mid _{S=0}$ = $-{\cal A}^2\leq 0$, showing level
attraction near $S=0$, which is weaker than that of a Poissonian
distribution for a finite number of levels but approaches the same in the
semi-classical limit. For large $S$ again our expression approaches
Poissonian distribution. To see this behaviour clearly we study from (\ref
{6-4}), dependence of $P(S)$ on length parameters and parameters $\kappa $%
~s. we will consider only three classes of the periodic orbits. Fig.6.1
shows the effect of variations in the length parameters and Fig.6.2 shows
the same for $\kappa _\alpha $'s, parameters chosen are same as that in case
of two-point cluster function. Again conclusions that can be drawn from
these figures are similar to that stated in the previous chapter. In the
high energy limit, again $P(S)$ distribution converges to a Poissonian
distribution. For finite energy levels however, $P(S)$ distribution shows
mixed behaviour of weaker level attraction as well as level repulsion with
respect to Poissonian and GOE respectively, for different values of $S$.
Note, a closeness of curve (d) in Fig. 6.1 to GOE for $S<1,$ though curve do
not rise as much as that of GOE one may easily misunderstood it as showing
GOE kind of behaviour.
\begin{figure}[htbp]
\begin{center}
{\epsfig{file=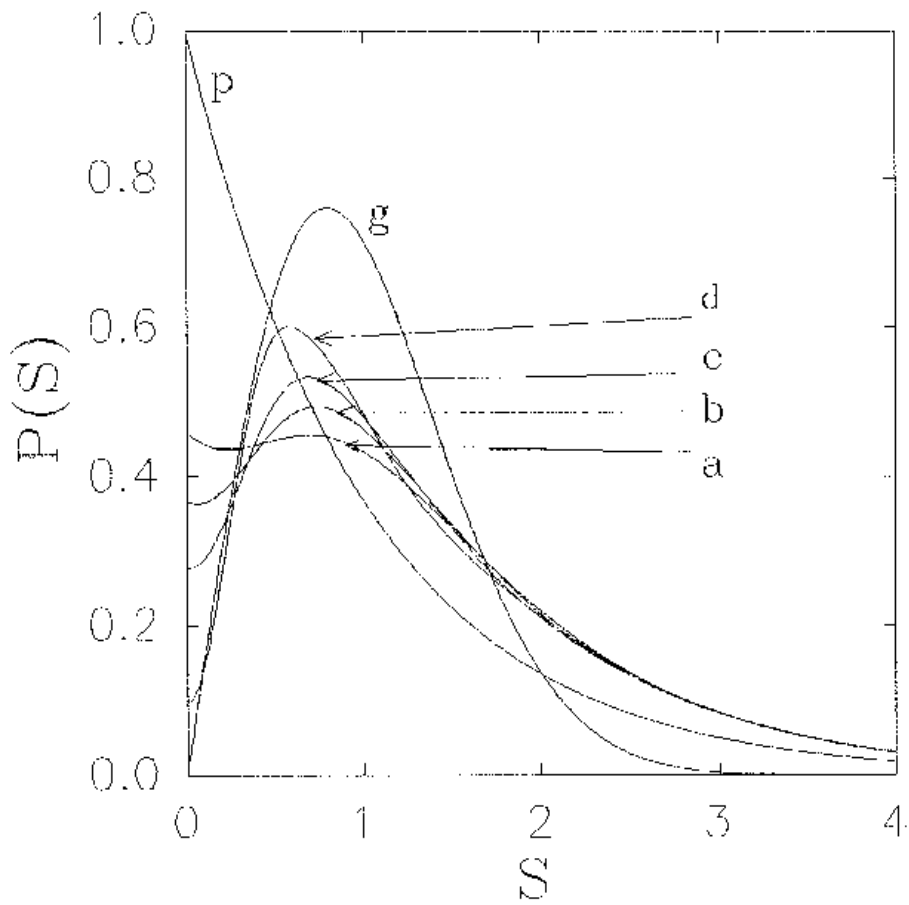,bbllx=80,bblly=170,bburx=560,bbury=700,height=6in}}
\caption{Nearest neighbour spacing distribution. Efect of variation of length parameters (see text)}
\end{center}
\end{figure}
\begin{figure}[htbp]
\begin{center}
{\epsfig{file=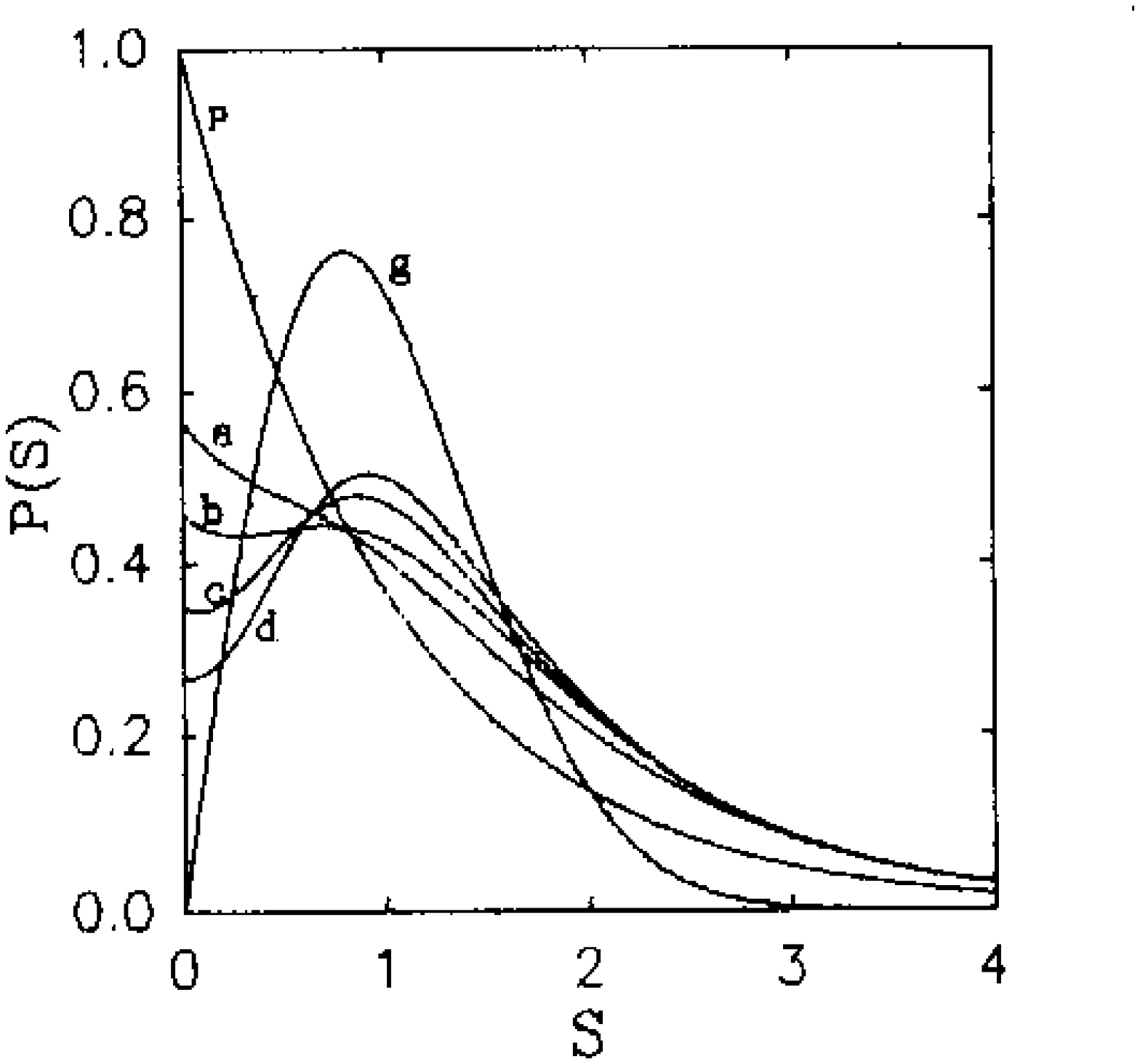,bbllx=80,bblly=170,bburx=560,bbury=700,height=6in}}
\caption{Nearest neighbour spacing distribution. 
Efect of variation of $\kappa$ parameters (see text)}
\end{center}
\end{figure}

In Fig. 6.3 and 6.4 we show plots of $P(S)$ distribution for complete
rhombus and even states of the rhombus. $P(S)$ for complete rhombus
approaches comparatively faster to Poisson distribution reasons are obvious
as it is a superposition of four different modes. In Fig. 6.4 for curve (a)
we choose levels $r_{\min }=47$ and $r_{\max }=370$, to compare our results
with that of numerical results of \cite{db-srj} for approximately same
number of levels. As stated earlier our results are in good agreement with
numerical results for $S<1$. This can be more clearly seen from Fig. 6.5
where we show cumulative $P(S)$ distribution. The curve (a) in this figure
is our results and numerical results are shown by triangles and dots (for
two different mode of even states). Results are in close agreement for $S<1$%
, however for $S>1$ deviation of our results is too large to be correct.
This behaviour of $P(S)$ distribution using diagonal approximation is also
in contradiction with recent studies(\cite{bogo1,grem}), where one get 
\begin{equation}
\label{6-7}P(S)=4S\exp (-2S) 
\end{equation}
Again for $S<<1$ our results (for finite number of levels) are in agreement
with eq. (\ref{6-7}) but deviates appreciably for $S>1$. This contradiction
may be removed by considering off-diagonal contributions. As the calculation
of $P(S)$ involves exponential of integration of $Y_2(S)$, any error in $Y_2$
will be amplified exponentially and effect of this error will be more for $%
S>1$. Further improvements are therefore necessary in (\ref{6-4}) to take
account of off-diagonal contributions.
\begin{figure}[htbp]
\begin{center}
{\epsfig{file=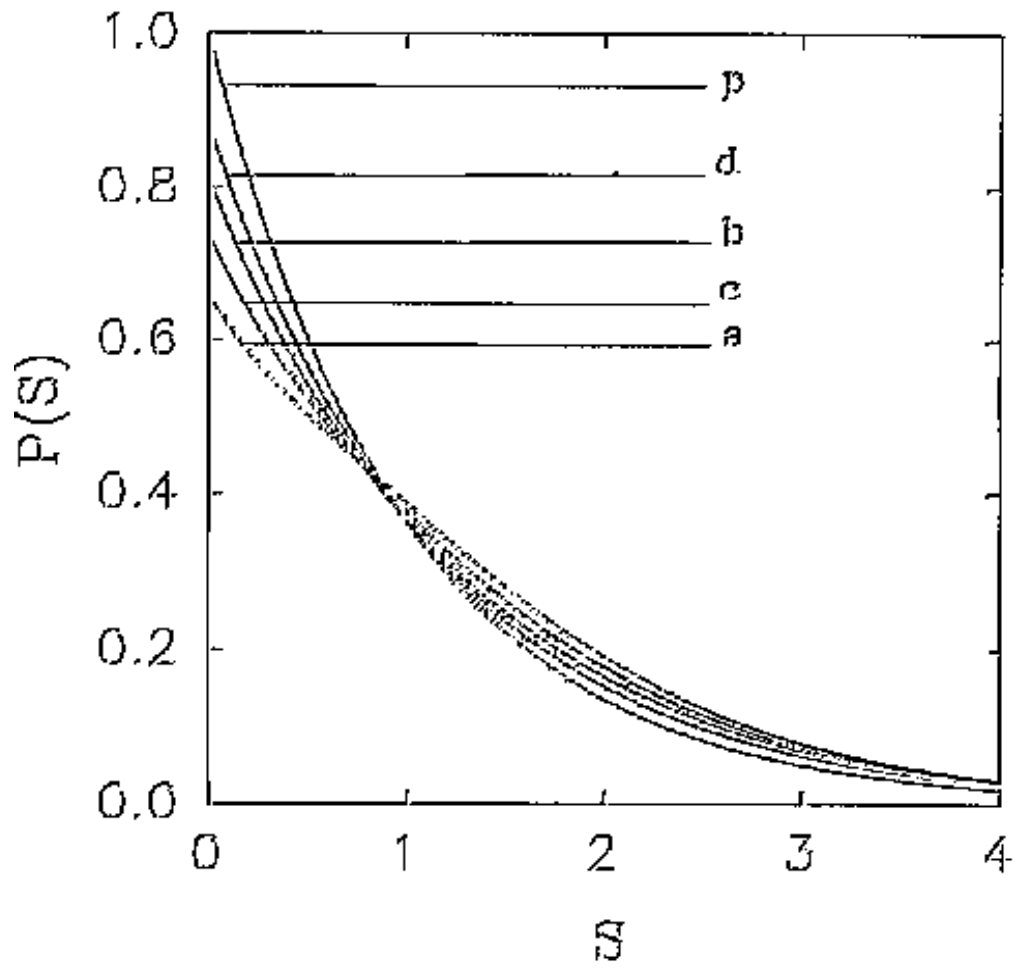,bbllx=80,bblly=170,bburx=560,bbury=700,height=6in}}
\caption{Nearest neighbour spacing distribution. 
complete $\pi/3$-rhombus billiard (see text)}

\end{center}
\end{figure}
\begin{figure}[htbp]
\begin{center}
{\epsfig{file=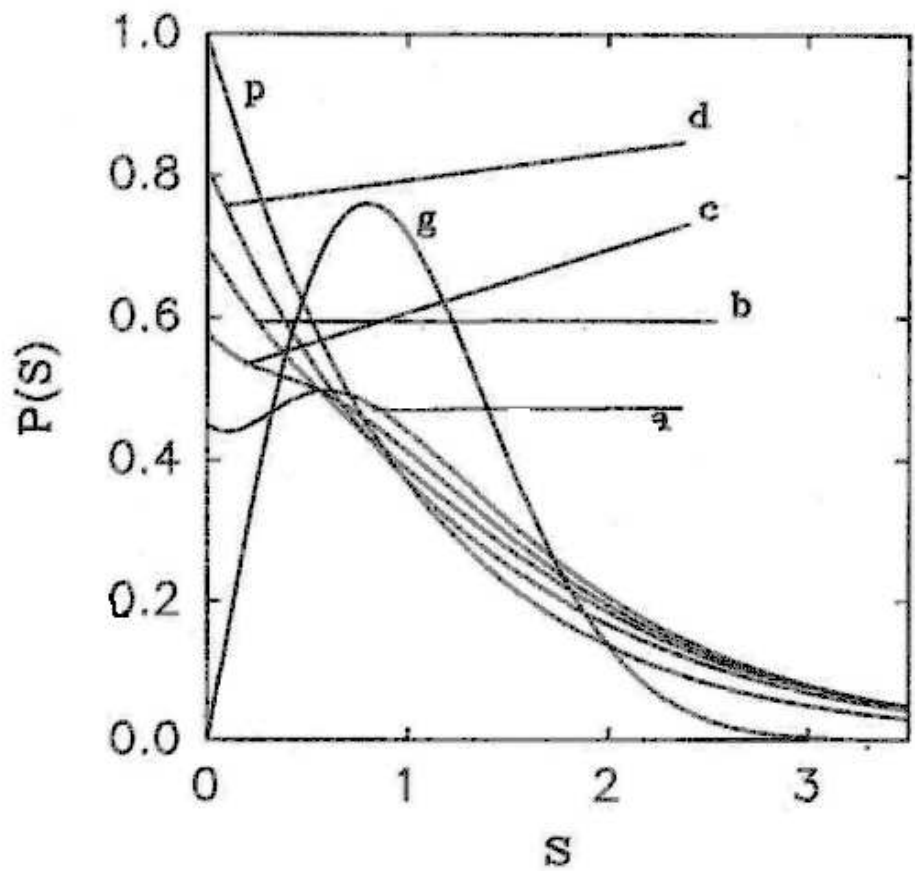,bbllx=80,bblly=170,bburx=560,bbury=700,height=6in}}
\caption{Nearest neighbour spacing distribution. 
pure rhombus modes of  $\pi/3$-rhombus billiard (see text)}
\end{center}
\end{figure}
\begin{figure}[htbp]
\begin{center}
{\epsfig{file=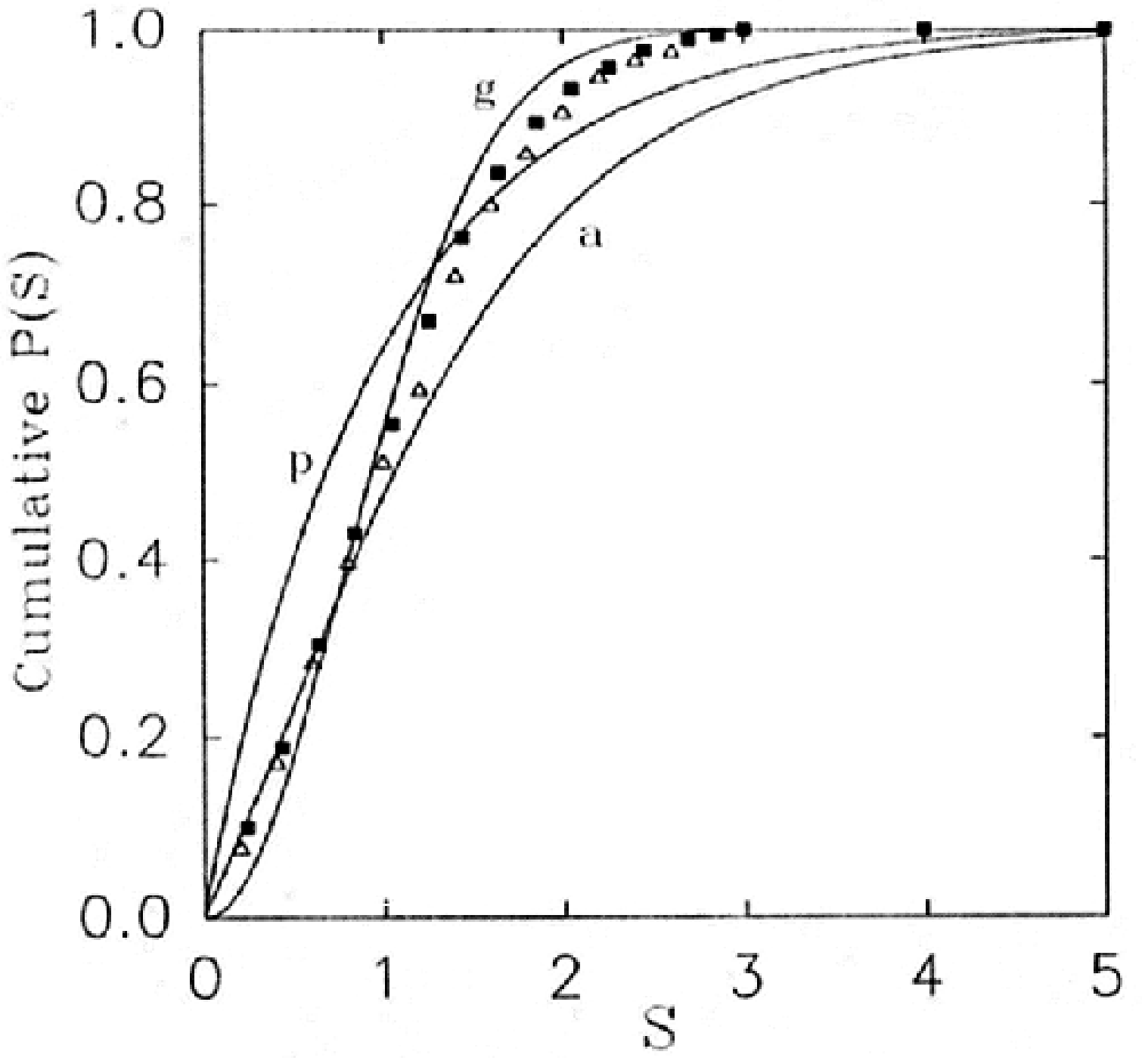,bbllx=80,bblly=170,bburx=560,bbury=700,height=6in}}
\caption{Cummulative nearest neighbour spacing distribution. 
 $\pi/3$-rhombus billiard (only even modes) (see text)}
\end{center}
\end{figure}

\section{The Number Variance}

In the preceding section we have discussed one of the important measure for
spectral fluctuations on the finer energy scales. On the intermediate scales
or on the scale of few mean level spacings, important measures are spectral
rigidity and number variance. Both of these statistic provide similar
information on spectral correlations. In this section we will discuss number
variance for the pseudointegrable billiards. The number variance $\Sigma
^2(L;r_0)$ is defined as the variance of the distribution of the number of
levels in intervals $[r_0,r_0+L]$, $n(L;r_0)$ = $N(r_0+L)-N(r_0)$, 
\begin{equation}
\label{6-8}\Sigma ^2(L;r_0)=<[n(L;r_0)-L]^2>, 
\end{equation}
where, $<\cdots >$ denotes a usual spectral averaging over $r_0$. A
completely random spectrum that follows Poissonian behaviour shows a linear
trend for a number variance, $\Sigma ^2(L;r_0)$=L, whereas on the other
hand, for the chaotic systems where levels are strongly correlated, a number
variance is asymptotically given by $\Sigma ^2(L;r_0)\sim (2/\pi
^2)ln(L)+.44 $ for $L>1$. The number variance $\Sigma ^2(L;r_0)$ depends on $%
Y_2(r)$ via relation (e.g., see \cite{del}) 
\begin{equation}
\label{6-9}\Sigma ^2(L;r_0)=L-2\ \int_0^L(L-s)Y_2(s)\ ds 
\end{equation}
Using results of last chapter, we can exactly integrate above equation to
get simple analytical form for $\Sigma ^2(L;r_0)$ as, 
\begin{equation}
\label{6-10}
\begin{array}{lll}
\Sigma ^2(L;r_0) & = & L-2\ \sum_\alpha \kappa _\alpha \left\{ 
\frac{r_{\max }}{\pi \Delta r}\left[ L{\rm Si}(\gamma _{\alpha ,\max }L)+%
\frac{L\sin (\gamma _{\alpha ,\max }L)}6\right. \right. \\  &  &  \\  
&  & + 
\frac{\gamma _{\alpha ,\max }L^2}6\cos (\gamma _{\alpha ,\max }L)+\frac
2{3\gamma _{\alpha ,\max }}\cos (\gamma _{\alpha ,\max }L) \\  &  &  \\  
&  & \left. + 
\frac{\gamma _{\alpha ,\max }^2L^3}6{\rm si}(\gamma _{\alpha ,\max }L)-\frac
2{3\gamma _{\alpha ,\max }}\right] \\  &  &  \\  
&  & - 
\frac{r_{\min }}{\pi \Delta r}\left[ L{\rm Si}(\gamma _{\alpha ,\min }L)+%
\frac{L\sin (\gamma _{\alpha ,\min }L)}6+\frac{\gamma _{\alpha ,\min }L^2}%
6\cos (\gamma _{\alpha ,\min }L)\right. \\  &  &  \\  
&  & \left. \left. +\frac 2{3\gamma _{\alpha ,\min }}\cos (\gamma _{\alpha
,\min }L)+\frac{\gamma _{\alpha ,\min }^2L^3}6{\rm si}(\gamma _{\alpha ,\min
}L)-\frac 2{3\gamma _{\alpha ,\min }}\right] \right\} 
\end{array}
\end{equation}
where again $\gamma _{\alpha ,\min (\max )}=2\pi l_{\alpha ,\min }/l_{H,\min
(\max )}.$ For small $L$, one can see that $\Sigma ^2(L;r_0)\sim L$ i.e. it
shows Poissonian behaviour, as $L$ increases, we can approximate {\rm Si}$%
(x)\sim \pi /2-\cos (x)/x$ and hence {\rm si}$(x)\sim \cos (x)/x$, $\Sigma
^2(L;r_0)$ starts deviating from the Poissonian and oscillates around a non-
universal value as%
$$
\begin{array}{ccc}
\Sigma ^2(L;r_0) & = & L-L\sum_\alpha \kappa _\alpha -2\sum_\alpha \kappa
_\alpha \left\{ 
\frac{r_{\max }}{\pi \Delta r}\left[ \frac{L\sin (\gamma _{\alpha ,\max }L)}%
6\right. \right. \\  &  &  \\  
&  & -\left. 
\frac{\cos (\gamma _{\alpha ,\max }L)}{3\gamma _{\alpha ,\max }}-\frac
2{3\gamma _{\alpha ,\max }}\right] \\  &  &  \\  
&  & \left. -\frac{r_{\min }}{\pi \Delta r}\left[ \frac{L\sin (\gamma
_{\alpha ,\min }L)}6-\frac{\cos (\gamma _{\alpha ,\min }L)}{3\gamma _{\alpha
,\min }}-\frac 2{3\gamma _{\alpha ,\min }}\right] \right\} 
\end{array}
$$
Thus the saturation value for very large number of levels (i.e. $\gamma
_\alpha \gg L$) is given by 
$$
\Sigma _\infty ^2(L;r_0)=2\sqrt{\frac{A_R}{\pi ^3}}\sum\limits_\alpha \frac{%
\kappa _\alpha }{l_{\min ,\alpha }}\left\{ \frac{r_{\max }^{3/2}-r_{\min
}^{3/2}}{\Delta r}\right\} 
$$
This behaviour is very much akin to that of the spectral rigidity and
saturation value is correctly given by $\Sigma _\infty \simeq 2\Delta
_{3\;\infty }$, where $\Delta _{3,\infty }$ is obtain directly (i.e. not
from $Y_2$) in the sext section. In the Fig. 6.6 and 6.7 we show this
behaviour as well as effect of variations in the length parameters and $%
\kappa $ parameters respectively as done earlier. Trend is same. Hence our
conclusions are also similar to the one stated earlier and hardly needs any
elaboration. In Fig. 6.8 and 6.9 we show more realistic cases namely
complete rhombus and even states of the rhombus respectively. In Fig. 6.9,
one can compare our results for energy level windows ($47,370$) (curve (a))
with numerical results of \cite{db}. Discounting oscillations of numerical
results agreement is good even for $r>1$. This is because, number variance
involves simple integration over $Y_2$ and any error due to off-diagonal
contributions is not as much amplified as that in the $P(S)\,$ distribution.
In the next section we will confirm this in the case of another spectral
measure known as spectral rigidity.
\begin{figure}[htbp]
\begin{center}
{\epsfig{file=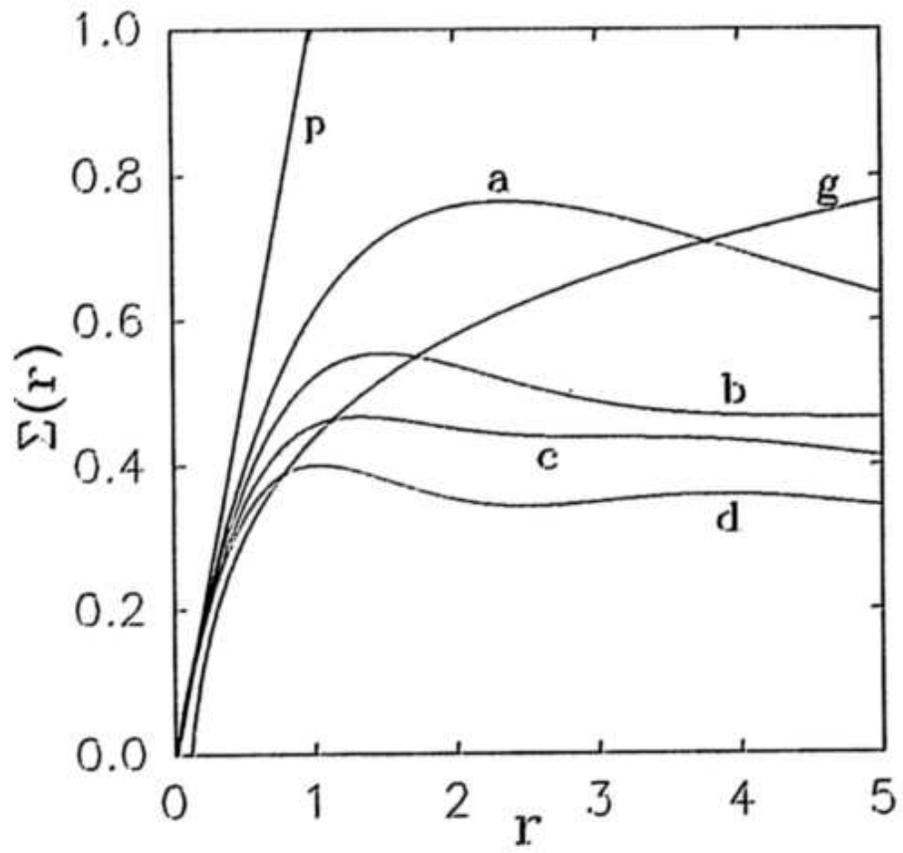,bbllx=80,bblly=170,bburx=560,bbury=700,height=6in}}
\caption{The number variance:  Efect of variation of length parameters.  (see text)}
\end{center}
\end{figure}
\begin{figure}[htbp]
\begin{center}
{\epsfig{file=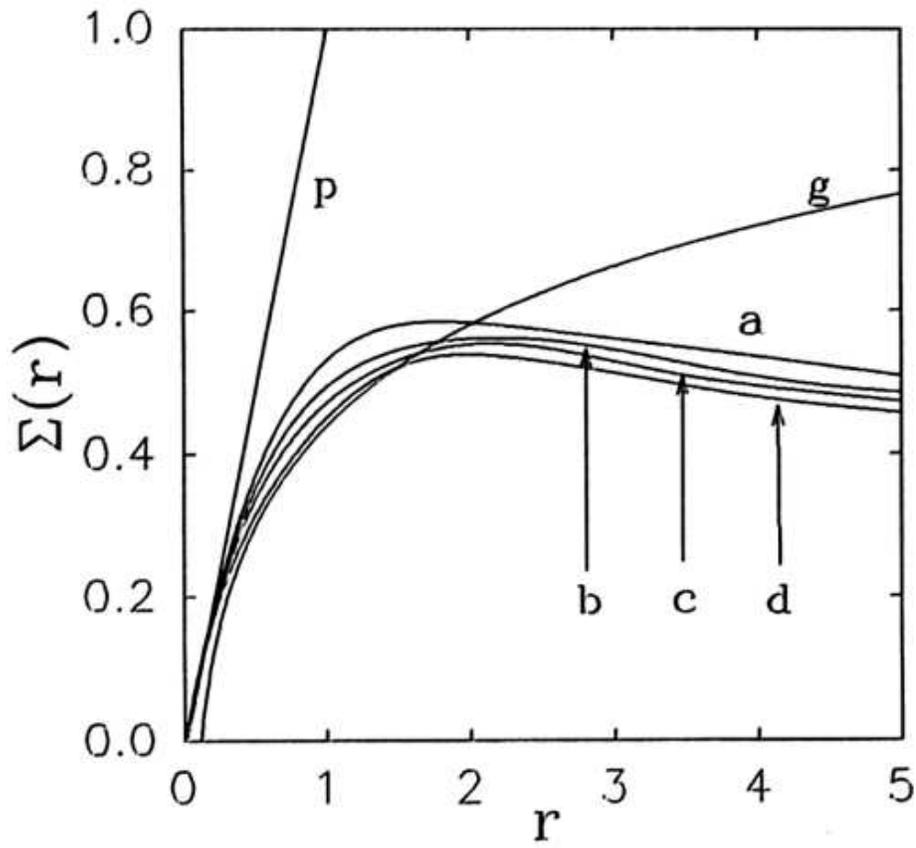,bbllx=80,bblly=170,bburx=560,bbury=700,height=6in}}
\caption{The number variance:  Efect of variation of $\kappa$ parameters.  (see text)}
\end{center}
\end{figure}
\begin{figure}[htbp]
\begin{center}
{\epsfig{file=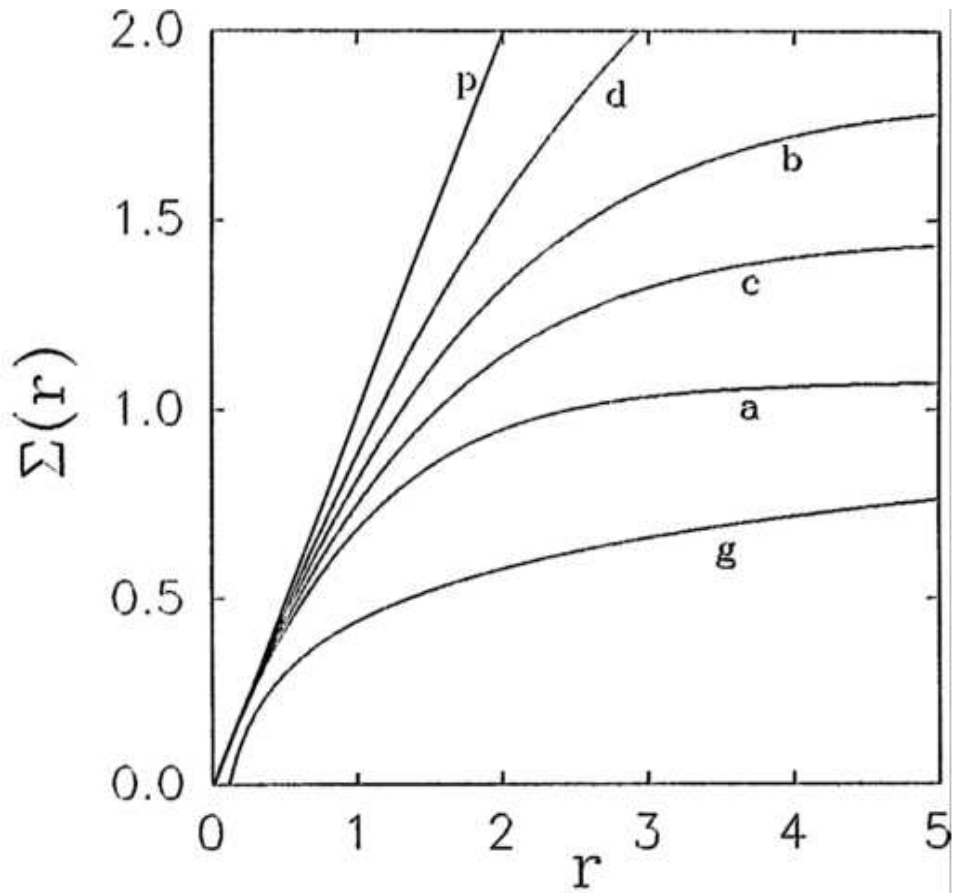,bbllx=80,bblly=170,bburx=560,bbury=700,height=6in}}
\caption{The number variance:complete $\pi/3$-rhombus billiard (see text)}    

\end{center}
\end{figure}
\begin{figure}[htbp]
\begin{center}
{\epsfig{file=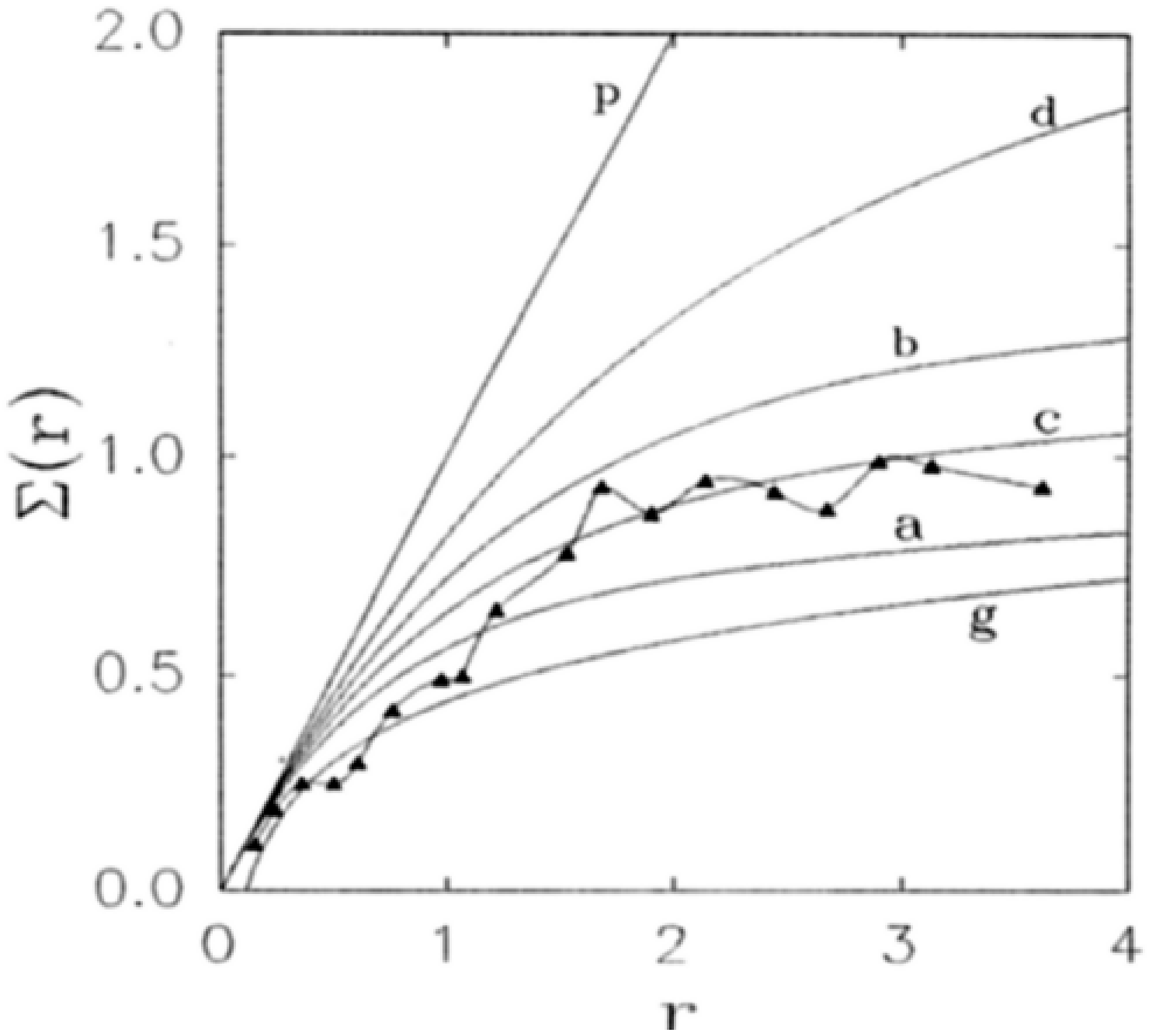,bbllx=80,bblly=170,bburx=560,bbury=700,height=5in}} 
\caption{The number variance: for even modes of $\pi/3$-rhombus billiard (see text)}
\end{center}
\end{figure}

\section{The Spectral Rigidity}

In this section we obtain \cite{6-rigi} the Dyson-Mehta $\Delta _3$%
-statistic for pseudointegrable billiards and show that it is nonuniversal
with a universal trend, also that this trend is similar to the one for
integrable billiards. We present a formula, based on exact semiclassical
calculations and the proliferation law of periodic orbits, which gives
rigidity for the entire range of L. To consolidate our theory, we discuss
several examples finding complete agreement with the numerical results , and
also the underlying fundamental reasons for the nonuniversality.

In their statistical theory of energy levels of complex systems, Dyson and
Mehta \cite{fd-mlm} proposed the $\Delta $-statistic to study spectral
fluctuations on the intermediate energy scale, the most popular being the $%
\Delta _3$-statistic defined as a {\it local average of the mean square
deviation of the spectral staircase from the best fitting straight line over
an energy range corresponding to }$L${\it \ mean level spacings.} 
\begin{equation}
\label{ri-1}\Delta _3(L)=\left\langle \min _{(A,B)}\frac{d_{av}(E)}%
L\int\limits_{-L/2d_{av}}^{L/2d_{av}}d\varepsilon \left[ N(E+\varepsilon
)-A-B\varepsilon \right] ^2\right\rangle 
\end{equation}
where $N(E)$ is spectral staircase introduced in chapter(1) (keeping in
consonance with existing literature we use symbol $L$ for correlation range
in this section, and not to be confused with symbol for side length used
earlier ). $\langle ...\rangle $ represents spectral averaging over energy
scale much larger than what is called as outer energy scale but much smaller
than classical scale\cite{mvb}, which is already discussed in the previous
chapter. Though, $\Delta _3(L)\,$ can also be expressed in terms of the
two-point correlation function, we prefer to use above direct definition to
obtain an expression for $\Delta _3$ in pseudo-integrable billiards.
Minimizing \ref{ri-1}over $A$ and $B$ we get 
\begin{equation}
\label{ri-2}
\begin{array}{lll}
\Delta _3(L) & = & \left\langle \left\{ 
\frac{d_{av}(E)}L\int\limits_{-L/2d_{av}}^{L/2d_{av}}d\varepsilon
N^2(E)-\left[ \frac{d_{av}(E)}L\int\limits_{-L/2d_{av}}^{L/2d_{av}}d%
\varepsilon N(E+\varepsilon )\right] ^2\right. \right. \\  &  &  \\  
&  & -12\left[ \left( \frac{d_{av}(E)}L\right)
^2\int\limits_{-L/2d_{av}}^{L/2d_{av}}\varepsilon d\varepsilon
N(E+\varepsilon )\right] ^2 
\end{array}
\end{equation}
For a Poisson spectrum (by which integrable systems are very well modeled) $%
\Delta _3(L)=L/15$. Consider another trivial example of a spectrum of the
harmonic oscillator in one dimension. Its spectrum is also called picket
fence spectrum due to equi-spacing between levels. $N(E)$ for such case is a
ideal staircase and obviously mean square deviation of a best fit from this
staircase is constant $(1/12)$ independent of $L$. For the chaotic systems
with time reversal symmetry which are modeled by GOE of random matrix
theory rigidity is given by $\Delta _3(L)=\ln L/\pi ^2-.007$. It may be
noted that for $L<<1$, the staircase nature of $N(E)\,$ leads to the limit $%
\Delta _3\rightarrow L/15$, whatever distribution(non-singular) the levels
have. We now proceed to obtain expression for spectral rigidity in
psedo-integrable billiards.

A asymptotic semiclassical approximation for $N(E)\,=\int\limits_0^Ed(E^{%
\prime })dE^{\prime }$ in case of pseudointegrable billiards is given by 
\begin{equation}
\label{ri-3}N(E)=\frac{E^{1/4}}{\sqrt{2\pi ^3\hbar }}\sum\limits_j\frac{A_j}{%
l_j^{3/2}}\cos \left( \frac{l_j\sqrt{E}}\hbar -\frac{3\pi }4\right) 
\end{equation}

Substituting this in eq. \ref{ri-2} we can obtain the rigidity as 
\begin{equation}
\label{ri-4}\Delta _3(L,E)=\frac{E^{1/2}}{4\pi ^3\hbar }\sum_j\sum_k\frac{%
A_jA_k}{l_j^{3/2}l_k^{3/2}}\cos \left\{ \frac{\sqrt{E}(l_j-l_k)}\hbar
\right\} G(y_j,y_k) 
\end{equation}
where 
\begin{equation}
\label{ri-5}G(y_j,y_k)=f(y_j-y_k)-f(y_j)f(y_k)-3f^{\prime }(y_j)f^{\prime
}(y_k) 
\end{equation}
and $f(y)=\sin y/y,\,\,\,\,\,\,y=Ll/(4d_{av}(E)E^{1/2}\hbar )=\pi Ll/l_H$.
All other symbols have their usual meaning.

The main questions that ensue from the numerical studies \cite
{ch-coh,db-srj,shu,ccg} are : a) are the levels of PIB uncorrelated and
mimic a Poisson process in a certain range of L, as seen in IB ?; b) is
there any saturation of rigidity in PIB if L exceeds the system dependent
range, as seen for the IB ?; c) what is the essential difference between
PIB, IB and chaotic billiards in terms of level correlations ?; d) can we
obtain a formula for the rigidity such that Poisson and non-Poisson results
follow in a natural way for the IB and PIB respectively ? . Here we answers
all these questions to a large (sometimes complete) extent. This success
holds due to the fact that the rigidity is a direct consequence of the
proliferation law with some simple, nontrivial modifications in the known
formalism shown below. Recalling discussion in the last chapter, we employ
the uniformity principle \cite{ha-ozo} and retain also, apart from the
diagonal, the off-diagonal part corresponding to the systematic degeneracies
in the lengths of po's (giving same contribution to the rigidity as the
diagonal terms). From the exact results on some of the PIBs obtained in
previous chapters, one can classify the bands of the po's in such a way that
the projective phase space area occupied by all periodic orbits in a given
class\ (defined by corresponding lattice points where po's closes) (say $%
\alpha $) is identical ($A_\alpha $). With this in mind, we can write $%
\Delta _3(L,E)$ as 
\begin{equation}
\label{ri-5-1}
\begin{array}{ccc}
\Delta _3(L,E) & = & \frac{E^{1/2}}{4\pi ^3\hbar }\left[ \sum_\alpha
g_\alpha ^2A_\alpha ^2\sum_j\frac{G(y_{\alpha ,j})}{l_{\alpha ,j}^3}\right.
\\  &  &  \\  
&  & +\left. \sum_\alpha \sum_\beta (1-\delta _{\alpha ,\beta })g_\alpha
g_\beta A_\alpha A_\beta \sum_j\sum_k\delta _{l_{\alpha ,j},l_{\beta ,k}}%
\frac{G(y_{\alpha ,j},y_{\beta ,k})}{l_{\alpha j\beta k}^{3/2l3/2}}\right] 
\end{array}
\end{equation}
where $G(y)=1-f^2(y)-3(\dot f(y))^2$ and Greek subscripts again denote
classes of periodic bands. In (\ref{ri-5-1}), $g_\alpha g_\beta $ denotes the
number of po's with the same action belonging to the class $\alpha (\beta )$%
, and $\delta _{i,j}$ is the usual Kronecker symbol. In the above equation
the summation $\sum_jG(y_{\alpha ,j})/l_{\alpha ,j}^3$ can be written as $%
\int_{y_{min}}^\infty \ dF_\alpha \ G(y_\alpha )/l_\alpha ^3$ in the
continuum limit due to the mathematical nature of the summand, where $%
dF_\alpha $ represents number of periodic orbits within length $l$ and $l+dl$%
. This $dF_\alpha $ can be deduced from the proliferation law (average or
asymptotic part) as we have done in the earlier chapters. We then have
(after unfolding spectrum via rescaled energies $r_0=Ed_{av}$) 
\begin{equation}
\label{ri-6}
\begin{array}{lll}
\Delta _3(L,r_0{\cal )} & = & \frac L{2\pi ^2A_R}[\sum_\alpha g_\alpha
^2A_\alpha ^2a_\alpha I_{1,\alpha }+\sum_\gamma \sum_\eta \delta _{l_\gamma
,l_\eta }g_\gamma g_\eta A_\gamma A_\eta a_\eta I_{1,\eta }] \\  
&  &  \\  
&  & +\frac{L^2}{8\pi ^{3/2}A_R^{3/2}r_0^{1/2}}[\sum_\alpha g_\alpha
^2A_\alpha ^2b_\alpha I_{2,\alpha }+\sum_\gamma \sum_\eta \delta _{l_\gamma
,l_\eta }g_\gamma g_\eta A_\gamma A_\eta b_\eta I_{2,\eta }] 
\end{array}
\end{equation}
where%
$$
I_{1,\alpha }=\int\limits_{y_{\min ,\alpha }}^\infty dy_\alpha
\,\,\,\,\,y_\alpha ^{-2}\,\,\,\,G(y_\alpha ) 
$$
and%
$$
I_{2,\alpha }=\int\limits_{y_{\min ,\alpha }}^\infty dy_\alpha
\,\,\,\,\,y_\alpha ^{-3}\,\,\,\,G(y_\alpha ) 
$$
both these can be evaluated easily with $y_{\min }=\pi L/L_{\max }$. And 
$$
L_{\max ,\alpha }=\sqrt{4\pi A_Rr_0/l_{\min ,\alpha }^2}=\frac{l_H}{l_{\min
,\alpha }} 
$$
where $l_H$ is Heisenberg length already introduced earlier. For small $%
y_{\min },\,\,\,\,I_1=2\pi /15,\,\,\,I_2=1/9$; and for large $y_{\min
},\,\,\,\,I_{1,\alpha }=L_{\max ,\alpha }/\pi L,$ $I_{2,\alpha }=L_{\max
,\alpha }^2/2\pi ^2L^2.$ For $L<\min _\alpha L_{\max ,\alpha }/\pi $, 
\begin{equation}
\label{ri-7}
\begin{array}{lll}
\Delta _3(L,r_0{\cal )} & = & \frac L{15\pi A_R}[\sum_\alpha g_\alpha
^2A_\alpha ^2a_\alpha +\sum_\gamma \sum_\eta \delta _{l_\gamma ,l_\eta
}g_\gamma g_\eta A_\gamma A_\eta a_\eta ] \\  
&  &  \\  
&  & +\frac{L^2}{72\pi ^{3/2}A_R^{3/2}r_0{\cal ^{1/2}}}[\sum_\alpha g_\alpha
^2A_\alpha ^2b_\alpha +\sum_\gamma \sum_\eta \delta _{l_\gamma ,l_\eta
}g_\gamma g_\eta A_\gamma A_\eta b_\eta ]. 
\end{array}
\end{equation}
For $L>>\max _\alpha L_{\max ,\alpha }/\pi ,$%
\begin{equation}
\label{ri-8}
\begin{array}{lll}
\Delta _3(L,r_0{\cal )} & = & \frac 1{2\pi ^3A_R}[\sum_\alpha g_\alpha
^2A_\alpha ^2a_\alpha L_{\max ,\alpha }+\sum_\gamma \sum_\eta \delta
_{l_\gamma ,l_\eta }g_\gamma g_\eta A_\gamma A_\eta a_\eta L_{\max ,\eta }]
\\  
&  &  \\  
&  & +\frac 1{16\pi ^{7/2}A_R^{3/2}{\cal r}_0^{1/2}}[\sum_\alpha g_\alpha
^2A_\alpha ^2b_\alpha L_{\max ,\alpha }^2+\sum_\gamma \sum_\eta \delta
_{l_\gamma ,l_\eta }g_\gamma g_\eta A_\gamma A_\eta b_\eta L_{\max ,\eta
}^2]. 
\end{array}
\end{equation}
It is important to note that minimum and maximum (over $\alpha $) $L_{\max }$
correspond respectively to the longest and shortest (over $\alpha $) orbits
of the set containing shortest periodic orbits of different $\alpha $'s. It
is the consequence of this observation that will lead us to understand the
fundamental distinction between the spectral correlations of integrable and
pseudointegrable billiards.

Ignoring $I_2$ for the sake of brevity, the formula valid for the entire
range of $L$ is given by (denoting $L_{\max ,\alpha }/\pi L=l_H/\pi Ll_{\min
,\alpha }$ by $\Lambda _\alpha $), 
\begin{equation}
\label{ri-9}
\begin{array}{lll}
\Delta _3(L,{\cal r}_0) & = & \frac L{2\pi }\sum_\alpha \kappa _\alpha
[\Lambda _\alpha -\frac 23\Lambda _\alpha ^3-\frac 3{10}\Lambda _\alpha ^5
\\  
&  &  \\  
&  & -\frac 2{15}\Lambda _\alpha \cos (2/\Lambda _\alpha )-\frac
1{15}\Lambda _\alpha ^2\sin (2/\Lambda _\alpha )+\frac 1{15}\Lambda _\alpha
^3\cos (2/\Lambda _\alpha ) \\  
&  &  \\  
&  & +\theta \frac 35\Lambda _\alpha ^4\sin (2/\Lambda _\alpha )+\frac
3{10}\Lambda _\alpha ^5\cos (2/\Lambda _\alpha )-\frac 4{15}si(2/\Lambda
_\alpha )]\Lambda 
\end{array}
\end{equation}
where ${\rm si}(x)=-\int_x^\infty dt\,\,t^{-1}\sin t$ and $\kappa _\alpha
=A_\alpha ^2g_\alpha ^2a_\alpha /\pi A_R$. Eq. (\ref{ri-9}) along with the
limiting results (\ref{ri-7}), (\ref{ri-8}) ; which clearly establishes
relation between $\Delta _3$ and information about the classical periodic
orbits such as proliferation law, band areas, degeneracies in lengths etc.
This result applies to all integrable and pseudointegrable billiards. To
understand the formulae better, we now propose to examine some paradigm
systems carefully, and subsequently compare the results with the known
numerical results.

In this regard, we consider the specific examples of an Incommensurate
Rectangle Billiard (IRB), the Single Slit Rectangle Billiard (SSRB) and the $%
\pi /3$-rhombus Billiard (RHB).

Firstly, let us consider the IRB with sides $({\cal L},\gamma {\cal L})$, $%
\gamma $ being an irrational number. The all periodic orbits fall in a
single class occupying projective phase space area $4A_R$ ($A_R=\gamma {\cal %
L}^2$), except the two shortest periodic orbit bands parallel to either pair
of sides of IRB. The area of these two shortest periodic bands is $2A_R$.
The proliferation law for the IRB can be easily found by employing the ideas
of stacking and replication, we get (counting all the repetitions of the
ppo's), 
\begin{equation}
\label{ri-10}F_{IRB}(l)=al^2+bl\ =\frac \pi {16{\cal L}^2}l^2+\frac{\gamma +1%
}{4\gamma {\cal L}}l. 
\end{equation}
Using a, $A(=4A_R)$ in (9) we get complete quantitative agreement with
results obtained earlier \cite{ccg} for $L<L_{\max }/\pi $. As observed in 
\cite{ccg}, the oscillations in $\Delta _3(L)$ are rather weak beyond the
''crossover regime''.

To get the correct saturation values of $\Delta _3(L)$, we have to consider $%
O(l)$ term in eq. (\ref{ri-10}). Taking account of this, in the region where 
$L<L_{\max }/\pi $, we get 
\begin{equation}
\label{ri-11}\Delta _3(L,{\cal r}_0)=\frac L{15}+\frac 1{9\sqrt{2}\pi
^{3/2}}\left( \frac{\gamma +1}\gamma \frac{L^2}{r_0^{1/2}}\right) 
\end{equation}
The second term is quite small as compared to the first one due to $r_0{\cal %
^{-1/2}}$ factor. For $L\gg L_{\max }/\pi $, on the other hand, we have 
\begin{equation}
\label{ri-12}\Delta _3(L,r_0{\cal )=}\frac{r_0^{1/2}}{2\pi ^{3/2}}\left( 1+%
\frac{\sqrt{\gamma }(\gamma +1)}{2\pi }\right) {\cal ,}{} 
\end{equation}
which is in very good agreement with the numerical results.

Our next example is the single slit rectangle billiard, which is a simple
variation of the barrier billiard already discussed in previous chapters.
Recall that this is an example of a PIB whose invariant surface is
topologically equivalent to a sphere with two handles (genus, $g=3$). The
law of proliferation is the same as $a_\alpha ^2l+b_\alpha l$. We can obtain 
$a_\alpha ,b_\alpha $ for different classes of bands in this system in
similar manner as in chapter (4).

With these, for $L<\min _\alpha (L_{\max ,\alpha }/\pi )=\sqrt{r_0/4\pi }$, 
\begin{equation}
\label{ri-13}\Delta _3(L,r_0)=\frac L{15}+\frac 1{18\sqrt{2\pi ^3}}\frac{L^2%
}{r_0^{1/2}} 
\end{equation}
and for $L>>\max _\alpha (L_{\max ,\alpha }/\pi )=\sqrt{8r_0/\pi }$, 
\begin{equation}
\label{ri-14}\Delta _3(L,r_0)\sim \left[ \frac{\sqrt{2}+13}{12\pi ^{3/2}}%
+\frac 9{8\sqrt{2\pi ^5}}\right] r_0^{1/2} 
\end{equation}

We will discuss these results after we present calculation for yet another
well studied system - the $\pi /3$-rhombus billiard. This is an almost
integrable system with an invariant integral surface of genus two. For this
system tessellation of the plane is not complete and results in more general
barrier structure \cite{e-f-v}. Again, recall that each trajectory from
origin to a coprime pairs, $(q,p)$ represents ppo ending at $c(q,p)$. We
then get for $L<\min _\alpha (L_{\max ,\alpha }/\pi )=\sqrt{2r_0/37\sqrt{3}%
{\cal \pi }}$, 
\begin{equation}
\label{ri-15}\Delta _3(L,r_0)\sim \frac L{15}+\frac 1{3^{7/4}2^{1/2}\pi
^{3/2}}\frac{L^2}{r_0^{1/2}} 
\end{equation}
and for $L>>\max _\alpha (L_{\max ,\alpha }/\pi )=\sqrt{2\sqrt{3}\pi r_0/3}$ 
\begin{equation}
\label{ri-16}\Delta _3(L,r_0)\sim .237\frac{r_0^{1/2}}{\pi ^{3/2}}+\sqrt{%
\frac{r_0/2}{3^{5/2}\pi ^5}} 
\end{equation}

From expressions and examples discussed above, one can clearly see that
there is an universal trend of $\Delta _3(L)$ with $L$ for integrable and
pseudointegrable billiards. More precisely, for $L<\min _\alpha (L_{\max
,\alpha }/\pi )$, the rigidity is very well approximated by $L/15$, and for $%
L>>\max _\alpha (L_{\max ,\alpha }/\pi )$ it saturates with a crossover
connecting these two limits smoothly. The extent of the crossover region is
given by the difference between $\min _\alpha $ and $\max _\alpha $ of $%
L_{\max ,\alpha }$, or in other words, depends on the spectrum of lengths of
shortest periodic orbits over $\alpha $. Nonuniversal aspects, thus, arise
due to nontrivial classification depending upon the degree of tessellation
of invariant surface in terms of a system-specific fundamental region. For
instance, in IRB, tessellation is complete and there is only one class of
bands ($\alpha =1$); the crossover region is expected to be of lesser extent
- a fact fully corroborated by the numerical experiments. In the SSRB, there
is a barrier (gap to barrier ratio is unity) in a rectangle which gives rise
to a periodic untessellated arabesque in terms of which classification is
facilitated - the number of bands here is seven. Similarly for the RHB, the
number of bands is eighteen. Importantly, it should be noted that the value
of $L$ at which the spectral rigidity deviates from the Poisson value of $%
L/15$, and the value at which saturation sets in, depends upon the lengths
of the shortest periodic orbits distributed over various classes admissible
in a given system. Indeed, this is the fundamental source of nonuniversality.

Let us discuss the numerical results on various pseudointegrable billiards.
Most of the studies have been on rhombus billiards \cite{db-srj}, square
torus billiard and its generalizations \cite{ch-coh} and singular billiards 
\cite{seba}. The analysis for the singular billiards was carried out and one
understands the level spacing statistics \cite{alb-seba}. The study of the
two-level cluster function (in particular $\Sigma ^2(L)$) does not give the
GOE result \cite{seba-w} although the level spacing is GOE raising, thereby,
a question currently beyond explanation. Therefore, we concentrate to
explain the results for non-singular systems.

Perhaps the paradigm pseudointegrable billiard is the RHB \cite{db-srj,shu}.
In both these studies, one can observe that the rigidity is intermediate to
Poisson and GOE. From our analysis, taking the energy and parameters from
these numerical work, it turns out that deviation from $L/15$ would occur at 
$L\sim 1$ and $2$ respectively. We illustrate this in Fig.6.10, where we
compare our analytical result with that of the numerical work \cite{db-srj}
and the agreement is clearly evident. Here also as in the case of number
variance errors due to neglecting off-diagonal contribution do not affect
much. The crossover values are also correctly predicted by our analysis. We
show behaviour of $\Delta _3$ for small values of $L$ in Fig.6.11, where
deviation from $L/15$is evident. In Fig.6.12 we show $\Delta _3$ for
complete range of $L,$ where the crossover region and the saturation can be
seen. Since the numerical results are not available for higher energy and
for larger range of $L$, the saturation cannot be clearly seen in the
numerical experiments. It is, therefore, desirable to carry out extensive
numerical work for higher energy and for larger range of $L$. The formulae (%
\ref{ri-7},\ref{ri-8}) provide the guidelines for choosing appropriate
number of levels to bring out all the salient features of the systems
discussed above. Our analysis also explains the results of \cite{ch-coh}
where one gets $L/15$ for very small values of $L$ and there is a saturation
regime. Unfortunately, because of constraints over levels available, the
belief of an intermediate behaviour between that of Poisson and GOE has been
pursued for quite sometime. Our analysis clearly reveals, that such a
behaviour does not exist and the spectral rigidity never becomes GOE.

The occurrence of periodic orbits in the bands is a likely reason for the
slow rise of $\Delta _3(L)$ in large $L$ region and overall stronger
fluctuations than the GOE result. A recent result on Stadium billiard
indicates this possibility too \cite{graf} - in this work, $\Delta _3(L)$ is
shown to be rising well above the GOE curve if the contribution of the
bouncing ball modes is taken into account. In chaotic systems like this
(also, e.g. the Sinai billiard) the analysis of bands can be carried out
using the above theory and it is expected that there exist a departure from
GOE as well as a rise in spectral fluctuations after some $L$ decided by the
length of the periodic orbits in the band. Recently, non-genericity of the
rigidity arising from banded orbits is discussed for the Stadium billiard 
\cite{sscl}.
\begin{figure}[htbp]
\begin{center}
{\epsfig{file=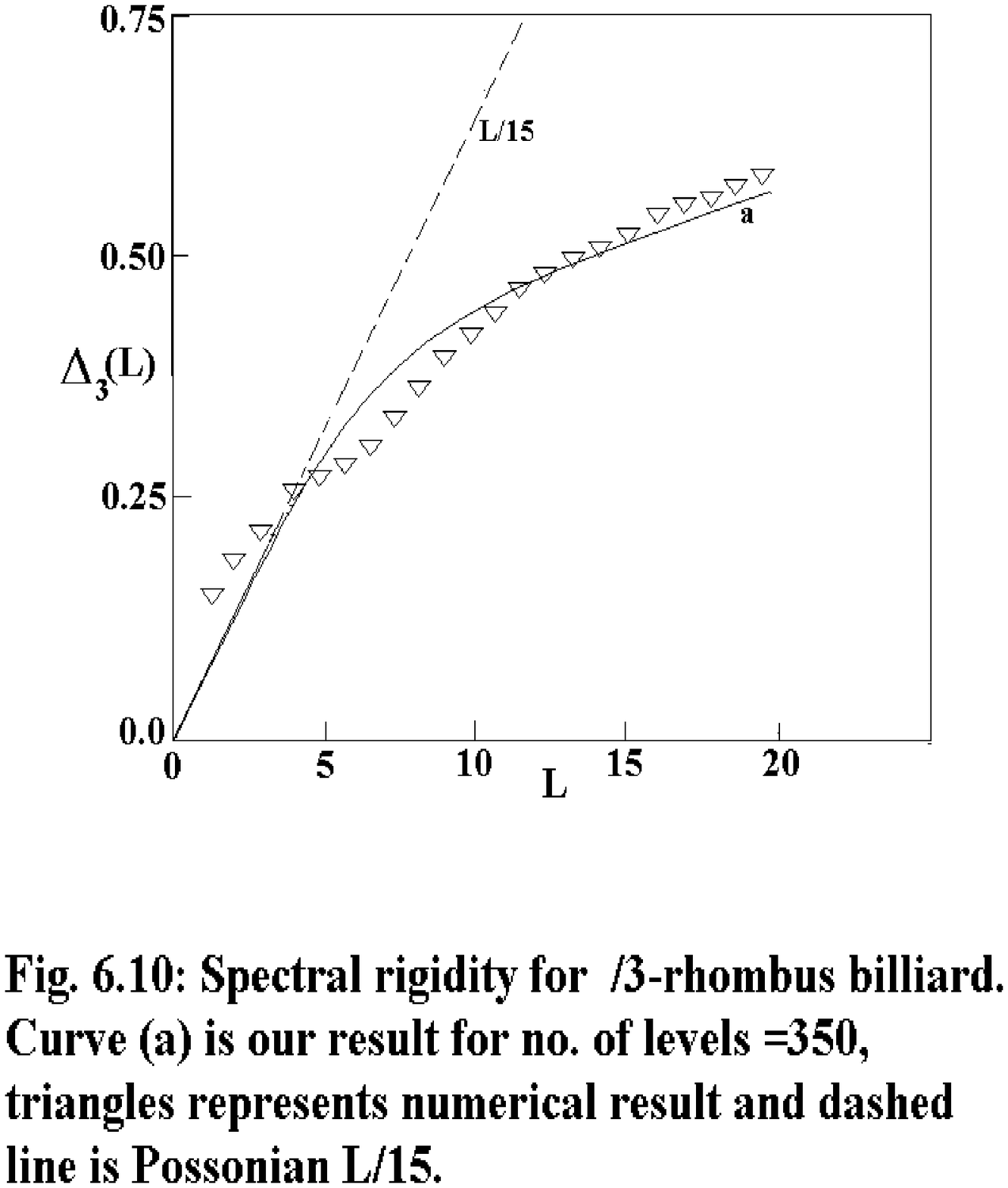,bbllx=80,bblly=150,bburx=580,bbury=770,height=6in}}
\end{center}
\end{figure}
\begin{figure}[htbp]
\begin{center}
{\epsfig{file=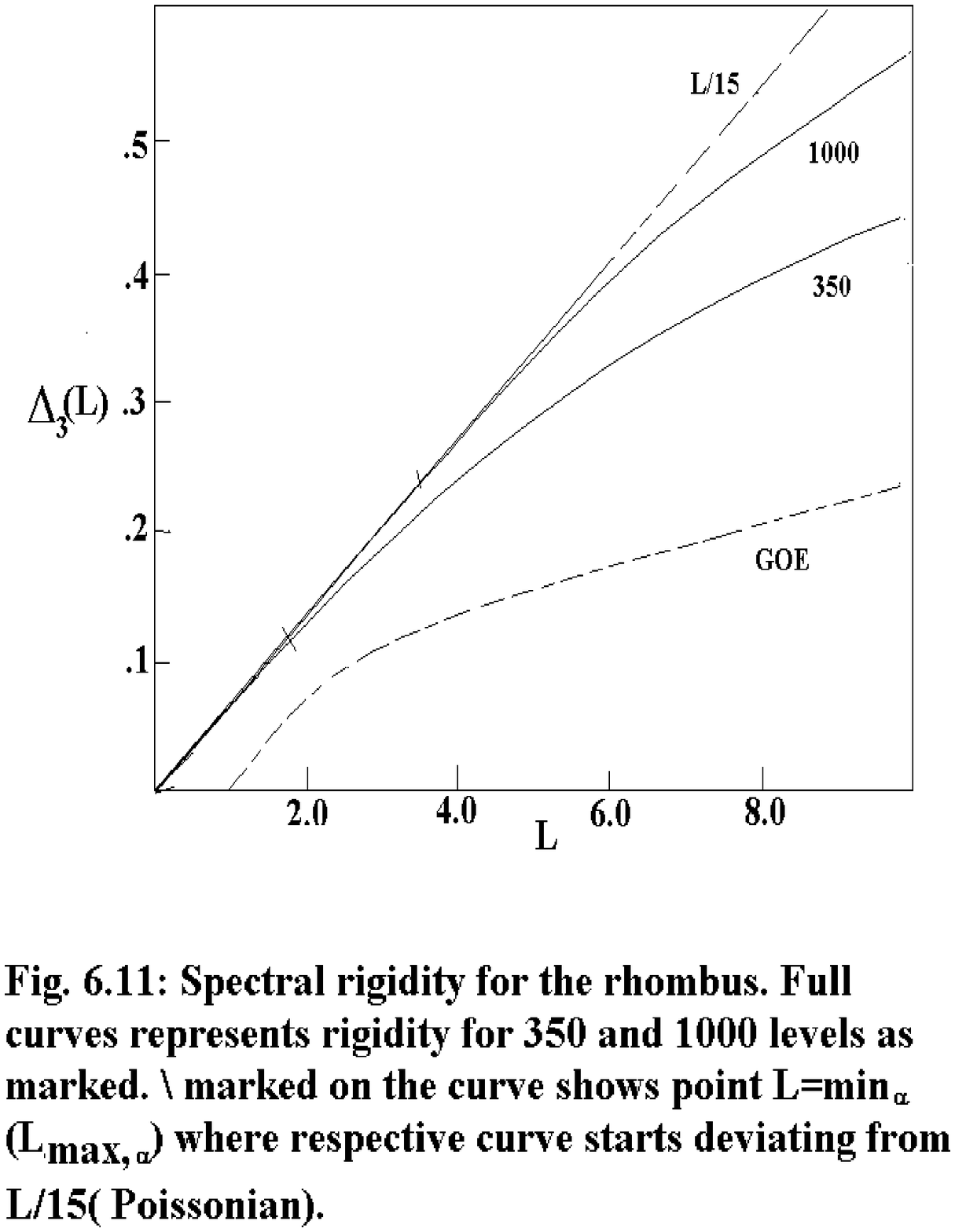,bbllx=80,bblly=120,bburx=580,bbury=770,height=6in}}
\end{center}
\end{figure}
\begin{figure}[htbp]
\begin{center}
{\epsfig{file=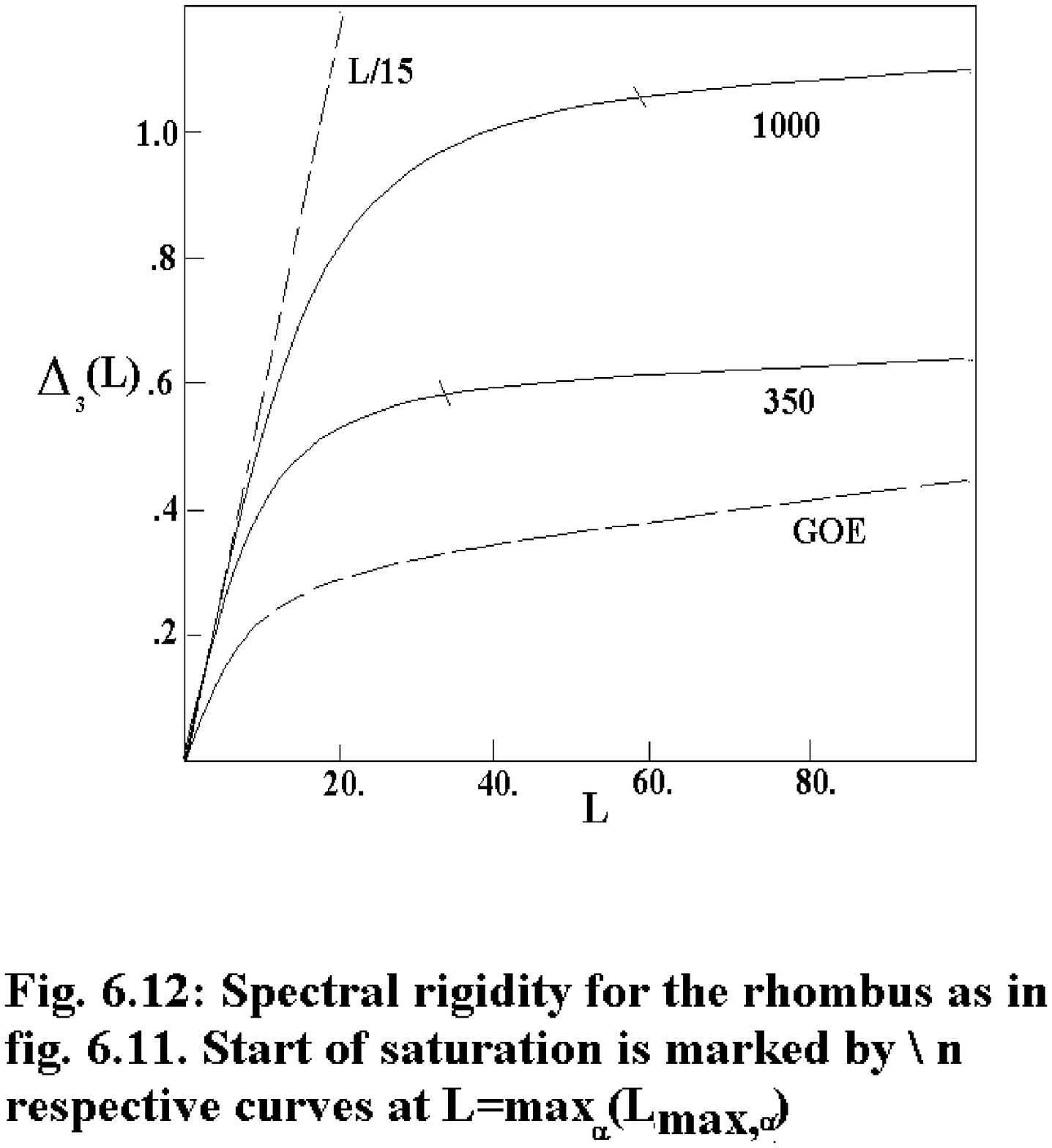,bbllx=80,bblly=170,bburx=560,bbury=700,height=6in}}
\end{center}
\end{figure}

In conclusion, we have obtained good approximation for the $\Delta _3$%
-statistic in an close analytical form for systems in which periodic orbits
of the marginal stability (in bands) occur, from which Poisson and
non-Poission results follow in a natural way. Answers to the basic questions
we have asked above are as follows: a) The levels of PIB are uncorrelated
and mimic a Poission process for $L<\min _\alpha (L_{\max ,\alpha })$ which
depends on shortest periodic orbit of a given system, hence nonuniversal
value. This condition also stipulate minimum number of energy levels that
one should be consider in the numerical experiment to observe this effect.
b) For $L>>\max _\alpha (L_{\max ,\alpha })$ which depends mainly on the
longest of the shortest p.o. among the different classes, the spectral
rigidity saturates to a nonuniversal value. c) The fluctuation properties of
PIB and IB differ essentially in the extent of transition region. In IB
transition region will be of less extent, since there is only one class of
p.o., deviation from a Poission and saturation is determined by same
p.o.(i.e. shortest one). In PIB because large number of classes of p.o. are
present, all shortest p.o. amongst the different classes play important role
in determining shape and extent of crossover region.

\chapter{Conclusions}

The work presented in this thesis provides first step to attain complete
understanding of spectral fluctuation properties in pseudointegrable systems
within semiclassical framework. It also supplements our knowledge about
underlying reasons for universal and non-universal properties of spectral
fluctuations. Furthermore, close analytical forms for various spectral
measures are obtained, which in our knowledge were not available before this
work (albeit under diagonal approximation only).

Semiclassical framework needs complete information about periodic orbits of
a system under consideration. In particular one needs to specify actions of
periodic orbits and their stability properties. For example, in the case of
pseudointegrable billiards we need to know lengths and band areas of the
periodic orbits in a given system. This is a herculean task, and there are
vary few systems about which complete information regarding this is
available. Different methods are developed to gather information about
periodic orbits in different kinds of system (e.g. symbolic dynamics). These
methods are some times system specific and sometimes more general. In
chapter three we have developed a methodology to enumerate and classify
periodic orbits of some pseudo-integrable billiards. Here we have exploited
important geometrical attributes of systems under considerations. To our
knowledge these are only pseudointegrable systems about which complete
information regarding periodic orbits is available. The same methodology can
be used to study large number of rational polygonal billiards in particular
one can apply technique to rectangular billiard with many slits(or barriers)
or L-shaped billiards or its further generalization. This is one of the open
problem in which we will be interested in future. This will also give us
opportunity to deal with higher genus pseudointegrable systems.

As it turns out from our analysis proliferation of periodic orbits (or
growth rate) in one of the most important factor in the semiclassical
methods. Spectral fluctuation properties indeed depends on the exact
asymptotic proliferation of the periodic orbits in the system. We obtained
this law in chapter four using probabilistic arguments. We proved that the
law of proliferation of periodic orbits in the pseudointegrable billiards
under consideration is quadratic in length, in fact , of the form $al^2+bl+c$%
. We also obtained coefficients $a,b,c$ explicitly using probabilistic
arguments.

A complete knowledge about periodic orbits enabled us to obtain analytical
expressions for various spectral measures. Most important of these is
two-level cluster function which is studied along with its Fourier transform
in chapter five. To obtain two level cluster function one have to evaluate
density-density correlation function first. The semiclassical approximation
for density correlation function leads us to summation over all periodic
orbits. This formulation does not explicitly brings out various factors that
affect spectral fluctuations in the system. Using simple but non trivial
arguments we convert this form of summation to the integration. It then
turns out that proliferation law play important role in this integration and
hence, in the spectral fluctuations. This conversion to integration enables
us to obtain analytical expressions for various spectral measures. This also
enables us to study effect of considering finite number of levels on the
spectral fluctuations. In particular spectral properties depends on
1)lengths of shortest periodic orbit occurring in different classes of
periodic orbits.  2)phase space areas occupied by classes of periodic orbits
along with respective coefficients of proliferation law which forms a factor 
$\kappa $ in our analysis. In particular we show importance of factor $%
l_{\alpha \min }/l_H$, where $l_H=\sqrt{4\pi A_Rr_0}$ is the Heisenberg
length. From our analysis it is clear that this ratio for $\max _\alpha
(l_{\alpha \min })\,$ should be $<<1$, for one to talk about asymptotic
properties ($\hbar \rightarrow 0$) of spectral fluctuations. For $\pi /3$%
-rhombus billiard this condition indicates that the number of levels
considered should be at orders of magnitude  larger than $1600$. For other
pseudointegrable system this number may still be much higher. Though it is
meaningful to study statistically large number of levels in the
non-asymptotic regions, this will bring not bring out universal properties
of that class of systems as spectral fluctuations are not free of system
dependent parameters. This chapter thus we are able to define classical
parameters that governs spectral fluctuations and extent of their influence.
In the case of form factor even diagonal approximation gives correct
asymptotic ($\tau >>1)$ form unlike that in chaotic systems where it was
found that diagonal approximation leads to linear rise in the form factor in
asymptotic region ($\tau >>1$), which was compensated by off-diagonal
contribution to give correct saturation. This indicates that off-diagonal
contribution in pseudointegrable systems can affect only transition region.
Further, we have shown that imprints of different classes of periodic orbits
can be seen in the behaviour of the form factor. This feature is still gone
unnoticed though many numerical experiments do have these imprints.

In chapter six we consider short and intermediate range spectral measures,
namely spacing distribution, number variance and spectral rigidity. We also
compare our results with numerical experiments available. We found that
intermediate range measures are well modeled by our theoretical
expressions. The departure from numerical results are not serious and may be
attributed to off-diagonal contributions. Spacing distribution however,
agree with numerical experiments only for short correlation range this is
due to fact that any error due to off-diagonal contributions is exponentially
amplified hence rises fast for larger correlation ranges. This can be
improved further by considering off-diagonal contributions.

We now enlist some open avenues that our work suggest

1) To obtain complete understanding off-diagonal contributions should be
taken into account explicitly. This is however difficult task. Recently
developed indirect method by Bogomolny to consider these contributions in
terms of diagonal contribution may proved to be fruitful in this regard.
Some preliminary study confirms our conclusions above. It also indicate that
approach to Poisson like behaviour is further slowed down by these
contributions.

2) As mentioned in the thesis trajectories of pseudointegrable billiards
resides on foliated surface of genus greater that two. This therefore do not
allow trajectories to explore phase available as in the case of chaotic
systems. The singular points in the phase space, however, results in
diffraction of trajectories that hits it. These diffracted orbits are
therefore not confined to one foliated surface but may visit many more via
multiple diffractions thus imitating chaotic behaviour. These should be
properly accounted for.

3) Yet, another kind of orbits which are on the edges of the band of
periodic orbits can give different contribution to periodic orbit sum hence
should be studied in detail.

4) Extension of methods developed here to higher genus systems is also a one
of the important problem.

5) Can one generalise these results on two dimensional polygonal billiard to
n-dimensional polyhedra.

6) Using semiclassical formalism and complete information about periodic
orbits can we develop a method to extract information about the
eigenfunctions for these systems.

7) One of the important applied field where study about polygonal billiards
can be used is semiconductor microstructures where dimensions of a system is
much smaller that the mean free paths of electrons (hence electron imitates
billiard like dynamics). To design and develop microstructure with desired
properties it is important to develop complete semicalssical understanding
of such systems.

I hope this is just a beginning of my persuasion of knowledge about
dynamical systems in general.
\end{document}